\documentclass{osa-article}

\journal{osajournal}

\articletype{Research Article}
\usepackage[utf8]{inputenc}
\usepackage[english]{babel}
\usepackage{fancyhdr}
\usepackage{lastpage}
\usepackage{geometry}
\usepackage{float}
\usepackage{tcolorbox}
\usepackage[hypcap=false]{caption}
\usepackage{url}
\begin{document}

\title{Space-time wave packets}

\author{Murat Yessenov,\authormark{1} Layton A. Hall,\authormark{1} Kenneth L. Schepler,\authormark{1} and Ayman F. Abouraddy\authormark{1,*}}

\address{\authormark{1}CREOL, The College of Optics \& Photonics, University of Central Florida, Orlando, FL 32816, USA}

\email{\authormark{*}raddy@creol.ucf.edu}

\begin{abstract}
`Space-time' (ST) wave packets constitute a broad class of pulsed optical fields that are rigidly transported in linear media without diffraction or dispersion, and are therefore propagation-invariant in absence of optical nonlinearities or waveguiding structures. Such wave packets exhibit unique characteristics, such as controllable group velocities in free space and exotic refractive phenomena. At the root of of these behaviors is a fundamental feature underpinning ST wave packets: their spectra are not separable with respect to the spatial and temporal degrees of freedom. Indeed, the spatio-temporal structure is endowed with non-differentiable angular dispersion, in which each spatial frequency is associated with a single prescribed wavelength. Furthermore, deviation from this particular spatio-temporal structure yields novel behaviors that depart from propagation invariance in a precise manner, such as acceleration with an arbitrary axial distribution of the group velocity, tunable dispersion profiles, and Talbot effects in space-time. Although the basic concept of ST wave packets has been known since the 1980's, only very recently has rapid experimental development emerged. These advances are made possible by innovations in spatio-temporal Fourier synthesis, thereby opening a new frontier for structured light at the intersection of beam optics and ultrafast optics. Furthermore, a plethora of novel spatio-temporally structured optical fields (such as flying-focus wave packets, toroidal pulses, and ST optical vortices) are now providing a swathe of surprising characteristics, ranging from tunable group velocities to transverse orbital angular momentum. We review the historical development of ST wave packets, describe the new experimental approach for their efficient synthesis, and enumerate the various new results and potential applications for ST wave packets and other spatio-temporally structured fields that are rapidly accumulating, before casting an eye on a future roadmap for this field. 
\end{abstract}

\newpage

\tableofcontents

\newpage

\section{Introduction}

Examples abound in optics where the spatial and temporal degrees of freedom (DoFs) are coupled, especially when considering ultrashort pulses \cite{Akturk2010JO}. However, such space-time couplings are typically considered nuisances to be tolerated or combated, drawbacks to be overcome, or curious features to be examined \cite{Pariente2016NP,Li2019OE,Wikmark2019PNAS,Leroux2020OE}; see the reviews in \cite{Dorrer2019JSTQE,Jeandet2021arxiv}. Nevertheless, space-time coupling can be viewed as a yet-to-be-exploited \textit{resource} in optics, which can enable novel phenomena to be observed in freely propagating optical fields and new opportunities to be harnessed in the interaction of light with matter and photonic devices. This reorientation of perspective is leading to rapidly evolving developments, exciting new results, and a burgeoning field of research that may be called \textit{space-time optics and photonics}, which constitutes a new frontier for classical optics and structured light. In this regard, the past few decades have witnessed the maturation of manipulating the \textit{temporal} DoF of optical fields as evinced by developments, for instance, in ultrafast pulse modulation \cite{Weiner2000RSI,Weiner2009Book} and optical combs \cite{Cundiff2003RMP,Picque2019NP}. Over the same period of time, manipulating the \textit{spatial} DoF has also experienced significant maturation, leading to the development of new classes of optical beams, such as beams endowed with orbital angular momentum \cite{Allen1992PRA,Yao2011AOP}, Airy beams \cite{Siviloglou2007OL,Siviloglou2007PRL,Efremidis2019Optica}, and other spatially structured optical fields \cite{Forbes2021NP}, which have found applications in microscopy \cite{Maurer2011LPR}, optical communications \cite{Wang2012NP,Bozinovic2013Science}, and micro-particle control \cite{Grier2003Nature}. The new frontier of space-time optics and photonics explores the unique consequences of manipulating the spatial and temporal DoFs \textit{jointly}, rather than modulating each separately. One characteristic in particular that exploits space-time coupling has been pursued for the past 40 years: propagation invariance of pulsed beams (or wave packets) in free space.

\begin{figure}
\begin{center}
\includegraphics[width=11.0cm]{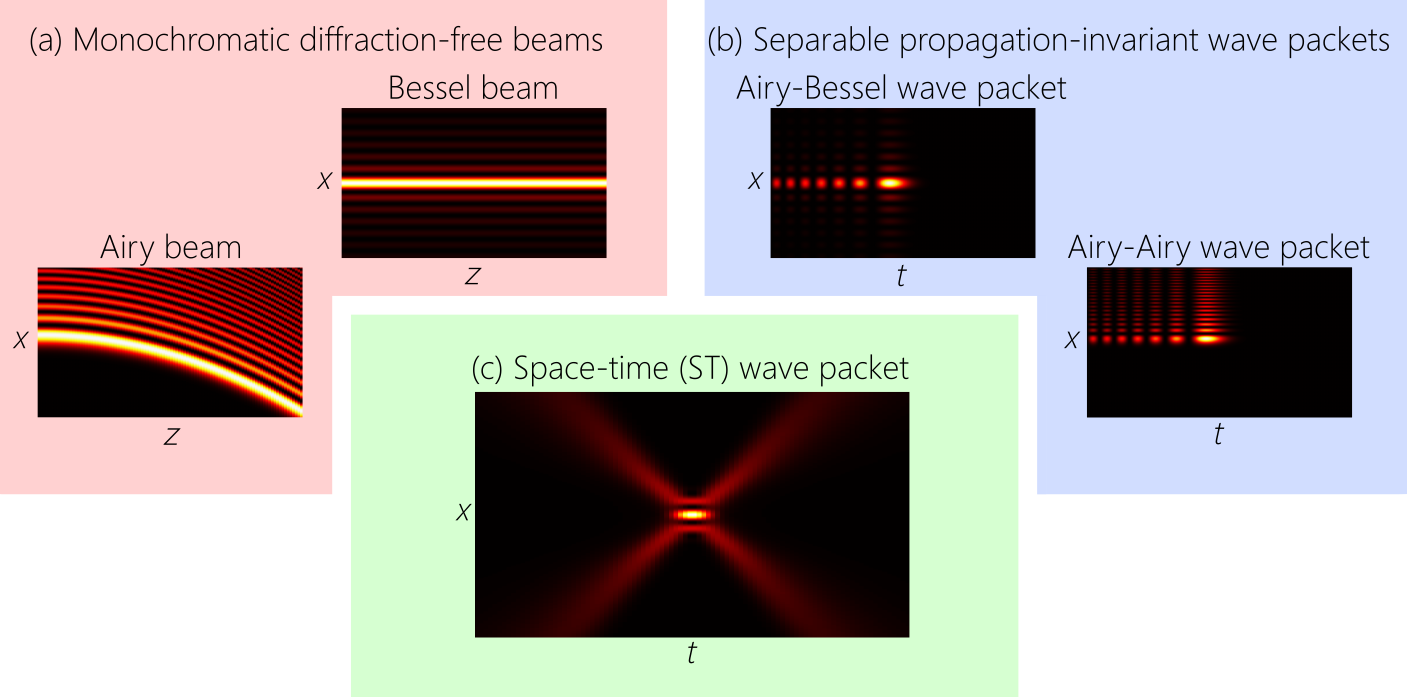}
\end{center}
\caption{(a) The evolution of the transverse intensity $I(x,z)$ for monochromatic zeroth-order-Bessel and Airy beams. Both are diffraction-free, but the latter travels on a parabolic trajectory. (b) Wave packets (pulsed beams) that are separable with respect to the spatial and temporal DoFs can be propagation invariant when the spatial profile is a Bessel or Airy beam and the temporal profile is an Airy pulse (in a dispersive medium); the Airy-Airy wave packet travels along a curved trajectory as in (a). (c) ST wave packets are propagation invariant in free space or dispersive media, are non-separable with respect to space and time, and travel along a straight line. In (b,c) we plot the spatio-temporal intensity profile $I(x,t)$ at a fixed axial plane z.}\label{Fig:Intro}
\end{figure}

The `diffraction-free' Bessel beam was introduced into optics to widespread attention in 1987 \cite{Durnin1987PRL,Durnin1987JOSAA}. A Bessel beam is a monochromatic optical field whose transverse spatial profile conforms to the Bessel function, which is a solution of the source-free Helmholtz equation \cite{Stratton1941Book}. Unlike traditional optical beams (e.g., the Gaussian beam) that undergo diffractive spreading upon free propagation, ideal Bessel beams travel without change in shape or scale; hence the term `diffraction-free'; see Fig.~\ref{Fig:Intro}(a,b). Less well-known is that a propagation-invariant \textit{pulsed} beam was proposed earlier in 1983 by James Brittingham and called a `focus-wave mode' (FWM), which propagates rigidly in free space without change in the shape or scale of its spatio-temporal profile at a group velocity of $c$ (the speed of light in vacuum) \cite{Brittingham1983JAP}. It is a curious accident in the history of optics that a propagation-invariant pulsed beam (the FWM) was discovered prior to its monochromatic limit (the Bessel beam).

Brittingham's FWM generated early excitement as a potential platform for directed energy, especially in light of the contemporaneous Strategic Defense Initiative (SDI; see \cite{Figueroa2014Book}, Ch.~2). Nevertheless, subsequent interest waned in comparison to the flourishing effort -- continuing to the present \cite{McGloin2005CP,Mazilu2010LPR,Vicente2021Optica} -- dedicated to studying Bessel beams. In large part, this asymmetry in interest is due to practical considerations; namely, the ease of producing Bessel beams and the corresponding difficulty of synthesizing FWMs \cite{Reivelt2000JOSAA,Reivelt2002PRE}. Indeed, producing FWMs reliably remains a problem that has yet to be resolved to this very day. In the decades since 1983, other propagation-invariant wave packets have been identified, chief among these is the X-wave \cite{Lu1992IEEETUFFC-Xwaves} that shares many of the characteristics of FWMs, but has the additional intriguing feature of its group velocity $\widetilde{v}$ potentially taking on \textit{superluminal} values (i.e., $\widetilde{v}\!>\!c$). However, despite successes in ultrasonic X-waves \cite{Lu1992IEEETUFFCexperimentalX}, synthesizing optical X-waves \cite{Saari1997PRL} and tuning their properties face some of the same challenges as FWMs. For example, producing a X-wave whose group velocity deviates from $c$ by even $1\%$ requires operating deep in the non-paraxial regime \cite{Yessenov2019PRA}. It is another curious -- and unfortunate -- accident in the history of optics that the first discovered examples of propagation-invariant wave packets (FWMs and X-waves) are the most difficult to synthesize, and the least versatile with respect to their propagation characteristics. Consequently, FWMs and X-waves have received only a fraction of the interest from the optics community at large as Bessel beams. A host of other propagation-invariant wave packets have been proposed over the years, some with exotic names such as slingshot mode \cite{Ziolkowski1993JOSAA} and splash mode \cite{Besieris1989JMP,ZamboniRached2002EPJD}, among many others \cite{Figueroa2014Book}. To the best of our knowledge, none of these has been convincingly demonstrated in the optical spectrum (although there have been in acoustics and ultrasonics \cite{Ziolkowski1989PRL,Lu1992IEEETUFFCexperimentalX}).

Recently, we have developed a class of \textit{easily synthesized} propagation-invariant wave packets that possess several crucial advantages over FWMs and X-waves, to which we have given the generic name `space-time' (ST) wave packets, and whose study has rapidly advanced over the course of the past few years \cite{Wong2017ACSP2,Porras2017OL,Efremidis2017OL}; see Fig.~\ref{Fig:Intro}(c). Prime amongst these characteristics is that their group velocity can be continuously tuned from subluminal to superluminal and even negative values, all in the same simple experimental arrangement, and all without leaving the paraxial regime. Indeed, values of $\widetilde{v}$ in the range from $30c$ to $-4c$ have been recorded in free space \cite{Kondakci2019NC}. In addition to propagation invariance and tunable group velocity, ST wave packets exhibit self-healing characteristics \cite{Kondakci2018OL}, can display precisely controllable departures from propagation invariance (e.g., acceleration \cite{Yessenov2020PRLaccel} with arbitrary axial distribution of the group velocity), are the basis for new ST Talbot effects \cite{Yessenov2020PRLveiled,Hall2021APLSTTalbot,Hall2021OLtempTalbot}, display a host of anomalous refraction phenomena \cite{Bhaduri2020NP,AllendeMotz2021OLisochronous}, and are a platform for propagation-invariant surface waves (e.g., surface plasmon polaritons \cite{Schepler2020ACSP}), among a litany of other emerging unique properties and applications \cite{Yessenov2019OPN,Yessenov2020NC,Shiri2020NC_Hybrid}. Most recently, bridges between ST wave packets and nanophotonics \cite{Guo2021Light} and device physics \cite{Shiri2020OL,Shiri2020NC_Hybrid,Guo2021PRR} are being established. Consequently, despite decades of work devoted to this area, it is still in many ways a young field with basic discoveries appearing only now in nascent form, especially on the experimental front. These recent successes motivate a re-canvasing of this field of research with an eye for future developments.

Despite the bewildering variety of propagation-invariant wave packets for which closed-form expressions have been found \cite{Figueroa2014Book}, they \textit{all} share a common foundational property: each spatial frequency (underlying the transverse spatial profile) is associated with only a single wavelength (underlying the temporal profile) -- \textit{independently} of the details of their spatio-temporal profile. In other words, a particular form of ST coupling undergirds the propagation invariance of these wave packets. Consequently, the functional form of their spatio-temporal spectra in general has a reduced dimensionality with respect to those of traditional pulsed beams where the spatial and temporal DoFs are typically independent of each other. That is, \textit{all} propagation-invariant wave packets have a similar underlying spatio-temporal spectral structure, and the appellation `ST wave packets' is therefore an apt and fitting name for the field as a whole. Indeed, the previous examples of FWMs and X-waves are also structured in space and time to ensure that each spatial frequency is assigned to a single wavelength, and are thus specific instances of ST wave packets. 

In addition to propagation-invariant ST wave packets, a flourishing area of research has emerged around a plethora of novel spatio-temporally structured wave packets that have been developed very recently. Some of these new wave packets share characteristics with ST wave packets; e.g., the group velocity of a flying-focus \cite{SaintMarie2017Optica,Froula2018NP} can be tuned in free space. Distinctive field geometries and topologies are involved in the structure of other wave packets, such as toroidal pulses \cite{Papasimakis16NM}, ST optical vortices \cite{Jhajj2016PRX,Hancock2019Optica,Chong2020NP}, and pulsed orbital angular momentum beams \cite{Pariente2015OL,Ornigotti2015PRL,Porras2019PRL,Zhao2020NC}. We envision that these spatio-temporally structured fields alongside ST wave packets will be the basis for the above-mentioned emerging area of research of ST optics and photonics. 

The goals of this Review are multifold. First, we provide a sketch of the historical development of propagation-invariant fields. Indeed, an extensive theoretical literature has accumulated that easily overwhelms the newcomer. Furthermore, this literature is intensely mathematical, sometimes with little concern for potential experimental implementation or connection to practical applicability. Therefore, rather than merely relating a chronology of apparently disconnected findings reported over the years, we formulate the historical sketch in tutorial fashion by first establishing the spatio-temporal spectral-support formalism pioneered in \cite{Donnelly1993ProcRSLA}, which unifies in a single framework all previous results. This allows synthesizing the multifarious examples available to date in a coherent fashion, and paves the way towards appreciating recent achievements and anticipating future breakthroughs. While it is laudable to find exact closed-form expressions for ST wave packets, this should not be a critical requirement, and is far from constituting a criterion for their intrinsic merit. In retrospect, some of the controversies in the early days of FWMs could have been easily resolved by relying on the strategy explicated here rather than tying the validity of ST wave packets to particular analytic expressions. As part of this survey, we elucidate the fundamental difficulties in synthesizing the early examples of FWMs and X-waves, and make the case that they in fact represent a `dead end' for optics (but not necessarily in ultrasonics or acoustics).

A second goal is to describe in detail the new experimental methodology for synthesizing ST wave packets, which combines elements from ultrafast pulse modulation \cite{Weiner2009Book} with those from Fourier-optics-based beam shaping \cite{Goodman2005Book} together in a spatio-temporal Fourier synthesizer based on spectral-phase modulation. Rather than modulating the spatial and temporal DoFs separately, this experimental strategy enables joint modulation of the spatio-temporal spectrum, which is key to the synthesis of ST wave packets. In this context, we review the conceptual advance made recently in understanding the factors determining the achievable limits with respect to propagation distance and group velocity.

Finally, we sketch the current field of play in this area and lay out a roadmap for further developments that are now within reach in light of these new capabilities. We hope that this tutorial format helps introduce the reader to the vast literature on the topic, and thus initiates the newcomer into this exciting emerging field of ST optics and photonics.

\subsection{Comparison to previous reviews}

There exist several reviews of localized waves and we describe here the most prominent examples briefly to situate our current review. Early papers by Ziolkowski \cite{Ziolkowski1985JMP,Ziolkowski1989PRA,Ziolkowski1989PRL,Ziolkowski1993JOSAA} and Shaarawi \cite{Shaarawi1995JOSAA,Shaarawi1995JMP,Shaarawi1996JOSAA,Shaarawi2000JPA} are dedicated to FWMs and related examples of propagation-invariant wave packets. This early work in the optical domain was solely theoretical. One of the earliest reviews after the first optical experimental demonstrations \cite{Saari1997PRL,Reivelt2000JOSAA,Reivelt2002OE,Reivelt2002PRE} is that by Reivelt and Saari in 2003 \cite{Reivelt2003arxiv}, which is a useful survey of the literature up to 2002 on coherent pulsed and partially coherent propagation-invariant fields, encompassing FWMs and X-waves.The review is clear with regards to the difficulty of producing FWMs in particular, and conveys appropriate skepticism with regard to their potential synthesis. Kiselev in 2007 \cite{Kiselev2007OS} provides a strictly mathematical review of localized waves.

The review by Turunen and Friberg in 2010 \cite{Turunen2010PO} covers monochromatic diffraction-free beams (emphasizing Bessel beams) and their pulsed and partially coherent counterparts. Although this review emphasizes theoretical developments, it nevertheless usefully surveys the synthesis approaches available at the time and highlights the lack of convenient general strategies for preparing ST wave packets. Furthermore, in addition to FWMs and X-waves, it  describes theoretically the cases of ST wave packets that we focus on here (but excludes wave packets with negative group velocity), although these latter wave packets had not been synthesized at the time. In general, the Review in \cite{Turunen2010PO} is closest in spirit to our Review. Finally, two books \cite{HernandezFigueroa2008Book,Figueroa2014Book} provide a broad survey of the theoretical and experimental developments from groups engaged worldwide with theoretical and experimental research on localized waves.

Some shared features emerge from these previous Reviews: (1) they are focused on propagation invariance as the most interesting consequence of ST-coupling; (2) they emphasize research on X-waves and FWMs; (3) they are mostly dedicated to theoretical work; and (4) little attention is paid to the practical aspects of feasible synthesis. In contrast, the current Review moves in a different direction. Most importantly, by placing a premium on physical realizability, X-waves and FWMs become considerably less interesting than so-called `baseband' ST wave packets that offer unprecedented versatile tunability of their characteristics, and yet can be readily produced in the paraxial regime with small bandwidths. In addition, experimental efforts are developing rapidly and are now keeping apace with theoretical efforts. Critically, propagation invariance is now only one consequence of precise spatio-temporal structuring of the field, in addition to realizing arbitrary dispersion profiles, accelerating wave packets, spectral axial encoding, and new interaction modalities with photonic devices. Furthermore, we encompass within this Review recent families of spatio-temporally structured optical fields that provide new and useful behaviors by virtue of their specific structure, although they are not necessarily propagation invariant.

\subsection{Plan of this Review}

This Review focuses on the newly emerging work carried out in the past five years with regards to ST wave packets and other spatio-temporally structured fields. However, it is critical for the reader to appreciate the previous work done in the area of FWMs and X-waves, which have received the most interest over the previous 4 decades, and the difficulties involved in their synthesis. Because previous reviews have usually addressed a more specialized technical community, there exists an entry barrier to this topic. To provide a convenient entry point for newcomers to this field, we formulate the first part of this Review as a tutorial (Sections~\ref{Section:Preliminaries}-\ref{Section:classification}). We open with an overview of the basic mathematical formalism used for analyzing ST wave packets (Section~\ref{Section:Preliminaries}), and we emphasize the usefulness of the geometric representation of their spectral support domain on the surface of the light-cone, which provides a simple visualization of the crucial aspects of ST wave packets. An examination of the conditions for achieving propagation invariance (Section~\ref{Section:conditionsforpropinv}) then leads to a classification of such fields into baseband ST wave packets, sideband ST wave packets, and X-waves (Section~\ref{Section:classification}). This groundwork enables us to present a historical sketch (Section~\ref{Section:HistoricalSketch}) for the developments and previous achievements in this area using the nomenclature outlined here. We hope that this formulation will provide the reader with an entry point to the vast literature on propagation-invariant wave packets. In Section~\ref{Section:dispelling_misconceptions} we dispel misconceptions regarding ST wave packets, and in Section~\ref{Section:Deadend} we outline the rationale for considering X-waves, FWMs, and sideband ST wave packets to be a dead end for optics.

In Section~\ref{Synthesis} we describe our general experimental procedure for the synthesis and characterization of ST wave packets, before surveying the results that have been achieved with this strategy (Section~\ref{Sec:PropCharacteristics} through Section~\ref{Sec:StateOfTheArt}). We the broaden our perspective to describe examples from the recent flourishing area of spatio-temporally structured optical fields; specifically flying-foci, toroidal pulses, pulsed OAM fields, and ST optical vortices (Section~\ref{Section:novelST-structures}). We subsequently highlight in Section~\ref{sec:STPhotonics} a few examples of the interaction of ST wave packets with photonic devices, which are ushering in the new field of ST photonics. Finally, we survey the related topics that we did \textit{not} cover in this Review (Section~\ref{Section:notcovered}), before giving a brief outlook on potential future developments (Section~\ref{Section:Roadmap}).

This Review is dedicated solely to realizations of ST wave packets in the optical regime, and we are therefore not concerned with microwaves \cite{Recami98AnnPhys,Mugnai2000PRL,Chiotellis18PRB,Chiotellis2020IEEETAP}, acoustics and ultrasonics \cite{Ziolkowski1989PRL,Ziolkowski1990,Lu1992IEEETUFFC-Xwaves,Lu1992IEEETUFFCexperimentalX}, gravitational waves \cite{Asenjo2021EPJC}, or particle physics \cite{Ziolkowski1989NPB,Shaarawi90JMP,Ziolkowski1995PRE}. The fundamental and practical concerns in optics are unique, and have not been adequately addressed, resulting in a paucity of experimental results despite the dearth of theoretical studies. We hope that the rapid developments over the past 5~years have obviated these difficulties.

\section{Preliminaries}\label{Section:Preliminaries}

Our goal in this Section is to provide physical insight and intuition for the representation and behavior of ST wave packets by relying on geometric arguments that -- despite their simplicity -- capture the basic physics of ST wave packets.

\subsection{Mathematical formulation}

\begin{figure}
\begin{center}
\includegraphics[width=6.0cm]{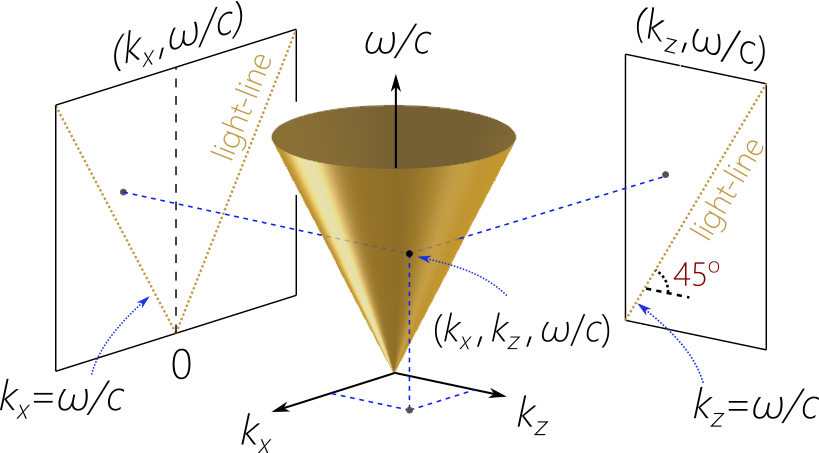}
\end{center}
\caption{The free-space dispersion relationship $k_{x}^{2}+k_{z}^{2}\!=\!(\tfrac{\omega}{c})^{2}$ is represented geometrically in the spatio-temporal spectral space by the light-cone. A monochromatic plane wave $e^{i(k_{x}x+k_{z}z-\omega t)}$ corresponds to a point of coordinates $(k_{x},k_{z},\tfrac{\omega}{c})$ on the light-cone surface.}\label{Fig:BasicLightCone}
\end{figure}

We refer to a pulsed beam generally as a \textit{wave packet}. We consider here scalar fields $E(x,y,z;t)$ in a Cartesian coordinate system $(x,y,z)$, where $z$ is the propagation direction, the transverse plane is spanned by $x$ and $y$, and $t$ is time. The wave-vector components $\vec{k}\!=\!(k_{x},k_{y},k_{z})$ and the angular frequency $\omega$ for a monochromatic plane wave in free space satisfy the dispersion relationship $k_{x}^{2}+k_{y}^{2}+k_{z}^{2}\!=\!(\tfrac{\omega}{c})^{2}$ which corresponds to the surface of a hypercone in 4D. We refer to $k_{z}$ as the axial wave number, and to the transverse components $k_{x}$ and $k_{y}$ of the wave vector as \textit{spatial} frequencies. To symmetrize the treatment of the spatial and temporal DoFs, we refer to $\omega$ henceforth as the \textit{temporal} frequency. To facilitate visualization of this geometric structure in 3D, we hold the field uniform along $y$, $E(x,y,z;t)\!\rightarrow\!E(x,z;t)$, whereupon $k_{y}\!=\!0$. The free-space dispersion relationship becomes $k_{x}^{2}+k_{z}^{2}\!=\!(\tfrac{\omega}{c})^{2}$, which corresponds geometrically to the surface of a cone in $(k_{x},k_{z},\tfrac{\omega}{c})$-space, referred to as the `light-cone'. This simplification facilitates an instructive geometric representation in Fourier space of optical fields in general, and ST wave packets in particular -- without loss of generality. A monochromatic plane wave $e^{i(k_{x}x+k_{z}z-\omega t)}$ is represented by a point of coordinates $(k_{x},k_{z},\tfrac{\omega}{c})$ on the light-cone surface [Fig.~\ref{Fig:BasicLightCone}]. Because any physically realizable optical field can be written as a superposition of such elementary plane waves after excluding any evanescent components, the spectral support domain of the field must correspond to some region on the light-cone surface. Indeed, one may say that optics `lives' on the light-cone!

We assume throughout that the slowly varying envelope and paraxial approximations are valid, and write the wave packet $E(x,z;t)\!=\!e^{i(k_{\mathrm{o}}z-\omega_{\mathrm{o}}t)}\psi(x,z;t)$ as the product of a slowly varying spatio-temporal envelope $\psi(x,z;t)$ and a carrier wave of frequency $\omega_{\mathrm{o}}$, where $k_{\mathrm{o}}\!=\!\tfrac{\omega_{\mathrm{o}}}{c}$ is the associated wave number. The envelope can be expressed as:
\begin{equation}\label{Eq:BasicIntegral}
\psi(x,z;t)=\iint\!dk_{x}d\Omega\,\,\widetilde{\psi}(k_{x},\Omega)e^{i\{k_{x}x+(k_{z}-k_{\mathrm{o}})z-\Omega t\}},
\end{equation}
where we have introduced for convenience a new temporal frequency variable $\Omega\!=\!\omega-\omega_{\mathrm{o}}$ measured with respect to $\omega_{\mathrm{o}}$, and the \textit{spatio-temporal spectrum} $\widetilde{\psi}(k_{x},\Omega)$ is the 2D Fourier transform of the initial wave packet $\psi(x,0;t)$ with respect to $x$ and $t$. This formula captures diffractive spreading and space-time coupling \cite{Saleh2007Book}. We could equivalently describe the spatio-temporal spectrum in terms of the axial wave number $k_{z}$ along with $\Omega$, rather than $k_{x}$ and $\Omega$. However, the variables $k_{x}$ and $\Omega$ are more convenient because they are under direct control experimentally. Practically, we can only change $k_{z}$ indirectly by tuning $k_{x}$ and/or $\omega$. 

\subsection{Representation of optical fields on the light-cone}

Any optical field can be represented on the light-cone surface by a domain corresponding to the support of the spatio-temporal spectrum $\widetilde{\psi}(k_{x},\Omega)$. Some general rules apply with respect to representations of the spectral support domain of optical fields on the surface of the light-cone $k_{x}^{2}+k_{z}^{2}\!=\!(\tfrac{\omega}{c})^{2}$.
\begin{enumerate}
    \item Because we exclude evanescent components, only points \textit{on the surface} are allowed. Consequently, the projection of the spectral support domain on the $(k_{x},\tfrac{\omega}{c})$ and $(k_{z},\tfrac{\omega}{c})$ planes must be above the light-lines $k_{x}\!=\!\tfrac{\omega}{c}$ and $k_{z}\!=\!\tfrac{\omega}{c}$, respectively \cite{Donnelly1993ProcRSLA}.
    \item We restrict the axial wave number to $k_{z}\!\geq\!0$ to be consistent with causal excitation into the region $z\!\geq\!0$ from sources at $z\!<\!0$ \cite{Heyman1987JOSAA,Heyman1989IEEE,Shaarawi1995JMP,Shaarawi1995OC,Yessenov2019PRA}.
    \item We employ only positive temporal frequencies $\omega\!>\!0$. We are typically interested in frequencies in the vicinity of a carrier frequency $\omega_{\mathrm{o}}$, $\omega\!=\!\omega_{\mathrm{o}}+\Omega$. The frequency $\Omega$ measured from $\omega_{\mathrm{o}}$ can thus be positive or negative.
    \item The spatial frequencies $k_{x}$ can take on positive or negative values.
\end{enumerate}
Although references to the light-cone representation, or at least to projections onto the $(k_{z},\tfrac{\omega}{c})$-plane, were made early on \cite{Besieris1989JMP}, the first decisive study highlighting the versatility of this approach for ST wave packets was made by Donnelly and Ziolkowski in \cite{Donnelly1993ProcRSLA}, and was subsequently used in \cite{Valtna2007OC,Saari2007LP,Saari2004PRE}. The recent literature now makes heavy use of this representation for visualization of the spectral support domain, developing new concepts, and providing useful insights \cite{Kondakci2016OE,Kondakci2017NP,Efremidis2017OL,Wong2017ACSP1}.

We henceforth eschew the usual reliance on particular field profiles associated with particular spectral amplitudes, and rely instead on the equivalency classes defined by the spectral support domain itself. We first examine a few basic field configurations as limiting cases to develop visual intuition regarding light-cone representations. 

\begin{figure}[t!]
  \begin{center}
  \includegraphics[width=12.0cm]{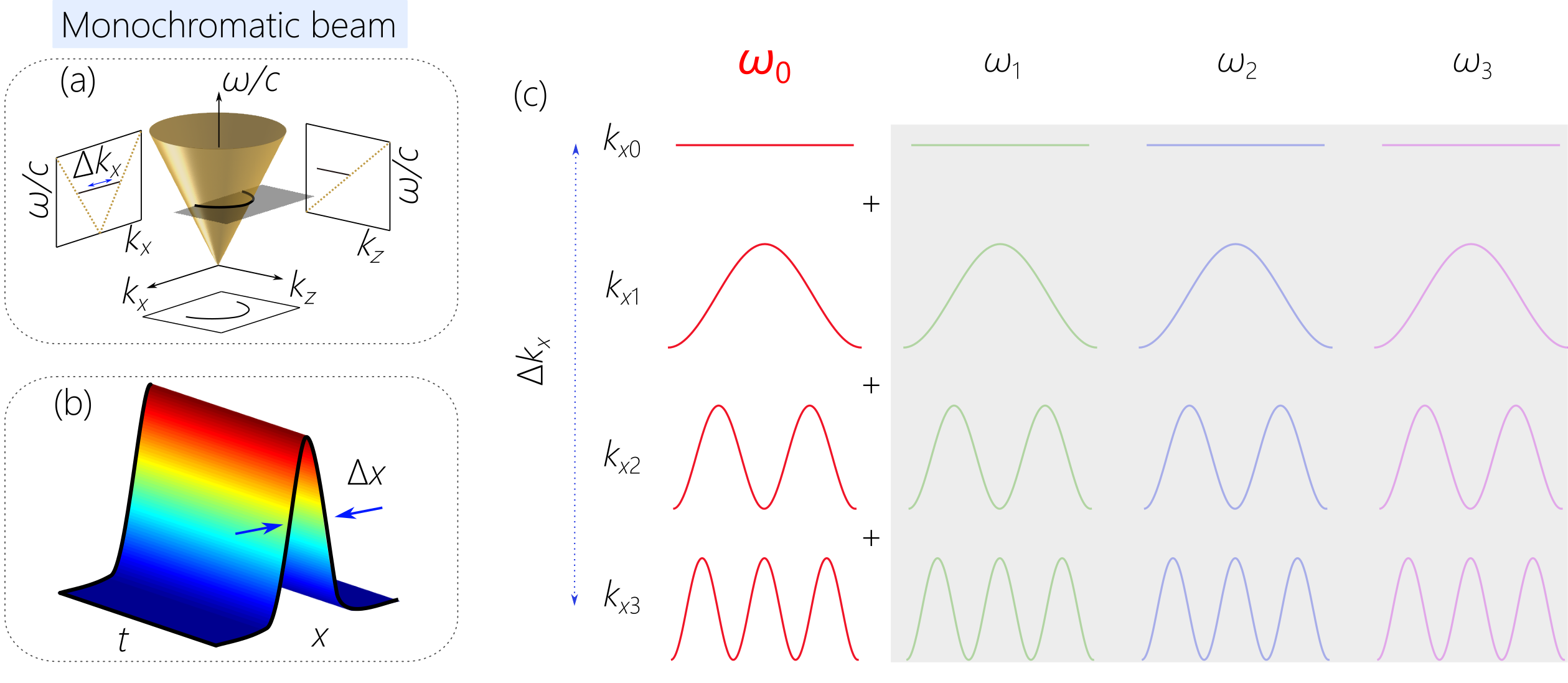}
  \end{center}
  \caption{Representation of the spatio-temporal spectrum of \textit{monochromatic beams}. (a) The spectral support domain on the light-cone surface is at its intersection with a horizontal iso-frequency plane $\omega\!=\!\omega_{\mathrm{o}}$. Spectral projections onto the $(k_{z},\tfrac{\omega}{c})$ and $(k_{x},\tfrac{\omega}{c})$ planes are straight lines, and onto the $(k_{x},k_{z})$-plane a semicircle after excluding $k_{z}\!<\!0$. (b) The spatio-temporal intensity profile $I(x,z=0;t)$. (c) The monochromatic beam is a superposition of spatial frequencies all at the same wavelength.}
  \label{Fig:MonochromaticBeam}
\end{figure}

\subsubsection{Monochromatic optical beams}

Because \textit{monochromatic} beams are defined by the constraint $\omega\!=\!\omega_{\mathrm{o}}$, their purely \textit{spatial} spectrum lies along the circle $k_{x}^{2}+k_{z}^{2}\!=\!k_{\mathrm{o}}^{2}$ at the intersection of the light-cone with a horizontal iso-frequency plane [Fig.~\ref{Fig:MonochromaticBeam}(a)]. Any such beam features a transverse spatial profile but no temporal linewidth [Fig.~\ref{Fig:MonochromaticBeam}(b)]. The field can be written as $E(x,z;t)\!=\!e^{i(k_{\mathrm{o}}z-\omega_{\mathrm{o}}t)}\psi(x,z)$, with the envelope given by:
\begin{equation}\label{Eq:MonochromaticBeamEnvelope}
\psi(x,z)=\int\!dk_{x}\widetilde{\psi}(k_{x})e^{ik_{x}x}e^{i(k_{z}-k_{\mathrm{o}})z};
\end{equation}
here the spatio-temporal spectrum from Eq.~\ref{Eq:BasicIntegral} was restricted as follows: $\widetilde{\psi}(k_{x},\Omega)\!\rightarrow\!\widetilde{\psi}(k_{x})\delta(\Omega)$, where the spatial spectrum $\widetilde{\psi}(k_{x})$ is the Fourier transform of $\psi(x,0)$ in Eq.~\ref{Eq:MonochromaticBeamEnvelope}. The inverse of the spatial bandwidth $\Delta k_{x}$ determines the initial transverse spatial width $\Delta x$ of the beam. Of course, the beam width increases with free propagation along $z$ because of diffraction.

The spectral projection onto the $(k_{z},\tfrac{\omega}{c})$-plane is a horizontal line $\omega\!=\!\omega_{\mathrm{o}}$, indicating that $\widetilde{v}\!=\!0$ \cite{Zamboni2004OE} and the absence of any temporal dynamics. The spectral projection onto the $(k_{x},\tfrac{\omega}{c})$-plane is also a horizontal line: all the spatial frequencies $k_{x}$ share the same temporal frequency $\omega_{\mathrm{o}}$. This condition is illustrated pictorially in Fig.~\ref{Fig:MonochromaticBeam}(c) where we depict spatial frequencies $k_{x}$ of increasing magnitude as sinusoids of decreasing period, and the temporal frequencies $\omega$ as different colors. In general, the spatio-temporal spectrum $\widetilde{\psi}(k_{x},\Omega)$ can be represented on a 2D grid with each cell in the grid occupied by a particular spatio-temporal frequency pair $(k_{x},\Omega)$: a specific sinusoid (for $k_{x}$) and color (for $\omega$). In the case of monochromatic beams considered here, only a single color is needed, and therefore only a column through the grid is required to represent such a field. It is clear that the spatio-temporal spectrum is separable with respect to the spatial and temporal DoFs, as can be seen from the substitution $\widetilde{\psi}(k_{x},\Omega)\!\rightarrow\!\widetilde{\psi}(k_{x})\delta(\Omega)$.

\begin{figure}[t!]
  \begin{center}
  \includegraphics[width=12.0cm]{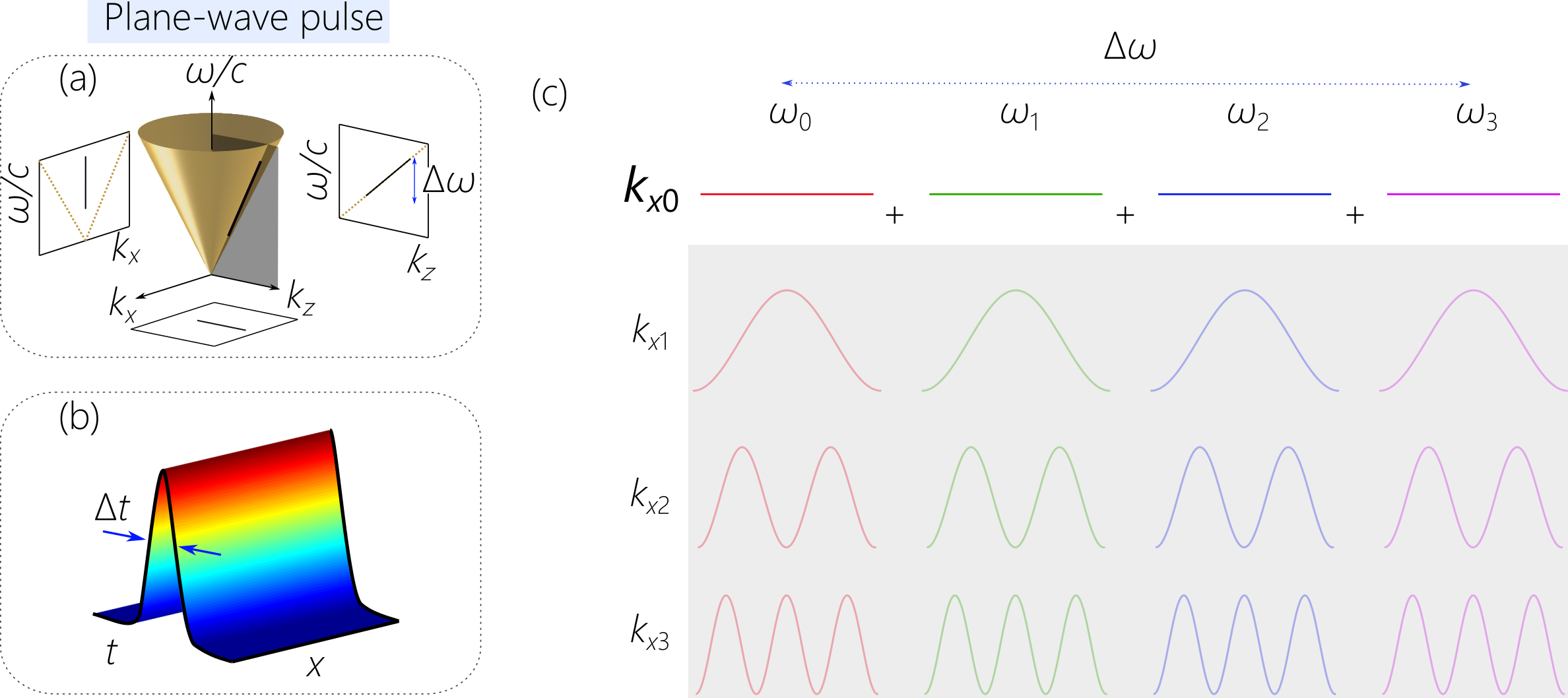}
  \end{center}
  \caption{Representation of the spatio-temporal spectrum of \textit{plane-wave pulses}. (a) The spectral support domain on the light-cone surface is its intersection with an iso-$k_{x}$ plane $k_{x}\!=\!0$. The spectral projection onto any plane is a straight line. (b) The spatio-temporal intensity profile $I(x,z\!=\!0;t)$. (c) A plane-wave pulse is a superposition of temporal frequencies (wavelengths) all at the same spatial frequency $k_{x}\!=\!0$.}
  \label{Fig:PulsedPlaneWave}
\end{figure}

\subsubsection{Plane-wave pulses}

Another paradigmatic optical field is the plane-wave pulse. We take here the simplest case corresponding to the constraint $k_{x}\!=\!0$ [Fig.~\ref{Fig:PulsedPlaneWave}(a)], so the pulse has no spatial features. The purely temporal spectrum of such a pulse lies at the intersection of the light-cone with the vertical iso-$k_{x}$ plane $k_{x}\!=\!0$. Such a pulse features a temporal linewidth but no spatial profile [Fig.~\ref{Fig:PulsedPlaneWave}(b)]. The field can be written as $E(x,z;t)\!=\!e^{i(k_{\mathrm{o}}z-\omega_{\mathrm{o}}t)}\psi(z;t)$, with the envelope given by:
\begin{equation}\label{Eq:PulsedPlaneWaveProfile}
\psi(z;t)=\int\!d\Omega\widetilde{\psi}(\Omega)e^{i(k_{z}-k_{\mathrm{o}})z}e^{-i\Omega t}=\psi(0;t-z/c);
\end{equation}
here the spatio-temporal spectrum from Eq.~\ref{Eq:BasicIntegral} is restricted as follows: $\widetilde{\psi}(k_{x},\Omega)\!\rightarrow\!\widetilde{\psi}(\Omega)\delta(k_{x})$, where the temporal spectrum $\widetilde{\psi}(\Omega)$ is the Fourier transform of $\psi(0;t)$ in Eq.~\ref{Eq:PulsedPlaneWaveProfile}, and $k_{z}\!=\!\omega/c$. The inverse of the temporal bandwidth $\Delta\omega$ determines the pulse linewidth $\Delta t$ [Fig.~\ref{Fig:PulsedPlaneWave}(b)].

The spectral projection onto the $(k_{z},\tfrac{\omega}{c})$-plane lies along the light-line $k_{z}\!=\!\tfrac{\omega}{c}$ [Fig.~\ref{Fig:PulsedPlaneWave}(a)], indicating a group velocity $\widetilde{v}\!=\!c$ and the absence of dispersion. The pulse therefore travels invariantly in free space. In the pictorial depiction in Fig.~\ref{Fig:PulsedPlaneWave}(c), only one spatial frequency $k_{x}\!=\!0$ is needed, but all the colors for $\omega$ are included. Thus, only a single row through the $(k_{x},\omega)$ grid is required to represent the plane-wave pulse. Again, the spatio-temporal spectrum is separable with respect to the spatial and temporal DoFs.

Common between these two examples -- the monochromatic beam and the plane-wave pulse -- is that one spectral DoF is limited by a strict constraint: $\omega\!=\!\omega_{\mathrm{o}}$ in the former and $k_{x}\!=\!0$ in the latter. Consequently, the spectral support domain for each on the light-cone surface is a 1D trajectory rather than a 2D domain, as further emphasized by the pictorial depictions in Fig.~\ref{Fig:MonochromaticBeam}(c) and Fig.~\ref{Fig:PulsedPlaneWave}(c).

\begin{figure}[t!]
  \begin{center}
  \includegraphics[width=12.0cm]{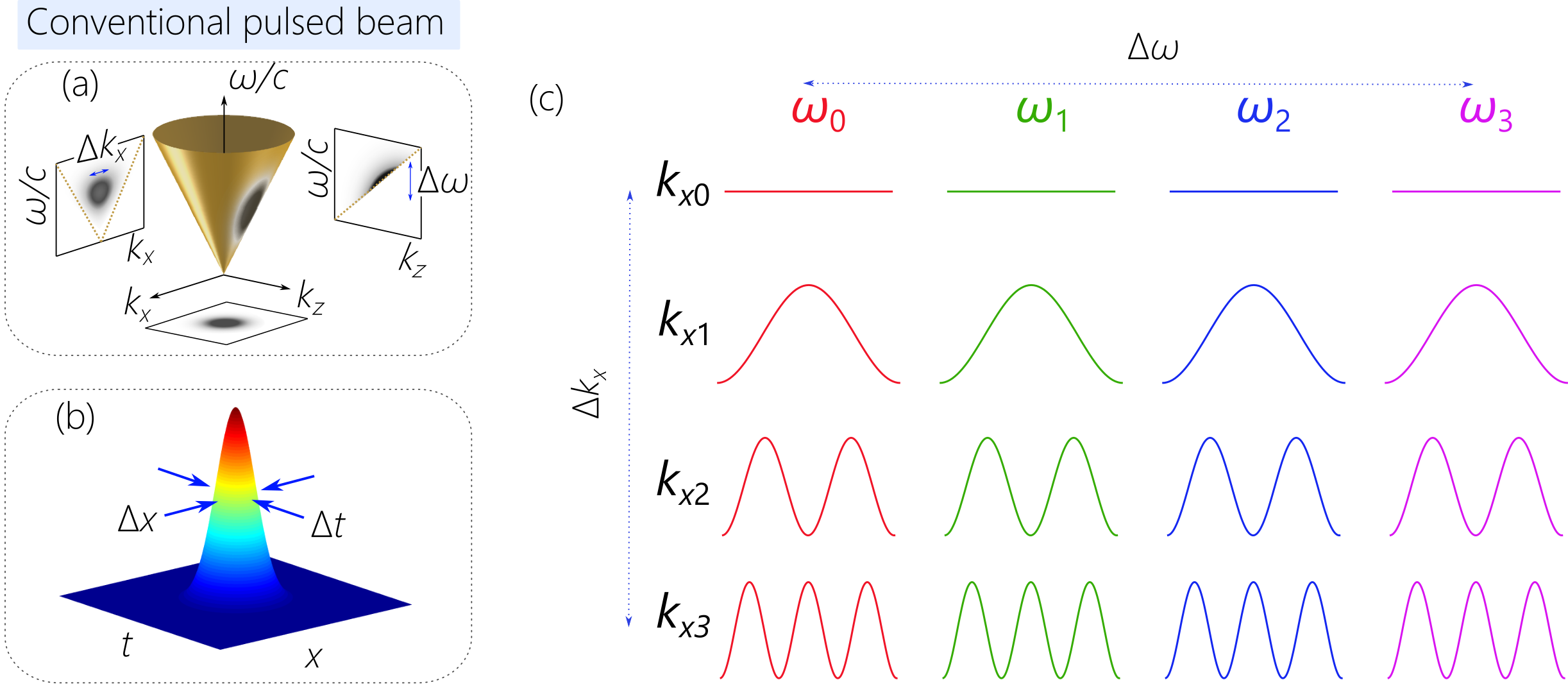}
  \end{center}
  \caption{Representation of the spatio-temporal spectrum of a \textit{conventional pulsed beam} with separable spatial and temporal DoFs. (a) The 2D spectral support domain on the light-cone surface for a pulsed beam. The spectral projections onto the $(k_{x},\tfrac{\omega}{c})$, $(k_{x},k_{z})$, and $(k_{z},\tfrac{\omega}{c})$ planes are all 2D domains. (b) The spatio-temporal intensity profile $I(x,z\!=\!0;t)$ of the separable pulsed beam for a double-Gaussian spatio-temporal spectrum. (c) The spatio-temporal spectrum as depicted is separable; every spatial frequency $k_{x}$ is associated with all temporal frequencies $\omega$, and vice versa.}
  \label{Fig:TraditionalPulsedBeam}
\end{figure}

\subsubsection{Conventional pulsed beams}

A conventional pulsed beam has finite spatial \textit{and} temporal bandwidths, and its spatio-temporal spectral support domain is thus represented by a 2D patch on the light-cone surface [Fig.~\ref{Fig:TraditionalPulsedBeam}(a)]. The spectral projections onto any of the planes $(k_{x},\tfrac{\omega}{c})$, $(k_{z},\tfrac{\omega}{c})$, or $(k_{x},k_{z})$ are also 2D domains. In most pulsed lasers $\widetilde{\psi}(k_{x},\Omega)$ separates into a product $\widetilde{\psi}(k_{x},\Omega)\!\rightarrow\!\widetilde{\psi}_{x}(k_{x})\widetilde{\psi}_{t}(\Omega)$; i.e., the spatial and temporal DoFs are in general separable. Although this is not always strictly the case, it nevertheless encompasses most practical scenarios. The field can be written as $E(x,z;t)\!=\!e^{i(k_{\mathrm{o}}z-\omega_{\mathrm{o}}t)}\psi(x,z;t)$ with the envelope given by Eq.~\ref{Eq:BasicIntegral}. The spatial and temporal bandwidths $\Delta k_{x}$ and $\Delta\omega$ determine the transverse beam width $\Delta x$ and the temporal linewidth $\Delta t$, respectively [Fig.~\ref{Fig:TraditionalPulsedBeam}(b)]. Free propagation leads to diffractive spreading of the transverse beam profile and potentially pulse deformation via space-time coupling. In Fig.~\ref{Fig:TraditionalPulsedBeam} we illustrate this case with an example of a double-Gaussian pulsed beam, whereupon $\widetilde{\psi}_{x}(k_{x})$ and $\widetilde{\psi}_{t}(\Omega)$ are both Gaussian functions (although the basic features are independent of the particular profiles chosen).

The pictorial depiction of the spatio-temporal spectrum in Fig.~\ref{Fig:TraditionalPulsedBeam}(c) reflects its 2D nature. Here a range of spatial-frequency sinusoids are included corresponding to a finite spatial bandwidth $\Delta k_{x}$, and a range of temporal-frequency colors corresponding to the finite temporal bandwidth $\Delta\omega$. All the cells in the $(k_{x},\omega)$ grid are occupied, signifying the separability of the spectrum in term of spatial and temporal DoFs.

\section{Conditions for propagation invariance}
\label{Section:conditionsforpropinv}

Except for the plane-wave pulse that has no transverse spatial profile, monochromatic and pulsed beams both undergo diffractive spreading with free propagation. Monochromatic beams having certain spatial profiles represent exceptions that travel without diffractive spreading and are usually dubbed `diffraction-free' beams \cite{McGloin2005CP,Mazilu2010LPR,Vicente2021Optica}. We establish here a connection between the general conditions for diffraction-free propagation of monochromatic beams and the conditions for propagation invariance of a pulsed beam. This connection is made particularly clear when examining the role of the dimensionality of the spectra involved.

\subsection{Diffraction-free beams in two transverse dimensions}

\begin{figure}[t!]
\begin{center}
  \includegraphics[width=6cm]{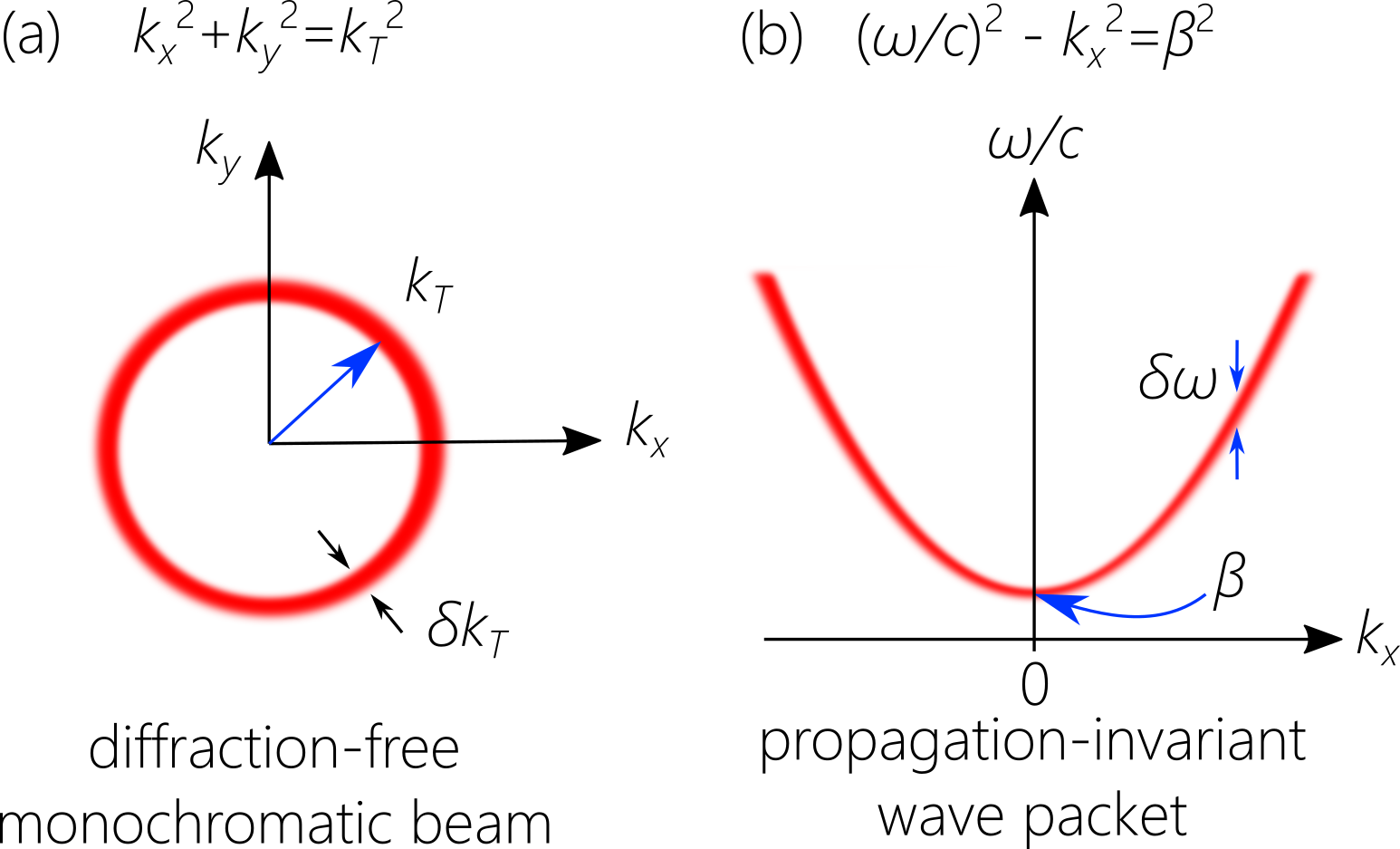}
  \caption{Diffraction-free monochromatic beam and propagation-invariant wave packet. Comparison of the conditions for achieving diffraction-free propagation of a monochromatic beam in two transverse dimensions and for the propagation invariance of a pulsed beam in one transverse dimension. (a) A monochromatic beam is diffraction-free when the spatial frequencies satisfy $k_{x}^{2}+k_{y}^{2}\!=\!k_{\mathrm{T}}^{2}$, with $k_{\mathrm{T}}$ a constant. (b) A pulsed beam is propagation invariant when $(\tfrac{\omega}{c})^{2}-k_{x}^{2}\!=\!\beta^{2}$, with $\beta$ a constant. Here, $\delta k_{\mathrm{T}}$ and $\delta\omega$ are the uncertainties associated with realistic experimental realizations.}
  \label{Fig:DiffractionFreeAndPropagationInvariant}
\end{center}
\end{figure}

We first elucidate why monochromatic diffraction-free beams require two transverse spatial dimensions for their realization. We say that a monochromatic beam is diffraction-free if its transverse spatial profile retains its shape and scale along the propagation axis. The Helmholtz equation for a monochromatic field at frequency $\omega_{\mathrm{o}}$ is $\{\nabla^{2}+k_{\mathrm{o}}^{2}\}E\!=\!0$, which yields the dispersion relationship $k_{x}^{2}+k_{y}^{2}+k_{z}^{2}\!=\!k_{\mathrm{o}}^{2}$. The field can be written as $E(x,y,z;t)\!=\!e^{i(k_{\mathrm{o}}z-\omega_{\mathrm{o}}t)}\psi(x,y,z)$, where $\psi(x,y,z)\!=\!\iint\!dk_{x}dk_{y}\widetilde{\psi}(k_{x},k_{y})e^{i\{k_{x}x+k_{y}y+(k_{z}-k_{\mathrm{o}})z\}}$; here $\widetilde{\psi}(k_{x},k_{y})$ is the 2D Fourier transform of $\psi(x,y,0)$, and $k_{z}^{2}\!=\!k_{\mathrm{o}}^{2}-k_{x}^{2}-k_{y}^{2}$. Because the axial phases $e^{ik_{z}z}$ will be in general different for each pair $(k_{x},k_{y})$, the plane waves underlying the beam dephase along $z$, which is manifested in diffractive spreading. Diffraction-free behavior is achieved when $k_{z}$ is a constant $k_{z}\!=\!\beta$, whereby $k_{x}^{2}+k_{y}^{2}\!=\!k_{\mathrm{o}}^{2}-\beta^{2}\!=\!k_{\mathrm{T}}^{2}$ is also a constant [Fig.~\ref{Fig:DiffractionFreeAndPropagationInvariant}(a)]. If all the plane waves underpinning the beam have transverse components $k_{x}$ and $k_{y}$ that satisfy this constraint, they will share the same axial wave number $E(x,y,z;t)\!=\!e^{i(\beta z-\omega_{\mathrm{o}}t)}\psi(x,y,0)$, and the transverse spatial profile evolves self-similarly except for an overall phase. Therefore, the spatial frequencies $(k_{x},k_{y})$ that lie on a circle of radius $k_{\mathrm{T}}$ give rise to a diffraction-free beam independently of the particular spectral amplitudes $\widetilde{\psi}(k_{x},k_{y})$ associated with these plane waves. Having the specific one-to-one relationship between $|k_{x}|$ and $|k_{y}|$ as defined by the circle $k_{x}^{2}+k_{z}^{2}\!=\!k_{\mathrm{T}}^{2}$ is essential for diffraction-free propagation because the axial phase factor gained by the $k_{x}$ component is then counterbalanced by that resulting from $k_{y}$. The special case of all the field amplitudes being identical in magnitude and phase yields the zeroth-order Bessel beam $\psi(x,y,0)\!=\!J_{0}(k_{\mathrm{T}}r)$, which takes the form of a central peak with surrounding concentric rings, each of which carries an equal amount of power to that in the central peak. As such, the intensity drops slowly from the center outwards. Other special functions yield diffraction-free beams in different coordinate systems \cite{Levy2016PO}. Furthermore, careful modulation of the field amplitudes and phases can produce a much broader family of field distributions with almost arbitrary profiles \cite{Bouchal2002OL,Lopez2010OL}. 

The fact that the spectral support domain projected onto the $(k_{x},k_{z})$-plane is a circle suggests that placing an annulus in the Fourier plane of a spherical lens illuminated with a plane wave will produce such a beam \cite{Durnin1987PRL}. Although straightforward, this spatial filtering approach is not power-efficient, and other so-called Bessel-beam generators have been developed \cite{McGloin2005CP,Turunen2010PO,Mazilu2010LPR}. However, the annulus-based spatial filter was instrumental as an inspiration for the development of X-waves as discussed below \cite{Saari1997PRL}.

All diffraction-free beams have infinite total power; i.e., $\iint\!dxdy|E(x,y,z)|^{2}$ is unbounded in any axial plane. Therefore, diffraction-free beams strictly speaking cannot be produced; however, realistic finite-energy realizations can exhibit extended propagation distances compared to conventional beams \cite{Durnin1988OL}. Such a finite-power beam is produced by a finite-width annular aperture, which relaxes the precise association between $k_{x}$ and $k_{y}$ [Fig.~\ref{Fig:DiffractionFreeAndPropagationInvariant}(a)]. That is, a spatial `uncertainty' is inevitable in a realistic diffraction-free beam, which determines its propagation distance.

\subsection{Why are there no 1D diffraction-free beams?}\label{Section:DiffractionFree1D}

All monochromatic diffraction-free beams require two transverse spatial dimensions for their realization. The profound impact of dimensionality on the prospects for diffraction-free propagation can be brought out by eliminating one transverse dimension (say $y$, whereupon $k_{y}\!=\!0$). We thus consider fields in the form of a light-sheet extended uniformly along $y$ but localized along $x$, which are described by the reduced-dimensionality dispersion relationship $k_{x}^{2}+k_{z}^{2}\!=\!k_{\mathrm{o}}^{2}$, and the field envelope in Eq.~\ref{Eq:MonochromaticBeamEnvelope}. 

Diffraction-free behavior requires that $k_{z}\!=\!\beta$ for all spatial frequencies $k_{x}$, whereupon $k_{x}^{2}\!=\!k_{\mathrm{o}}^{2}-\beta^{2}\!=\!k_{\mathrm{T}}^{2}$, and thus $k_{x}\!=\!\pm k_{\mathrm{T}}$. In other words, only the trivial solutions corresponding to a plane wave $e^{i(\pm k_{\mathrm{T}}x+\beta z)}$ or a cosine wave $e^{i\beta z}\cos{k_{\mathrm{T}}x}$ are diffraction-free in one transverse dimension -- neither of which is localized along $x$. Indeed, even a Bessel beam $J_{0}(k_{\mathrm{T}}x)$ along $x$ is \textit{not} diffraction-free in contrast to its 2D counterpart. From a different perspective, because the axial phase factor resulting from $k_{x}$ is not compensated by that from $k_{y}$, only a single $k_{x}$ can be used in constructing the beam and still maintain diffraction-free propagation. For this fundamental reason, there are no diffraction-free monochromatic light sheets or surface plasmon polaritons (which are surface waves with one transverse dimension), with the exception of the cosine plasmon \cite{Lin2012PRL}. The existence of diffraction-free beams in 2D is therefore predicated on the availability of two free spectral variables $k_{x}$ and $k_{y}$.

Incidentally, a unique exception to this statement exists if we relax the definition of diffraction-free propagation by allowing for the axially invariant profile to travel along a curved trajectory rather than a straight line \cite{Siviloglou2007OL,Siviloglou2007PRL,Hu2012Book,Efremidis2019Optica}. In other words, we have $\psi(x,z)\!=\!\psi(x-x_{\mathrm{o}}(z),0)$, where $x_{\mathrm{o}}(z)$ is a fixed point in the spatial profile that is transversely displaced as the field propagates along $z$. It can be shown that the Airy function is the unique solution subject to these constraints in the paraxial regime \cite{Berry1979AMP,Unnikrishnan1996AJP}. The Airy beam is therefore the unique profile providing diffraction-free light-sheets \cite{Vettenburg2014NM,Piksarv17SR,Subedi2021OE} and surface plasmon polaritons \cite{Salandrino2010OL,Minovich2011PRL,Li2011PRL,Zhang2011OL,Minovich2014LPR,Wang2017OE}, although both travel along curved trajectories. The uniqueness of the 1D Airy wave packet also implies that it is the only propagation-invariant (but accelerating) plane-wave pulse in a dispersive medium \cite{Chong2010NP,Kaminer2011OE}. Furthermore, approximately diffraction-free beams can be crafted in 1D  \cite{Rosen1995OL,Eliyahu1995OL}, which would then also correspond to approximately dispersion-free pulses. Of course, diffraction-free 1D beams can be supported in the appropriate nonlinear medium, such as a photorefractive planar waveguide  \cite{Christodoulides1996OL}

\subsection{Why can pulsed beams propagate invariantly in one transverse dimension?}\label{Section:PropagationInvarianceIn1D}

While there are no diffraction-free monochromatic light sheets, propagation-invariant \textit{pulsed} beams or wave packets \textit{do exist} in only one transverse dimension \cite{Kondakci2016OE,Parker2016OE,Kondakci2017NP,Kondakci2018PRL}. We define a propagation-invariant wave packet as a pulsed field whose spatio-temporal profile shape and scale do not change with free propagation except for a group delay along the propagation axis and possibly an overall phase.

Starting with the wave equation $\{\partial_{x}^{2}+\partial_{z}^{2}-c^{-2}\partial_{t}^{2}\}E(x,z;t)\!=\!0$, we obtain the dispersion relationship $k_{x}^{2}+k_{z}^{2}\!=\!(\tfrac{\omega}{c})^{2}$,  and the envelope of the field $E(x,z;t)\!=\!e^{i(k_{\mathrm{o}}z-\omega_{\mathrm{o}}t)}\psi(x,z;t)$ is given by Eq.~\ref{Eq:BasicIntegral}. One simple scenario to realize propagation invariance is obtained by setting $k_{z}\!=\!\beta$ \cite{Kondakci2016OE,Parker2016OE,Kondakci2019ACSP}, whereupon $E(x,z;t)\!=\!e^{i(\beta z-\omega_{\mathrm{o}}t)}\psi(x,0;t)$; i.e., the wave-packet envelope is independent of $z$. Here $k_{x}$ and $\omega$ are related through $\omega\!=\!\omega(k_{x})\!=\!c\sqrt{\beta^{2}+k_{x}^{2}}$, and we thus have a one-to-one relationship between $\omega$ and $|k_{x}|$, $\widetilde{\psi}(k_{x},\Omega)\!\rightarrow\!\widetilde{\psi}(k_{x})\delta(\Omega-\Omega(k_{x}))$, where $\Omega(k_{x})\!=\!\omega(k_{x})-\omega_{\mathrm{o}}$. The axial phase factor resulting from $k_{x}$ is compensated by the phase factor from $\Omega$. In other words, the temporal frequency $\omega$ in the pulsed beam in one spatial dimension replaces the spatial frequency $k_{y}$ in the monochromatic beam in two spatial dimensions. In this sense, the condition for propagation invariance of a pulsed beam with one transverse dimension is analogous to the condition for diffraction-free propagation of a monochromatic beam having two transverse dimensions. In the latter, a constraint is imposed upon $k_{x}$ and $k_{y}$, and in the former a constraint is imposed upon $k_{x}$ and $\Omega$. However, because of the different roles played by space and time in the wave equation, the circular relationship $k_{x}^{2}+k_{y}^{2}\!=\!k_{\mathrm{T}}^{2}$ [Fig.~\ref{Fig:DiffractionFreeAndPropagationInvariant}(a)] is replaced by the hyperbolic relationship $(\tfrac{\omega}{c})^{2}-k_{x}^{2}\!=\!\beta^{2}$ [Fig.~\ref{Fig:DiffractionFreeAndPropagationInvariant}(b)].

A further similarity between these two different scenarios is that the energy of this propagation-invariant wave packet $\iint\!dxdt|\psi(x,z;t)|^{2}$ is infinite in any axial plane, just as the power is infinite for the monochromatic diffraction-free beam. In both cases, the strict delta-function association between the two spectral coordinates underpins this divergent infinity. Introducing a `spectral uncertainty' $\delta\omega$ [Fig.~\ref{Fig:DiffractionFreeAndPropagationInvariant}(b)] in the association between $k_{x}$ and $\omega$ renders both the energy and the propagation distance finite. The example considered here demonstrates that propagation-invariant wave packets do indeed exist in one transverse spatial dimension, whereas diffraction-free monochromatic beams do \textit{not}. However, it is important to note that this particular example where $k_{z}\!=\!\beta$ does not exhaust all the possible scenarios of propagation-invariant wave packets, and the general case will be discussed below.

The analogy between the spectra in Fig.~\ref{Fig:DiffractionFreeAndPropagationInvariant}(a) and Fig.~\ref{Fig:DiffractionFreeAndPropagationInvariant}(b) suggests a general strategy for synthesizing propagation-invariant wave packets by selecting the appropriate spatio-temporal frequency pairs $(k_{x},\omega)$ just as pairs of spatial frequencies $(k_{x},k_{y})$ are selected via spatial filtering in the Fourier domain to produce a diffraction-free beam. However, whereas purely spatial filtering in the Fourier domain is straightforward, \textit{spatio-temporal filtering} remains a challenge and previous attempts at such filtering have not yielded positive results \cite{Dallaire2009OE,Jedrkiewicz2013OE}. We discuss in Section \ref{Synthesis} our recently developed energy-efficient phase-only spectral modulation methodology to address this challenge.

\begin{figure}[t!]
  \begin{center}
  \includegraphics[width=6.5cm]{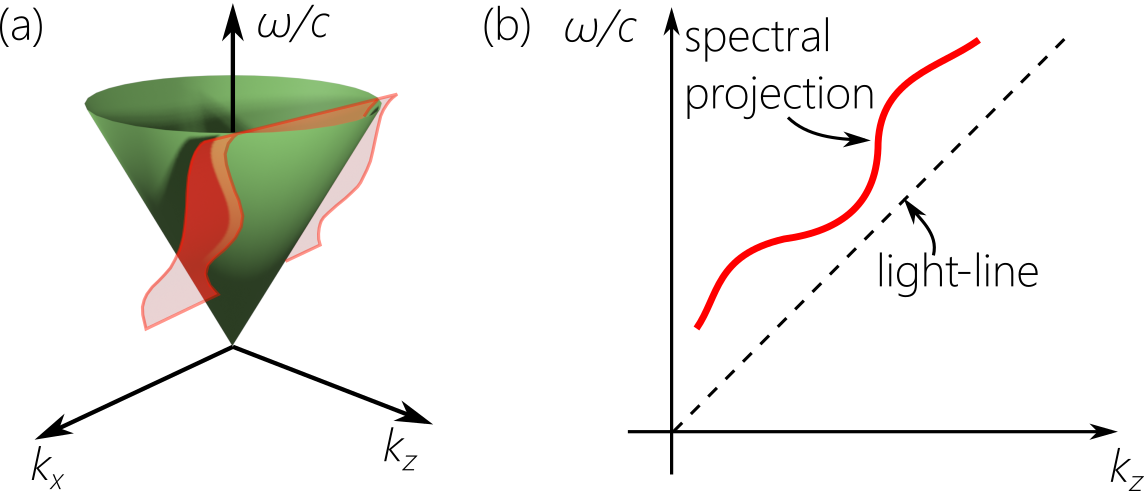}
   \caption{Example of the representation for a wave packet with diffraction-free time-averaged intensity, but whose spatio-temporal envelope is not necessarily propagation-invariant. (a) The spectral support domain is the intersection of the light-cone with a planar curved surface that is parallel to the $k_{x}$-axis. (b) As long as there is a one-to-one relationship between $k_{z}$ and $\omega$, the diffraction-free behavior $I(x,z)\!=\!I(x,0)$ of the time-averaged intensity is independent of the precise shape of the spectral projection onto the $(k_{z},\tfrac{\omega}{c})$-plane.\label{Fig:RepresentationForTimeAveragedDiffractionFree}}
  \end{center}
\end{figure}

\subsection{Conditions for propagation invariance of the time-averaged intensity}

We described above a particular example of a propagation-invariant wave packet for which $k_{z}\!=\!\beta$. We proceed to discuss the more general requirements for a wave packet to propagate invariantly. First, we consider propagation invariance of the \textit{time-averaged intensity} (or equivalently the energy distribution) $I(x,z)\!=\!\int\!dt|\psi(x,z;t)|^{2}\!=\!I(x,0)$  \cite{Porras2003PREBessel-X}. Starting with Eq.~\ref{Eq:BasicIntegral} for the most general wave-packet envelope, we have
\begin{equation}
I(x,z)=\iint\!dk_{x}dk_{x}'G(k_{x},k_{x}';z)e^{i(k_{x}-k_{x}')x},
\end{equation}
where $G(k_{x},k_{x}';z)\!=\!\int\!d\Omega\widetilde{\psi}(k_{x},\Omega)\widetilde{\psi}^{*}(k_{x}',\Omega)e^{i(k_{z}-k_{z}')z}$, $k_{z}\!=\!\sqrt{(\omega/c)^{2}-k_{x}^{2}}$, and $k_{z}'\!=\!\sqrt{(\omega/c)^{2}-k_{x}'^{2}}$. Satisfying the diffraction-free requirement $I(x,z)\!=\!I(x,0)$ for any spatio-temporal spectrum $\widetilde{\psi}(k_{x},\Omega)$, necessitates that $G(k_{x},k_{x}';z)\!=\!G(k_{x},k_{x}';0)$.

\begin{figure}[t!]
  \begin{center}
  \includegraphics[width=10cm]{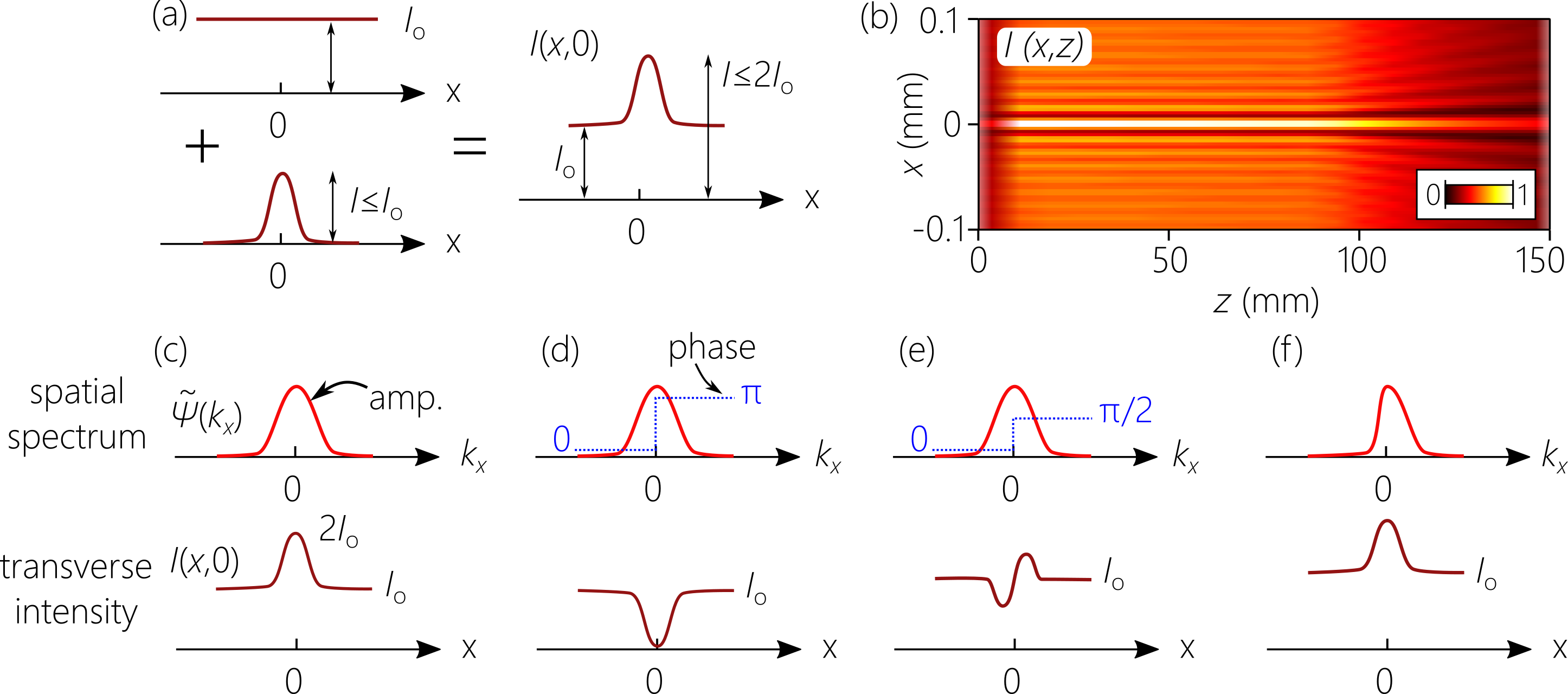}
   \caption{Structure of the diffraction-free intensity $I(x,z)$ for a pulsed beam. (a) According to Eq.~\ref{Eq:TwoTermsForDiffFree}, the intensity $I(x,0)$ at $z\!=\!0$ comprises a narrow localized spatial feature atop a pedestal. (b) The axial evolution of the transverse intensity distribution. We have included a spectral uncertainty of $\delta\lambda\!=\!25$~pm that results in a finite propagation distance, and a spectral tilt angle $\theta\!=\!50^{\circ}$. (c-f) Structure of the transverse intensity distribution for different spatial spectra: $\widetilde{\psi}(k_{x})$ is (c) even; (d) odd; (e) has a $\tfrac{\pi}{2}$-phase step across $k_{x}\!=\!0$; and (f) asymmetric distribution; i.e., a superposition of even and odd functions. The resulting intensity $I(x,0)$ corresponding to these spectra has (c) a high-visibility peak; (d) a high-visibility dip; (e) a low-visibility peak-dip feature; and (f) a low-visibility peak.}
   \label{Fig:StructureOfTimeAveragedIntensity}
  \end{center}
\end{figure}

This requires that $\widetilde{\psi}(k_{x},\Omega)$ be composed of pairs of spatial and temporal frequencies for which $\widetilde{\psi}(k_{x},\Omega)\!=\!\widetilde{\psi}(k_{x})\delta(\Omega-\Omega(|k_{x}|))$; i.e., each temporal frequency $\Omega$ can be related to only one pair of spatial frequencies $\pm k_{x}$. The specific functional form of $\Omega(|k_{x}|)$ is \textit{not} relevant; it is necessary only that a one-to-one relationship exists between $|k_{x}|$ and $\Omega$. The spectral support domain for a wave packet satisfying this condition lies at the intersection of the light-cone with a planar curved surface that is parallel to the $k_{x}$-axis and orthogonal to the $(k_{z},\tfrac{\omega}{c})$-plane [Fig.~\ref{Fig:RepresentationForTimeAveragedDiffractionFree}(a)]. The spectral projection onto the $(k_{z},\tfrac{\omega}{c})$-plane is a 1D curve [Fig.~\ref{Fig:RepresentationForTimeAveragedDiffractionFree}(b)]. In this case $G(k_{x},k_{x}';z)\!=\!|\widetilde{\psi}(k_{x})|^{2}\delta(k_{x}-k_{x}')+\mathrm{Re}\{\widetilde{\psi}(k_{x})\widetilde{\psi}^{*}(-k_{x})\}\delta(k_{x}+k_{x}')$, and there are no axial dynamics:
\begin{equation}\label{Eq:TwoTermsForDiffFree}
I(x,z)=\int\!dk_{x}|\widetilde{\psi}(k_{x})|^{2}+\mathrm{Re}\!\int\!dk_{x}\widetilde{\psi}(k_{x})\widetilde{\psi}^{*}(-k_{x})e^{i2k_{x}x}=I(x,0).
\end{equation}
The time-averaged intensity profile is the sum of a background pedestal $\int\!dk_{x}|\widetilde{\psi}(k_{x})|^{2}$ atop of which there is a localized diffraction-free feature $\int\!dk_{x}\widetilde{\psi}(k_{x})\widetilde{\psi}^{*}(-k_{x})e^{i2k_{x}x}$ whose width is the inverse of the spatial bandwidth and whose maximum height is equal to or less than that of the pedestal \cite{Kondakci2019OL}. The maximum height of the localized feature occurs when $\widetilde{\psi}(k_{x})$ is an even function. When $\widetilde{\psi}(k_{x})$ is odd, a dip is produced in the intensity profile rather than a peak (a so-called `dark beam' \cite{Ponomarenko2007OL,Yessenov2019Optica,Zhu2019OL}); see Fig.~\ref{Fig:StructureOfTimeAveragedIntensity}. The factor-of-2 in the exponent $e^{i2k_{x}x}$ in the second term in Eq.~\ref{Eq:TwoTermsForDiffFree} implies that the width of the diffraction-free spatial feature is half that expected from the width of its spatial spectrum.

All this presumes a perfect association between the spatial and temporal frequencies in the wave packet spectrum. A departure from this precise association, as occurs in the presence of spectral uncertainty [Fig.~\ref{Fig:DiffractionFreeAndPropagationInvariant}(b)] not only results in a finite propagation distance, but also reduces the height of the background pedestal. Ultimately, when the spatio-temporal spectrum becomes separable with respect to $k_{x}$ and $\omega$, the pedestal vanishes. In other words, observing the pedestal in the time-averaged intensity of wave packets in one transverse dimension attests to the tight association between $k_{x}$ and $\omega$.

Of course, if the wave packet envelope is propagation invariant, then the associated time-averaged intensity will also be diffraction-free. The converse, however, does not hold. When the \textit{time-averaged intensity} is diffraction-free, we cannot expect the underlying spatio-temporal profile of the propagating wave packet to be necessarily propagation invariant. Indeed, as long as a one-to-one relationship exists between the spatial and temporal frequencies, the wave packet may undergo dramatic spatio-temporal dynamics axially with free propagation that remain veiled from our view when the time-averaged intensity alone is observed. 

As a particular example, consider a wave packet that is propagation-invariant in a dispersive medium. Such a wave packet would necessitate a particular functional form for the tight association between $k_{x}$ and $\Omega$ in its spatio-temporal spectrum. Of course, such a wave packet would undergo temporal dispersive spreading in free space. Nevertheless, its time-averaged intensity profile would be \textit{diffraction-free} and contain a constant pedestal. In previous work on the synthesis of 1D wave packets predicted to be propagation-invariant in media having anomalous\cite{Dallaire2009OE} and normal \cite{Jedrkiewicz2013OE} GVD, the pedestal was not visible and a diffraction-free time-averaged intensity was not verified. This indicates that the spectral uncertainty was likely too large for the diffraction-free behavior to emerge (Section~\ref{Section:PropagationDistance}).

\subsection{Requirements for propagation invariance of the envelope}

Propagation invariance of the spatio-temporal envelope is a more stringent constraint than diffraction-free propagation of the time-averaged intensity. Besides the one-to-one association between $|k_{x}|$ and $\omega$ necessary for the latter, a particular functional form for this association must additionally be imposed. We take propagation invariance to imply that $\psi(x,z;t)\!=\!\psi(x,0;t-z/\widetilde{v})$; that is, the wave-packet envelope propagates invariantly at a group velocity $\widetilde{v}$, not necessarily equal to $c$, and that the field takes the form $E(x,z;t)\!=\!e^{i\varphi(z)}E(x,0;t-z/\widetilde{v})$. This condition implies that $\varphi(z)\!=\!(k_{z}-\tfrac{\omega}{\widetilde{v}})z$. However, because $\varphi$ is independent of $k_{z}$ and $\omega$, we have the general constraint:
\begin{equation}\label{Eq:GeneralKz}
k_{z}=k_{\mathrm{a}}+\frac{\omega}{\widetilde{v}},
\end{equation}
where $k_{\mathrm{a}}$ is a constant. Consequently, the phase $\varphi(z)\!=\!k_{\mathrm{a}}z$ is linear in $z$.

\begin{figure}[t!]
\begin{center}
 \includegraphics[width=13cm]{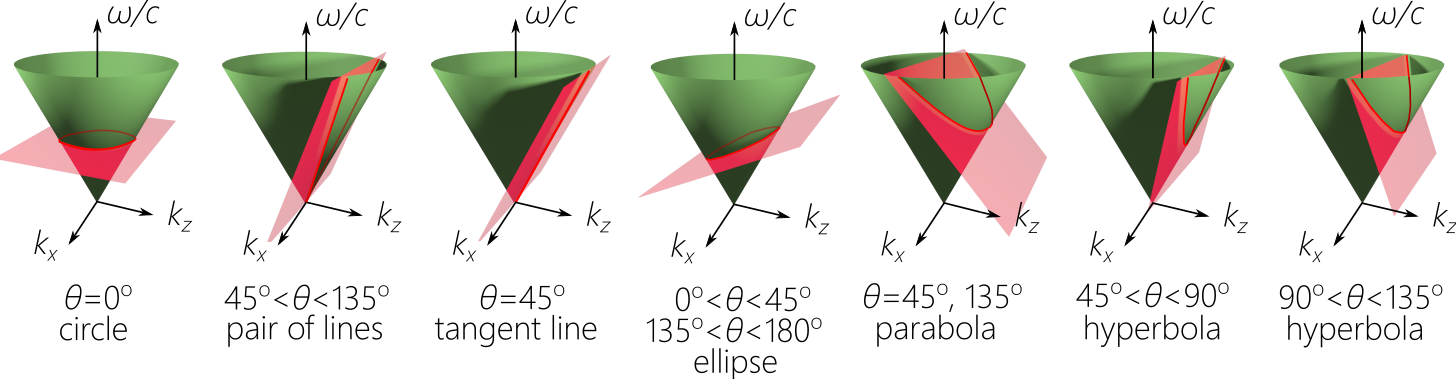}
 \caption{The variety of spectral support domains corresponding to the conic sections resulting from the intersection of the light-cone with a plane are depicted. Each of these cases is consistent with a propagation-invariant ST wave packet.} \label{Fig:ConicSections}
\end{center}
\end{figure}

\begin{figure}\begin{center}[t!]
  \includegraphics[width=11cm]{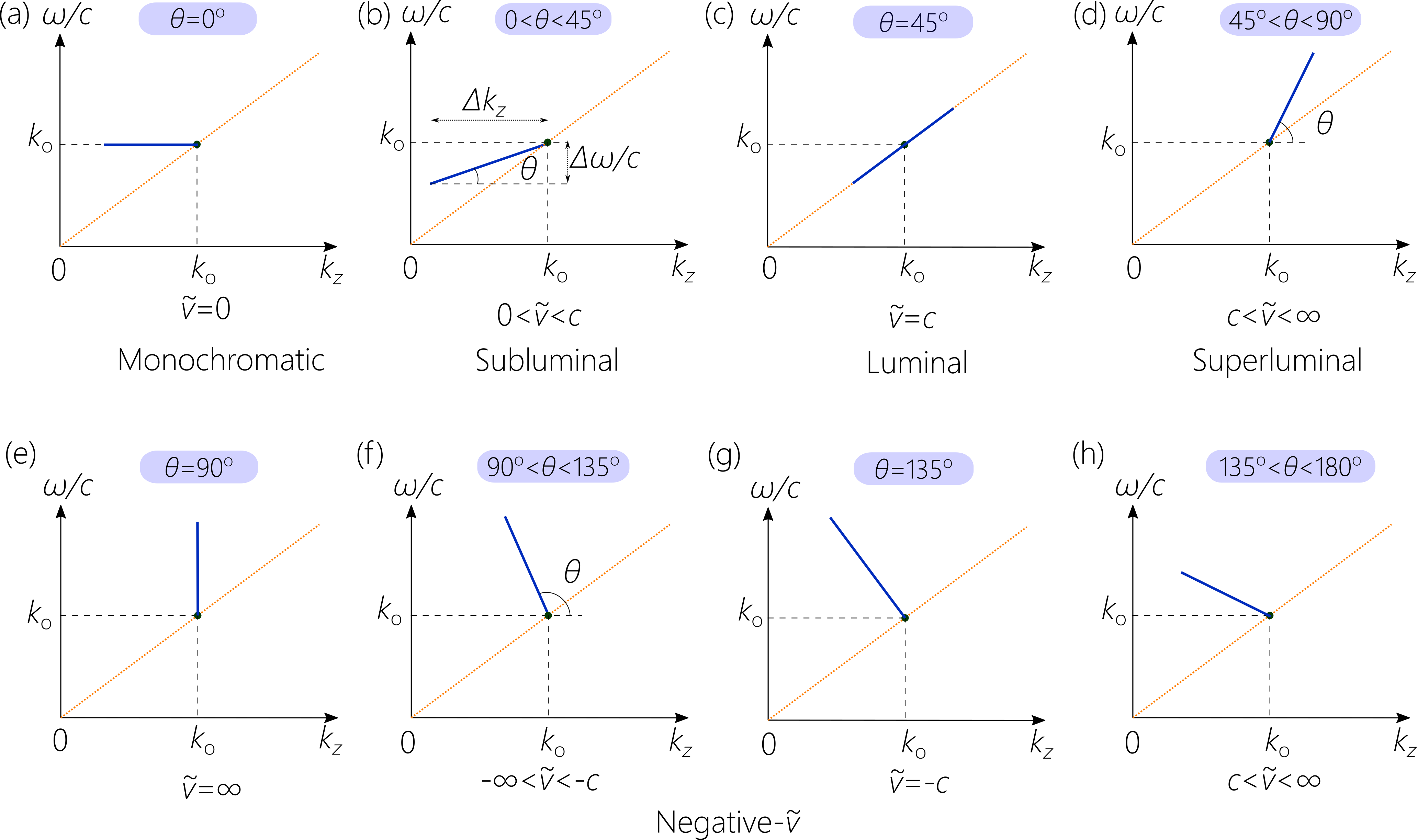}
    \caption{Spectral projections of baseband ST wave packets onto the $(k_{z},\tfrac{\omega}{c})$-plane as $\theta$ increases from $0^{\circ}$ to $180^{\circ}$; see Fig.~\ref{Fig:ConicSections} for the corresponding 3D representation. Each projection is associated with a different conic section on the surface of the light-cone; see Table~\ref{Table:BasebandClasses}.}
    \label{Fig:SpectralProjectionsOntoKzOmegaPlane}
\end{center}
\end{figure}

The result in Eq.~\ref{Eq:GeneralKz} implies that the projection of the spectral support domain onto the $(k_{z},\tfrac{\omega}{c})$-plane is a straight line. For this to be true, the spectral support domain on the surface of the light-cone must be restricted to its intersection with a plane $k_{z}\!=\!k_{\mathrm{a}}+\tfrac{\omega}{\widetilde{v}}$ that is parallel to the $k_{x}$-axis and makes an angle $\theta$ with respect to the $k_{z}$-axis, where $\widetilde{v}\!=\!c\tan{\theta}$. We refer to $\theta$ as the spectral tilt angle. The spectral support domain is thus a conic section whose nature depends on the values of the two parameters $k_{\mathrm{a}}$ and $\theta$; see Fig.~\ref{Fig:ConicSections}, Fig.~\ref{Fig:SpectralProjectionsOntoKzOmegaPlane}, and Table~\ref{Table:BasebandClasses}. The spectral trajectory on the light-cone surface is one-dimensional, so that $\omega$ is uniquely related to $|k_{x}|$ for any values of $k_{\mathrm{a}}$ and $\theta$. This construction thus guarantees a diffraction-free time-averaged intensity. In addition, the field is also propagation-invariant in the sense defined above. We call propagation-invariant pulsed beams in general \textit{ST wave packets}. Because ST wave packets are pulsed beams, they have finite spatial \textit{and} temporal bandwidths, in common with conventional wave packets having 2D spectral support domains on the surface of the light-cone. However, more akin to the simpler examples of monochromatic beams [Fig.~\ref{Fig:MonochromaticBeam}] or plane-wave pulses [Fig.~\ref{Fig:PulsedPlaneWave}], ST wave packets have 1D spectral support domains. The distinctive reduced-dimensionality spectral representation associated with finite spatial \textit{and} temporal bandwidths underpins the unique characteristics of ST wave packets.

\section{Classification of propagation-invariant wave packets}
\label{Section:classification}

Three distinct scenarios emerge corresponding to the regimes $k_{\mathrm{a}}\!>\!0$, $k_{\mathrm{a}}\!=\!0$, and $k_{\mathrm{a}}\!<\!0$, all subject to the constraint $0\!<\!k_{z}\!<\!\tfrac{\omega}{c}$ for each $\omega$. As we vary $\theta$ in these three regimes, a variety of conditions arise that define distinct families of ST wave packets. Visualizing the representation of the ST wave packet on the light-cone surface is particularly instructive here. In all these scenarios, the spectral support domain is at the intersection of the light-cone with a plane that is parallel to the $k_{x}$-axis and makes an angle $\theta$ with respect to the $k_{z}$-axis, with $\widetilde{v}\!=\!c\tan{\theta}$ for any $k_{\mathrm{a}}$. Figure~\ref{Fig:ConicSections} enumerates all the possible intersections of such a plane with the light-cone from a geometric perspective: a circle, a pair of straight lines, a single tangential line, an ellipse, a parabola, or a hyperbola. Each geometry is associated with at least one family of propagation-invariant ST wave packets as we proceed to show.

\subsection{Baseband ST wave packets: Positive $k_{\mathrm{a}}$}

We first consider $k_{\mathrm{a}}\!>\!0$, so that $k_{z}\!=\!k_{\mathrm{a}}+\tfrac{\omega}{\widetilde{v}}\!\leq\!\tfrac{\omega}{c}$, where the equality sign corresponds to the intersection point with the light-line $k_{z}\!=\!\tfrac{\omega}{c}$ that occurs when $k_{x}\!=\!0$. We denote the frequency $\omega\!=\!\omega_{\mathrm{o}}$ at this intersection point, so that $k_{\mathrm{a}}\!=\!k_{\mathrm{o}}-\tfrac{\omega}{\widetilde{v}}$, and:
\begin{equation}\label{Eq:BasebandPlane}
k_{z}=k_{\mathrm{o}}+\frac{\Omega}{\widetilde{v}}.
\end{equation}
The \textit{spatial} spectrum is centered around $k_{x}\!=\!0$, corresponding to $\omega\!=\!\omega_{\mathrm{o}}$ or $\Omega\!=\!0$. This is the condition for realizing a so-called `baseband' ST wave packet. This nomenclature is borrowed from radio engineering to signify that the spatial spectrum includes the origin $k_{x}\!=\!0$, and that the spectrum is typically concentrated in its vicinity. Most importantly in the optical domain, this condition signifies that such wave packets can be readily produced in the paraxial regime. Only one parameter remains to be varied: the spectral tilt angle $0^{\circ}\!<\!\theta\!<\!180^{\circ}$, as shown in Fig.~\ref{Fig:SpectralProjectionsOntoKzOmegaPlane}.

By writing the field as $E(x,z;t)\!=\!e^{i(k_{\mathrm{o}}z-\omega_{\mathrm{o}}t)}\psi(x,z;t)$, the wave-packet envelope can be expressed as follows: 
\begin{equation}
\psi(x,z;t)=\int\!dk_{x}\widetilde{\psi}(k_{x})e^{i(k_{x}x-\Omega(k_{x})\{t-z/\widetilde{v}\})}=\psi(x,0;t-z/\widetilde{v}).
\end{equation}
Here we take $k_{x}$ as the independent variable, and $\Omega\!=\!\Omega(k_{x})$ is a function of $k_{x}$. We may equivalently take $\Omega$ as the independent variable and express $k_{x}$ as a function of $\Omega$. The relationship between $k_{x}$ and $\Omega$ is determined from the geometry of the spectral support domain on the light-cone surface that lies at its intersection with a plane $\mathcal{P}_{\mathrm{B}}(\theta)$ defined by Eq.~\ref{Eq:BasebandPlane}, and the conic sections associated with the different angular spans of $\theta$ in Fig.~\ref{Fig:SpectralProjectionsOntoKzOmegaPlane} are enumerated in Table~\ref{Table:BasebandClasses}. We divide the various families of baseband ST wave packets into three classes: subluminal when $0^{\circ}\!<\!\theta\!<\!45^{\circ}$, superluminal when $45^{\circ}\!<\!\theta\!<\!90^{\circ}$, and negative-$\widetilde{v}$ when $\theta\!>\!90^{\circ}$.

\begin{table}[t!]
\caption{Classes of baseband ST wave packets following the enumeration in Fig.~\ref{Fig:SpectralProjectionsOntoKzOmegaPlane}.}
\label{Table:BasebandClasses}
\begin{center}
\begin{tabular}{|p{0.5cm}||p{2.2cm}|p{2.2cm}|p{2.2cm}|}
 \hline
  & $\theta$ & $\widetilde{v}$ & conic section\\
 \hline\hline
 (a) & $0^{\circ}$ & 0 & circle\\ \hline
 (b) & $0^{\circ}\!<\!\theta\!<\!45^{\circ}$ & $0\!<\!\widetilde{v}\!<\!c$ & ellipse\\ \hline
 (c) & $45^{\circ}$ & $c$ & tangent line\\ \hline
 (d) & $45^{\circ}\!<\!\theta\!<\!90^{\circ}$ & $c\!<\!\widetilde{v}\!<\!\infty$ & hyperbola\\ \hline
 (e) & $90^{\circ}$ & $\infty$ & hyperbola\\ \hline
 (f) & $90^{\circ}\!<\!\theta\!<\!135^{\circ}$ & $-\infty\!<\!\widetilde{v}\!<\!-c$ & hyperbola\\ \hline
 (g) & $135^{\circ}$ & $-c$ & parabola\\\hline
 (h) & $135^{\circ}\!<\!\theta\!<\!180^{\circ}$ & $-c\!<\!\widetilde{v}\!<\!0$ & ellipse\\ \hline
 \end{tabular}
 \end{center}
\end{table}

Some general features are shared by all baseband ST wave packets. In the limit of small spatial and temporal bandwidths, $\Delta k_{x}\!\ll\!k_{\mathrm{o}}$ and $\Delta\omega\!\ll\!\omega_{\mathrm{o}}$, respectively, the spectral support domain projected onto the $(k_{x},\tfrac{\omega}{c})$-plane can be approximated by a parabola:
\begin{equation}\label{Eq:BasebandParabola}
\frac{\Omega}{\omega_{\mathrm{o}}}=\frac{k_{x}^{2}}{2k_{\mathrm{o}}^{2}(1-\cot{\theta})}.
\end{equation}
The curvature of the parabola is inversely proportional to the quantity $1-\widetilde{n}$ in free space, where $\widetilde{n}\!=\!\cot{\theta}$. In a non-dispersive medium of refractive index $n$, this quantity becomes $n(n-\widetilde{n})$ \cite{Bhaduri2020NP}. When investigating the refraction of ST wave packets (Section~\ref{Section:Refraction}), we refer to this quantity as the spectral curvature. This parabolic approximation also points to a relationship between the spatial and temporal bandwidths. If the spatial frequency $k_{x}\!=\!0$ is associated with the temporal frequency $\omega\!=\!\omega_{\mathrm{o}}$, and the maximum spatial frequency $k_{x}\!=\!\Delta k_{x}$ is associated with the frequency $\omega\!=\!\omega_{\mathrm{o}}+\Delta\omega$, then we have:
\begin{equation}\label{Eq:SpatialAndTemporalBandwidthBaseband}
\frac{\Delta\omega}{\omega_{\mathrm{o}}}=\frac{1}{2|1-\cot{\theta}|}\left(\frac{\Delta k_{x}}{k_{\mathrm{o}}}\right)^{2}.
\end{equation}
The temporal bandwidth $\Delta\omega$ increases quadratically with $\Delta k_{x}$, $\Delta\omega\!\propto\!(\Delta k_{x})^{2}$. Furthermore, the proportionality constant is enhanced as $\theta$ deviates away from $45^{\circ}$. However, by exploiting the concept of `spectral recycling' this proportionality between $\Delta\omega$ and $(\Delta k_{x})^{2}$ can be alleviated \cite{Hall2021PRArecycling}. In this strategy, rather than associating each $k_{x}$ to a single $\omega$, the spectral bandwidth $\Delta\omega$ is divided into segments, and each $k_{x}$ is associated with one frequency in \textit{each} subsection; i.e., each $k_{x}$ is `recycled' with multiple temporal frequencies \cite{Hall2021PRArecycling}.

\subsubsection{Subluminal ST wave packets}\label{Subluminal_ST}

\begin{figure}[t!]
  \begin{center}
  \includegraphics[width=12.0cm]{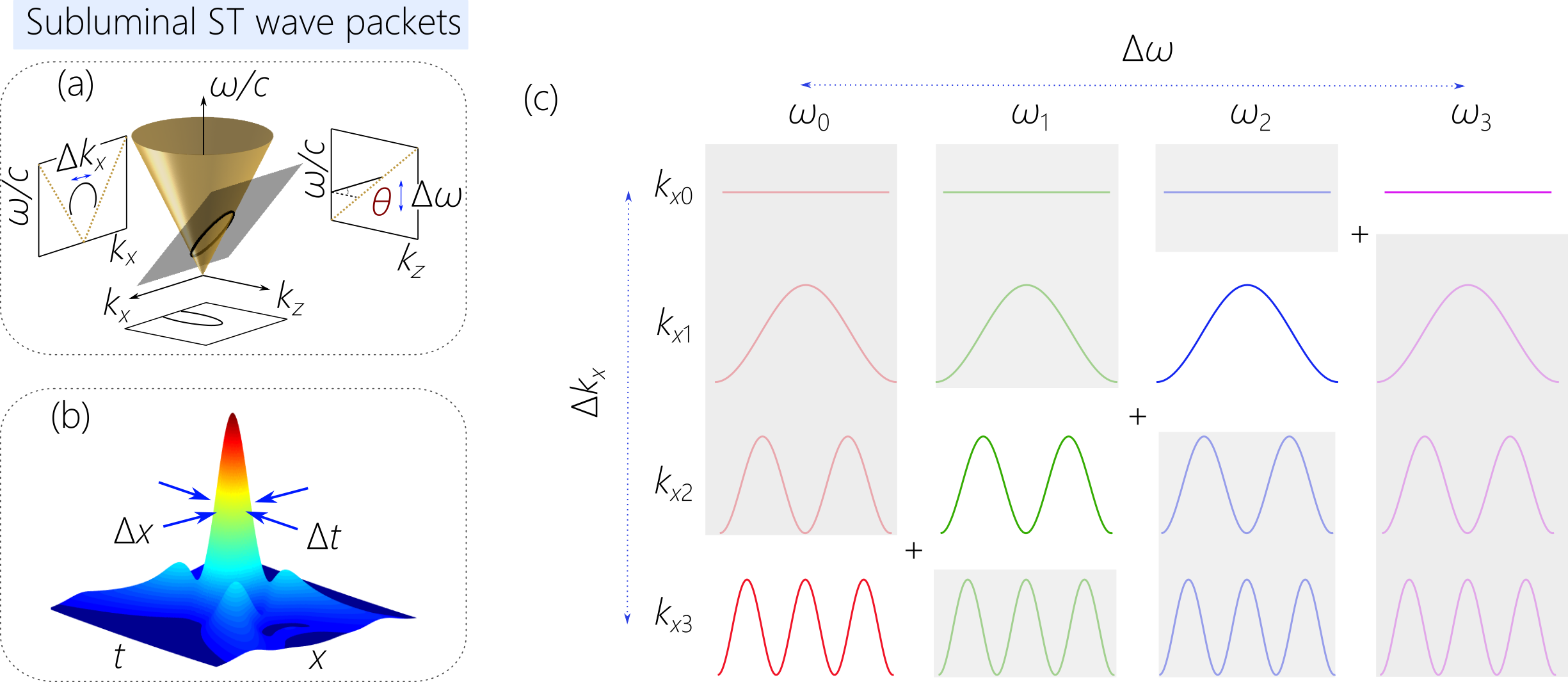}
  \end{center}
  \caption{Representation of the spatio-temporal spectrum of \textit{subluminal} ST wave packets. (a) The spectral support domain on the surface of the light-cone lies at the intersection of the light-cone with a spectral plane with a spectral tilt angle $\theta\!<\!45^{\circ}$. Spectral projections onto the $(k_{z},\tfrac{\omega}{c})$-plane is a straight line, and on the $(k_{x},\tfrac{\omega}{c})$ and $(k_{x},k_{z})$ planes are ellipses. (b) The spatio-temporal intensity profile $I(x,0;t)$. (c) The subluminal ST wave packet is a superposition of spatial frequencies, each associated with a single wavelength. The spatial frequency \textit{decreases} with the temporal frequency. The monochromatic plane waves obscured behind a gray foreground are eliminated from the wave packet angular spectrum of the conventional pulsed beam.}
  \label{Fig:SubluminalSTWavePacket}
\end{figure}

For subluminal baseband ST wave packets, the spectral projection onto the $(k_{z},\tfrac{\omega}{c})$-plane is the straight line in Eq.~\ref{Eq:BasebandPlane} but with $\theta$ restricted to the range $0^{\circ}\!<\!\theta\!<\!45^{\circ}$ [Fig.~\ref{Fig:SpectralProjectionsOntoKzOmegaPlane}(b)], such that $0\!<\!\widetilde{v}\!<\!c$ and $\widetilde{n}\!>\!1$. In the $(k_{z},\tfrac{\omega}{c})$-plane, this linear spectral projection intersects with the light-line at the point $(k_{z},\tfrac{\omega}{c})\!=\!(k_{\mathrm{o}},k_{\mathrm{o}})$. Within this geometrical construction, $\omega_{\mathrm{o}}$ is the \textit{maximum} possible frequency to be included in the spectrum, $\omega\!\leq\!\omega_{\mathrm{o}}$. Because $0^{\circ}\!<\!\theta\!<\!45^{\circ}$, the intersection of the plane $\mathcal{P}_{\mathrm{B}}(\theta)$ with the light-cone is an ellipse [Fig.~\ref{Fig:SubluminalSTWavePacket}(a)], whose projection onto the $(k_{x},\tfrac{\omega}{c})$-plane is described by the equation:
\begin{equation}
\frac{1}{k_{1}^{2}}\left(\frac{\omega}{c}-k_{2}\right)^{2}+\frac{k_{x}^{2}}{k_{3}^{2}}=1,
\end{equation}
where the parameters $k_{1}$, $k_{2}$, and $k_{3}$ are defined as follows:
\begin{equation}
\label{k1,k2,k3-def}
\frac{k_{1}}{k_{\mathrm{o}}}=\left|\frac{\tan{\theta}}{1+\tan{\theta}}\right|,\,\,\,\frac{k_{2}}{k_{\mathrm{o}}}=\frac{1}{|1+\tan{\theta}|},\,\,\,\frac{k_{3}}{k_{\mathrm{o}}}=\sqrt{\left|\frac{1-\tan{\theta}}{1+\tan{\theta}}\right|}.
\end{equation}
This ellipse simplifies to the parabola in Eq.~\ref{Eq:BasebandParabola} in the vicinity of the point $(k_{x},\tfrac{\omega}{c})\!=\!(0,k_{\mathrm{o}})$ for small bandwidths. Of course, when $\theta\!\rightarrow\!0$ [Fig.~\ref{Fig:SpectralProjectionsOntoKzOmegaPlane}(a)], we reach the limit of a monochromatic beam [Fig.~\ref{Fig:MonochromaticBeam}].

We cannot take the ellipse in its entirety to contribute to the angular spectrum. Only the portion of the ellipse corresponding to $k_{z}\!>\!0$ is physically accessible. Furthermore, within the paraxial regime only the portion of the ellipse in the vicinity of the point $(k_{x},\tfrac{\omega}{c})\!=\!(0,k_{\mathrm{o}})$ is utilized. We plot in Fig.~\ref{Fig:SubluminalSTWavePacket}(b) the intensity distribution for a generic subluminal ST wave packet, which has a characteristic X-shaped profile.

It is sometimes thought that propagation-invariant subluminal ST wave packets are \textit{not} X-shaped, in contrast to their superluminal counterparts \cite{ZamboniRached2008PRA}. This is not true in general. We will elaborate on this misconception in Section~\ref{Section:AreXWavesXshaped}. In general, the synthesis of subluminal ST wave packets received scant attention and was not attempted experimentally until our recent work \cite{Kondakci2017NP,Kondakci2019NC} (see Section~\ref{Section:HistoricalSketch}). The main reason for this is that \textit{baseband} ST wave packets did not receive the same level of interest as X-waves and sideband ST wave packets.

Although the subluminal ST wave packet is itself a pulsed beam, the dimension of its spectral support domain is reduced with respect to that of a conventional pulsed beam as a consequence of the strict association between the spatial and temporal frequencies. This condition is depicted pictorially in Fig.~\ref{Fig:SubluminalSTWavePacket}(c) using the grid of spatial and temporal frequencies used above in Fig.~\ref{Fig:MonochromaticBeam}(c), Fig.~\ref{Fig:PulsedPlaneWave}(c), and Fig.~\ref{Fig:TraditionalPulsedBeam}(c). The spatio-temporal spectrum is a slice through this grid. Because $\omega_{\mathrm{o}}$ is the maximum allowed temporal frequency and is associated with $k_{x}\!=\!0$, $k_{x}$ increases with decreasing $\omega$ for subluminal ST wave packets. Here each $\omega$ is associated with a single $k_{x}$, so that the spatial and temporal DoFs are no longer separable.

\subsubsection{Superluminal ST wave packets}

\begin{figure}[t!]
  \begin{center}
  \includegraphics[width=12.0cm]{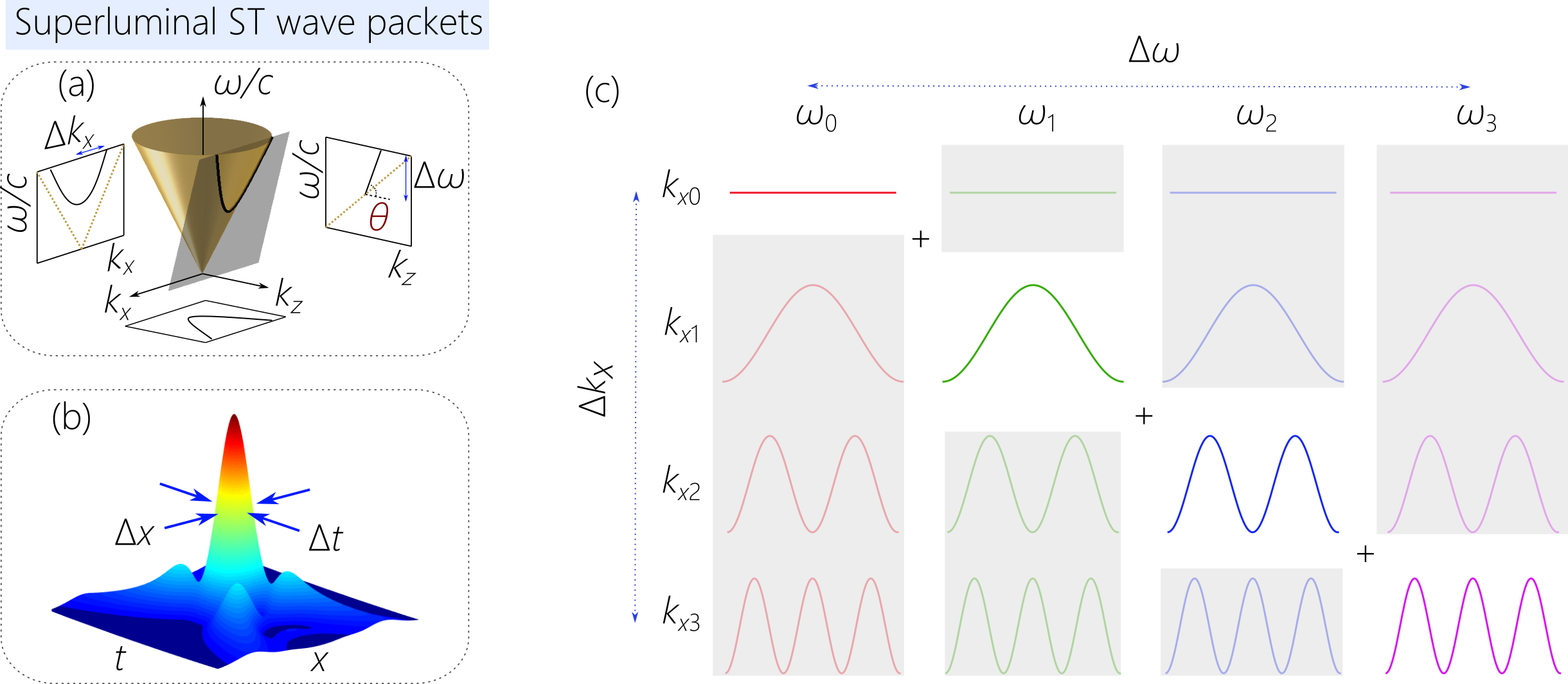}
  \end{center}
  \caption{Representation of the spatio-temporal spectrum of \textit{superluminal} ST wave packets. (a) The spectral support domain on the surface of the light-cone lies at the intersection of the light-cone with a spectral plane with a spectral tilt angle $\theta\!>\!45^{\circ}$. Spectral projections onto the $(k_{z},\tfrac{\omega}{c})$-plane is a straight line, and on the $(k_{x},\tfrac{\omega}{c})$ and $(k_{x},k_{z})$ planes are hyperbolas when $45^{\circ}\!<\!\theta\!<\!135^{\circ}$, parabola when $\theta\!=\!135^{\circ}$, and ellipses when $\theta\!>\!135^{\circ}$. (b) The spatio-temporal intensity profile $I(x,0;t)$. (c) The superluminal ST wave packet is a superposition of spatial frequencies, each associated with a single wavelength. The spatial frequency \textit{increases} with the temporal frequency. The monochromatic plane waves obscured behind a gray foreground are eliminated from the wave packet angular spectrum.}
  \label{Fig:SuperluminalSTWavePacket}
\end{figure}

The study of superluminal wave packets in general has been more active than that of their subluminal counterparts, but has primarily focused on X-waves and other wave packets \cite{Recami1998PA,Shaarawi2000JPA,Shaarawi2000JPA2,ZamboniRached2002EPJD,ZamboniRached2003OptCommun,Recami2003IEEEJSTQE} and not on superluminal baseband ST wave packets (with the exceptions in \cite{Valtna2007OC,Saari2007LP,ZamboniRached2009PRA}). For such wave packets, the spectral projection onto the $(k_{z},\tfrac{\omega}{c})$-plane is the straight line in Eq.~\ref{Eq:BasebandPlane} but with $\theta$ restricted to the range $45^{\circ}\!<\!\theta\!<\!90^{\circ}$, whereupon $c\!<\!\widetilde{v}\!<\!\infty$, and $\widetilde{n}\!<\!1$. The spectral support domain on the light-cone is its intersection with the plane $\mathcal{P}_{\mathrm{B}}(\theta)$; however, this intersection is now a hyperbola (rather than an ellipse) whose projection onto the $(k_{x},\tfrac{\omega}{c})$-plane is described by the equation:
\begin{equation}
\frac{1}{k_{1}^{2}}\left(\frac{\omega}{c}-k_{2}\right)^{2}-\frac{k_{x}^{2}}{k_{3}^{2}}=1,
\end{equation}
where $k_{1}$, $k_{2}$, and $k_{3}$ are given in Eq.~\ref{k1,k2,k3-def}. In contrast to the subluminal case, the temporal and spatial bandwidths can in principle extend indefinitely. Indeed, the plane $\mathcal{P}_{\mathrm{B}}(\theta)$ does not cross $k_{z}\!=\!0$, so that the wave packet is causal [Fig.~\ref{Fig:SpectralProjectionsOntoKzOmegaPlane}(d)] no matter how large the bandwidth is \cite{ZamboniRached2009PRA}. Nevertheless, this hyperbola can once again be approximated by the parabola in Eq.~\ref{Eq:BasebandParabola} in the vicinity of $k_{x}\!=\!0$. At the transition from subluminal to superluminal regimes at $\theta\!=\!45^{\circ}$, $\mathcal{P}_{\mathrm{B}}$ is tangential to the light-cone, and the wave packet degenerates into a plane-wave pulse [Fig.~\ref{Fig:PulsedPlaneWave} and Fig.~\ref{Fig:SpectralProjectionsOntoKzOmegaPlane}(c)].

We plot in Fig.~\ref{Fig:SuperluminalSTWavePacket}(a) the hyperbolic spectral support domain for a superluminal ST wave packet and its projections onto the three spectral planes. The projections onto the $(k_{x},\tfrac{\omega}{c})$ and $(k_{x},k_{z})$ planes are both hyperbolas. The spatio-temporal profile for a generic wave packet [Fig.~\ref{Fig:SuperluminalSTWavePacket}(b)] is X-shaped and resembles that of a subluminal ST wave packet [Fig.~\ref{Fig:SubluminalSTWavePacket}(b)], which emphasizes that there need be no major distinction between the profiles of these two classes of wave packets in the paraxial regime. Similarly to subluminal ST wave packets, the dimension of the spectral support domain for superluminal ST wave packets is reduced with respect to that for conventional wave packets. This condition is depicted pictorially in Fig.~\ref{Fig:SuperluminalSTWavePacket}(c), where the spatio-temporal spectrum is once again a slice through the $(k_{x},\omega)$ grid. However, because $\omega_{\mathrm{o}}$ is the minimum temporal frequency and is associated with $k_{x}\!=\!0$, $k_{x}$ \textit{increases} with increasing $\omega$. 
\subsubsection{Negative-$\widetilde{v}$ ST wave packets}

If we consider propagation-invariant ST wave packets whose spectral support domain is at the intersection of the light-cone surface with $\mathcal{P}_{\mathrm{B}}(\theta)$ when $90^{\circ}\!<\!\theta\!<\!180^{\circ}$ [Fig.~\ref{Fig:SpectralProjectionsOntoKzOmegaPlane}(f-h)], then the group velocity is negative $-\infty\!<\!\widetilde{v}\!<\!0$, $\widetilde{n}\!<\!0$. This configuration may raise concerns with respect to relativistic causality, and it has been consequently the least studied -- if not completely ignored -- class of ST wave packets; see \cite{Zapata2006OL} for an exception, although this example of negative-$\widetilde{v}$ is \textit{not} propagation invariant. However, there is no violation of relativistic causality here as explicated in \cite{Porras2005OL,Saari2018PRA,Saari2019PRAenergyflow}. At the transition from superluminal positive-$\widetilde{v}$ to negative-$\widetilde{v}$ at $\theta\!=\!90^{\circ}$, we formally have $\widetilde{v}\!=\!\infty$ [Fig.~\ref{Fig:SpectralProjectionsOntoKzOmegaPlane}(e)], corresponding to the example considered in Section~\ref{Section:PropagationInvarianceIn1D}.

\begin{figure}
\begin{center}
  \includegraphics[width=8cm]{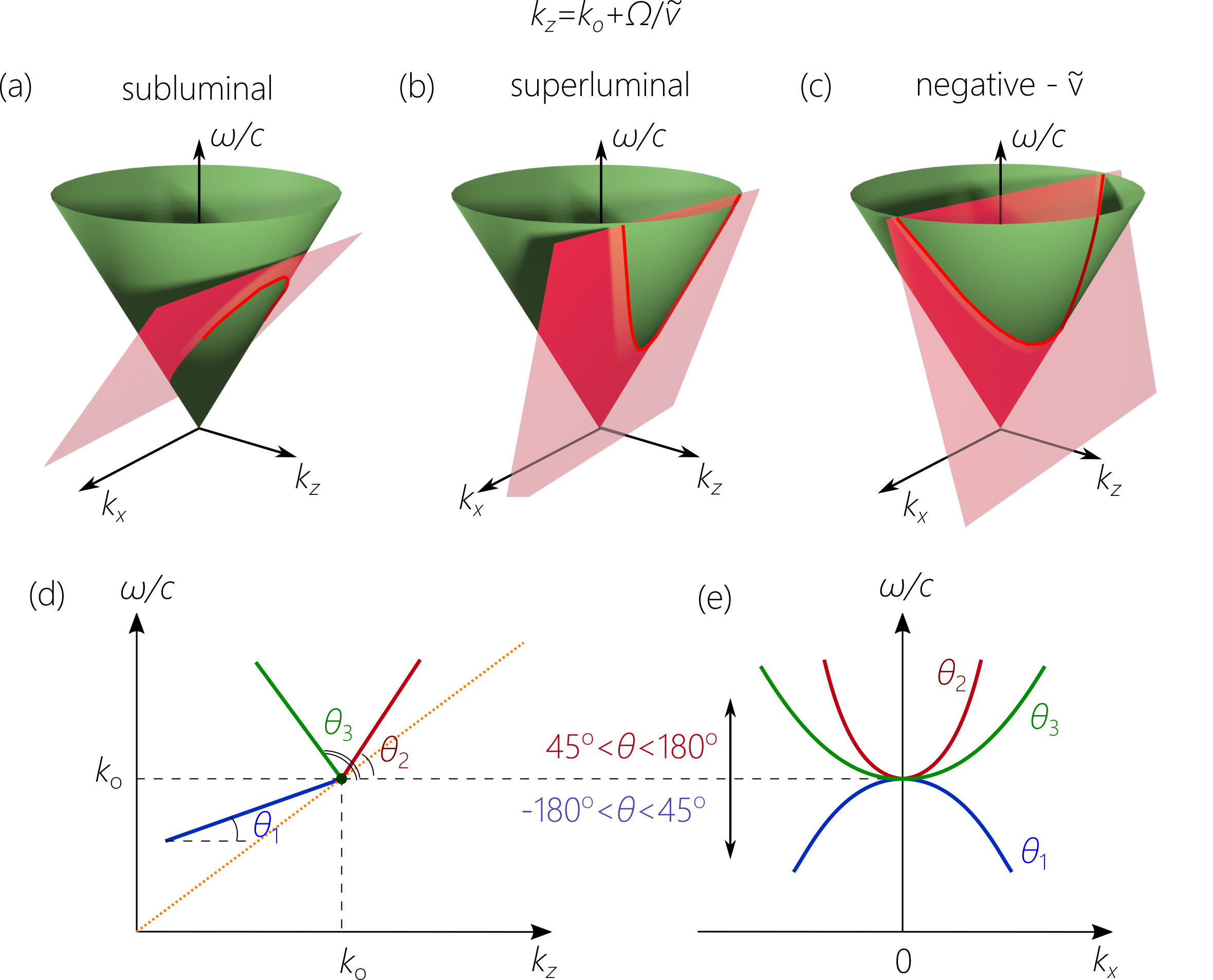}
    \caption{Summary of baseband ST wave packets. (a-c) Representation of the spectral support domain on the light-cone surface for baseband ST wave packets: (a) subluminal, (b) superluminal with positive-$\widetilde{v}$, and (c) negative-$\widetilde{v}$. (d) Spectral projections onto the $(k_{z},\tfrac{\omega}{c})$-plane for the baseband ST wave packets in (a-c), and (e) onto the $(k_{x},\tfrac{\omega}{c})$-plane. The spectral tilt angle $\theta_{1}$ is that for a subluminal ST wave packet, $\theta_{2}$ for superluminal, and $\theta_{3}$ for negative-$\widetilde{v}$.}
    \label{Fig:BasebandSummary}
\end{center}
\end{figure}

The spectral support domain can take on multiple forms. First, when $90^{\circ}\!<\!\theta\!<\!135^{\circ}$ and $|\widetilde{v}|\!>\!c$ [Fig.~\ref{Fig:SpectralProjectionsOntoKzOmegaPlane}(f)], the spectral support domain is a hyperbola that mirrors the positive-$\widetilde{v}$ superluminal regime $45^{\circ}\!<\!\theta\!<\!90^{\circ}$. Note, however, a critical distinction between the positive-$\widetilde{v}$ ($45^{\circ}\!<\!\theta\!<\!90^{\circ}$) and negative-$\widetilde{v}$ ($90^{\circ}\!<\!\theta\!<\!135^{\circ}$) superluminal regimes. Unlike the hyperbola for the superluminal baseband ST wave packet where there is no upper limit on bandwidth \cite{ZamboniRached2009PRA,Yessenov2019PRA}, there \textit{does} exist an upper limit on the usable bandwidth with negative-$\widetilde{v}$ wave packets when $\mathcal{P}_{\mathrm{B}}(\theta)$ reaches $k_{z}\!=\!0$. Consequently, the spectrum must lie within the range $\omega_{\mathrm{o}}\!<\!\omega\!<\!\omega_{\mathrm{o}}(1+|\tan{\theta}|)$. The second form of the spectral support domain occurs when $\theta\!=\!135^{\circ}$ and $\widetilde{v}\!=\!-c$ [Fig.~\ref{Fig:SpectralProjectionsOntoKzOmegaPlane}(g)], whereupon the spectral support domain is a parabola. This case does \textit{not} mirror that of the plane-wave pulse at $\theta\!=\!45^{\circ}$. Moreover, there is here also a maximum exploitable bandwidth that is reached when $k_{z}\!=\!0$. Third, when $135^{\circ}\!<\!\theta\!<\!180^{\circ}$ and $|\widetilde{v}|\!<\!c$ [Fig.~\ref{Fig:SpectralProjectionsOntoKzOmegaPlane}(h)], the spectral support domain is an ellipse that mirrors the positive-$\widetilde{v}$ subluminal regime $0^{\circ}\!<\!\theta\!<\!45^{\circ}$. 

The three general families of baseband ST wave packets are summarized in Fig.~\ref{Fig:BasebandSummary}: subluminal baseband ST wave packets when $0^{\circ}\!<\!\theta\!<\!45^{\circ}$ whose spectral support domain is an ellipse [Fig.~\ref{Fig:BasebandSummary}(a)]; superluminal baseband ST wave packets when $45^{\circ}\!<\!\theta\!<\!90^{\circ}$ whose spectral support domain is a hyperbola [Fig.~\ref{Fig:BasebandSummary}(b)]; and negative-$\widetilde{v}$ baseband ST wave packets when $90^{\circ}\!<\!\theta\!<\!180^{\circ}$ whose spectral support domain is a hyperbola ($90^{\circ}\!<\!\theta\!<\!135^{\circ}$), a parabola ($\theta\!=\!135^{\circ}$), or an ellipse ($135^{\circ}\!<\!\theta\!<\!180^{\circ}$) [Fig.~\ref{Fig:BasebandSummary}(c)]. In all cases, the projection onto the $(k_{z},\tfrac{\omega}{c})$-plane is a straight line making an angle $\theta$ with the $k_{z}$-axis, and that onto the $(k_{x},\tfrac{\omega}{c})$-plane is approximately a parabola for small bandwidths in the vicinity of $k_{x}\!=\!0$, whose sign of curvature switches at $\theta\!=\!45^{\circ}$.

\subsection{X-waves: $k_{\mathrm{a}}\!=\!0$}

\begin{figure}\begin{center}
  \includegraphics[width=10cm]{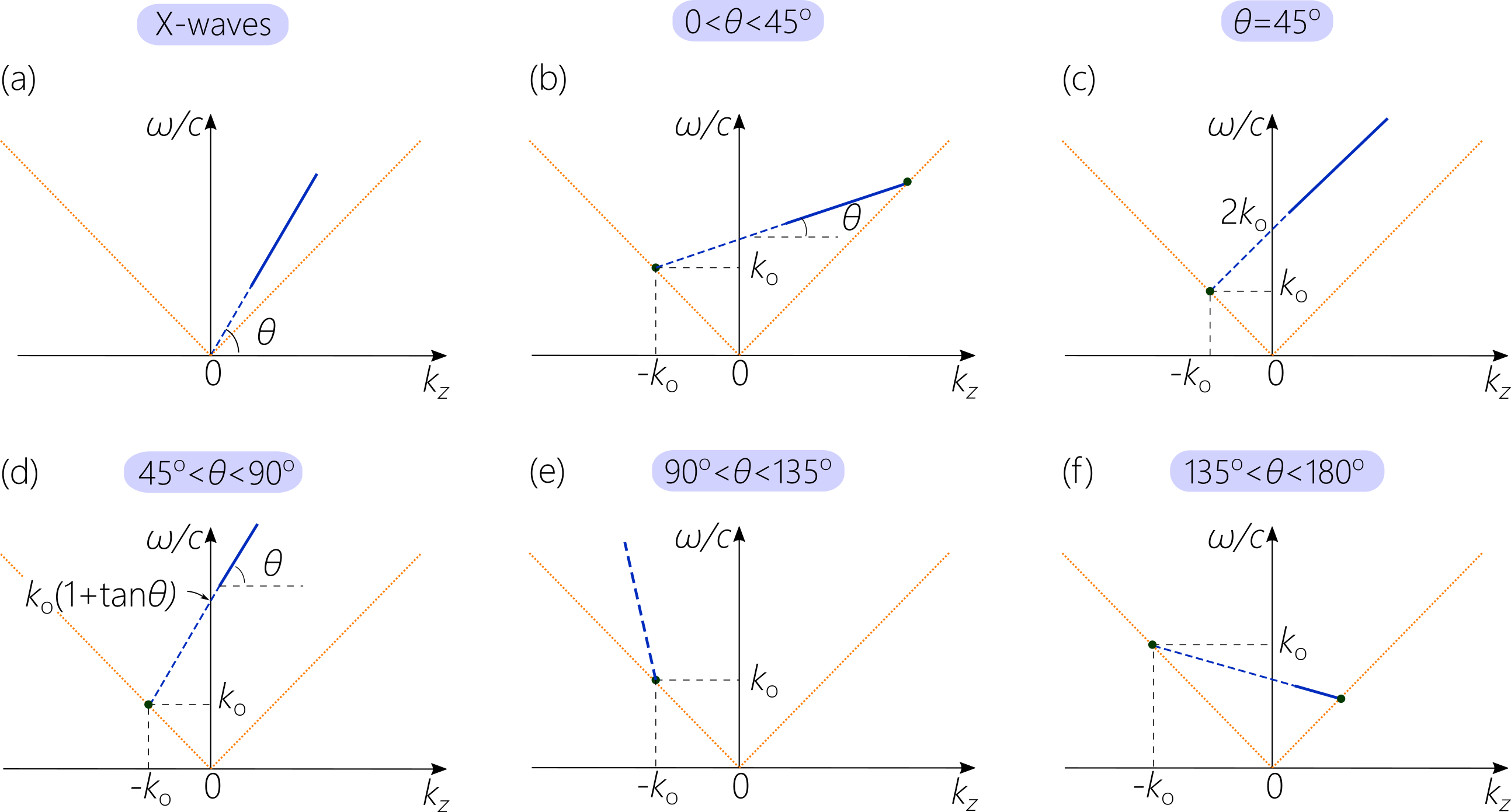}
    \caption{Spectral projections of X-waves and sideband ST wave packets onto the $(k_{z},\tfrac{\omega}{c})$-plane as $\theta$ increases from $0^{\circ}$ towards $180^{\circ}$. Not all of these projections are unique with respect to baseband ST wave packets; some are redundant, as enumerated in Table~\ref{Table:SidebandClasses}.}
    \label{Fig:SpectralProjectionsXSideband}
\end{center}
\end{figure}

By setting $k_{\mathrm{a}}\!=\!0$ in Eq.~\ref{Eq:GeneralKz}, we obtain the linear relationship:
\begin{equation}\label{Eq:XWavePlane}
k_{z}=\frac{\omega}{\widetilde{v}}.
\end{equation}
Such a projection corresponds to a spectral support domain at the intersection of the light-cone with a plane passing through the origin $(k_{x},k_{z},\tfrac{\omega}{c})\!=\!(0,0,0)$; see Fig.~\ref{Fig:SpectralProjectionsXSideband}(a). Because the spectral projection must be above the light-line (i.e., $k_{z}\!\leq\!\tfrac{\omega}{c}$) and $k_{z}\!>\!0$, the spectral tilt angle is limited here to the range $45^{\circ}\!<\!\theta\!<\!90^{\circ}$. This restriction implies that $c\!<\!\widetilde{v}\!<\!\infty$ and $\widetilde{n}\!<\!1$. In other words, X-waves are only \textit{superluminal}, with only positive-valued $\widetilde{v}$ allowed. Wave packets satisfying these criteria are known as X-waves. The name arises from the characteristics X-shape of the wave packet in a plane containing the $z$-axis. However, we now know that this characteristic X-shaped profile is \textit{not} unique to X-waves, and is in fact shared by \textit{all} propagation-invariant wave packets in the paraxial regime (see Section~\ref{Section:AreXWavesXshaped} for further details). It is therefore crucial to retain the various names of ST wave packets as indicators of their spectral support domain rather than their profile shapes.

The spectral support domain is the intersection of the light-cone with a plane $\mathcal{P}_{\mathrm{X}}(\theta)$ given by Eq.~\ref{Eq:XWavePlane}, which is parallel to the $k_{x}$-axis and makes an angle $\theta$ with the $k_{z}$-axis, so that the group velocity is $\widetilde{v}\!=\!c\tan{\theta}$. Because $\mathcal{P}_{\mathrm{X}}(\theta)$ passes through the origin $(k_{x},k_{z},\tfrac{\omega}{c})\!=\!(0,0,0)$, it intersects with the light-cone in a pair of straight lines rather than a conic section [Fig.~\ref{Fig:SpectralProjectionsXSideband}(a) and Fig.~\ref{Fig:SummarySideband}(a)]. Consequently, $k_{x}$ and $k_{z}$ are both linear in $\omega$: $k_{x}\!=\!\tfrac{\omega}{c}\sin{\varphi_{\mathrm{o}}}$ and $k_{z}\!=\!\tfrac{\omega}{c}\cos{\varphi_{\mathrm{o}}}$, where $\cos{\varphi_{\mathrm{o}}}\!=\!\cot{\theta}$. The ensuing mathematical simplification makes analytical studies of X-waves particularly straightforward in comparison to baseband or sideband ST wave packets. The particular spectral support domain of X-waves leads to unique features. For example, the phase velocity is the same as the group velocity: $v_{\mathrm{ph}}\!=\!\widetilde{v}\!=\!c/\cos{\varphi_{\mathrm{o}}}\!=\!c\tan{\theta}$. Consequently, both the envelope and the carrier move together rigidly; i.e., no phase dynamics ensue from free propagation $E(x,z;t)\!=\!E(x,0;t-z/\widetilde{v})$, and the entire field (and not only the envelope) is propagation-invariant.

Since their introduction into ultrasonics in 1992 \cite{Lu1992IEEETUFFC-Xwaves,Lu1992IEEETUFFCexperimentalX}, and into optics in 1996 \cite{Saari1996Proc}, X-waves have likely been the most widely studied class of propagation-invariant wave packets. The well-known analytical expression for X-waves assumes a spectrum extending down to $\omega\!=\!0$, and so is of little practical utility in the optical regime, unlike its ultrasonics counterpart. Instead, a finite temporal bandwidth centered at a carrier frequency $\omega_{\mathrm{o}}$ should be considered. This type of X-wave was the first propagation-invariant wave packet to be realized in optics in the pioneering work of Peeter Saari and dubbed a Bessel-X pulse \cite{Saari1997PRL,Saari1997LP,Reivelt2003arxiv}. The analytical simplicity of the X-wave notwithstanding, such a wave packet is considerably less useful than baseband ST wave packets, as we discuss in more detail in Section~\ref{Section:Deadend}.

\subsection{Sideband ST wave packets: Negative $k_{\mathrm{a}}$}

\begin{table}[t!]
\caption{Classes of sideband ST wave packets (in addition to X-waves) following the enumeration in Fig.~\ref{Fig:SpectralProjectionsXSideband}.}
\label{Table:SidebandClasses}
\begin{center}
\begin{tabular}{|p{0.5cm}||p{2.2cm}|p{2.0cm}|p{2.4cm}|p{1.5cm}|}
 \hline
  & $\theta$ & $\widetilde{v}$ & conic section & status\\
 \hline\hline
 (a) & $45^{\circ}\!<\!\theta\!<\!90^{\circ}$ & $c\!<\!\widetilde{v}\!<\!\infty$ & two straight lines & unique\\ \hline
 (b) & $0^{\circ}\!<\!\theta\!<\!45^{\circ}$ & $0\!<\!\widetilde{v}\!<\!c$ & ellipse & redundant\\ \hline
 (c) & $45^{\circ}$ & $c$ & parabola & unique\\ \hline
 (d) & $45^{\circ}\!<\!\theta\!<\!90^{\circ}$ & $c\!<\!\widetilde{v}\!<\!\infty$ & hyperbola & unique\\ \hline
 (e) & $90^{\circ}\!\leq\!\theta\!\leq\!135^{\circ}$ & --- & --- & forbidden\\ \hline
 (f) & $135^{\circ}\!<\!\theta\!<\!180^{\circ}$ & $-c\!<\!\widetilde{v}\!<\!0$ & ellipse & redundant\\ \hline
  \end{tabular}
 \end{center}
\end{table}

A third family of ST wave packets arises when $k_{\mathrm{a}}$ is negative. We rewrite Eq.~\ref{Eq:GeneralKz} as $k_{z}\!=\!-k_{\mathrm{a}}+\tfrac{\omega}{\widetilde{v}}$, with $k_{\mathrm{a}}$ positive-valued. Such a line will intersect with the light-line $k_{z}\!=\!-\tfrac{\omega}{c}$ at a point that we identify with $\omega\!=\!\omega_{\mathrm{o}}$ and $k_{z}\!=\!-k_{\mathrm{o}}$, whereupon $k_{\mathrm{a}}\!=\!k_{\mathrm{o}}+\tfrac{\omega_{\mathrm{o}}}{\widetilde{v}}$ and
\begin{equation}\label{Eq:PlaneForSideband}
k_{z}\!=\!-k_{\mathrm{o}}+\frac{\Omega}{\widetilde{v}}.
\end{equation}
The intersection point with the light-line $k_{z}\!=\!-\tfrac{\omega}{c}$, which is associated with the spatial frequency $k_{x}\!=\!0$ and temporal frequency $\omega_{\mathrm{o}}$, lies in the forbidden region $k_{z}\!<\!0$, and is therefore excluded from the spectral support domain along with the whole portion of the spatial spectrum associated with $k_{z}\!\leq\!0$. Moreover, the low-frequency domain in the vicinity of $k_{x}\!=\!0$ is excluded from such ST wave packets, which instead include only high spatial frequencies. In analogy with the nomenclature of radio engineering, we call these pulsed beams `sideband' ST wave packets. The synthesis of these wave packets is thus considerably more challenging than baseband ST wave packets or X-waves.

The spectral support domain lies at the intersection of the light-cone with a plane $\mathcal{P}_{\mathrm{S}}(\theta)$ given by Eq.~\ref{Eq:PlaneForSideband}, which -- similarly to its counterparts $\mathcal{P}_{\mathrm{B}}(\theta)$ for baseband ST wave packets and $\mathcal{P}_{\mathrm{X}}(\theta)$ for X-waves -- is parallel to the $k_{x}$-axis and makes and angle $\theta$ with the $k_{z}$-axis, such that the resulting wave packet has a group velocity $\widetilde{v}\!=\!c\tan{\theta}$. Although $\theta$ can formally span the range from $0^{\circ}$ to $180^{\circ}$, only certain angular ranges are unique with respect to the previously considered cases; see Fig.~\ref{Fig:SpectralProjectionsXSideband}(b-f). The spans $0^{\circ}\!<\!\theta\!<\!45^{\circ}$ and $135^{\circ}\!<\!\theta\!<\!180^{\circ}$ both coincide with the baseband ST wave packets over the same ranges, where the spectral support domain is an ellipse [Fig.~\ref{Fig:SpectralProjectionsXSideband}(b,f)]. Furthermore, the span $90^{\circ}\!<\!\theta\!<\!135^{\circ}$ is forbidden in its entirety because $k_{z}\!<\!0$ for all $\omega$ in this range [Fig.~\ref{Fig:SpectralProjectionsXSideband}(e)]. This leaves us with only the span $45^{\circ}\!\leq\!\theta\!<\!90^{\circ}$ [Fig.~\ref{Fig:SpectralProjectionsXSideband}(c,d)] that yields unique classes of sideband ST wave packets; see Table~\ref{Table:SidebandClasses} for an enumeration of these different cases. 

First, when $\theta\!\rightarrow\!45^{\circ}$ and $\widetilde{v}\!\rightarrow\!c$ [Fig.~\ref{Fig:SpectralProjectionsXSideband}(d)], the spectral projection onto the $(k_{x},\tfrac{\omega}{c})$-plane is given by the parabola:
\begin{equation}\label{Eq:FWMParabola}
\frac{\omega}{c}=k_{\mathrm{o}}+\frac{k_{x}^{2}}{4k_{\mathrm{o}}},
\end{equation}
which corresponds to Brittingham's FWM \cite{Brittingham1983JAP}. Second, for sideband ST wave packets with $45^{\circ}\!<\!\theta\!<\!90^{\circ}$ and $c\!<\!\widetilde{v}\!<\!\infty$ (i.e., superluminal), the spectral projection onto the $(k_{x},\tfrac{\omega}{c})$-plane is a hyperbola:
\begin{equation}
\frac{1}{k_{1}'^{2}}\left(\frac{\omega}{c}+k_{2}'\right)^{2}-\frac{k_{x}^{2}}{k_{3}'^{2}}=1,
\end{equation}
with the parameters $k_{1}'$, $k_{2}'$, and $k_{3}'$ given by:
\begin{equation}
\frac{k_{1}'}{k_{\mathrm{o}}}=\left|\frac{\tan{\theta}}{1-\tan{\theta}}\right|,\frac{k_{2}'}{k_{\mathrm{o}}}\frac{1}{|1-\tan{\theta}|},\frac{k_{3}'}{k_{\mathrm{o}}}=\sqrt{\left|\frac{1+\tan{\theta}}{1-\tan{\theta}}\right|}.
\end{equation}
These superluminal sideband ST wave packets are sometimes known as focused-X-waves \cite{Shaarawi2003JOSAA,ZamboniRached2004JOSAA}.

Because $k_{x}\!=\!0$ and the low-frequency portion of the spatial spectrum is excluded on physical grounds, we expect these wave packets to be difficult to synthesize. Indeed, as we show below, to achieve even a small deviation in $\widetilde{v}$ from $c$, we must operate at large numerical apertures far from the paraxial regime. It is in this sense an unfortunate historical accident -- from the experimental perspective -- that these difficult-to-synthesize propagation-invariant wave packets were the first to be discovered theoretically.

\begin{figure}\begin{center}
  \includegraphics[width=8cm]{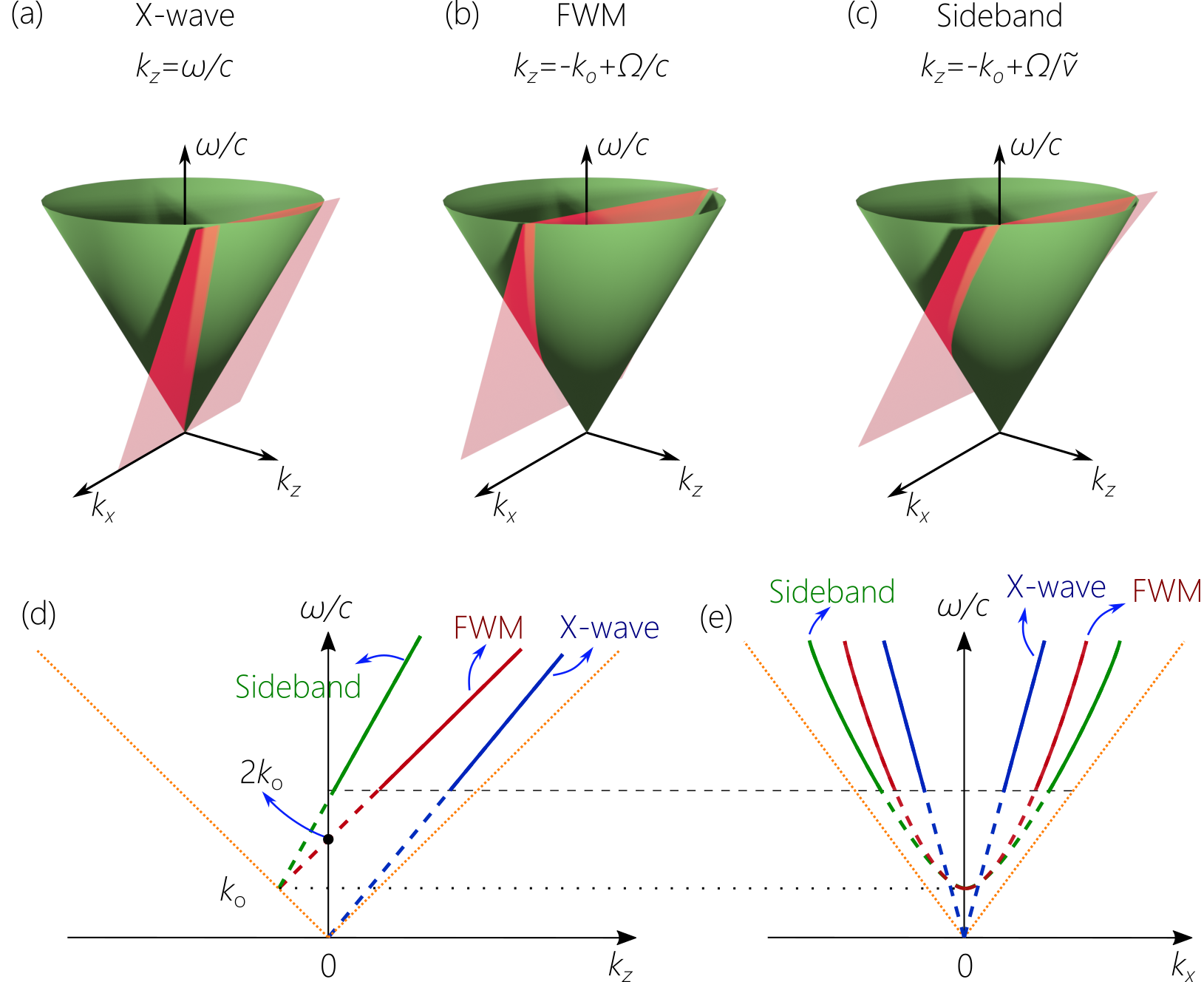}
    \caption{Summary of sideband ST wave packets. (a-c) Spectral support domains on the surface of the light-cone for (a) X-waves, (b) FWMs, and (c) sideband ST wave packets. (d) Spectral projections onto the $(k_{z},\tfrac{\omega}{c})$-plane and (e) the $(k_{x},\tfrac{\omega}{c})$-plane for the wave packets in (a-c).}
    \label{Fig:SummarySideband}
\end{center}
\end{figure}

We summarize in Fig.~\ref{Fig:SummarySideband} the various scenarios of X-waves [Fig.~\ref{Fig:SummarySideband}(a)], FWMs [Fig.~\ref{Fig:SummarySideband}(b)], and sideband ST wave packets [Fig.~\ref{Fig:SummarySideband}(c)]. Although FWMs are a special limit of sideband ST wave packets corresponding to $\theta\!=\!45^{\circ}$, it is nonetheless useful to consider them separately: first, for their historical significance and consequently the literature that has accumulated regarding them; and, second, they warrant separate treatment because their projection onto the $(k_{z},\tfrac{\omega}{c})$-plane is parallel to the light-line. In all the cases of FWMs and sideband ST wave packets, the spatial frequencies in the vicinity of $k_{x}\!=\!0$ are physically excluded. In the case of X-waves, although spatial frequencies in the vicinity of $k_{x}\!=\!0$ are not excluded strictly speaking, they are nevertheless inaccessible in optics because they are associated with temporal frequencies in the vicinity of $\omega\!=\!0$ [Fig.~\ref{Fig:SummarySideband}(d,e)].

\section{Historical development of ST wave packets}\label{Section:HistoricalSketch}

A large (predominantly theoretical) literature has accumulated over the past 4 decades in the field of `localized waves'. The shear size of this literature can represent an obstacle to the newcomer without proper critical evaluation. Indeed, much of this literature is now of historical interest as the field embarks on a new era made possible by recent experimental and conceptual developments. We provide here an overview of the literature from this standpoint.

\begin{figure}[t!]
\begin{center}
\includegraphics[width=11.0cm]{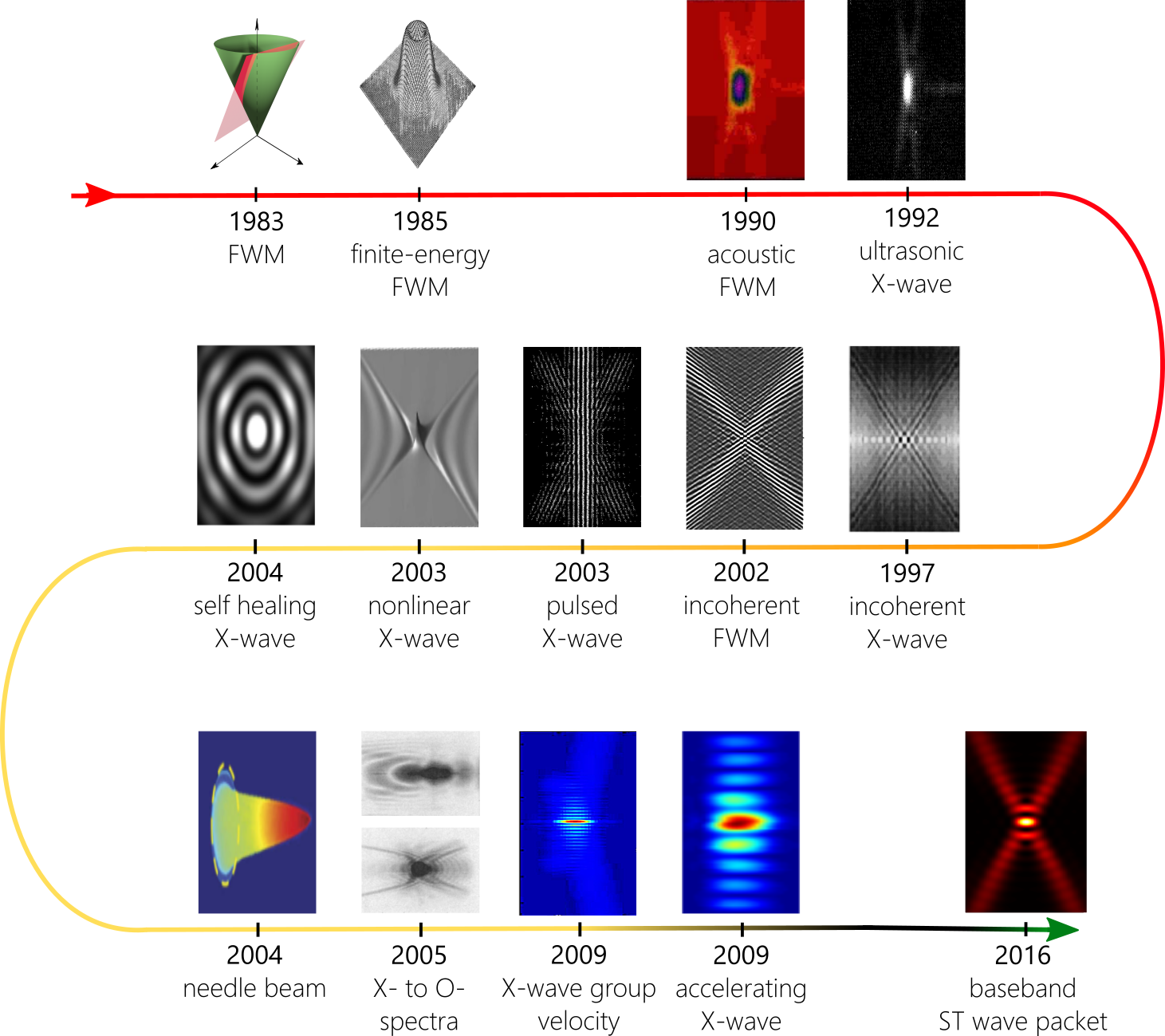}
\end{center}
\caption{A timeline of significant events in the development of propagation-invariant wave packets. In chronological order: Brittingham's FWM (1983) \cite{Brittingham1983JAP}; finite-energy FWM (1985) \cite{Sezginer1985JAP}; synthesis of acoustic FWMs (1990) \cite{Ziolkowski1990} and ultrasonic X-waves (1992) \cite{Lu1992IEEETUFFCexperimentalX}; producing incoherent optical X-waves (1997) \cite{Saari1997PRL} and incoherent optical FWMs (2002) \cite{Saari2004PRE}; nonlinear X-waves (2003) \cite{Conti2003PRL} and coherent optical X-waves (2003) \cite{Grunwald2003PRA}; observing X-wave self-healing (2004) \cite{Grunwald2004CLEO}; producing a needle beam (2004) \cite{Grunwald2004SPIE}; observing a nonlinear transition from X- to O-shaped spectra in water (2005) \cite{Porras2005OL}; measuring the group velocity of X-waves \cite{Bowlan2009OL}; and accelerating X-waves (2009) \cite{ValtnaLukner2009OE}. Baseband ST wave packets have been pursued since 2016 \cite{Kondakci2016OE}.}\label{Fig:history}
\end{figure}

It is useful to adopt a chronological periodization to establish some signposts in the landscape [Fig.~\ref{Fig:history}]; the endpoints of these periods are of course porous. A pre-history prior to 1983 features early premonitions of propagation-invariant wave packets, preceding the first period 1983--1996 that witnessed concerted theoretical research on luminal FWMs. Although the X-wave was introduced into ultrasonics during this period, no \textit{optical} experiments were conducted. The second period 1996--2003 witnessed the pioneering efforts of Peeter Saari and co-workers on synthesizing X-waves at optical frequencies \cite{Saari1996Proc,Saari1997PRL,Saari1997LP}, and subsequently FWMs \cite{Reivelt2000JOSAA,Reivelt2002PRE,Reivelt2002PRE2,Reivelt2003arxiv} -- both of which made use of incoherent light rather than laser pulses. The third period 2003--2010 was in many regards the heyday of localized waves in which progress developed along several lines of inquiry: pulsed X-waves were produced and characterized, nonlinear optical effects yielded X-shaped spectra, and the propagation invariance in linear dispersive media was studied. Research on localized waves waned after $\sim\!2010$. However, interest in ST wave packets has witnessed a rebirth since 2016, with developments now emerging at a rapid pace.

\subsection{Pre-history: --1983}

It is rare for the origin of a major scientific discovery to be clearly tied to a precisely defined point in time. Case in point, although the Bessel beam in its current form was introduced into optics in 1987 \cite{Durnin1987JOSAA,Durnin1987PRL}, there nevertheless exists a prehistory of early intimations that extend as far back as Airy in 1841 and Lord Rayleigh in 1872 (see \cite{Figueroa2014Book}, Ch.~17, for historical details). Similarly for ST wave packets; although the first widely known example is Brittingham's FWM\cite{Brittingham1983JAP}, there indeed exist earlier precursors that are worth mentioning. For example, it was recognized in the well-known textbook on mathematical physics by Courant and Hilbert \cite{Courant1966Book} (Vol.~2, page~760) that Maxwell's equations admit propagation-invariant wave-packet solutions (see also the even earlier description by Bateman \cite{Bateman1915Book}), which are termed `undistorted progressive waves'. Research on propagation-invariant wave packets continued with interest in their implications for special relativity and quantum electrodynamics, mostly in the context of exploring hypothetical superluminal particles known as `tachyons' \cite{Barut1982NC,Barut1990PLA,Barut1990FPL,Barut1993PLA}.

However, the most significant result in this pre-history is Mackinnon's proposal in 1978 of a subluminal non-dispersive ST wave packet \cite{Mackinnon1978FP} based on relativistic de Broglie waves (rather than non-relativistic Schr{\"o}dinger waves) for a massive particle. This `Mackinnon wave packet' went initially unrecognized by the optics community until after the development of FWMs. Uniquely, the spatio-temporal profile of the Mackinnon wave packet is circularly symmetric in space and time, in contrast to all other ST wave packets that are X-shaped \cite{Recami1998PA} in free space in the paraxial regime (see Section~\ref{Section:dispelling_misconceptions} for further details). No direct observation of a circularly symmetric spatio-temporal profile has been reported for propagation-invariant wave packets in free space to date.    

\subsection{First Period: 1983--1996}

In 1983, James Brittingham took the optics community by surprise by identifying a mathematical function that satisfies Maxwell's equations and represents a pulsed beam that travels rigidly in free space at a group velocity $\widetilde{v}\!=\!c$ while overcoming diffraction and dispersion. This so-called focus-wave mode (FWM) is a sideband ST wave packet with $\theta\!=\!45^{\circ}$. Brittingham's result appears to contradict our expectations regarding the inevitability of diffractive spreading with free propagation. Immediately following this breakthrough, a meeting was organized by the U.S. Department of Defense and held at MIT to verify the published result. It was perhaps thought that FWMs could be a platform for directed energy applications in the era of the Strategic Defense Initiative (SDI). One outcome of the meeting was the realization that the FWM, although a valid solution of Maxwell's equations, nevertheless requires infinite energy for its realization\cite{Sezginer1985JAP}. However, finite-energy realizations with significantly enhanced propagation distances can be produced by introducing finite spectral uncertainty [Fig.~\ref{Fig:DiffractionFreeAndPropagationInvariant}(b)]. Furthermore, the vectorial solution found by Brittingham gave way to simpler scalar formulations \cite{Belanger1984JOSAA,Sezginer1985JAP}. The reader is referred to Chapter~2 of \cite{Figueroa2014Book} for further interesting historical details of the early development of FWMs.

Brittingham discovered the FWM by tinkering with solutions to Maxwell's equations and did not provide intuition with regard to the underlying physics, or a methodology for constructing wave packets with similar characteristics. Subsequently, a variety of approaches were introduced to physically ground FWMs: the FWM envelope was obtained from a modified paraxial equation \cite{Belanger1984JOSAA}; the FWM was found to result from implementing a Lorentz transformation on a monochromatic beam\cite{Belanger1986JOSAA}; and the FWM was identified as the wave packet emitted by a charged particle source at imaginary coordinates \cite{Ziolkowski1985JMP}. 

Ziolkowski, Besieris, Shaarawi, and co-workers produced a large body of theoretical work during this period on various aspects of FWMs \cite{Ziolkowski1985JMP,Ziolkowski1988SPIE,Ziolkowski1989PRA,Ziolkowski91PRA,Ziolkowski1991ProcIEEE}, including proposals for their preparation \cite{Shaarawi89JAP,Hernandez1992JASA,Ziolkowski1992IEEETAP,Ziolkowski1993JOSAA,Shaarawi1995JOSAA,Shaarawi1996JOSAA} (see Section~\ref{Synthesis}), propagation in optical fibers \cite{Vengsarkar1992JOSAA}, and a theoretical tool called the `bidirectional traveling-wave representation' for representing the spatio-temporal spectrum of FWMs \cite{Besieris1989JMP}. In lieu of the spectral variables $(k_{x},k_{z},\tfrac{\omega}{c})$ and the constraint $k_{x}^{2}+k_{z}^{2}\!=\!(\tfrac{\omega}{c})^{2}$, the bidirectional wave representation makes use of transformed positive-valued spectral variables $\alpha\!=\!\tfrac{1}{2}(\tfrac{\omega}{c}+k_{z})$ and $\beta\!=\!\tfrac{1}{2}(\tfrac{\omega}{c}-k_{z})$ along with the constraint $\alpha\beta\!=\!\tfrac{1}{4}k_{x}^{2}$. In physical space, the corresponding variables are $(x,z+ct,z-ct)$, the latter two of which correspond to propagation along the negative and positive $z$-axis, respectively. This representation offers advantages in deriving closed-form expressions for FWMs having particular spectral weights, and facilitates introducing a form of `spectral uncertainty' in the parameter $\beta$ that renders the energy of the wave packet finite. Nevertheless, the abstract nature of this approach can mask the intuitive geometric picture we introduced  in Section~\ref{Section:Preliminaries}. Indeed, these early studies provided unlikely estimates of the propagation distance (e.g., 940 km in free space for a FWM of spatial width 10~$\mu$m \cite{Ziolkowski2020EuCAP-FWM}, and 40,000~km in an optical fiber\cite{Vengsarkar1992JOSAA}).

The burgeoning early effort on FWMs elicited a critical response from Heyman, Felson, and co-workers who noted that typical FWMs included contributions from negative-$k_{z}$ components that are not compatible with causal excitation, and which should therefore be excluded on physical grounds \cite{Heyman1987JOSAA}. This `causality debate' led to multiple papers arguing for \cite{Heyman1987JOSAA,Heyman1989IEEE,Heyman1993Chapter} and against \cite{Hillion1992IEEETAP,Hillion1994IEEETAP} this premise (see Section~\ref{Section:CausalityDebates} for further details). 

The entire body of work accumulated in this period was theoretical. With the exception of the acoustic realization in \cite{Ziolkowski1989PRL,Hernandez1992JASA}, no optical experiments were carried out. Despite the tremendous effort previously dedicated to FWMs and related sideband ST wave packets, it is unlikely that this class of propagation-invariant wave packets will be useful in practical applications because of the large numerical aperture required to produce a wave packet whose characteristics (except for propagation invariance) deviate in a meaningful way from a conventional wave packet (Section~\ref{Section:Deadend}). 

\subsection{Second Period: 1996--2003}

Whereas research in the first period 1983--1996 focused on \textit{luminal} wave packets, the second period witnessed interest shifting to \textit{superluminal} wave packets. Remarkably, there were no attempts to produce propagation-invariant wave packets optically in the intervening period from 1983 when Brittingham first proposed the FWM to 1997 when Saari carried out his pioneering experiments on X-waves. The X-wave (so-called because of its characteristic X-shaped profile in a meridional plane) was introduced theoretically and demonstrated experimentally in ultrasonics by Lu and Greenleaf in 1992 \cite{Lu1992IEEETUFFC-Xwaves,Lu1992IEEETUFFCexperimentalX}. The original formulation of the X-wave assumed spectral amplitudes extending down to the DC component $\omega\!=\!0$. Whereas such a field configuration may be approximated in acoustics and ultrasonics, it is not realistic in optics. Instead, a modification of the X-wave having a finite spectral width centered at an optical frequency (usually called a Bessel-X pulse) was developed theoretically \cite{Saari1996Proc} and then demonstrated experimentally by Saari in 1997 \cite{Saari1997PRL}, which is the first propagation-invariant optical wave packet. From a theoretical perspective, it was shown that the bidirectional traveling-wave representation can be used to produce closed-form expressions for X-waves after implementing appropriate Lorentz boosts to the spatio-temporal variables \cite{RECAMI2009bookchapter}.

Fortunately, the synthesis of optical X-waves is relatively straightforward. Writing $k_{x}(\omega)\!=\!k\sin{\varphi(\omega)}$, we see from Eq.~\ref{Eq:XWavePlane} that X-waves have a constant propagation angle $\varphi(\omega)\!=\!\varphi_{\mathrm{o}}$ (the so-called cone angle or axicon angle) across the entire spectrum. The X-wave is thus free of angular dispersion, which suggests that such a wave packet can be produced by directing a pulse through a Bessel-beam generator. However, Saari recognized early on \cite{Saari1996Proc} an awkward aspect of optical X-waves: observing the characteristic X-shaped profile requires ultrashort pulses ($<\!20$~fs pulse widths), whereas longer pulses only produce a Bessel beam modulated longitudinally with the pulse profile (i.e., the spatio-temporal profile is approximately separable in space and time), with only faint X-shaped tails far away from the beam center. Consequently, the first experimental demonstration made use of broadband incoherent light (rather than a pulsed laser) from a high-pressure Xe-arc lamp with coherence time $\sim\!3$~fs at $\lambda_{\mathrm{c}}\!=\!600$~nm. The spectral tilt angle corresponded to $\widetilde{v}\!=\!1.00018c$ ($\Delta\widetilde{v}\!\!=\!1.8\times10^{-4}c$). Only later were optical X-waves produced using coherent pulsed sources \cite{Grunwald2003PRA}. Although the measurements in \cite{Saari1997PRL} suggested that $\widetilde{v}\!>\!c$, precise measurements of $\widetilde{v}$ were only carried out more than a decade later \cite{Bowlan2009OL}.

These considerations throw doubt on the first results reported on Bessel-X waves in a dispersive medium where the pulse width was $\sim\!200$~fs \cite{Sonajalg1996OL,Sonajalg1997OL}, and which preceded the report in \cite{Saari1997PRL}. The spatio-temporal profile was not reported in \cite{Sonajalg1997OL} and only on-axis ($x\!=\!0$) data is provided. It is likely that the wave packet produced was a pulsed Bessel beam (Saari states in \cite{Figueroa2014Book}, Ch.~4, that a version of the focused-X-wave was produced). In any event, this first attempt \cite{Sonajalg1997OL} -- which remains the only attempt at synthesizing a wave packet in free space to achieve propagation invariance in a dispersive medium -- was quickly superseded by the more conclusive results in \cite{Saari1997PRL}. 

Despite the rapid developments in superluminal X-waves, interest in FWMs did not die out. Indeed, Saari and Reivelt conducted a study of the feasibility of synthesizing optical FWMs \cite{Reivelt2000JOSAA,Reivelt2002PRE2}, and an initial experimental realization was reported using incoherent light along one transverse dimension \cite{Reivelt2002PRE}. No subsequent experimental efforts following up on this pioneering attempt have been reported. The review by Reivelt and Saari \cite{Reivelt2003arxiv} describing their results concerning X-waves and FWMs can be viewed as the culmination of this second period.

\subsection{Third Period: 2003--2010}

\subsubsection{Pulsed X-waves and needle beams}

Following the assessment by Saari that the characteristic spatio-temporal profile of X-waves can be observed with only ultrashort pulse widths ($<\!20$~fs) \cite{Saari1996Proc}, Grunwald, Pich{\'e}, and co-workers produced the first pulsed optical X-waves. Initial experiments examined the spatial and temporal aspects of the wave packets separately \cite{Piche1999SPIE,Grunwald2000OL,Grunwald2001JOSAA}. Using pulses of width $<\sim\!10$~fs, X-waves were synthesized using thin-film axicons with small cone angles and characterized in space-time \cite{Grunwald2002OSA,Grunwald2002SPIE,Grunwald2003CLEO,Grunwald2003PRA,Grunwald2003OL}. The group velocity of such wave packets is extremely close to $c$ $(\theta\!\rightarrow\!45^{\circ}$), and no attempt was made at measuring its value.

Such short broadband pulses introduce a host of experimental difficulties. Although issues related to dispersion and aberrations in the synthesis system can be alleviated by relying on thin-film components \cite{Grunwald2004OEng,Grunwald2007Book}, there remain challenges in observing the spatio-temporal profile. Subsequently, spatially truncated versions of these wave packets were developed by placing a hard aperture in the beam path whose edge coincides with the first minimum of the underlying Bessel beam structure or a small-aperture axicon with ultrasmall conical angles ($\sim\!0.01^{\circ}$) was used to produce the X-wave. In both cases, only a bright central peak is obtained. Measurements reveal that this truncation produces a stable `needle beam' for an extended propagation distance with no side `fringes' \cite{Grunwald2004SPIE,Grunwald2006CLEO,Bock2013AS}. This unique feature allows for images to be transmitted with no diffraction by pixellating the transverse profile and utilizing an array of needle beams, one at each pixel \cite{Bock2009OE,Grunwald2012JEOSRP,Bock12OE}. A recent review \cite{Grunwald2020APX} surveys the state-of-the-art on needle beams.

Subsequently, $\widetilde{v}$ for an X-wave produced by a coherent pulsed laser was measured by Saari, Trebino, and co-workers \cite{Bowlan2009OL}, and was found to be $\widetilde{v}\!\approx\!1.00012c$. This value is in agreement with estimates from other groups using different measurement techniques \cite{Bonaretti2009OE,Kuntz2009PRA}. We discuss in Section~\ref{Section:Deadend} why a significant departure of $\widetilde{v}$ for an X-wave from $c$ is not feasible in the paraxial regime. The measurement approach in \cite{Bowlan2009OL} enabled for a sequence of beautiful experiments to record spatial phenomena involving pulsed light, such as the observation of Bessel beams produced by circular gratings \cite{Lohmus2012OL}, thereby confirming boundary diffraction waves \cite{Maggi1888AdMI,Rubinowicz1957Nature,Horvath2001PRE}, and observing the Arago spot resulting from a circular aperture \cite{Saari2010OE}. Furthermore, the same technique was utilized to observe accelerating and decelerating wave packets \cite{ValtnaLukner2009OE} produced via the strategy proposed by M. Clerici \textit{et al} \cite{Clerici2008OE}. However, only minute changes in $\widetilde{v}$ were observed (Section~\ref{Sec:Acceleration}).

\subsubsection{Nonlinear X-waves}

A major development in this period was the discovery that a variety of nonlinear effects give rise spontaneously to X-shaped \textit{spectra} \cite{DiTrapani2003PRL,Conti2003PRL,Jedrkiewicz2003PRE}. The spectra were measured in $(\varphi,\lambda)$-space, where $\varphi$ is the propagation angle with respect to the $z$-axis, and is thus related to the spatial frequency $k_{x}\!=\!\tfrac{\omega}{c}\sin{\varphi}$. The spectra are in fact parabolic and centered at $\varphi\!=\!0$, thus corresponding to \textit{baseband} ST wave packets, although they were confusingly referred to as nonlinear X-waves. These experiments thus established a new methodology for producing ST wave packets via nonlinear optical interactions. However, such wave packets are not propagation invariant for any significant distance because of the large spectral uncertainty $\delta\lambda$ resulting from the phase-matching process involved. Although the full bandwidth produced is usually broad ($\Delta\lambda\!\sim\!100$~nm \cite{Faccio2006PRL,Faccio2007OE}), the spectral uncertainty is also very large ($\delta\lambda\!\sim\!1$~nm). As we show in Section~\ref{Section:PropagationDistance}, the propagation distance of a ST wave packet is \textit{not} governed by the ratio $\tfrac{\Delta\lambda}{\delta\lambda}$, but is instead determined by the absolute value of $\delta\lambda$. The large $\delta\lambda$ characteristic of nonlinear X-waves therefore precludes propagation invariance over any significant distance in free space. Additionally, the dual parabolic dispersion curves observed in the spatio-temporal spectra correspond in free space to a \textit{pair} of baseband ST wave packets, one subluminal and the other superluminal. Axial walk-off between these two pulses offer a challenge to reconstruction of the wave packet's spatio-temporal profile in free space. Finally, nonlinear X-waves are expected to be dispersion-free \textit{in the dispersive nonlinear medium}, and are thus dispersive in free space once they exit the nonlinear medium. Whereas the time-averaged intensity of the  nonlinear X-wave remains diffraction-free in free space (albeit for a very short distance because of the large spectral uncertainty), the spatio-temporal profile will undergo rapid dispersive spreading in time over this distance. 

Nevertheless, these findings sparked research on propagation-invariant wave packets in dispersive media \cite{Porras2001OL,Porras2003OL,Porras2003PREBessel-X,Longhi2003PRE,Longhi2004OL,Porras2004,Porras2007JOSAB}, much of which awaits experimental confirmation. These studies have indicated that besides the more common X-shaped wave packets, O-shaped propagation-invariant counterparts also exist, which are related to Mackinnon's wave packet. Measurements have provided evidence for a transition from X- to O-shaped \textit{spectra} (but not the spatio-temporal profiles) by tuning the pump wavelength through a zero-dispersion wavelength from the regime of normal material GVD to the anomalous regime \cite{Porras2005OL}. However, subsequent work has shown that O-waves can indeed exist in the presence of either normal or anomalous GVD \cite{Malaguti2008OL,Malaguti2009PRA}. These O-waves remain yet to be observed.

\subsubsection{Theoretical results}

Further theoretical results were reported in  this period with respect to superluminal wave packets \cite{ZamboniRached2002EPJD,Recami2003IEEEJSTQE,ZamboniRached2003OptCommun,ZamboniRached2004JOSAA,ZamboniRached2006JOSAA,ZamboniRached2009PRA}. Additionally, the first papers examining subluminal ST wave packets appeared \cite{ZamboniRached2008PRA}, although no experiments attempted verifying these predictions \cite{Turunen2010PO}. Furthermore, the propagation of X-waves in waveguide structures was investigated theoretically \cite{ZamboniRached2001PRE,ZamboniRached2002PRE,ZamboniRached2003PRE}.

The negative-$\widetilde{v}$ baseband ST wave packet is notable for its conspicuous absence from the theoretical literature. In general, this case has been missed by previous classifications propagation-invariant wave packets; it was excluded from \cite{Turunen2010PO} and is absent from \cite{Figueroa2014Book} (Ch.~17). Porras \textit{et al}. \cite{Zapata2006OL} described a field configuration endowed with angular dispersion that has a negative group velocity, but the wave packet is dispersive and not propagation invariant. A useful explanation for why such wave packets do not violate relativistic causality is provided in \cite{Zapata2006OL}. Basically, there is a built-in latency in synthesizing such wave packets after which the wave packet is formed and then travels backwards. No information is thus transmitted beyond the bound set by relativistic causality.

\subsubsection{Baseband ST wave packets}

In retrospect, a useful development in this period that did not receive sufficient attention was the gradual emergence of baseband ST wave packets. Unfortunately, none of these proposed wave packets were pursued experimentally. It appears that the first consideration was by Besieris and Shaarawi \cite{Besieris2004OE}, followed by Saari in \cite{Valtna2007OC,Saari2007LP} where he noted the similarity between the dispersion relationship in the superluminal regime and that of a standard grating at grazing angles. Subsequently, superluminal baseband ST wave packets were examined in \cite{ZamboniRached2009PRA}, where the motivation was to find a propagation-invariant wave packet that is completely free of any acausal components; that is, $k_{z}\!>\!0$ for all $\omega$.

\subsection{After 2010}

Research on propagation-invariant wave packets after 2010 went through a dry spell. Attempts at producing wave packets with one transverse dimension that are propagation-invariant in media with anomalous \cite{Dallaire2009OE} and normal \cite{Jedrkiewicz2013OE} GVD were reported, but without verifying propagation invariance. Subsequent work was stimulated by the introduction of Airy beams \cite{Siviloglou2007OL,Siviloglou2007PRL}, pulsed Airy beams \cite{Saari2008OE,Saari2009LP,Kaganovsky2011JOSAA,Piksarv2013OL,Valdmann2014OL}, and Airy pulses \cite{Chong2010NP,Piksarv2012OE}. In part, research was stimulated along the lines motivated by the Airy-Bessel wave packet reported in 2010 \cite{Chong2010NP}. After many years of slow experimental progress, this paper demonstrated an easily synthesizable wave packet that is propagation invariant and yet \textit{separable} with respect to space and time (Bessel-beam spatial profile and Airy-pulse temporal profile). Strictly speaking, this wave packet is propagation invariant only in a dispersive medium, and it does not travel at a fixed group velocity. However, only limited deformation and changes in $\widetilde{v}$ occur in free space. In many ways, the Airy-Bessel wave packet was an antidote to the lack of previous experimental progress in the area of localized waves. Starting from 2016 \cite{Kondakci2016OE,Parker2016OE}, interest was revived in ST wave packets and continues to grow. We survey this more recent work starting in Sections~\ref{Sec:PropCharacteristics}. 

\section{Dispelling misconceptions regarding ST wave packets}
\label{Section:dispelling_misconceptions}

The appellation `X-wave' initially referred to the X-shaped profile of this wave packet \cite{Lu1992IEEETUFFC-Xwaves}. However, this terminology can be confusing; e.g., are X-waves the only X-shaped ST wave packets? We address this question and others in this Section.

\subsection{Are only X-waves X-shaped?}\label{Section:AreXWavesXshaped}

\begin{figure}[t!]
  \begin{center}
  \includegraphics[width=10cm]{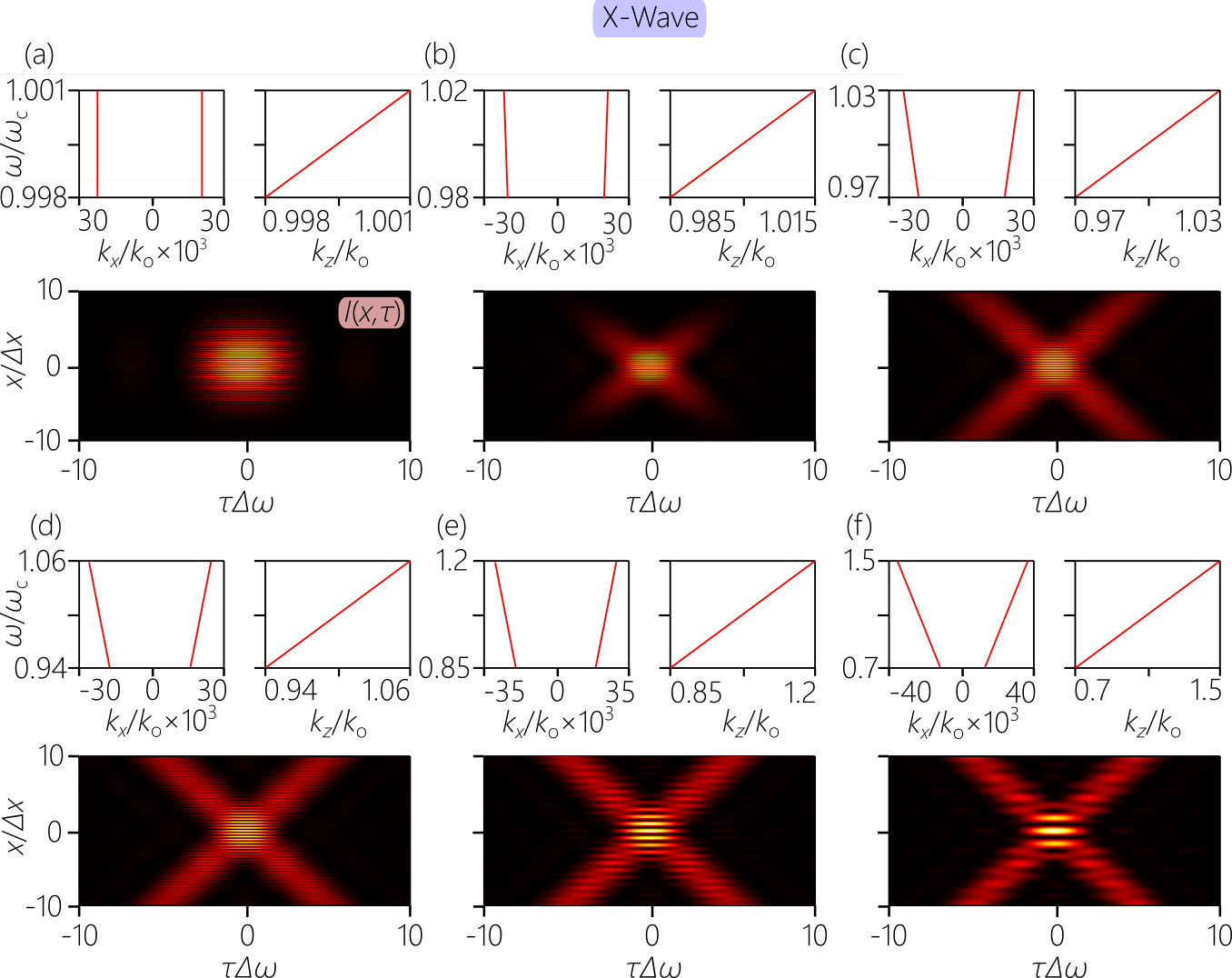}
  \end{center}
  \caption{Calculated spatio-temporal profiles for an X-wave with $\theta\!=\!45.001^{\circ}$ as the bandwidth $\Delta\omega$ increases. We plot for each case the spectral projections $\widetilde{\psi}(k_x,\omega)$ and $\widetilde{\psi}(k_z,\omega)$, and the intensity profile $I(x;\tau)$ at $z\!=\!0$; $\tau$ is scaled by $\Delta\omega$ and $x$ by $\Delta x\!=\!c\cot{\varphi_{\mathrm{o}}}/\Delta\omega$ (the lobe size of the X-wave). The bandwidths are: (a) $\Delta\omega\!=\!0.003\omega_\mathrm{c}$, (b) $0.03\omega_\mathrm{c}$, (c) $0.06\omega_\mathrm{c}$, (d) $0.12\omega_\mathrm{c}$, (e) $0.35\omega_\mathrm{c}$, and (f) $0.7\omega_\mathrm{c}$.}
  \label{Fig:subluminal_xtoo}
\end{figure}

There have been persistent misconceptions regarding the spatio-temporal profile of various ST wave packets (assumed to have a flat spectral phase). To dispel these misconceptions, we compare here the spatio-temporal profiles of X-waves, FWMs, and baseband ST wave packets. We first plot in Fig.~\ref{Fig:subluminal_xtoo} the evolution of the spatio-temporal profile for an X-wave (defined by the constraint in Eq.~\ref{Eq:XWavePlane}) as its bandwidth $\Delta\omega$ is increased at a fixed central frequency $\omega_{\mathrm{c}}$. Initially, at a narrow bandwidth of $\Delta\omega\!\approx\!0.002\omega_{\mathrm{c}}$, the profile is approximately separable with respect to the spatial and temporal DoFs [Fig.~\ref{Fig:subluminal_xtoo}(a)]. Increasing $\Delta\omega$ leads to the gradual emergence of an X-shaped profile [Fig.~\ref{Fig:subluminal_xtoo}(b)], which becomes clear when $\Delta\omega\!>\!0.05\omega_{\mathrm{o}}$ [Fig.~\ref{Fig:subluminal_xtoo}(c-f)] and maintained with subsequent increases in $\Delta\omega$. At $\lambda_{\mathrm{c}}\!=\!800$~nm, this corresponds to a bandwidth $\Delta\lambda\!>\!40$~nm, and thus a pulsewidth $\Delta t\!<\!24$~fs. As such, the X-shaped profile of an X-wave is only observable with ultrashort pulses as noted in \cite{Saari1996Proc}. Note however that the transverse field along $x$ is modulated by a periodic cosine wave (alternatively, by a Bessel beam when both transverse dimensions are included). We discuss in Section~\ref{Section:Deadend} the conditions that must be fulfilled to observe a purely X-shaped profile.

\begin{figure}[t!]
  \begin{center}
  \includegraphics[width=10cm]{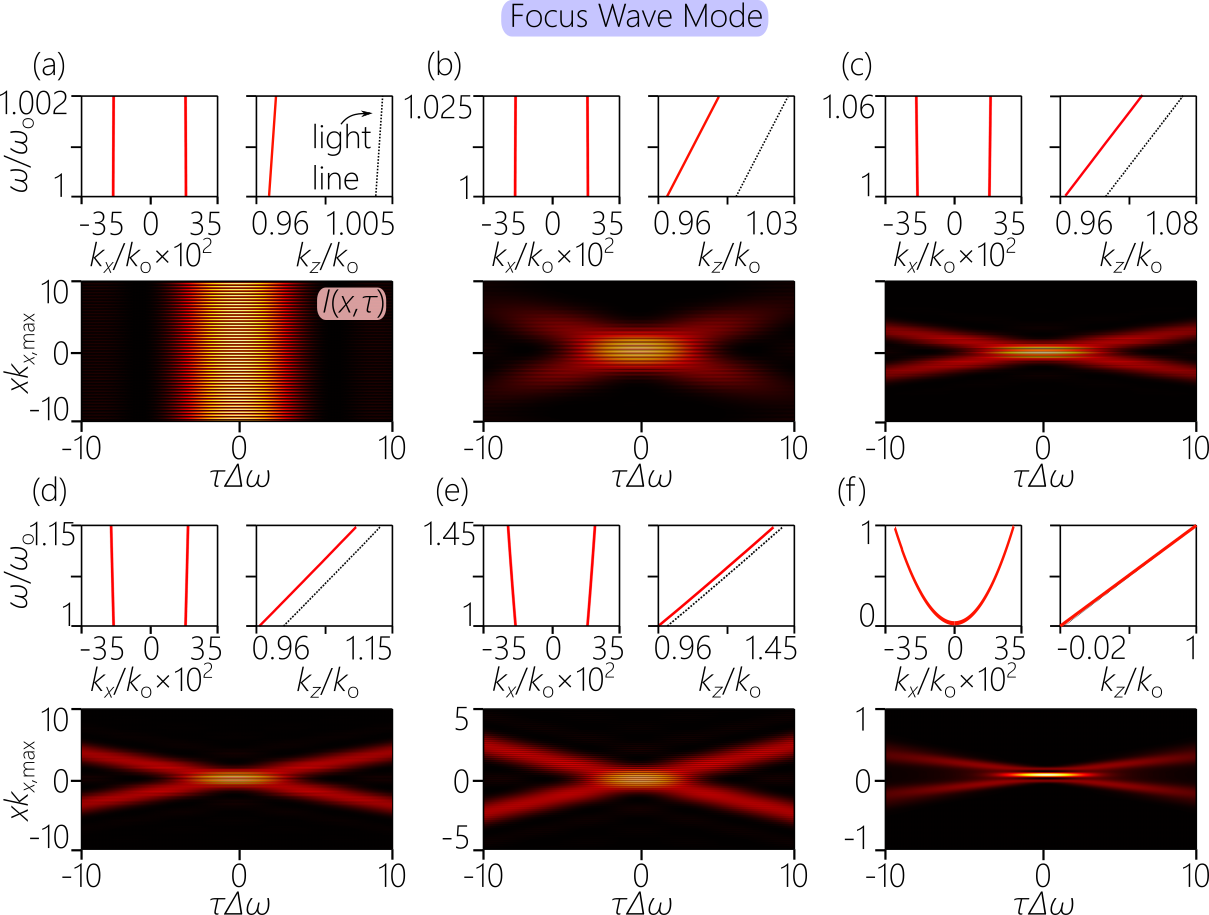}
  \end{center}
  \caption{Calculated spatio-temporal profiles for a luminal FWM ($\theta\!=\!45^{\circ}$) as $\Delta\omega$ increases, plotted similarly to Fig.~\ref{Fig:subluminal_xtoo}, and $\omega_\mathrm{o}\!=\!0.02\omega_{\mathrm{c}}$. The bandwidths are (a) $\Delta\omega\!=\!0.002\omega_\mathrm{c}$, (b) $0.025\omega_\mathrm{c}$, (c) $0.06\omega_\mathrm{c}$, (d) $0.15\omega_\mathrm{c}$, (e) $0.45\omega_\mathrm{c}$, and (f) $\omega_\mathrm{c}$, the latter including the $k_{z}\!<\!0$ components.}
  \label{Fig:luminal_xtoo}
\end{figure}

Much of the same evolution of the spatio-temporal profile ensues for FWMs with increasing bandwidth $\Delta\omega$ centered at $\omega_{\mathrm{c}}\!\gg\!\omega_{\mathrm{o}}$. The approximately separable spatio-temporal profile for a narrow bandwidth at $\Delta\omega\!=\!0.002\omega_{\mathrm{c}}$ [Fig.~\ref{Fig:luminal_xtoo}(a)] gives way to an X-shaped profile at larger bandwidths [Fig.~\ref{Fig:luminal_xtoo}(b-f)]. Once again, we observe the above-mentioned transverse spatial oscillations. Extending the bandwidth to encompass non-physical, backward-propagating $k_{z}\!<\!0$ components can lead to changes in the profile shape that may obfuscate the essentially X-shaped profile. Similar behavior is observed for sideband ST wave packets.

\begin{figure}[t!]
  \begin{center}
  \includegraphics[width=10cm]{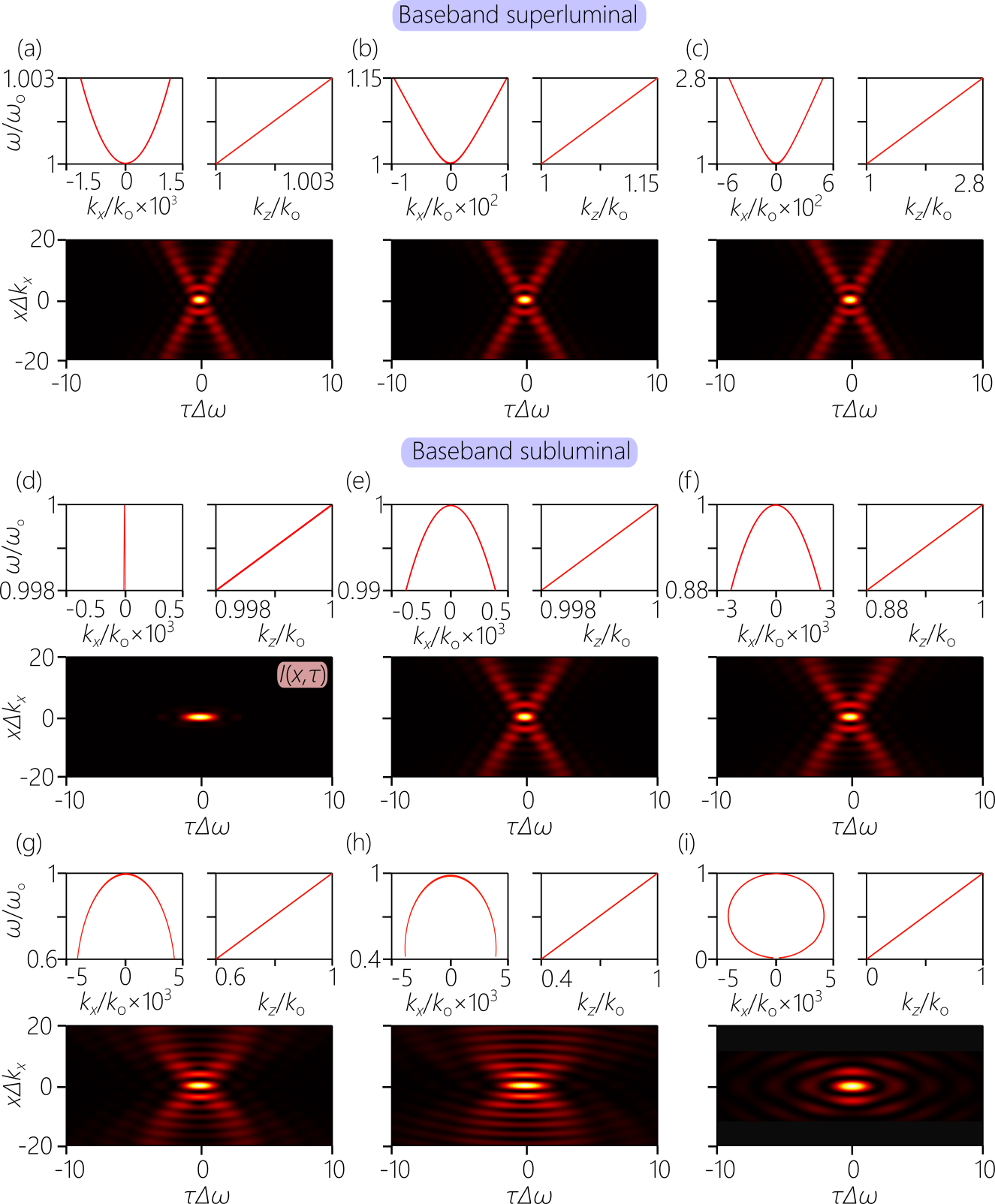}
  \end{center}
  \caption{Calculated spatio-temporal profiles for baseband ST wave packets as $\Delta\omega$ increases, plotted similarly to Fig.~\ref{Fig:subluminal_xtoo}. (a-c) Superluminal ST wave packets with $\theta\!=\!45.001^{\circ}$. (d) A separable spatio-temporal spectrum. (e-i) Subluminal ST wave packets with $\theta\!=\!44.999^{\circ}$. The bandwidths are (a) $\Delta\omega\!=\!0.003\omega_\mathrm{o}$, (b) $0.15\omega_\mathrm{o}$, and (c) $1.8\omega_\mathrm{o}$ for the superluminal range; and (d) $0.002\omega_\mathrm{o}$, (e) $0.002\omega_\mathrm{o}$, (f) $0.12\omega_\mathrm{o}$, (g) $0.4\omega_\mathrm{o}$, (h) $0.6\omega_\mathrm{o}$, and (i) $0.95\omega_\mathrm{o}$ for the subluminal range.}
  \label{Fig:xtoo}
\end{figure}

Although there are similarities between the profile for superluminal \textit{baseband} ST wave packets to those for X-waves and FWMs, striking contrasts emerge in the plots in Fig.~\ref{Fig:xtoo}(a-c). First, the X-shape emerges immediately even for extremely narrow bandwidths, in contrast to X-waves and FWMs. For example, at the same narrow bandwidth $\Delta\omega\!=\!0.002\omega_{\mathrm{o}}$, the X-shape has fully formed for the baseband ST wave packet, but has yet to emerge for the X-wave and the FWM. Second, we note a remarkable stability in the X-shaped profile of the baseband ST wave packet with increased bandwidth. Indeed, the profile is \textit{unchanged} with $\Delta\omega$ except for the different scale for space and time. Associated with this profile stability is the lack of transverse interference fringes within the beam center. The observed structure is due to the selection of a flat square spectrum leading to a sinc-function profile, which can be smoothed out by using a Gaussian or other smooth spectrum. We will comment further on these features below in Section~\ref{Section:Deadend}. Finally, the same trend occurs for \textit{subluminal baseband} ST wave packets, as shown in Fig.~\ref{Fig:xtoo}(d-g) for $\Delta\omega\!<\!0.6\omega_{\mathrm{o}}$.

From Figs.~\ref{Fig:subluminal_xtoo}-\ref{Fig:xtoo} we conclude that X-waves are \textit{not} the only X-shaped propagation-invariant wave packets. Indeed, within the paraxial domain, \textit{all} propagation-invariant wave packets are generally X-shaped, as long as the spectral phase is flat. In the case of baseband ST wave packets, the X-shape emerges even for narrow spectra and remain stable with changing bandwidth. It is much more challenging to observe the X-shaped profile for an X-wave or FWM, which necessitate exploiting an exorbitantly large bandwidth. Modulating the spectral phase enables modifying the spatio-temporal profile (Section~\ref{Sec:PropInvarianceAiry} and Section~\ref{Sec:StateOfTheArt}). It is therefore critical to retain the various distinct terms (X-waves, FWMs, baseband, sideband) for the particular spectral support domain for each family of ST wave packets rather than using `X-wave' as a generic blanket term.

\subsection{Are superluminal ST wave packets X-shaped but their subluminal counterparts O-shaped?}

It is sometimes stated that the X-shaped profile is a consequence of the superluminal group velocity of X-waves, whereas subluminal wave packets are O-shaped \cite{ZamboniRached2008PRA}. Indeed, E. Recami bases this claim on a particular extension of special relativity \cite{Barut1982NC,Figueroa2014Book}. As seen from the plots in Fig.~\ref{Fig:subluminal_xtoo} for superluminal X-waves, Fig.~\ref{Fig:luminal_xtoo} for luminal FWMs, and Fig.~\ref{Fig:xtoo}(a-c) for superluminal baseband ST wave packets, the profiles are indeed all X-shaped. However, even for \textit{subluminal} baseband ST wave packets, the spatio-temporal profile is X-shaped for small bandwidths as shown in Fig.~\ref{Fig:xtoo}(d-g), where the profile follows the same pattern of their superluminal counterparts. Nevertheless, for extremely large bandwidths where the spectrum encompasses the entire ellipse at the intersection of the light-cone with $\mathcal{P}_{\mathrm{B}}(\theta)$ when $0^{\circ}\!<\!\theta\!<\!45^{\circ}$, an O-shaped profile emerges [Fig.~\ref{Fig:xtoo}(h,i)], corresponding to the Mackinnon wave packet \cite{Mackinnon1978FP}.

The circularly symmetric wave packet is not constructed from plane-waves restricted to the vicinity of $k_{x}\!=\!0$, but requires integrating the spectrum over the \textit{entire} ellipse at the intersection of the light-cone with the spectral plane $k_{z}\!=\!k_{\mathrm{o}}+\Omega/\widetilde{v}$ [Fig.~\ref{Fig:SubluminalSTWavePacket}]. Consequently, there is a significant contribution from plane waves having $k_{z}\!<\!0$, which are not causal. This can be minimized in one of two ways. In the first approach we take $\theta\!\rightarrow\!45^{\circ}$, in which case only a small part of the spectrum extends beyond $k_{z}\!=\!0$, and the negative-$k_{z}$ components can be eliminated altogether from the spectrum without impacting the resulting O-shaped profile. However, this requires an exorbitantly large bandwidth $\Delta\omega$, and a numerical aperture approaching 1, deep within the non-paraxial regime. Reducing $\theta$ in turn reduces the required bandwidth, but the numerical aperture remains large and the negative-$k_{z}$ portion of the spectrum increases, thereby reverting from the O-shaped profile back to its X-shaped counterpart \cite{ZamboniRached2008PRA}. It is therefore unlikely that O-shaped optical ST wave packets will be observed in free space (the situation is altogether different in linear dispersive media \cite{Malaguti2008OL,Malaguti2009PRA}). To the best of our knowledge, subluminal propagation-invariant wave packets were not demonstrated experimentally until our recent work in \cite{Kondakci2017NP,Kondakci2019NC,Yessenov2019OE,Yessenov2019PRA}.

\subsection{Causality debates}\label{Section:CausalityDebates}

One of the pitfalls of the over-reliance on closed-form expressions for propagation-invariant wave packets became clear in the early development of FWMs when a debate emerged regarding their causality. Felson \textit{et al}. pointed out that the expression for FWMs includes contributions from backward-propagating plane-wave components having $k_{z}\!<\!0$, which are not consistent with causal excitation and propagation. The abstract of \cite{Heyman1987JOSAA} famously stated: ``... The spectral decomposition in the spatial wave-number domain reveals that the FWM is synthesized by both forward- and backward-propagating plane waves ... Asymptotic considerations show that the dominant mechanism is constructive interference of the backward-propagating waves ... The conclusions cast doubt on the possibility of embedding the FWM within a \textit{causal} excitation scheme.'' According to Ziolkowski in \cite{Figueroa2014Book} (Chapter~2), this statement convinced the US National Science Foundation to decline funding this research area.

\begin{figure}[t!]
  \begin{center}
  \includegraphics[width=11cm]{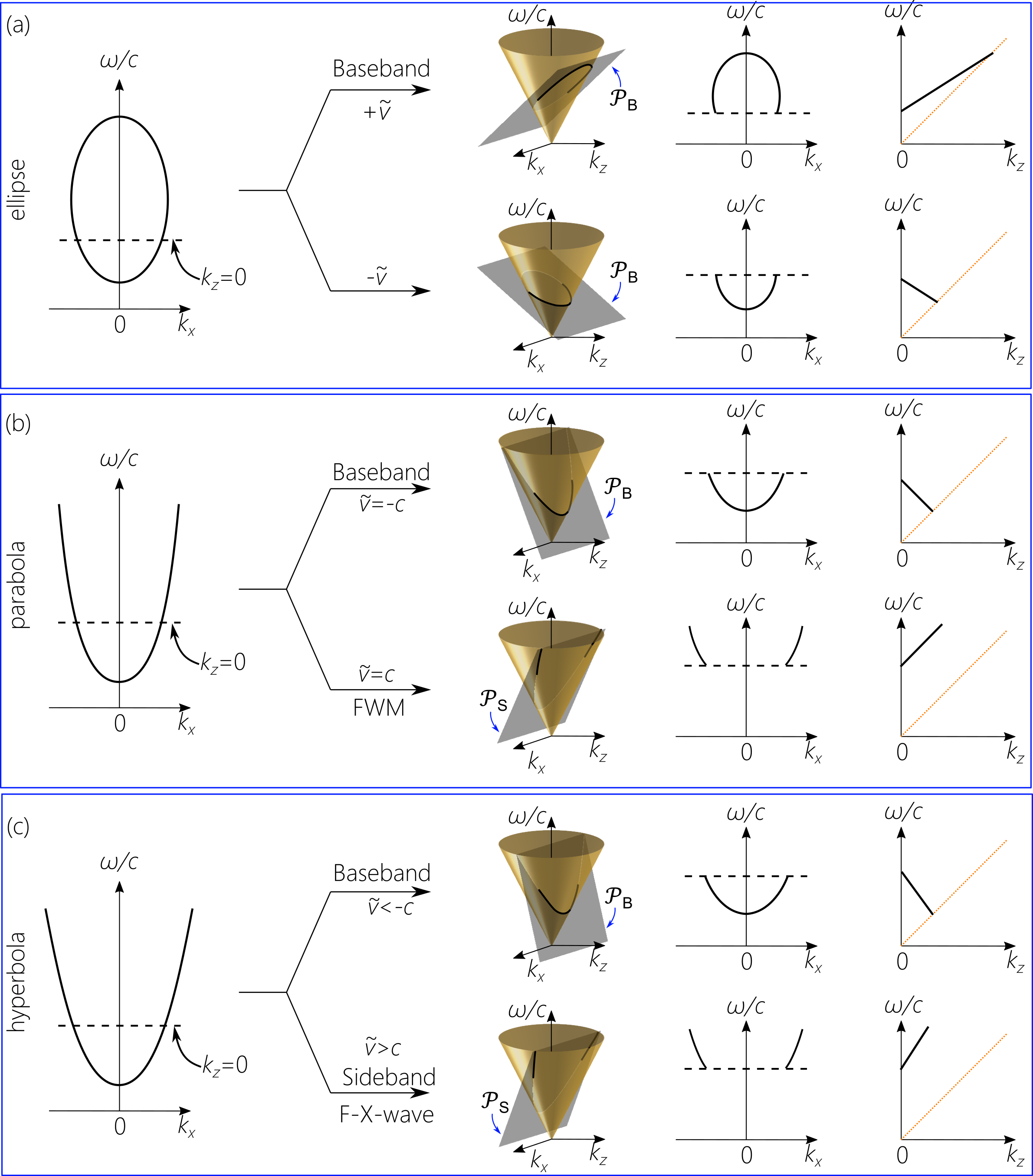}
  \end{center}
  \caption{(a) An ellipse at the intersection of the light-cone with a plane, projected onto the $(k_{x},\tfrac{\omega}{c})$-plane. Only one portion of the ellipse is causal at a time: the upper half when $0^{\circ}\!<\!\theta\!<\!45^{\circ}$ corresponding to a baseband subluminal ST wave packet, and the bottom half when $135^{\circ}\!<\!\theta\!<\!180^{\circ}$ corresponding to a baseband negative-$\widetilde{v}$ ST wave packet. (b) A parabola at the intersection of the light-cone with a plane, projected onto the $(k_{x},\tfrac{\omega}{c})$-plane. Only one portion of the parabola is causal at a time: the lower half when $\theta\!=\!135^{\circ}$ corresponding to a baseband negative-$\widetilde{v}$ ST wave packet ($\widetilde{v}\!=\!-c$), and the upper half when $\theta\!=\!45^{\circ}$ corresponding to a sideband FWM ($\widetilde{v}\!=\!c$). (c) A hyperbola at the intersection of the light-cone with a plane, projected onto the $(k_{x},\tfrac{\omega}{c})$-plane. Only one portion of the hyperbola is causal at a time: the lower half when $90^{\circ}\!<\!\theta\!<\!135^{\circ}$ corresponding to a baseband negative-$\widetilde{v}$ ST wave packet, and the upper half when $45^{\circ}\!<\!\theta\!<\!90^{\circ}$ corresponding to a superluminal sideband ST wave packet.}
  \label{Fig:Spectral_lines}
\end{figure}

The response at the time was to modify the structure of the FWM to `minimize' the non-causal components while retaining a closed-form expression for the FWM. This can be done, for example, by reducing $\omega_{\mathrm{o}}$ in Eq.~\ref{Eq:FWMParabola}, whereupon the spectral support domain approaches the light-line. In our opinion, and with the benefit of hindsight, this emphasis on closed-form expressions was misplaced, and ultimately hindered further developments in this field. The dilemma regarding the causality of FWMs could have easily been resolved by eliminating the $k_{z}\!<\!0$ regime altogether from the integrals, even if no closed-form expression is found \cite{Turunen2010PO}. That is, finding an analytical expression should not be the sole criterion for identifying a new propagation-invariant wave packet. Ironically, a closed-form expression was indeed identified years later with only $k_{z}\!>\!0$ \cite{Sheppard2008OE} in terms of the Lommel functions \cite{born2019principles} commonly used in theoretical modelling of optical microscopy. The attitude that we adopt and promote here is that the various names such as FWM and X-wave should \textit{not} be associated with particular closed-form expressions, but should instead be assigned to the underlying spatio-temporal spectral support domain, which is the crucial structural feature distinguishing ST wave packets independently of their particular spectral field amplitudes.

Additional perspective on the issue of causality of ST wave packets can be gleaned from Fig.~\ref{Fig:Spectral_lines}. In general, the intersection of a plane with the light-cone is an ellipse [Fig.~\ref{Fig:Spectral_lines}(a)], a parabola [Fig.~\ref{Fig:Spectral_lines}(b)], or a hyperbola [Fig.~\ref{Fig:Spectral_lines}(c)]. These conic sections all appear in the spectral projection onto the $(k_{x},\tfrac{\omega}{c})$-plane. However, the entire conic section is not necessarily compatible with the causal restriction $k_{z}\!>\!0$. We can then divide the conic section into two parts separated at $k_{z}\!=\!0$. Each part of the conic section can then be associated with a \textit{different} spectral plane.

Consider the ellipse in Fig.~\ref{Fig:Spectral_lines}(a): the lower part of the ellipse when $k_{z}\!>\!0$ is associated with the plane $\mathcal{P}_{\mathrm{B}}(\theta)$, $0^{\circ}\!<\!\theta\!<\!45^{\circ}$, corresponding to a subluminal baseband ST wave packet; whereas the upper part of the ellipse $k_{z}\!>\!0$ is associated with $\mathcal{P}_{\mathrm{B}}(\theta')$, $\theta'\!=\!180^{\circ}-\theta$, and thus corresponds to a negative-$\widetilde{v}$ baseband ST wave packet. For the parabola in Fig.~\ref{Fig:Spectral_lines}(b): the lower part is associated with the plane $\mathcal{P}_{\mathrm{B}}(\theta)$, $\theta\!=\!135^{\circ}$, and thus corresponds to a baseband ST wave packet with $\widetilde{v}\!=\!-c$; whereas the upper part to the plane $\mathcal{P}_{\mathrm{S}}(\theta)$, $\theta\!=\!45^{\circ}$, corresponding to a FWM with $\widetilde{v}\!=\!c$. Finally, for the hyperbola in Fig.~\ref{Fig:Spectral_lines}(c): the lower part is associated with the plane $\mathcal{P}_{\mathrm{B}}(\theta)$, $90^{\circ}\!<\!\theta\!<\!135^{\circ}$, and thus corresponds to a negative-$\widetilde{v}$ baseband ST wave packet; whereas the upper part is associated with $\mathcal{P}_{\mathrm{S}}(\theta')$, $45^{\circ}\!<\!\theta'\!<\!90^{\circ}$ and thus corresponds to a superluminal sideband ST wave packet (a so-called focused X-wave \cite{Shaarawi2003JOSAA,ZamboniRached2004JOSAA}).

\subsection{Nomenclature}

From the discussion above, we can make the following observations:
\begin{enumerate}
    \item X-waves are \textit{not} the only X-shaped ST wave packet. Indeed, all propagation-invariant wave packets are X-shaped in free space in the paraxial regime as long as their spectral phase is flat.
    \item Not all propagation-invariant wave packets are X-shaped; exceptions include Mackinnon's wave packet (or O-waves) and the Bessel-Airy wave packet. Furthermore, radically different profiles can result from modulating the spectral phase (Section~\ref{Sec:PropInvarianceAiry} and Section~\ref{Sec:StateOfTheArt}).
    \item Some wave packets may initially be X-shaped, but are in fact \textit{not} propagation invariant; e.g., tilted pulse fronts (TPFs) with symmetrized spectra (conventional TPFs are a single branch from this potentially X-shaped profile).
\end{enumerate}

Based on these observations, we advocate for a precise and consistent nomenclature that helps delineate the various classes of propagation-invariant wave packets independently of their particular spatio-temporal profiles. First, we suggest the use of the term `ST wave packet' for any pulsed field in which a one-to-one relationship is introduced between $\omega$ and $k_{z}$ (or $\omega$ and $k_{x}$). It will be understood that in almost cases, a linear relationship between $\omega$ and $k_{z}$ is targeted in order to achieve propagation invariance. If not understood from the context, the descriptor `propagation-invariant' can be added. The term `X-wave' should \textit{not} be associated with the X-shaped profile of a wave packet or a closed-form expression. Instead, we retain this term for any wave packet in which $k_{x}\!\propto\!\omega$ \textit{and} $k_{z}\!\propto\!\omega$. Based on the nature of the conic section at the intersection of the light-cone with a plane, the wave packets should always be explicitly identified with their class: baseband, sideband, or X-wave. In the case of sideband ST wave packets, we can identify them as superluminal or FWM (luminal); and in the case of baseband ST wave packets, we can identify subluminal, superluminal, and negative-$\widetilde{v}$ regimes.

Propagation-invariant wave packets, have been previously studied under the general name of `localized waves', `diffraction-free' beams, `non-diffracting' beams, `limited-diffraction' waves, or other related terms. Such names accentuate the propagation invariance of these wave packets. We use instead the name `ST wave packets', and more generally ST optics and photonics. There are several reasons why we find adopting this general term more appropriate. First, as mentioned above, propagation-invariance follows from a particular spatio-temporal spectral structure, which provides a unifying framework to investigate and classify all propagation-invariant wave packets. Second, more generally, manipulation of the spatio-temporal optical field structure can lead to new behaviors in wave packets that are \textit{not} propagation invariant, such as axial acceleration, programmable axial spectral encoding, and group velocity dispersion experienced in free space (Section~\ref{Sec:ModifyingTheAxialProp}). Such characteristics represent deviations from propagation-invariance, yet they are all a consequence of sculpting the spatio-temporal spectrum. Third, even incoherent fields can be structured in space and time along the same lines \cite{Yessenov2019Optica}. Although such fields may be localized in space, they are \textit{not} localized in time, and as such the appellation `localized waves' is \textit{not} appropriate. Finally, there is an interesting and instructive connection between ST wave packets and relativistic transformations of traditional optical fields that was recognized early on \cite{Belanger1986JOSAA,Saari2004PRE,Longhi2004OE}, which lends support to the aptness of the term ST wave packets.


\section{Are X-waves, sideband ST wave packets, and FWMs a dead end for optics?}\label{Section:Deadend}

Despite the substantial body of theoretical work on FWMs and X-waves over the past 4 decades, the pace of experimental progress with regard to their synthesis and applications in the optical regime was slow, and ultimately came to a standstill. For example, although it is predicted that X-waves may take on arbitrary superluminal group velocities, the realized values are typically $\widetilde{v}\!\sim\!1.0001c$ \cite{Bonaretti2009OE,Bowlan2009OL}. Moreover, it was recognized early on that observing the X-shaped spatio-temporal profile requires extremely short pulse widths \cite{Saari1996Proc}, and the first demonstration of an X-wave thus made use of broadband \textit{incoherent} light (rather than an ultrashort pulse) of coherence time $\sim\!3$~fs \cite{Saari1997PRL}. Subsequent experiments making use of $\sim\!10$~fs pulses faced challenges even in recording the X-shaped profile \cite{Grunwald2003PRA}. Measurements of the group velocity relied on 30-fs pulses \cite{Bowlan2009OL}, in which case only faint tails of the X-branches are resolved away from the wave packet center. At the center, the wave packet is approximately a pulsed Bessel beam that is basically separable with respect to $x$ and $t$ \cite{Bowlan2009OL}. Furthermore, with regards to optical FWMs, there have been no conclusive observations beyond the initial investigations by Reivelt and Saari \cite{Reivelt2002PRE}, despite the almost 40~years that have elapsed since Brittingham initiated this whole field of study. To the best of our knowledge, there have been no observations of superluminal sideband ST wave packets of any kind. This paucity of previous experimental results is contrasted to the rapid progress over the past 5 years with baseband ST wave packets.

Why has there been such a wide disconnect between theory and experiment regarding X-waves, FWMs, and sideband ST wave packets over such an extended period of time? We argue here that these particular classes of ST wave packets are a `dead end' for optics; that is, progress -- in principle -- cannot be made in synthesizing these particular optical wave packets in a meaningful way. The central reason is that tuning the characteristics of these classes of ST wave packets away from those of conventional pulsed beams requires either an extremely broad temporal bandwidth, a very large numerical aperture, and in most cases both. Whereas these constraints are not deal-breakers in acoustics or ultrasonics, they are unbridgeable obstacles in the optical regime.

\subsection{Challenges faced by X-waves}

It is straightforward to demonstrate why any significant deviation in $\widetilde{v}$ from $c$ for an X-wave requires operating deep within the non-paraxial regime. Consider an X-wave whose temporal spectrum is centered at $\omega_{\mathrm{c}}$, at which point we have $k_{\mathrm{c}x}^{2}+k_{\mathrm{c}z}^{2}\!=\!k_{\mathrm{c}}^{2}$, where $k_{\mathrm{c}x}\!=\!k_{\mathrm{c}}\sin{\varphi_{\mathrm{x}}}$ and $k_{\mathrm{c}z}\!=\!k_{\mathrm{c}}\cos{\varphi_{\mathrm{x}}}$, $k_{\mathrm{c}}\!=\!\omega_{\mathrm{c}}/c$, and $\varphi_{\mathrm{x}}$ is the propagation angle with respect to the $z$-axis. For X-waves, $\varphi_{\mathrm{x}}$ is fixed for all $\omega$. To realize a superluminal group velocity $\widetilde{v}\!=\!c\tan{\theta}$ requires that $\cos{\varphi_{\mathrm{x}}}=\cot{\theta}$, and the associated numerical aperture is $\mathrm{NA}\!=\!\sin{\varphi_{\mathrm{x}}}$. Because only superluminal group velocities are allowed, we express the spectral tilt angle as $\theta\!=\!45^{\circ}+\delta$, where $\delta$ is a positive angle formally in the range $0^{\circ}\!<\!\delta\!<\!45^{\circ}$, thus $\mathrm{NA}\!=\!\tfrac{2\sqrt{\tan{\delta}}}{1+\tan{\delta}}$ and $\widetilde{v}\!=\!c\tfrac{1+\tan{\delta}}{1-\tan{\delta}}$. It is clear that only small angles $\delta$ are compatible with paraxial propagation. For example, $\delta\!=\!0.1^{\circ}$ results in $\mathrm{NA}\!=\!0.083$, $\varphi_{\mathrm{x}}\!\approx\!4.8^{\circ}$, and $\widetilde{v}\!=\!1.0035c$; whereas $\delta\!=\!5^{\circ}$ results in $\mathrm{NA}\!=\!0.544$, $\varphi_{\mathrm{x}}\!\approx\!33^{\circ}$, and $\widetilde{v}\!=\!1.19c$. In other words, any deviation of $\widetilde{v}$ from $c$ for an X-wave requires operating deep within the non-paraxial regime. For realistic scenarios where $\delta$ is small, we have $\mathrm{NA}\!\approx\!2\sqrt{\delta}$ and $\widetilde{v}\!\approx\!(1+2\delta)c$. This is consistent with the reported measured values of $\widetilde{v}$. 

\begin{figure}[t!]
  \begin{center}
  \includegraphics[width=11cm]{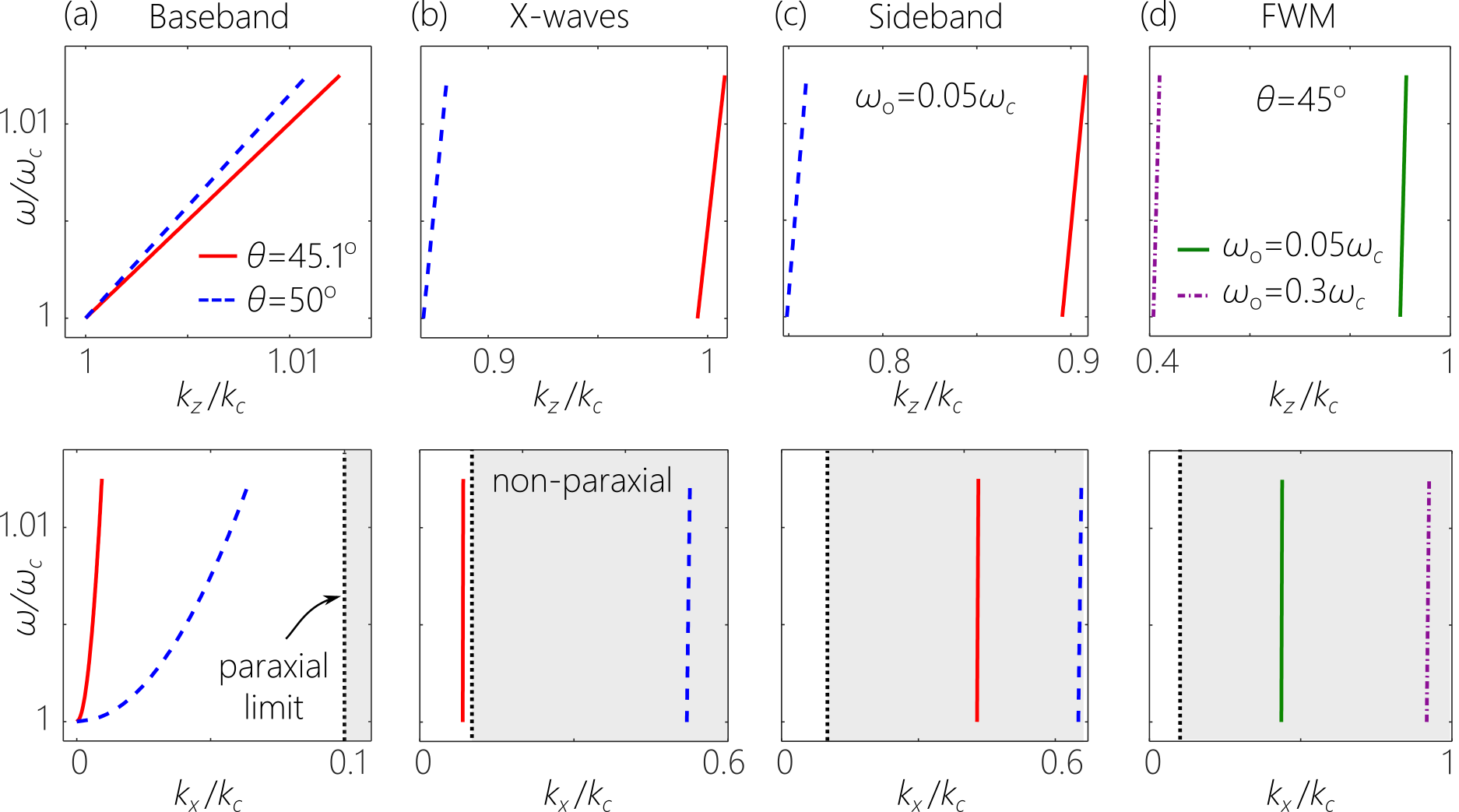}
  \end{center}
  \caption{Realizing ST wave packets in the paraxial domain (defined as $k_{x}\!<\!0.1k$). In each case we plot the spectral projections onto the $(k_{z},\tfrac{\omega}{c})$ and $(k_{x},\tfrac{\omega}{c})$ planes for two wave packets having $\theta\!=\!45.1^{\circ}$ and $\theta\!=\!50^{\circ}$. (a) Baseband ST wave packets; both are in the paraxial regime. (b) X-waves; $\theta\!=\!45.1^{\circ}$ is already close to the paraxial limit, while $\theta\!=\!50^{\circ}$ is well within the non-paraxial regime. (c) Sideband ST wave packets; both examples are in the non-paraxial regime. (d) FWM at $\theta\!=\!45^{\circ}$ while varing $\omega_{\mathrm{o}}$; both examples are in the non-paraxial regime. Bandwidth throughout is $\Delta\omega\!=\!0.0125\omega_{\mathrm{c}}$ ($\Delta\lambda\!=\!8$~nm at $\lambda_{\mathrm{c}}\!=\!800$~nm).}
  \label{Fig:Dead-end}
\end{figure}

A further constraint results from the proportionality between the spatial and temporal bandwidths, $\Delta k_{x}$ and $\Delta\omega$, respectively. The relationship $k_{x}\!=\!\tfrac{\omega}{c}\sin{\varphi_{\mathrm{x}}}$ imposes the constraint $c\Delta t\!=\!2\sqrt{\delta}\Delta x$, between the spatial and temporal widths, $\Delta x$ and $\Delta t$, respectively. At $\delta\!=\!0.1^{\circ}$, a beam of width $\Delta x\!=\!10$~$\mu$m requires a pulse of width $\Delta t\!\approx\!3$~fs. That is, mildly localizing an X-wave in space requires a prohibitively large temporal bandwidth. Furthermore, the required pulsewidth is independent of the central wavelength $\lambda_{\mathrm{c}}$, so that it becomes practically impossible to produce a localized X-wave at long wavelengths in the mid-infrared. Increasing $\delta$ reduces this burden only slightly (at $\delta\!=\!1^{\circ}$, the required pulse width is $\Delta t\!\approx\!9$~fs for $\Delta x\!=\!10$~$\mu$m), but at the cost of a higher NA; see Fig.~\ref{Fig:Dead-end}(b).

Finally, a different constraint emerges when examining the spatio-temporal profile of the X-wave. Because the spatial spectrum is centered at $k_{\mathrm{c}x}\!\approx\!4\pi\sqrt{\delta}/\lambda_{\mathrm{c}}$, the transverse profile of the X-wave is modulated with spatial fringes of period $L\!\approx\!\lambda_{\mathrm{c}}/(4\sqrt{\delta})$. For a single fringe to occur within the central lobe of the X-wave, an added constraint is imposed on the X-wave localization. At $\delta\!=\!0.1^{\circ}$, we have $L\!\approx\!5$~$\mu$m at $\lambda_{\mathrm{c}}\!\sim\!800$~nm, thus requiring a pulse of width $\Delta t\!\approx\!1.5$~fs. Making use of larger pulse widths leads to spatial fringes in the central lobe and the spatial and temporal DoFs of the X-waves are practically separable. With the inclusion of both transverse spatial dimensions, this spatial modulation takes the form of a Bessel beam profile, and the expected X-shaped profile is instead separable in space and time: a Bessel beam modulated by the temporal pulse profile, with faint tails of X-branches far from the beam center \cite{Saari1996Proc}. In other words, observing an X-shaped profile in which a single fringe is confined to the central lobe places extreme constraints on the required bandwidth and/or numerical aperture.

\subsection{Challenges faced by sideband ST wave packets}

The restrictions described above on the synthesis of X-waves are further exacerbated for sideband ST wave packets whose spatial and temporal spectra are centered at $k_{\mathrm{c}x}\!=\!k_{\mathrm{c}}\sin{\varphi_{\mathrm{s}}}$ and $\omega_{\mathrm{c}}$, respectively, equal to those of an X-wave of the same $\widetilde{v}$ and $\omega_{\mathrm{c}}$. Here $\varphi_{\mathrm{s}}$ is the angle made by the frequency component at $\omega\!=\!\omega_{\mathrm{c}}$ with the $z$-axis. Unlike X-waves where $\varphi_{\mathrm{x}}$ is independent of $\omega$, $\varphi_{\mathrm{s}}$ is $\omega$-dependent. It is straightforward to show that $\cos{\varphi_{\mathrm{s}}}\!=\!(1-\tfrac{\omega_{\mathrm{o}}}{\omega_{\mathrm{c}}})\cos{\varphi_{\mathrm{x}}}-\tfrac{\omega_{\mathrm{o}}}{\omega_{\mathrm{c}}}$, so that $\cos{\varphi_{\mathrm{s}}}\!\leq\!\cos{\varphi_{\mathrm{x}}}$ and $\varphi_{\mathrm{s}}\!\geq\!\varphi_{\mathrm{x}}$. Equality is achieved as $\omega_{\mathrm{o}}\!\rightarrow\!0$ ($\omega_{\mathrm{o}}\!\ll\!\omega_{\mathrm{c}}$), in which case the sideband ST wave packet approaches an X-wave). In other words, the NA needed for the sideband ST wave packet is larger than that for an X-wave at the same $\widetilde{v}$ and $\omega_{\mathrm{c}}$. Within the paraxial regime, we have $\mathrm{NA}\!\approx\!2\sqrt{\delta+\tfrac{\omega_{\mathrm{o}}}{\omega_{\mathrm{c}}}}$. It is expected that the NA will be larger when considering the maximum temporal frequency in the spectrum. The proportionality between the spatial and temporal widths of the sideband ST wave packet is $c\Delta t\!\approx\!2\Delta x\sqrt{\delta+\tfrac{\omega_{\mathrm{o}}}{\omega_{\mathrm{c}}}}$, so that the pulse width required is larger to realize the same $\Delta x$ as the corresponding X-wave. This relaxing of the required temporal bandwidth $\Delta\omega$ comes of course at the price of a larger NA; see Fig.~\ref{Fig:Dead-end}(c).  

\subsection{Challenges faced by FWMs}

Luminal FWMs ($\theta\!=\!45^{\circ}$ and $\widetilde{v}\!=\!c$) are a limiting case of sideband ST wave packets. Nevertheless, their historical importance and some of their unique characteristics warrant a separate treatment. Consider a FWM whose spatial and temporal spectra are centered at $k_{\mathrm{c}x}\!=\!k_{\mathrm{c}}\sin{\varphi_{\mathrm{f}}}$ and $\omega_{\mathrm{c}}$, respectively, where $\varphi_{\mathrm{f}}$ is the angle made by the frequency component at $\omega\!=\!\omega_{\mathrm{c}}$ with the $z$-axis; in general, $\varphi_{\mathrm{f}}$ is $\omega$-dependent. Here $\cos{\varphi_{\mathrm{f}}}\!=\!1-2\tfrac{\omega_{\mathrm{o}}}{\omega_{\mathrm{c}}}$ and $\mathrm{NA}=\sin{\varphi_{\mathrm{f}}}\!=\!2\sqrt{\tfrac{\omega_{\mathrm{o}}}{\omega_{\mathrm{c}}}(1-\tfrac{\omega_{\mathrm{o}}}{\omega_{\mathrm{c}}})}$. It is clear that the NA will be extremely large unless $\omega_{\mathrm{o}}\!\ll\!\omega_{\mathrm{c}}$; see Fig.~\ref{Fig:Dead-end}(d). For example, $\omega_{\mathrm{o}}\!=0.01\!\omega_{\mathrm{c}}$ leads to $\mathrm{NA}\!=\!0.2$ ($\varphi_{\mathrm{f}}\!\approx\!12^{\circ}$) and $\omega_{\mathrm{o}}\!=0.1\!\omega_{\mathrm{c}}$ to $\mathrm{NA}\!=\!0.6$ ($\varphi_{\mathrm{f}}\!\approx\!37^{\circ}$). We can thus use the approximation $\mathrm{NA}\!\approx\!2\sqrt{\tfrac{\omega_{\mathrm{o}}}{\omega_{\mathrm{c}}}}$.

Examining an example of a FWM developed by R. Ziolkowski using this approach is instructive. This example is a FWM having a single central peak (no transverse spatial fringes), of spatial width $\Delta x\!\sim\!10$~$\mu$m and temporal width $\Delta t\!\sim\!100$~fs\cite{Ziolkowski1989PRA,Ziolkowski2020Conf}. Both values of $\Delta x$ and $\Delta t$ appear reasonable, and it may be expected that they do not place strenuous requirements on the experimental resources needed to synthesize this wave packet. However, a closer look at this example reveals that such a pulse can be constructed only at lower frequencies $\sim\!2$~THz (the duration of one cycle is $\approx\!500$~fs) and not in the optical domain. This field configuration for a FWM thus corresponds to a sub-cycle pulse, which is a challenging task. In addition, the spectrum in this example extends from $\sim\!2$~THz all the way down to $\omega_{\mathrm{o}}$, and hence includes non-causal components $k_{z}\!<\!0$. This example, while positing an intriguing field configuration, clearly demonstrates the difficulties involved in synthesizing FWMs, and sideband ST wave packets in general.

\subsection{Baseband ST wave packets resolve these challenges}

The restrictions outlined above for X-waves, sideband ST wave packets, and FWMs are obviated when using baseband ST wave packets. Uniquely, because baseband ST wave packets have a spatial spectrum centered at $k_{x}\!=\!0$ (no spatial-frequency `carrier'), there are no transverse spatial fringes along $x$ in their profile, which is therefore always X-shaped (in absence of spectral phase modulation), even when large spatial widths $\Delta x$ (small spatial bandwidths) are used, in contradistinction to all three classes examined above. Second, the spatial and temporal bandwidth are related through Eq.~\ref{Eq:SpatialAndTemporalBandwidthBaseband}, so that we have $\mathrm{NA}\!\approx\!\sqrt{2\tfrac{\Delta\omega}{\omega_{\mathrm{o}}}(1-\cot{\theta})}$. The NA can be retained low for any group velocity $\widetilde{v}\!=\!c\tan{\theta}$ if we reduce the temporal bandwidth $\Delta\omega$; see Fig.~\ref{Fig:Dead-end}(a).

These characteristics of ST wave packets allow us to explore realms that are entirely inaccessible to other families of propagation-invariant wave packets. For example, a wave packet with $\theta\!\rightarrow\!90^{\circ}$ (which is impossible to realize with X-waves or sideband ST wave packets) can be readily realized in the paraxial domain with only picosecond pulses! Indeed, at $\lambda_{\mathrm{o}}\!=\!800$~nm and $\Delta\lambda\!=\!2$~nm (pulsewidth $\approx\!0.5$~ps), only $\mathrm{NA}\!=\!0.126$ is needed to achieve negative-$\widetilde{v}$ of $\widetilde{v}\!=\!-1.732c$ ($\theta\!=\!120^{\circ}$). Furthermore, baseband ST wave packets offer the most spatially localized profiles for a given temporal bandwidth and NA.

Historically, the elegant closed-form analytical expressions that can be easily developed for X-waves and FWMs have made them attractive targets for theoretical investigations, whereas baseband ST wave packets have not yet offered the same alluring analytical expressions. However, the fundamental difficulties faced by FWMs, X-waves, and sideband ST wave packets as outlined above with regards to the required bandwidth and numerical aperture are resolved by the use of \textit{baseband} ST wave packets, which therefore offer a much more fertile ground for future experimental investigations.

\section{Synthesis and characterization of ST wave packets}\label{Synthesis}

Now that we have established that sideband ST wave packets, FWMs, and X-waves do not offer a path forward in optics, we focus from this point onward on the synthesis of \textit{baseband} ST wave packets, the new phenomena that have been realized with them, and their potential applications. 

\subsection{Historical background on the spatio-temporal synthesis problem}

\begin{figure}[t!]
  \begin{center}
  \includegraphics[width=9cm]{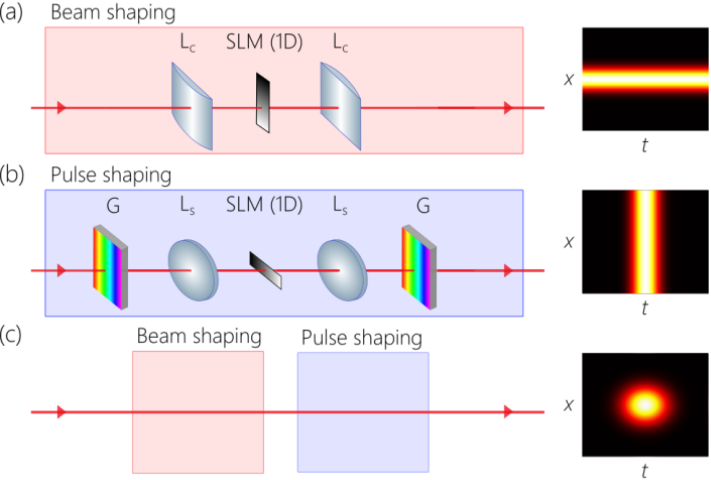}
  \end{center}
  \caption{(a) A monochromatic beam shaped along 1D using a SLM in a conventional Fourier optics system. (b) Conversely, a pulse can be shaped through spectral phase modulation via a SLM between two gratings. (c) Combining the beam and pulse shaping systems from (a) and (b) in succession produces a pulsed beam with separable DoFs. }
  \label{Fig:ModulationSchemes}
\end{figure}

Despite the initial interest generated by Brittingham's theoretical breakthrough with the FWM \cite{Brittingham1983JAP}, no proposals emerged for its optical synthesis, and the first experimental tests utilized acoustic waves instead \cite{Ziolkowski1989PRL}. Indeed, the synthesis of ST wave packets in general introduces a challenge because the spatial and temporal DoFs of the field are \textit{not} separable. In producing a monochromatic beam, the spatial beam profile can be manipulated directly in physical space or its spatial spectrum can be modulated in the Fourier domain \cite{Goodman2005Book}; see Fig.~\ref{Fig:ModulationSchemes}(a). This is the arena of Fourier optics \cite{Goodman2005Book}, beam shaping \cite{Maurer2011LPR}, and more recently `structured light' \cite{Forbes2021NP,Rubinsztein_Dunlop_2016JoO}. The result is a monochromatic beam in which all the spatial frequencies share the same temporal frequency [Fig.~\ref{Fig:MonochromaticBeam}]. Alternatively, ultrafast pulses can be modulated in time to produce a broad range of temporal waveforms. Because we cannot modulate a pulse directly on femtosecond time scales, spectral phase modulation has emerged as the best approach for ultrafast pulse synthesis \cite{Weiner2000RSI,Weiner2009Book,Weiner2011OC}; see Fig.~\ref{Fig:ModulationSchemes}(b). In this methodology, the spectrum of the ultrafast pulse is first resolved spatially, and each temporal frequency is addressed independently, typically via a SLM. This produces a plane-wave pulse in which all the temporal frequencies share the same spatial frequency [Fig.~\ref{Fig:PulsedPlaneWave}], or all share the same spatial profile (e.g., a pulsed optical-fiber mode). Of course, a \textit{pulsed beam} can be prepared by cascading two systems, one that sculpts the spatial profile, and the other modifies the temporal waveform [Fig.~\ref{Fig:ModulationSchemes}(c)]. The result is a conventional pulsed beam in which the spatial and temporal DoFs are separable [Fig.~\ref{Fig:TraditionalPulsedBeam}]. Therefore, this strategy can\textit{not} yield a ST wave packet, where \textit{joint} manipulation of the spatial and temporal DoFs is needed.

In general, ST wave packets require introducing precise \textit{angular dispersion} \cite{Torres2010AOP,Fulop2010Review} into a pulsed optical field; i.e., each wavelength must be directed to travel at a prescribed angle with respect to the propagation axis. The only exception is the X-wave that is entirely free of angular dispersion, and whose wavelengths all travel at a common angle. In absence of a `universal angular-dispersion synthesizer', distinct experimental configurations have consequently been proposed for each class of ST wave packets. We classify here the diverse methodologies pursued to date to synthesize propagation-invariant wave packets:

(1) \textit{Antenna arrays}: By making use of localized emitters on a spatial grid with each emitter fed with an ultrafast signal, the radiation patterns from these emitters can be designed to interfere and produce the wave packet. Despite many theoretical explorations\cite{Ziolkowski1989PRA,Ziolkowski1988SPIE,Ziolkowski1991ProcIEEE}, this strategy has not been put to test in optics, and is more suitable for radio or acoustic waves\cite{Ziolkowski1990}.

(2) \textit{Modulating aperture}: Proposals have advanced the idea of a time-varying aperture modulating an incident field \cite{Shaarawi1995JMP,Shaarawi1995JOSAA}. This strategy is unlikely to succeed in optics because of the prohibitively large modulation speeds required.

(3) \textit{Bessel-beam generators}: Since their introduction in 1987, Bessel beams have met with widespread interest, which has been facilitated by the ease of synthesizing them. Bessel-beam generators include \cite{Herman1991JOSAA}: an annular amplitude filter combined with a converging spherical lens \cite{Durnin1987PRL}; axicons or conical lenses \cite{Sonajalg1997OL,Saari1997PRL,Piche1999SPIE ,Sheppard1977Optik}; computer-generated holograms \cite{Turunen1988AO,Vasara1989JOSAA}, among others \cite{Turunen2010PO}. These systems are usually designed for monochromatic light and they do not all produce the same ST wave packet structure when fed with an ultrafast pulse. The first demonstration of an optical X-wave utilized the annular amplitude filter approach, which imposes a fixed propagation angle $\varphi_{\mathrm{x}}$ with respect to the propagation axis for all frequencies, so that $k_{x}\!\propto\!\omega$ and $k_{z}\!\propto\!\omega$. This remains a specialized approach, not suitable as a universal angular-dispersion synthesizer.

(4) \textit{Wave front shaping}: Proposals in \cite{Valtna2007OC,Turunen2010PO} suggest combining diffractive devices (such as gratings) with dispersive devices (such as axicons constructed of a material with prescribed wavelength-dependent refractive index), to construct a system endowed with a sufficient number of controllable parameters that allow tuning of the angular-dispersion profile introduced into the field. Besides the attempt at synthesizing FWMs along these general lines \cite{Reivelt2000JOSAA,Reivelt2002PRE2,Reivelt2002PRE}, this approach has not been put to test.

\begin{figure}[t!]
  \begin{center}
  \includegraphics[width=9cm]{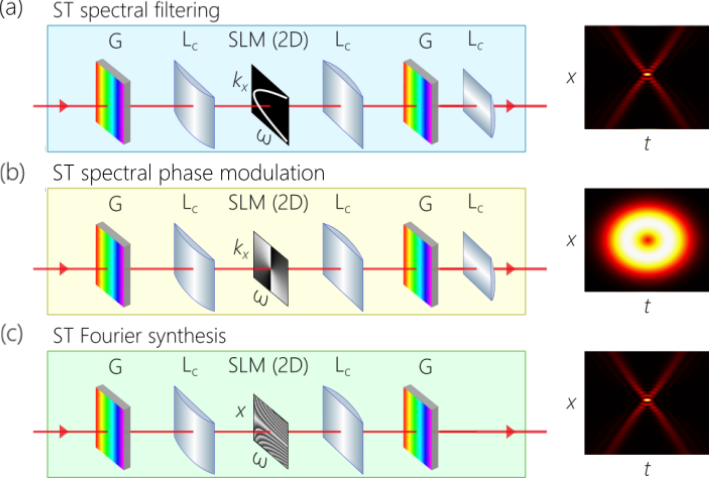}
  \end{center}
  \caption{(a) ST spectral filtering whereby an amplitude mask is placed in the spatio-temporal spectral domain $(k_{x},\lambda)$. (b) ST spectral phase modulation, wherein the amplitude mask in (a) is replaced by a phase mask. (c) ST wave packets are synthesized by combining spatial and temporal modulation in a single step; a phase mask is placed at the hybrid plane $(x,\lambda)$.}
  \label{Fig:ModulationSchemes1}
\end{figure}

(5) \textit{Spatio-temporal spectral filtering}: Because propagation invariance requires a tight association between spatial and temporal frequencies, a possible strategy is to \textit{select} the particular monochromatic plane waves identified by the spatial and temporal frequencies $(k_{x},\tfrac{\omega}{c})$ out of the separable spatio-temporal spectrum of a generic pulsed beam. This strategy can be viewed as a spatio-temporal analog of the annular-filter Bessel-beam generator [Fig.~\ref{Fig:DiffractionFreeAndPropagationInvariant}]. Starting with a pulsed plane-wave, a grating spreads the temporal spectrum in space, and an amplitude filter carves out the target spatio-temporal spectrum, before a grating reconstitutes the pulse and a Fourier transform lens converts the spatial spectrum into physical space [Fig.~\ref{Fig:ModulationSchemes1}(a)]. The plane at which the filter is placed is that of $(k_{x},\tfrac{\omega}{c})$, so that the field produced at the system exit results from a spatio-temporal Fourier transform.

This approach has been exploited for wave packets having one transverse dimension that are prepared for propagation-invariance in dispersive media \cite{Dallaire2009OE,Jedrkiewicz2013OE}, but were not successful: propagation invariance in the appropriate dispersive media was not attempted, diffraction-free behavior of the time-averaged intensity was not confirmed, and the necessary background pedestal [Fig.~\ref{Fig:StructureOfTimeAveragedIntensity}] was not observed, thus indicating the likelihood of a large spectral uncertainty $\delta\omega$ that precludes observing diffraction-free behavior. This is likely due to the finite physical size of the spatial filter utilized. Decreasing the width of the filter improves the spectral uncertainty but reduces the system throughput, and vice versa, which is the main drawback of this strategy (as with the annular spatial filter approach to generating Bessel beams). As we show below, the propagation distance depends on the absolute value of $\delta\omega$, which places a severe constraint on the energy efficiency of such an arrangement. Furthermore, this scheme cannot be extended to two transverse spatial dimensions. To the best of our knowledge, this approach has not been re-attempted since the reports in \cite{Dallaire2009OE,Jedrkiewicz2013OE}.

(6) \textit{Spatio-temporal spectral phase modulation}. This approach [Fig.~\ref{Fig:ModulationSchemes1}(b)] generalizes the conventional ultrafast pulse modulation scheme in Fig.~\ref{Fig:ModulationSchemes}(a) to the spatio-temporal domain by replacing the 1D SLM by a 2D SLM. The plane of the SLM is that of $(k_{x},\tfrac{\omega}{c})$, similarly to the ST spectral filtering technique in Fig.~\ref{Fig:ModulationSchemes1}(a), so that a lens is added to convert the field from the spectral $k_{x}$-domain to the physical $x$-domain. This approach was introduced to modulate the spatio-temporal spectral phase as a generalization of the traditional spectral phase modulation scheme used for ultrafast pulse shaping \cite{Wefers1996OL,Koehl1998OC,Feurer2002OL,Vaughan03OL}. Because this approach associates a finite spatial bandwidth to each wavelength, it does not yield propagation-invariant wave packets. Nevertheless, the authors in \cite{Vaughan03OL} demonstrated an impressive range of sophisticated spatio-temporal field structures that are very challenging to produce in any other way; see Fig.~\ref{fig:Other_groups3}(c) below. This approach has been recently revived to produce pulsed spatio-temporally structured fields in which OAM is introduced in the space-time domain rather than in the transverse spatial profile, thus leading to so-called transverse OAM \cite{Hancock2019Optica,Chong2020NP}; see Section~\ref{Sec:OpticalVortices} for details.

(7) \textit{Spatio-temporal Fourier synthesis}: The spatio-temporal spectral filtering approach highlights that the first step must be one of high-resolution spectral analysis because spectral uncertainty is the limiting factor in the performance of ST wave packets \cite{Yessenov2019OE,Bhaduri2019OL}. The missing ingredient is that the selected spatial frequency must be implemented in an energy-efficient manner by modulating the field at each wavelength spatially to produce a prescribed spatial frequency; see Fig.~\ref{Fig:ModulationSchemes1}(c). We proceed to discuss this approach in more detail.

\subsection{Spatio-temporal Fourier synthesis}\label{Section:SynthesisSetup}

\begin{figure}[t!]
  \begin{center}
  \includegraphics[width=10cm]{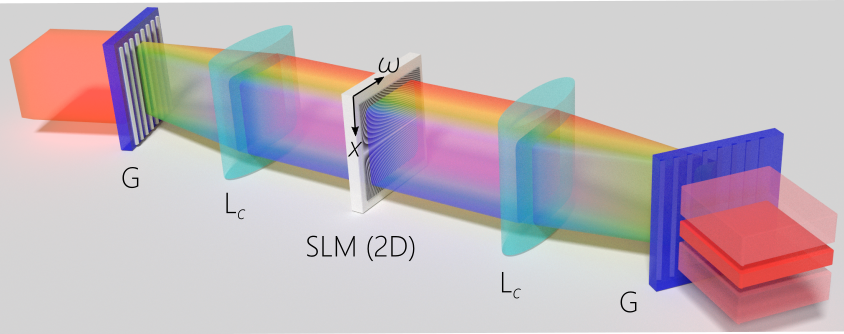}
  \end{center}
  \caption{ST wave packets synthesis setup. G: Diffraction grating; L$_{\mathrm{c}}$: cylindrical lens; SLM: spatial light modulator.}
  \label{Fig:STsetup}
\end{figure}

Because the unique properties of ST wave packets stem from the tight association between the spatial and temporal frequencies, the experimental arrangement for the spatio-temporal Fourier synthesis strategy [Fig.~\ref{Fig:ModulationSchemes1}(c)] starts from a generic pulsed beam (with separable spatial and temporal DoFs) and carefully assigns to each temporal frequency $\omega$ in the pulse a prescribed spatial frequency $k_{x}(\omega)$; see Fig.~\ref{Fig:STsetup}. This task is accomplished by the following procedure:
\begin{enumerate}
    \item \textit{High-resolution spectral analysis}. Because the spectral uncertainty is a critical parameter that must be minimized, the first step in synthesizing an ST wave packet \textit{must} be high-resolution spectral analysis, so that each wavelength can be modulated independently. Therefore, we start by analyzing the spectrum of a plane-wave pulse with a diffraction grating and then collimating the spectrally resolved spectrum with a cylindrical lens. The wavelengths are arranged continuously in columns at the focal plane of the lens. We do \textit{not} rely on the grating to introduce angular dispersion into the field as is typically done in the case of tilted pulse fronts \cite{Torres2010AOP,Fulop2010Review}. Instead, we make use of the grating to only resolve the spectrum; the spectral structure is modified in the next step.  
    \item \textit{Spatial-phase modulation of the spectrally resolved wave front}. Once the temporal spectrum has been resolved and collimated, each wavelength can be manipulated independently (within the limit set by the spectral resolution of the grating). By placing a device that spatially modulates the wave-front phase at the focal plane of the lens (e.g., a SLM \cite{Kondakci2017NP,Kondakci2019NC,Bhaduri2020NP} or phase plate \cite{Kondakci2018OE,Bhaduri2019OL,Yessenov2020OSAC}), we can impart a \textit{spatial phase distribution} to each wavelength along the direction orthogonal to the spread spectrum. It is critical to note here the distinction between this approach [Fig.~\ref{Fig:ModulationSchemes1}(c)] and the standard $4f$ pulse shaper used in ultrafast optics [Fig.~\ref{Fig:ModulationSchemes}(b)]. In the traditional pulse shaper, each wavelength is imparted a phase, and thus a 1D phase modulator suffices \cite{Weiner2000RSI,Weiner2009Book}. In our case, we provide \textit{a spatial phase pattern to each wavelength}, and thus a 2D SLM is needed. 
    \item \textit{Wave-packet reconstitution}. The wave packet is reconstituted by directing the phase-modulated spectrum to a diffraction grating identical to the one used for spectral analysis. If a reflective device imparts the spatial phase distribution, the retro-reflected field retraces its path backwards through the collimating lens to the grating, where the wavelengths are brought together to form the ST wave packet. Alternatively, for a transmissive SLM or phase plate, the spectrally resolved wave front is directed via a second lens and thence to a second diffraction grating, both identical to those in the path of the incident field [Fig.~\ref{Fig:STsetup}]. In this way, the ST wave packet is reconstituted after the second grating. Phase plates can help overcome some of the current technical limitations of SLMs: they can handle higher-power levels, can be exploited at extreme wavelengths \cite{Yessenov2020OSAC}, have superior spatial resolution (and thus can handle narrow bandwidths \cite{Kondakci2018OE}), larger numerical aperture (thus can handle large bandwidths \cite{Kondakci2018OE}), and higher diffraction efficiency.
\end{enumerate}

\begin{figure}[t!]
  \begin{center}
  \includegraphics[width=10cm]{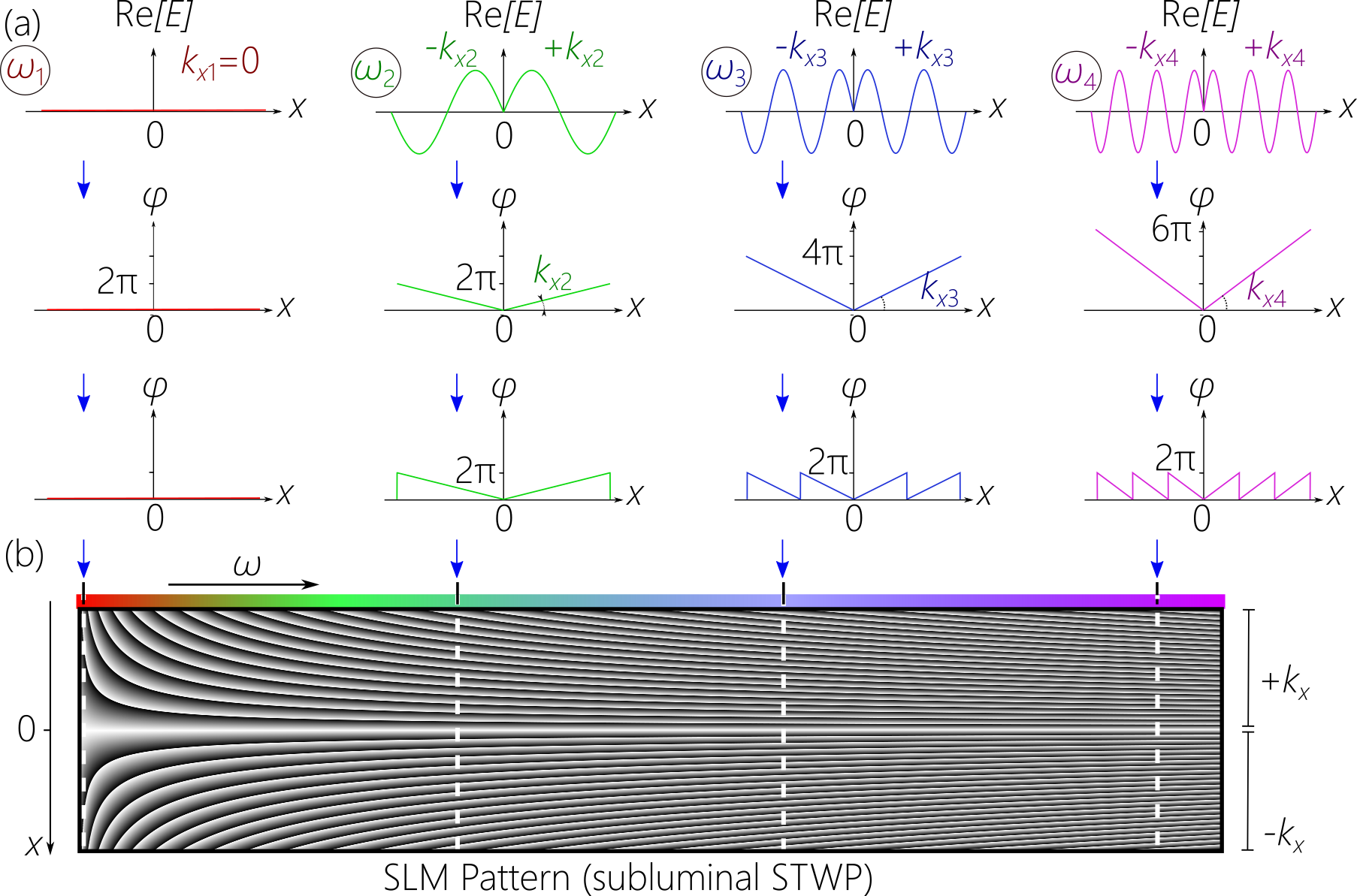}
  \end{center}
  \caption{Design of the phase pattern to synthesize ST wave packets imparted by a SLM or phase plate to the spectrally resolved wave front. (a) The real parts of the electric field $\mathrm{Re}{\{E\}}$ corresponding to pure spatial frequencies $E\!=\!e^{\pm i k_{x}x}$. We plot $\mathrm{Re}\{{e^{ik_{x}x}\}}$ when $x\!>\!0$ and $\mathrm{Re}{\{e^{-ik_{x}x}\}}$ when $x\!<\!0$; the phase distributions associated with the field, and the phase distributions in (b) after wrapping the phase to a maximum of $2\pi$. (b) The phase distributions in (a) assembled on the SLM to produce the requisite phase pattern.}
  \label{Fig:DesignOfThePhaseDistribution}
\end{figure}

To synthesize a propagation-invariant ST wave packet, the phase pattern on the SLM or phase plate for each wavelength takes the form of a linearly varying phase; i.e., a phase ramp. When the phase is implemented modulo $2\pi$, the phase modulation takes the form of a saw-tooth pattern, the slope of which determines the spatial frequency $k_{x}(\lambda)$ assigned to that particular wavelength $\lambda$; see Fig.~\ref{Fig:DesignOfThePhaseDistribution}(a). A different perspective to understand this strategy is that we \textit{deflect} each wavelength at a different angle $\varphi(\lambda)$ with respect to the initial propagation axis. From yet another perspective, the linear phase distribution corresponding to a particular spatial frequency imparted to each wavelength is reminiscent of the action of a grating. However, whereas a grating provides a \textit{fixed} change in transverse wave number to \textit{all} wavelengths, the arrangement described here provides a \textit{different} change in transverse wave number to each wavelength. Such a system therefore corresponds to a grating with a wavelength-dependent ruling period. Equivalently, we provide a separate grating to each wavelength. Consequently, this experimental strategy provides a much larger number of parameters that can be tuned to sculpt the targeted ST wave packets. Furthermore, symmetric deflection angles $\pm\varphi(\lambda)$ can be imparted to each wavelength, in contrast to a grating or a prism that provides a single deflection angle. This is achieved by dividing the SLM area into two halves, with each column providing symmetric angles $\varphi(\lambda)$ and $-\varphi(\lambda)$; see Fig.~\ref{Fig:DesignOfThePhaseDistribution}(b).

\subsection{Characterization of ST wave packets}

\begin{figure}
    \centering
    \includegraphics[width=11cm] {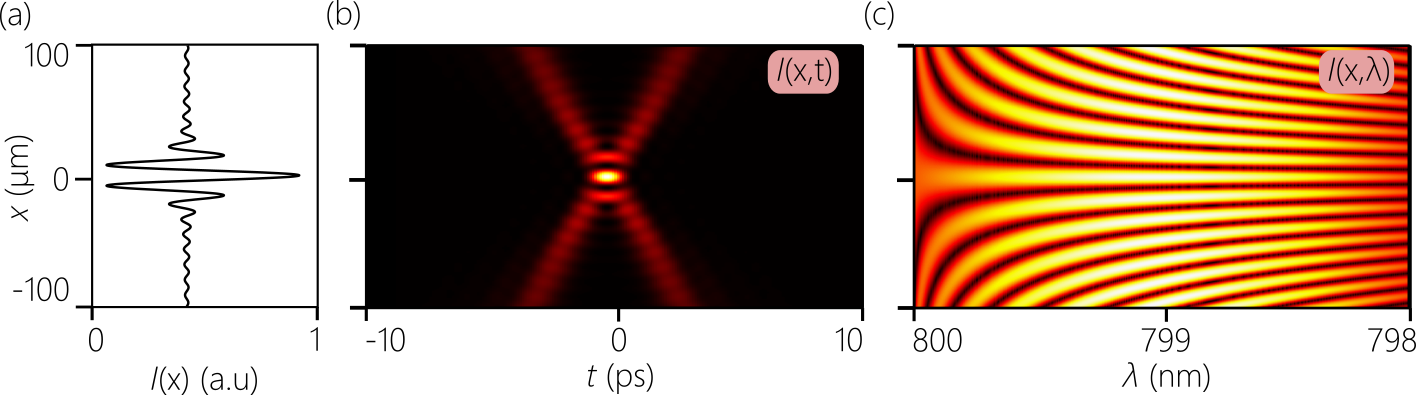} 
    \caption{Comparison of (a) the time-averaged intensity $I(x)$, (b) the spatio-temporal profile $I(x;t)$, and (c) the spatially resolved spectrum $I(x,\lambda)$, all at a fixed axial plane $z$ for a ST wave packet having $\theta\!=\!50^{\circ}$, $\lambda_{\mathrm{o}}\!=\!800$~nm, and bandwidth $\Delta\lambda\!=\!2$~nm.  }
    \label{fig:spectrum}
\end{figure}

\begin{figure}[t!]
  \begin{center}
  \includegraphics[width=10cm]{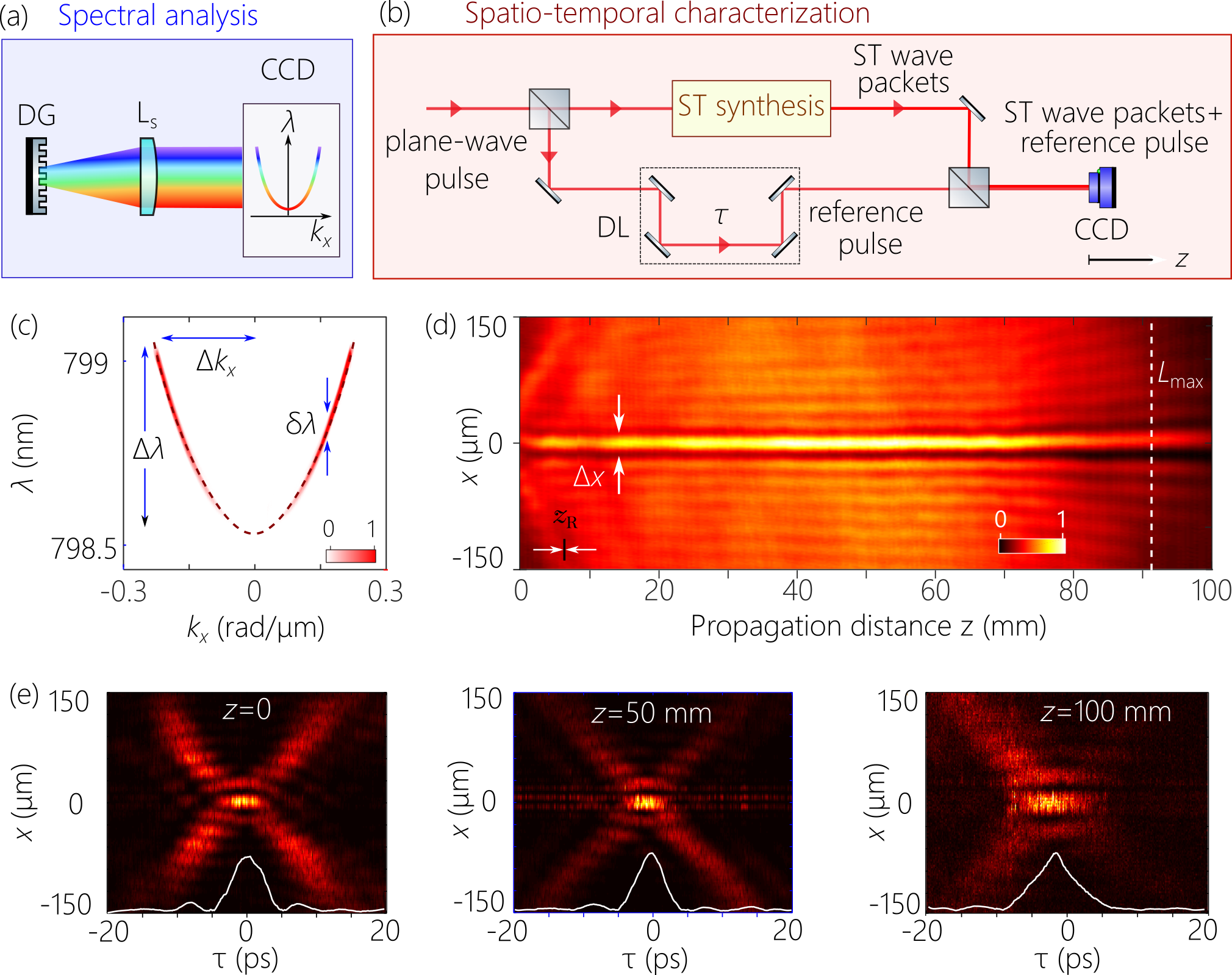}
  \end{center}
  \caption{(a) A spatio-temporal spectral analysis setup extracts the spectral projection onto the $(k_{x},\lambda)$-plane. (b) The configuration for characterization of spatio-temporal intensity of ST wave packets. The initial laser pulse traverses the reference arm. (c) Measured spatio-temporal spectral intensity $|\widetilde{\psi}(k_{x},\lambda)|^{2}$ for a subluminal ($\theta\!=\!35^{\circ}$; a section from an ellipse) ST wave packet. The spectrum appears approximately as a parabola because of the limited bandwidth ($\Delta\lambda\!\approx\!0.4$~nm). (d) Time-averaged intensity $I(x,z)$ of the subluminal ST wave packet from (c). (e) Time-resolved intensity $I(x,z,\tau)$ at $z\!=\!0$, 50, and 100~mm.}
  \label{Fig:DataFigure}
\end{figure}

ST wave packets can be characterized in three different domains to ascertain that the desired structure has been synthesized and that the targeted behavior has been realized: the time-averaged intensity [Fig.~\ref{fig:spectrum}(a)], the spatio-temporal profile [Fig.~\ref{fig:spectrum}(b)], and the spectra domain [Fig.~\ref{fig:spectrum}(c)]. It is crucial that any report of a ST wave packet combine all three measurements (although some may be more challenging than others) to benchmark and certify their characteristics, and thus systematize the comparison between different realizations.
\begin{enumerate}
    \item \textit{Spectral domain}. Measuring the overall spectrum of the ST wave packet (integrated over its area) does \textit{not} suffice to characterize the spectral structure of the wave packet. Indeed, the spectrum changes subtly across its spatial profile, as shown in Fig.~\ref{fig:spectrum}(c). It is therefore imperative to examine the spatio-temporal spectral structure by performing a double Fourier transform: a temporal Fourier transform using a grating that resolves the temporal spectrum, and a spatial Fourier transform using a lens in a $2f$-configuration that resolves the spatial spectrum [Fig.~\ref{Fig:DataFigure}(a)]. From this spatio-temporal spectrum projected onto the $(k_{x},\lambda)$-plane, we can immediately identify the temporal bandwidth $\Delta\lambda$, the spatial bandwidth $\Delta k_{x}$, and the spectral uncertainty $\delta\lambda$ (if sufficient spectral resolution is available in the measurement device) [Fig.~\ref{Fig:DataFigure}(c)]. Furthermore, the sign of the spectral curvature identifies whether the wave packet is subluminal ($\theta\!<\!45^{\circ}$) or superluminal ($\theta\!>\!45^{\circ}$), and the magnitude of the curvature helps identify the group velocity $\widetilde{v}\!=\!c\tan{\theta}$. This spectral analysis has been performed routinely in experiments generating ST wave packets via nonlinear optical interactions \cite{DiTrapani2003PRL,Conti2003PRL} and more recently in producing ST wave packets via spatio-temporal Fourier synthesis \cite{Kondakci2017NP,Kondakci2019NC,Yessenov2020NC}.
    \item \textit{Intensity domain}. A detector that is slow with respect to the pulsewidth, such as a CCD camera, records only the time-averaged intensity $I(x,z)\!=\!\int\!dt|\psi(x,z;t)|^{2}$, which corresponds to the wave-packet energy. Although such a measurement does not resolve the spatio-temporal profile of the wave packet, it can nevertheless help estimate its propagation distance. By axially scanning a CCD camera along $z$ [Fig.~\ref{Fig:DataFigure}(b) after blocking the reference arm], we can confirm the diffraction-free behavior and also measure the background pedestal height as an indicator of the spectral uncertainty; see Fig.~\ref{Fig:DataFigure}(d). Such measurements have been routinely performed for linear ST wave packets, but rarely for nonlinear X-waves. Combined with the spatio-temporal spectral measurements, they provide a wealth of information about the ST wave packet.
    \item \textit{Spatio-temporal domain}. Obtaining the spatio-temporal field or intensity profiles is typically the most challenging measurement domain. Several approaches have been followed so far. One simple strategy is off-axis holography or linear interferometry whereby the wave packet is made to interfere with a plane-wave pulse, one that is usually derived from the same source from which the ST wave packet itself is derived [Fig.~\ref{Fig:DataFigure}(b)]. Conceptually, the ST synthesis system is placed in one arm of a two-path interferometer, while the initial plane-wave pulse travels in a reference arm containing an optical delay. When the two wave packets overlap in space and time, spatially resolved fringes are observed, from whose visibility the wave-packet envelope can be reconstructed. Of course, this approach reveals only the intensity profile, and the pulse phase goes undetected, but this is usually sufficient to evaluate the wave packet profile; see Fig.~\ref{Fig:DataFigure}(e). An alternative approach relies on spectral interferometry \cite{Dallaire2009OE,Kondakci2017NP}, and full-field measurements have also been carried out \cite{Bowlan2009OL}. These measurements have been routinely performed for linear ST wave packets but rarely for their nonlinear counterparts. These measurements are also required to estimate the group velocity directly in physical space, which can then corroborate the value estimated from the curvature of the spatio-temporal spectrum.
\end{enumerate}

As mentioned above, it is crucial that all three measurements be combined as a benchmark for evaluating the synthesis efficacy and verifying the properties of the ST wave packets. More advanced characterization tools can be of course employed to extract the complex (magnitude and phase) spatio-temporal \cite{Akturk2010JO,Dorrer2019JSTQE,Alonso2020CP,Jolly2020JO} or spatio-spectral \cite{Borot2018OE,Jolly2021OSAC} field configuration.
 
\section{Propagation characteristics}\label{Sec:PropCharacteristics}

In this Section we focus on the characteristics of freely propagating baseband ST wave packets, highlighting their field structure and propagation invariance, the tunability of their group velocity, their diffraction-free propagation distance in relationship to the spectral uncertainty, and their self-healing properties.

\subsection{Propagation invariance and time diffraction}\label{Sec:PropInvarianceAiry}

The propagation invariance of ST wave packets has been confirmed \cite{Kondakci2017NP} using the synthesis and characterization configurations described in the previous Section. In the first experimental observation of a baseband ST wave packet, the time-averaged intensity display spatial widths of $\Delta x\!=\!14$~$\mu$m ($\theta\!=\!53^{\circ}$) and $\Delta x\!=\!7$~$\mu$m ($\theta\!=\!90^{\circ}$) were 100~mm and 25~mm, respectively. The spatio-temporal profile was obtained via spectral interferometry \cite{Dallaire2009OE}, but linear interferometry was subsequently adopted as a measurement approach \cite{Kondakci2019NC,Bhaduri2019Optica}. An example showing the recommended characterization of a ST wave packet is shown in Fig.~\ref{Fig:DataFigure} for a spectral tilt angle $\theta\!=\!35^{\circ}$. From the spatio-temporal spectrum in Fig.~\ref{Fig:DataFigure}(c) we have $\Delta\lambda\!=\!0.4$~nm and $\Delta k_{x}\!=\!0.2$~rad/$\mu$m, which point to a transform-limited on-axis pulse of width $\Delta\tau\!=\!4.3$~ps and beam width $\Delta x\!=\!15.6$~$\mu$m at the pulse center. The \textit{sign} of the curvature implies that the ST wave packet is subluminal (the opposite curvature implies that the ST wave packet is superluminal); the \textit{magnitude} of the curvature is consistent with $\theta\!=\!35^{\circ}$ and thus $\widetilde{v}\!\approx\!0.7c$; and the `thickness' of the spatio-temporal spectrum is the spectral uncertainty $\delta\lambda$.

The time-averaged intensity $I(x,z)$ [Fig.~\ref{Fig:DataFigure}(d)] shows the diffraction-free distance along which the underlying spatio-temporal intensity profile is maintained. In general, the transverse spatial profile remains invariant until some distance $L_{\mathrm{max}}$ is reached, after which the profile spreads rapidly \cite{Kondakci2016OE}. As we show below, $L_{\mathrm{max}}$ is related to the spectral uncertainty $\delta\lambda$ and the spectral tilt angle $\theta$. The spatio-temporal profile is reconstructed via linear interferometry [Fig.~\ref{Fig:DataFigure}(e)]. The on-axis pulsewidth $\Delta\tau\!=\!4.3$~ps and beam width at the pulse center is $\Delta x\!=\!15.6$~$\mu$m, both of which agree with the temporal and spatial bandwidths in Fig.~\ref{Fig:DataFigure}(c). The wave packet profile deforms once the diffraction-free length is exceeded. Finally, $\widetilde{v}$ is estimated by measuring the group delay accrued by the ST wave packet with $z$. The value of $\widetilde{v}$ obtained in this manner is in agreement with that predicted based on the spectral tilt angle $\theta$ extracted from Fig.~\ref{Fig:DataFigure}(c).

A useful perspective for understanding the propagation invariance of ST wave packets is that of `time diffraction' \cite{Moshinsky1952,Longhi2004OE,Porras2017OL}. Within this framework, the \textit{spatio-temporal} profile $I(x,0;t)$ is a scaled version of the \textit{spatial} diffraction pattern associated with a monochromatic beam sharing the same spatial bandwidth \cite{Longhi2004OE,Porras2017OL,Kondakci2018PRL}. Consider a monochromatic beam at $\omega\!=\!\omega_{\mathrm{o}}$ having a spatial bandwidth $\widetilde{\psi}(k_{x})$, whereupon its spatial profile evolves as follows in the paraxial regime: $\psi(x,z)\!=\!\int\!dk_{x}\widetilde{\psi}(k_{x})e^{ik_{x}x}\exp{\{-i\tfrac{k_{x}^{2}}{2k_{\mathrm{o}}}z\}}$, whereas for a ST wave packet we have $\psi(x,0;t)\!=\!\int\!dk_{x}\widetilde{\psi}(k_{x})e^{ik_{x}x}\exp{\{-i\tfrac{k_{x}^{2}}{2k_{\mathrm{o}}}\tfrac{ct}{1-\widetilde{n}}\}}$, so that $z\!\rightarrow\!\tfrac{ct}{1-\widetilde{n}}$; and when $\theta\!=\!90^{\circ}$, $\widetilde{n}\!\rightarrow\!0$ and $z\!\rightarrow\!ct$.

\begin{figure}[t!]
  \begin{center}
  \includegraphics[width=9cm]{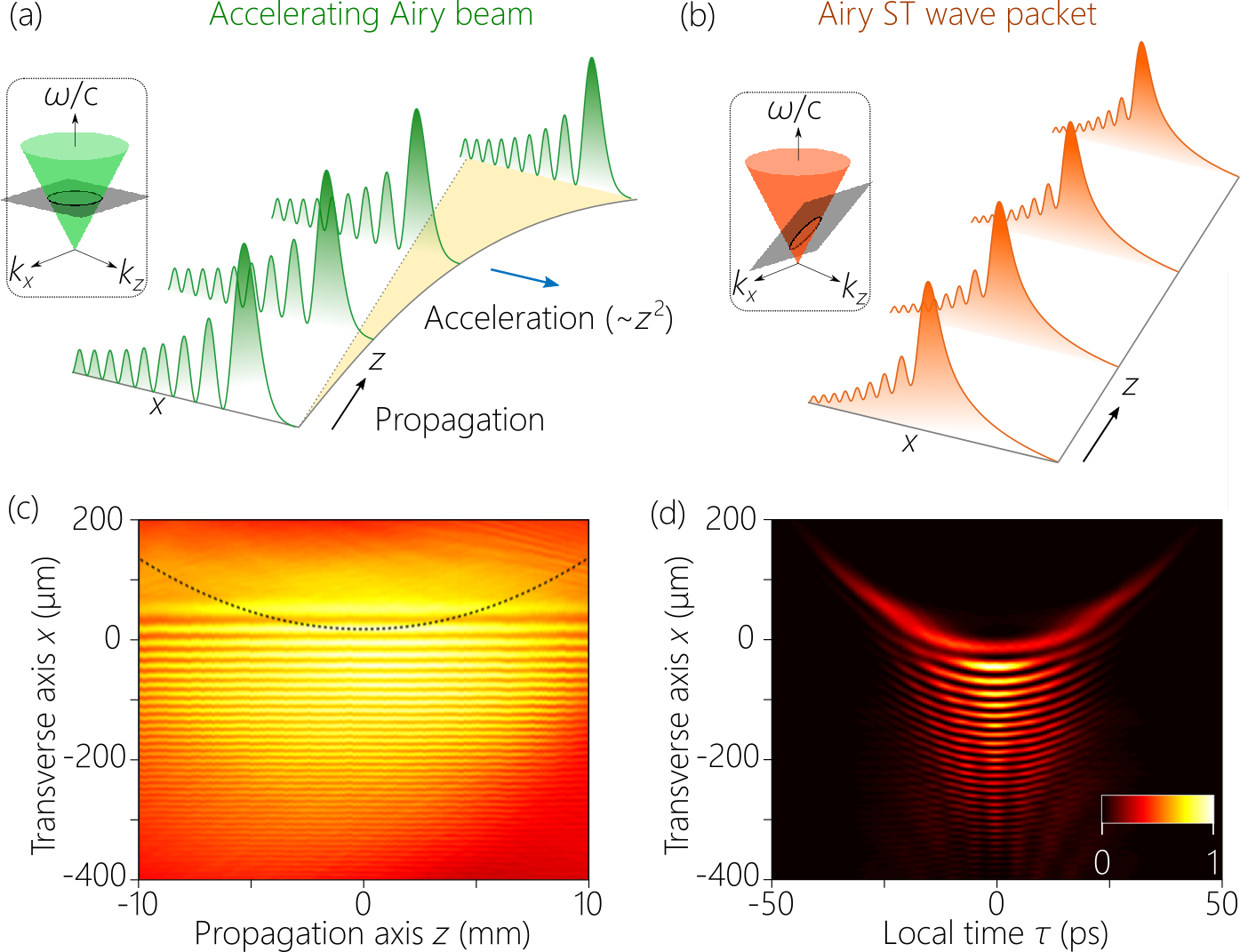}
  \end{center}
  \caption{(a) Intensity of a monochromatic Airy beam accelerating along a parabolic trajectory in the $(x,z)$-plane. The spectrum lies at the intersection of the light-cone with a horizontal iso-$\omega$-plane (inset). (b) The spatio-temporal spectrum of a ST Airy wave packet is at the intersection of the light-cone with a plane. The ST Airy wave packet does not exhibit the expected acceleration and instead travels in a straight line. We plot here the time-averaged intensity. (c) Measured transverse time-averaged intensity profile along the $z$-axis. The dashed line corresponds to the parabolic trajectory of a monochromatic Airy beam having the same beam waist. (d) Measured spatio-temporal intensity profile obtained at a fixed $z$; see \cite{Kondakci2018PRL}.}
  \label{Fig:AiryST}
\end{figure}

A particularly intriguing profile that brings out the unique aspects of this concept is the Airy beam. As discussed in Section~\ref{Section:DiffractionFree1D}, the Airy beam is the only diffraction-free 1D beam, except that its entire profile follows a parabolic trajectory with free propagation rather than a straight line [Fig.~\ref{Fig:AiryST}(a)]. In that sense, such a beam is not strictly speaking propagation invariant. Now, consider using the spatial spectrum of the Airy beam to construct a ST wave packet in which we then associate with each spatial frequency $k_{x}$ a single $\omega$ according to Eq.~\ref{Eq:BasebandParabola} to produce an Airy ST wave packet [Fig.~\ref{Fig:AiryST}(b)]. At $t\!=\!0$ the profile $I(x,0;0)$ is that of an Airy beam. According to the discussion above concerning time diffraction, the spatio-temporal profile at a fixed axial plane $z$ (say $z\!=\!0$) should take the form of the curved trajectory of the Airy beam -- but in the $(x,t)$ domain [Fig.~\ref{Fig:AiryST}(d)] rather than in physical space $(x,z)$ [Fig.~\ref{Fig:AiryST}(c)]. That is, the Airy ST wave packet accelerates in the local frame of the ST wave packet, but the time-averaged intensity reveals no sign of this acceleration. Instead, the intensity profile is diffraction-free and propagates in a straight line [Fig.~\ref{Fig:AiryST}(c)]. In other words, the spatio-temporal coupling underlying the ST wave packet arrested the acceleration in physical space and transformed it into the space-time domain. Consequently, the Airy ST wave packet has been `straightened out' \cite{Saenz2017NP} and now travels in a straight line while accelerating in the local frame of the wave packet.

Note that the spatio-temporal profile here does \textit{not} take the usual X-shaped form as a consequence of the cubic spectral phase introduced to produce the Airy structure. This is a first example of the recently developed generalized notion of propagation-invariant space-time caustics \cite{Wong21OE}.

\subsection{Tunable group velocity}

Previous attempts at tuning the group velocity of so-called localized waves produced only minute deviations from $c$. Although $\widetilde{v}$ for an X-wave can -- in principle -- take on arbitrary superluminal values, the observed values have been $1.00022c$ \cite{Bonaretti2009OE}, $1.00012c$ \cite{Bowlan2009OL}, and$1.00015c$ \cite{Kuntz2009PRA}. As pointed out in \cite{Turunen2010PO}, there has been much less work done historically on subluminal propagation-invariant wave packets in comparison to their superluminal counterparts. Indeed, there are no experimental reports of their synthesis prior to our recent work in \cite{Kondakci2017NP,Kondakci2019NC,Yessenov2019OE,Yessenov2019PRA}.  Measurements of diffracted pulses that are \textit{not} propagation invariant revealed $\widetilde{v}\!\approx\!0.999c$ \cite{Lohmus2012OL,Piksarv2012OE}, and recent work with broadband Bessel beams reported $\widetilde{v}\!=\!0.99999c$ \cite{Giovannini2015Science}, which are not propagation invariant; see also \cite{Bouchard2016Optica,Lyons2018Optica}. Furthermore, there have been no previous attempts at producing negative-$\widetilde{v}$ wave packets.

\begin{figure}[t!]
  \begin{center}
  \includegraphics[width=10cm]{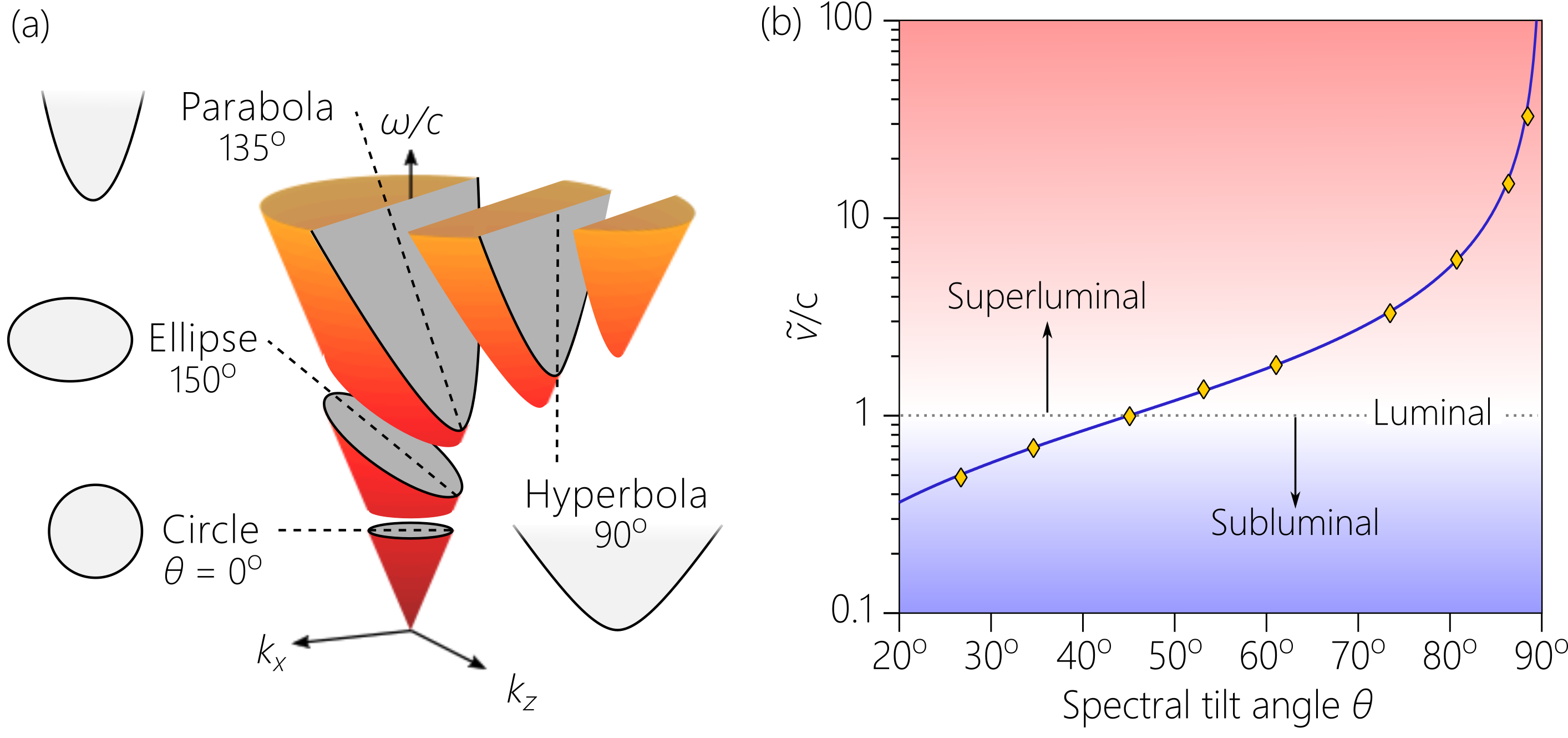}
  \end{center}
  \caption{Tuning the group velocity of a ST wave packet. (a) Changing the spectral tilt angle $\theta$ changes the conic section at the intersection of the light-cone with a plane making an angle $\theta$ with the $k_{z}$-axis, and thus changes the group velocity $\widetilde{v}\!=\!c\tan{\theta}$; see \cite{Yessenov2019OPN}. (b) Measured $\widetilde{v}$ while varying $\theta$. The data fits the predicted $\widetilde{v}\!=\!c\tan{\theta}$ dependence; see \cite{Kondakci2019NC}.}
  \label{Fig:TunableGroupVelocity}
\end{figure}

In contrast, baseband ST wave packets offer unique characteristics as a platform for varying $\widetilde{v}$. Crucially, ST wave packets are agnostic with respect to the subluminal, superluminal, and negative-$\widetilde{v}$ regimes. All that is needed is to change the spatio-temporal spectral structure to tune $\theta$ [Fig.~\ref{Fig:TunableGroupVelocity}(a)] by varying the phase distribution on the SLM in the spatio-temporal Fourier synthesis approach [Fig.~\ref{Fig:DesignOfThePhaseDistribution}]. As a result, $\widetilde{v}$ can be tuned over a very wide range from $30c$ to $-4c$ in \cite{Kondakci2017NP}; see Fig.~\ref{Fig:TunableGroupVelocity}(b). This is a major departure from previous reports. First, only a single setup is needed to span these values of $\widetilde{v}$, whereas different setups were proposed for the distinct regimes of $\widetilde{v}$. Second, the span of values exceeds previous reports by at least 4--5 orders of magnitude. Third, this tuning range is independent of beam size or pulsewidth, and the velocity can be tuned continuously while remaining in the paraxial regime.

\begin{figure}[t!]
  \begin{center}
  \includegraphics[width=11cm]{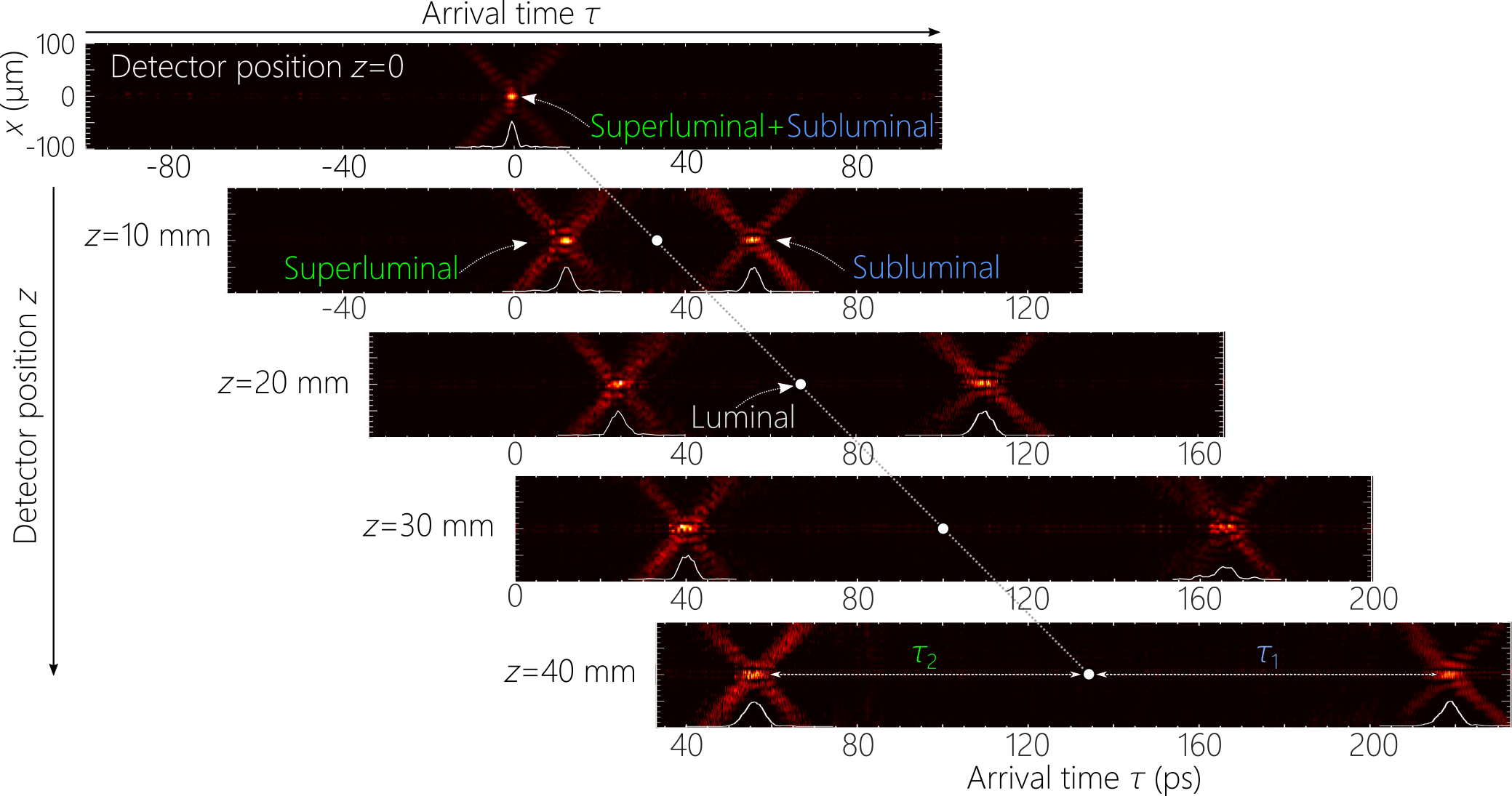}
  \end{center}
  \caption{Propagation of two ST wave packets in free space at different group velocities: a subluminal wave packet with $\theta_{1}\!=\!30^{\circ}$ ($\widetilde{v}\!\approx\!0.577c$) and a superluminal wave packet with $\theta_{2}\!=\!70^{\circ}$ ($\widetilde{v}\!\approx\!2.75c$). The panels display the arrival times of the two wave packets at different axial planes $z$. At $z\!=\!0$, the two wave packets overlap in space and time; at $z\!=\!40$~mm, the superluminal wave packet arrives at $\tau_1\!\approx\!56$~ps while the subluminal wave packet arrives at $\tau_2\!\approx\!218$~ps; see \cite{Yessenov2020NC}.}
  \label{Fig:NCPropagationDistnace}
\end{figure}

Figure~\ref{Fig:NCPropagationDistnace} shows a revealing example that highlights this unique feature. Two wave packets are prepared in different but adjacent spectral windows that are equal in bandwidth. The spatio-temporal profiles are initially identical and are launched while overlapping in space and time at $z\!=\!0$. Nevertheless, because each wave packet has a different group velocity (one subluminal and the other superluminal), a relative group delay develops between them as they travel along $z$ in free space, so that they no longer overlap as they did at $z\!=\!0$. Indeed, the maximum relative group delay that the two wave packets can sustain before they deform is much larger than the on-axis pulse width of either wave packet. 

\subsection{Propagation distance and the spectral uncertainty}\label{Section:PropagationDistance}

We have described so far an idealized scenario where a `delta-function' association is assumed between the spatial and temporal frequencies; that is, the spatio-temporal spectrum is restricted as follows: $\widetilde{\psi}(k_{x},\Omega)\!\rightarrow\!\widetilde{\psi}(k_{x})\delta(\Omega-\Omega(k_{x}))$. Each spatial frequency $\pm k_{x}$ is assigned \textit{exactly} to a single temporal frequency $\Omega(k_{x})$, and the wave packet propagates invariantly for an indefinite distance. However, such idealized pulsed-field configurations are unattainable in practice because they have infinite energy \cite{Sezginer1985JAP}. This conclusion applies to all ideal ST wave packets including X-waves, baseband, and sideband ST wave packets. In realistic fields produced by finite resources, the delta-function association is relaxed, and we have instead $\widetilde{\psi}(k_{x},\Omega)\!\rightarrow\!\widetilde{\psi}(k_{x})h(\Omega-\Omega(k_{x}))$, where $h$ is a narrow function of spectral width $\delta\omega$, which we call the \textit{spectral uncertainty}. By reducing $\delta\omega$, the ideal field structure is approached, whereas increasing $\delta\omega$ towards the full temporal bandwidth $\Delta\omega$ results in a conventional pulsed beam that is separable with respect to space and time.

The sources of spectral uncertainty differ from one configuration to another. In our spatio-temporal Fourier synthesis approach, there are three main factors contributing to $\delta\omega$: (1) The dominant factor is the spectral resolution of the grating, which sets a baseline for $\delta\omega$. The spectral resolution from the grating is determined by the product of the grating width and the ruling density. We therefore expand the initial beam using a telescope to fill the grating aperture. (2) The apertures in the system (e.g., the lens aperture) increase $\delta\omega$ as a result of diffractive coupling between $\omega$ and $k_{x}$. (3) The pixel size of the SLM or any other device used to modulate the wave front sets a lower limit on the bandwidth that can be associated with each $k_{x}$. This limit can be improved upon by increasing the focal length of the collimating lens between the grating and the SLM.

\begin{figure}[t!]
  \begin{center}
  \includegraphics[width=11cm]{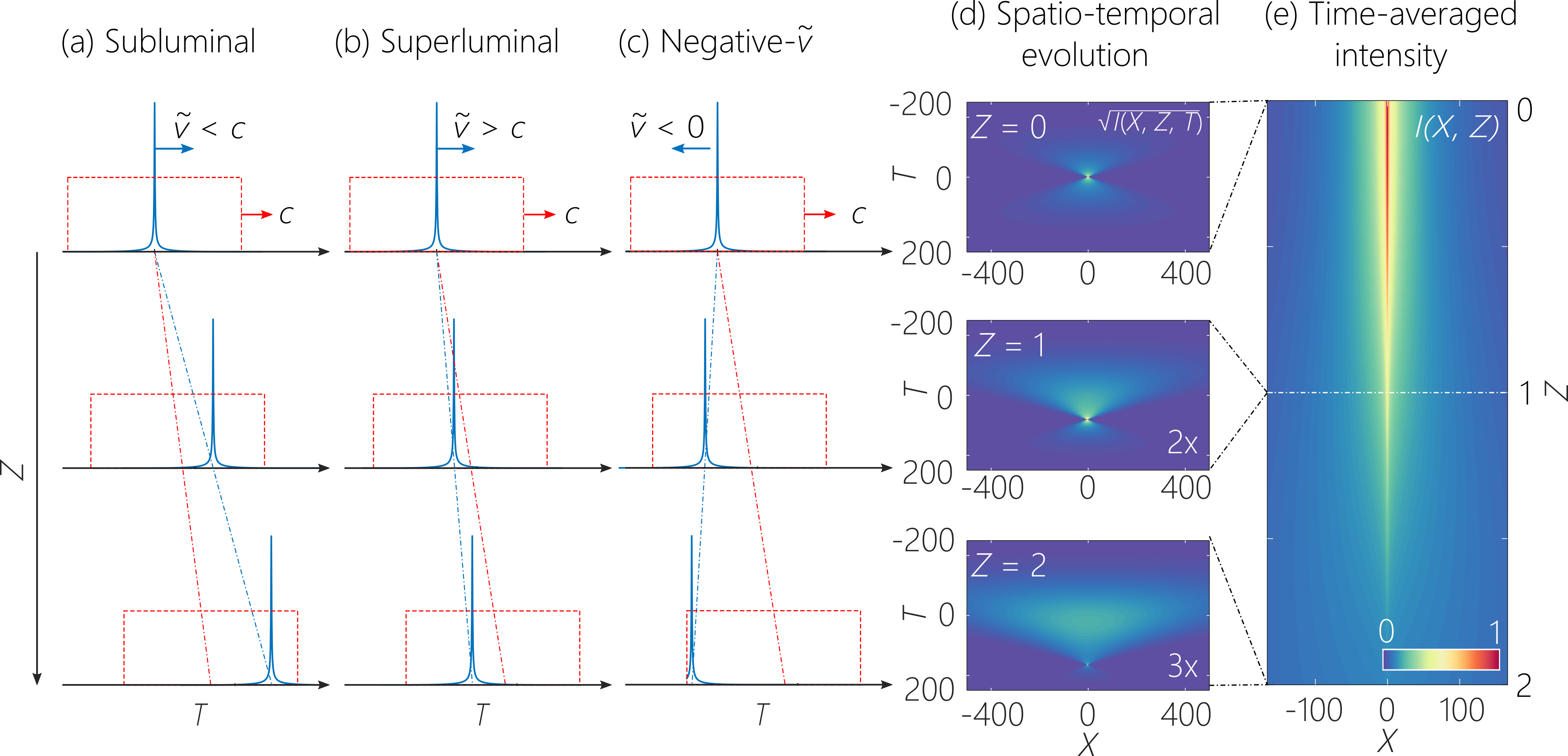}
  \end{center}
  \caption{A finite-energy ST wave packet is the product of an ideal ST wave packet (narrow solid blue pulse) traveling at $\widetilde{v}$ and a pilot envelope (wide dashed red pulse) traveling at $c$: (a) superluminal; (b) superluminal; and (c) negative-$\widetilde{v}$. (d) The spatio-temporal intensity profile at different $z$ showing evolution from a symmetric to an asymmetric wave packet. (e) The time-averaged intensity profile $I(x,z)$, showing the axial locations where the snapshots in (d) are calculated; see \cite{Yessenov2019OE}.}
  \label{Fig:Pilot}
\end{figure}

By including the finite spectral uncertainty $\delta\omega$, the energy of the ST wave packet is rendered finite. However, the wave packet is no longer strictly propagation-invariant, and its profile instead changes slowly for some finite distance, followed by rapid broadening. We define the propagation distance $L_{\mathrm{max}}$ as the axial length after which the on-axis ($x\!=\!0$) time-averaged intensity drops to half its initial value: $I(0,L_{\mathrm{max}})\!=\!\tfrac{1}{2}I(0,0)$. An expression for $L_{\mathrm{max}}$ can be derived from a simple model of the field of a baseband ST wave packet. We make use of a Gaussian spatial spectrum $\widetilde{\psi}(k_{x})\!\propto\!\exp(-\tfrac{k_{x}^{2}}{2(\Delta k_{x})^{2}})$ and a Gaussian function for the spectral uncertainty function $h(k_{x},\Omega)\!\propto\!\exp(-\tfrac{\Omega^{2}}{2(\delta\omega)^{2}})$, and obtain a spatio-temporal intensity profile $I(x,z;t)\!=\!|\psi(x,z;t)|^{2}\!=\!I_{\mathrm{ST}}(x,z;t)I_{\mathrm{p}}(x,z;t)$ that is given by a product of two terms: (1) an ideal ST wave packet that is indefinitely propagation invariant, travels at a group velocity $\widetilde{v}$, has an on-axis pulsewidth $\Delta\tau_{\mathrm{ST}}\!\sim\!\tfrac{1}{\Delta\omega}$, and is given by: 
\begin{equation}
I_{\mathrm{ST}}(x,z;t)=\frac{2\sqrt{\pi}\Delta k_{x}}{\sqrt{1+[\Delta\omega(t-z/\widetilde{v})]^{2}}}\exp{\left\{-\frac{x^{2}(\Delta k_{x})^{2}}{1+[\Delta\omega(t-z/\widetilde{v})]^{2}}\right\}}=I_{\mathrm{ST}}(x,0;t-z/\widetilde{v});
\end{equation}
and (2) a broad plane-wave pulse $I_{\mathrm{p}}(x,z;t)$ of temporal linewidth $\Delta\tau_{\mathrm{p}}\!\sim\!\tfrac{1}{\delta\omega}$, propagating at a group velocity $c$, and given by:
\begin{equation}
I_{\mathrm{p}}(x,z;t)=2\sqrt{\pi}\delta\omega\exp{\left\{-\left(t-\frac{z}{c}\right)^{2}(\delta\omega)^{2}\right\}}=I_{\mathrm{p}}(x,0;t-z/c).
\end{equation}

Figure~\ref{Fig:Pilot} provides an illustration of the relationship between the pilot envelope $I_{\mathrm{p}}$ that propagates at $c$ and the ideal ST wave packet $I_{\mathrm{ST}}$ that propagates at $\widetilde{v}$. In the subluminal regime, the ST wave packet travels slower than the pilot whose trailing edge catches up with it [Fig.~\ref{Fig:Pilot}(a)]. In the superluminal [Fig.~\ref{Fig:Pilot}(b)] and negative-$\widetilde{v}$ [Fig.~\ref{Fig:Pilot}(c)] regimes, the ST wave packet catches up with the leading edge of the pilot. In all cases, the ST wave packet cannot exceed the boundaries set by the broad pilot envelope without deformation [Fig.~\ref{Fig:Pilot}(d,e)]. Because the pilot envelope travels at a group velocity $c$, the ST wave packet cannot convey information from the source faster than $c$ regardless of the value of $\widetilde{v}$. There is a built-in latency in the synthesis system with regards to the emergence of the leading edge of the pilot envelope, which cannot be overtaken by the ST wave packet. See also the analysis in \cite{Shaarawi2000JPA,Saari2018PRA,Saari2019PRAenergyflow} that confirm from different perspectives that superluminal and even negative-valued $\widetilde{v}$ do not violate relativistic causality.

\begin{figure}[t!]
  \begin{center}
  \includegraphics[width=11cm]{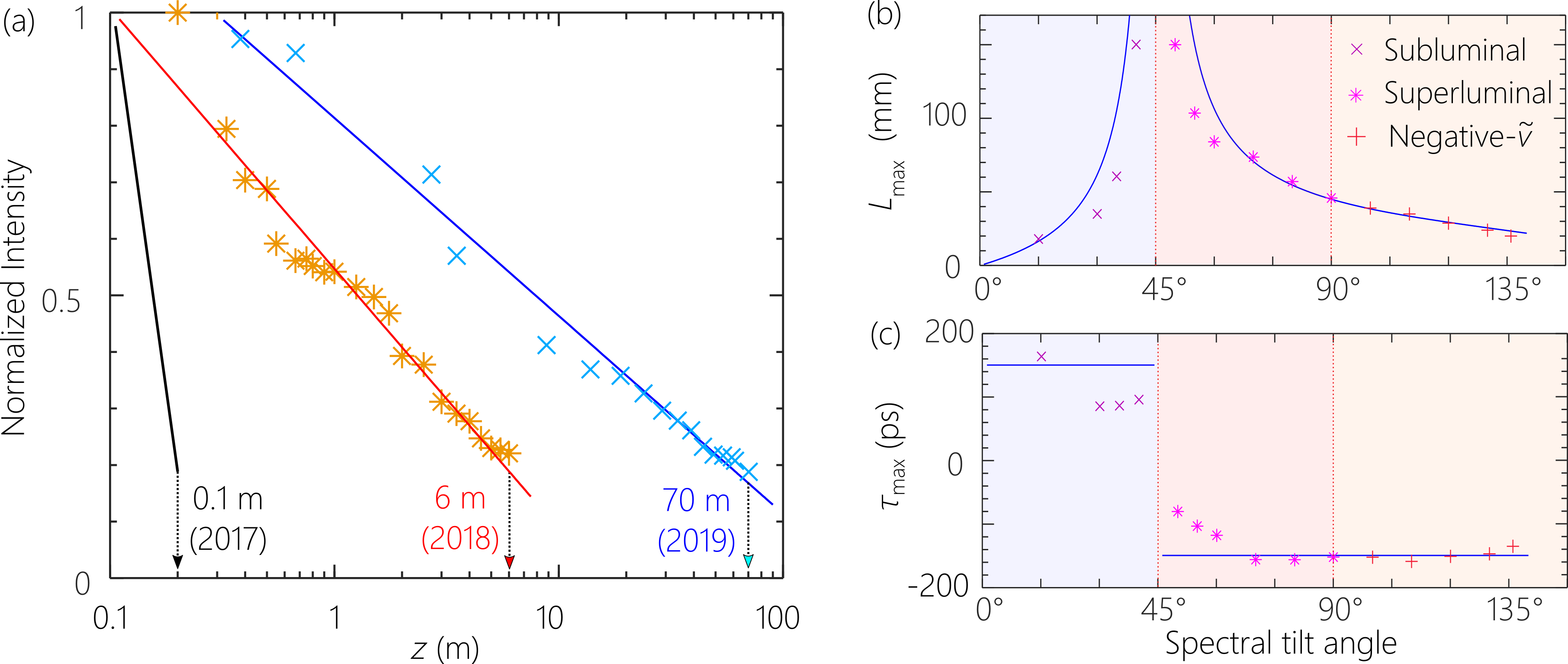}
  \end{center}
  \caption{Propagation distance for ST wave packets. (a) Measured propagation distance $L_{\mathrm{m}}$ for different $\theta$ at fixed $\delta\lambda\approx14$~pm. Theoretical curve is $L_{\mathrm{max}}\!=\!L_{\mathrm{p}}/|1-\cot{\theta}|$, with $L_{\mathrm{p}}\!=\!45$~mm; see \cite{Bhaduri2018OE,Bhaduri2019OL}. (b) Maximum DGD as a function of $\theta$ at fixed $\delta\lambda\approx14$~pm; see \cite{Yessenov2019OE}.}
  \label{Fig:PropagationDistance}
\end{figure}

From the above, we obtain a formula for the maximum propagation distance $L_{\mathrm{max}}$:
\begin{equation}\label{Eq:LMax}
L_{\mathrm{max}}\sim\frac{c}{\delta\omega}\frac{1}{|1-\cot{\theta}|}.
\end{equation}
That is, $L_{\mathrm{max}}$ decreases as $\theta$ deviates away from $45^{\circ}$ (the luminal condition), and increases rapidly as $\theta\!\rightarrow\!45^{\circ}$. The demonstration in \cite{Bhaduri2018OE} made use of $\theta\!=\!44.98^{\circ}$ for $\sim\!6$~m and  $\theta\!=\!45.00011^{\circ}$ in \cite{Bhaduri2019OL} for $\sim\!70$~m; see Fig.~\ref{Fig:PropagationDistance}(a). It is important to note that $\theta$ is \textit{not} an angle in physical space. It simply determines the rate of change of $k_{x}$ with $\omega$, and $\theta\!\rightarrow\!45^{\circ}$ merely indicates that $k_{x}$ changes slowly with $\omega$ \cite{Bhaduri2018OE}. The critical feature of Eq.~\ref{Eq:LMax} for $L_{\mathrm{max}}$ that has not been sufficiently appreciated, is that $L_{\mathrm{max}}$ depends on the absolute value of $\delta\omega$ and \textit{not} on its ratio to the bandwidth $\Delta\omega$. Increasing the temporal bandwidth $\Delta\omega$ increases the accompanying spatial bandwidth $\Delta k_{x}$ and thus reduces the transverse spatial width $\Delta x$ of the wave packet profile, but does \textit{not} change $L_{\mathrm{max}}$. In that sense, the enhancement ratio of $L_{\mathrm{max}}$ to the Rayleigh range of a conventional beam of equal spatial width is governed by the ratio $\tfrac{\Delta\omega}{\delta\omega}$. However, the absolute value of the propagation distance $L_{\mathrm{max}}$ is determined by $\delta\omega$ alone \cite{Bhaduri2018OE,Bhaduri2019OL,Yessenov2019OE}.

As an illustration, take $\theta\!=\!50^{\circ}$; $\delta\lambda\!=\!20$~pm results in $L_{\mathrm{max}}\!\sim\!100$~mm for \textit{any} bandwidth, whereas $\delta\lambda\!=\!1$~nm yields $L_{\mathrm{max}}\!\sim\!2$~mm. A spectral uncertainty of 20~pm can be easily produced by the system in Fig.~\ref{Fig:ModulationSchemes}(f), which can thus render access to a broad range of values of $\theta$ for baseband ST wave packets. On the other hand, the measured spectra for ST wave packets produced by nonlinear optical interactions \cite{Day2004PT,Faccio2007OE} reveal a large spectral uncertainty (on the order of nanometers). Even though the total bandwidth $\Delta\omega$ in those studies was extremely large, the large spectral uncertainty precludes observing axial propagation invariance.

From the expression for $L_{\mathrm{max}}$, it can be shown that the maximum relative group delay $\Delta\tau_{\mathrm{max}}$ that a ST wave packet can undergo with respect to a reference pulse traveling at $c$ is $\Delta\tau_{\mathrm{max}}\!\sim\!\tfrac{1}{\delta\omega}$. Because typically $\Delta\omega\!\gg\!\delta\omega$, this relative delay is much larger than the on-axis pulsewidth of the ST wave packet. As such, the delay-bandwidth product defined as the product of the temporal bandwidth of the ST wave packet and the maximum delay it undergoes (with respect to a luminal reference pulse) before deformation -- or, equivalently, the ratio of the delay to the on-axis pulsewidth -- can be much greater than 1. Indeed, this product can approach values as large as $\sim\!100$ \cite{Yessenov2019OE,Yessenov2020NC}, thus presenting an interesting contrast with slow-light systems \cite{Boyd2009Science} where this product is $\sim\!1$, which suggests potential applications in all-optical buffers \cite{Boyd2006OPN}.

\subsection{Self-healing}

\begin{figure}[t!]
  \begin{center}
  \includegraphics[width=10cm]{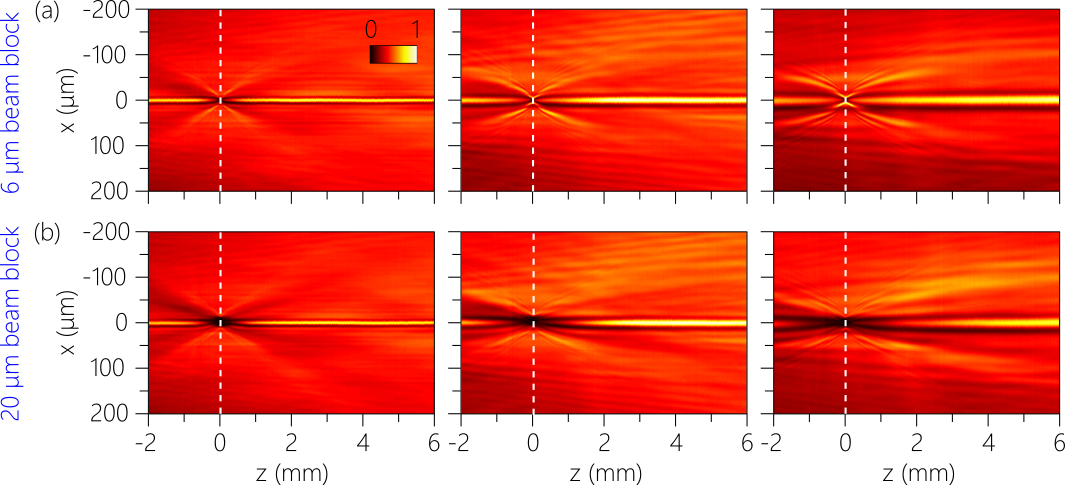}
  \end{center}
  \caption{Self-healing of ST wave packets after traversing an amplitude obstruction at fixed temporal bandwidth $\Delta\lambda\!\approx\!0.8$~nm. Each column corresponds to a different FWHM spatial width: $\Delta x\!=\!7$, 12, and 23~$\mu$m, and spectral tilt angles $\theta\!=\!90^{\circ}$, $53.1^{\circ}$, and $48.4^{\circ}$, respectively. Measured time-averaged intensity distribution at axial positions $-2\!<\!z\!<\!6$~mm ($z\!=\!0$ is the location of the obstruction) for obstruction widths of (a) 6~$\mu$m and (b) 20~$\mu$m.}
  \label{Fig:Self-healing}
\end{figure}

Monochromatic diffraction-free beams exhibit an intriguing behavior known as `self-healing', which refers to the rapid reconstruction of their initial field profile after traversing a perturbing obstacle, typically in the form of an amplitude obstruction. This property has been well-documented for Bessel \cite{Bouchal1998OC} and Airy beams \cite{Broky2008OE}, and can have important consequences for their utility in biomedical imaging \cite{Fahrbach2010NP,Vettenburg2014NM,Piksarv17SR}, among other applications. The propagation invariance of ST wave packets leads one to anticipate that similar behavior extends to them. To the best of our knowledge, self-healing has not been tested for FWMs, but has been confirmed with ultrashort-pulse X-waves \cite{Grunwald2004CLEO}. Self-healing of \textit{baseband} ST wave packets has been recently verified \cite{Kondakci2018OL}. Figure~\ref{Fig:Self-healing} shows samples from an experimental confirmation of the self-reconstruction of ST wave packets after traversing opaque obstructions (lithographically patterned 100-nm-thick chrome on a soda-lime glass substrate) of various widths placed in the path of ST wave packets of varying transverse spatial widths. The temporal bandwidth $\Delta\lambda$ is held fixed and the different spatial widths $\Delta x$ are realized by changing the spectral tilt angle $\theta$. It is readily observed that the profiles of the ST wave packets are rapidly recovered after scattering from the obstruction. Wave packets with smaller $\Delta x$ recover quicker, and the wave packets recover quickest from smaller obstructions. More work is needed to study the impact of \textit{phase} scatterers on self-healing as a first step towards assessing the impact of turbulence on ST wave packets.

\section{Modifying the axial propagation of ST wave packets}\label{Sec:ModifyingTheAxialProp}

Besides the familiar propagation-invariant ST wave packets, the spatio-temporal Fourier synthesis strategy enables exercising precise control over the axial evolution of one characteristic of the wave packet while maintaining all the others fixed, thus providing new behaviors. Examples include introducing arbitrary group-velocity dispersion (GVD) in free space, ST wave packets that accelerate or decelerate, and those with axially encoded spectrum on-axis. Another form of axial dynamics is induced by spectral discretization, whereupon axial revivals of the initial spatio-temporal profile occur periodically -- a ST Talbot effect.

\subsection{Dispersive ST wave packets and non-differentiable angular dispersion}\label{Sec:DispersiveWPs}

\begin{figure}[t!]
  \begin{center}
  \includegraphics[width=8.0cm]{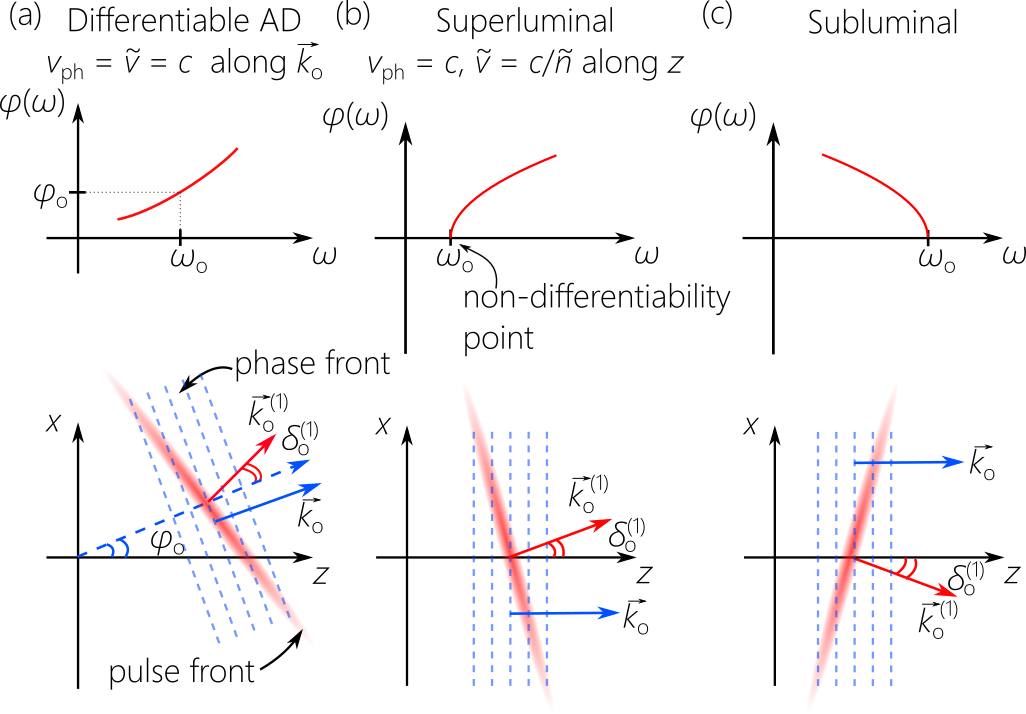}
  \end{center}
  \caption{Concept of non-differentiable angular dispersion with the angular dispersion spectrum $\varphi(\omega)$ and the associated pulse front. (a) After the introduction of angular dispersion, each frequency $\omega$ travels at a different angle $\varphi(\omega)$ with respect to the $z$-axis. (b) Non-differentiable angular dispersion associated with a superluminal ST wave packet. (c) Same as (b), but for a subluminal ST wave packet.}
  \label{Fig:nondiff}
\end{figure}

Angular dispersion (AD) refers to the frequency-dependent propagation angle $\varphi(\omega)$ -- measured here with respect to the $z$-axis -- introduced into a plane-wave pulse after traversing dispersive or diffractive devices such as prisms or gratings \cite{Fulop2010Review,Torres2010AOP}; see Fig.~\ref{Fig:nondiff}(a). As such, this concept can be traced back to Newton's experiments on spectral analysis with prisms \cite{Sabra1981Book}. However, the advent of lasers in the 1960's led to a re-conceptualization of AD in the context of pulsed optical fields. The most salient feature in a pulsed field after introducing AD is that the \textit{pulse front} (the plane of constant amplitude) is tilted with respect to the \textit{phase front} (the plane of constant phase). The optical field thus takes the form of a tilted pulse front (TPF) with respect to the propagation direction \cite{Hebling1996OQE}; that is, a single branch of an X-shaped spatio-temporal profile. TPFs have found a broad range of applications in traveling-wave optical amplification \cite{Bor1983APB,Klebniczki1988APB,Hebling1989OL,Hebling1991JOSAB}, dispersion compensation \cite{Martinez1984JOSAA,Fork1984OL,Gordon1984OL,Szatmari1996OL}, pulse compression \cite{Bor1985OC,Lemoff1993OL,Kane1997JOSABgrism,Kane1997JOSAB4thorder}, increasing the phase-matching bandwidth in nonlinear optics \cite{Martinez1989IEEE,Szabo1990APB,Szabo1994APB,Richman1998OL,Richman1999AO}, and in the generation of THz pulses \cite{Hebling2002OE,Nugraha2019OL,Wang2020LPR}. 

It is clear that AD also underpins ST wave packets \cite{Hall2021OL-AD-PFT}: each wavelength is associated with a single spatial frequency and hence a single propagation angle. From this perspective, ST wave packets can be viewed as a species of TPFs [Fig.~\ref{Fig:nondiff}(b,c)], except that they typically have a symmetric spatio-temporal spectrum, and thus have an X-shaped profile, whereas TPFs constitute one branch from the X-profile because of the asymmetric spectral structure produced by prisms and gratings. Previously, the distinction between TPFs and propagation-invariant wave packets hinged on the differences in geometry and dimensionality: the AD in TPFs is introduced in only one transverse dimension (similar to the formalism used here where the coordinate $y$ is dropped), whereas X-waves and FWMs are typically described in a cylindrically symmetric 3D configuration.  These superficial distinctions have limited the cross-pollination between the fields of propagation-invariant wave packets and TPFs (see \cite{Porras2003PREBessel-X,Kondakci2019ACSP} for exceptions). However, it has recently been appreciated that several characteristics of ST wave packets fly in the face of the well-established consequences of AD for TPFs. Therefore, there must be fundamental differences between the type of AD incorporated into conventional TPFs and that in ST wave packets.

To explicate this contrast, we first state the standard results of the conventional theory of AD. By expanding $\varphi(\omega)$ in a Taylor expansion around a carrier frequency $\omega_{\mathrm{o}}$: $\varphi(\omega)\!=\!\varphi(\omega_{\mathrm{o}}+\Omega)\!=\!\varphi_{\mathrm{o}}+\varphi_{\mathrm{o}}^{(1)}\Omega+\tfrac{1}{2}\varphi_{\mathrm{o}}^{(2)}+\cdots$, the AD coefficients $\{\varphi_{\mathrm{o}}^{(n)}\}$ can be related to those of the expansion of the longitudinal wave number $k_{z}(\omega)\!=\!k_{z}^{(0)}+k_{z}^{(1)}\Omega+\tfrac{1}{2}k_{z}^{(2)}\Omega^{2}\dots$, respectively \cite{Porras2003PREBessel-X}, yielding the axial phase velocity $v_{\mathrm{ph}}$ and group velocity $\widetilde{v}$ :
\begin{equation}\label{Eq:GroupVelocityGeneral}
v_{\mathrm{ph}}=\frac{\omega_{\mathrm{o}}}{k_{z}^{(0)}}=\frac{c}{\cos{\varphi_{\mathrm{o}}}},\;\;\;\; \widetilde{v}=\frac{1}{k_{z}^{(1)}}=\frac{c}{\cos{\varphi_{\mathrm{o}}}-\omega_{\mathrm{o}}\varphi_{\mathrm{o}}^{(1)}\sin{\varphi_{\mathrm{o}}}};
\end{equation}
and the axial GVD coefficient:
\begin{equation}\label{Eq:GVDGeneral}
c\omega_{\mathrm{o}}k_{z}^{(2)}=-(\omega_{\mathrm{o}}\varphi_{\mathrm{o}}^{(1)})^{2}\cos{\varphi_{\mathrm{o}}}-(\omega_{\mathrm{o}}^{2}\varphi_{\mathrm{o}}^{(2)}+2\omega_{\mathrm{o}}\varphi_{\mathrm{o}}^{(1)})\sin{\varphi_{\mathrm{o}}}.
\end{equation}

The conventional theory of AD makes a set of \textit{definitive predictions} that have been borne out in experiments over the past 5 decades of research into TPFs:
\begin{enumerate}
    \item For on-axis propagation $\varphi_{\mathrm{o}}\!=\!0$, we have $v_{\mathrm{ph}}\!=\!\widetilde{v}\!=\!c$. That is, the phase \textit{and} the group velocities are always equal and luminal along the propagation axis.
    \item The pulse front is tilted by an angle $\delta_{\mathrm{o}}^{(1)}$ with respect to the phase front, where $\tan{\delta_{\mathrm{o}}^{(1)}}\!=\!\omega_{\mathrm{o}}\varphi_{\mathrm{o}}^{(1)}$ [Fig.~\ref{Fig:nondiff}(a)], which is independent of the pulse bandwidth.
    \item Large deviation of $\widetilde{v}$ from $c$ typically occurs at large propagation angles $\varphi_{\mathrm{o}}$. 
    \item GVD always accompanies TPFs; i.e., there are no GVD-free TPFs in presence of AD.
    \item On-axis ($\varphi_{\mathrm{o}}\!=\!0$), the field can experience 
    \textit{only} anomalous GVD, $c\omega_{\mathrm{o}}k_{z}^{(2)}\!=\!-(\omega_{\mathrm{o}}\varphi_{\mathrm{o}}^{(1)})^{2}$ \cite{Martinez1984JOSAA}.
    \item Higher-order dispersion terms are always present.
\end{enumerate}

All these statements appear to have iron-clad theoretical justification, and they have indeed been taken to be universal consequences of introducing AD into a pulsed field \cite{Fulop2010Review,Torres2010AOP}. Nevertheless, baseband ST wave packets that travel on-axis $\varphi_{\mathrm{o}}\!=\!0$ \textit{contradict these statements}:
\begin{enumerate}
    \item Here $v_{\mathrm{ph}}\!=\!c$, but $\widetilde{v}\!=\!c/\widetilde{n}\!\neq\!c$ can take on any value in principle \cite{Kondakci2019NC}.
    \item In the vicinity of $\omega_{\mathrm{o}}$ the pulse front tilt angle is \textit{bandwidth-dependent}, $\tan{\delta_{\mathrm{ST}}^{(1)}}\!\propto\!\tfrac{1}{\sqrt{\Delta\omega}}$ \cite{Hall2021OL-AD-PFT}.
    \item Large deviations in $\widetilde{v}$ from $c$ can be realized in the paraxial regime \cite{Kondakci2019NC}.
    \item Propagation-invariant ST wave packets are GVD-free (Eq.~\ref{Eq:BasebandPlane}) \cite{Yessenov2019OE}.
    \item Both normal \textit{and} anomalous GVD can be produced symmetrically on the same footing \cite{Hall2021NGVDrealizing}.
    \item Arbitrary dispersion profiles can be readily realized involving any magnitude, sign, or order  or superposition of several dispersion orders, each controlled independently \cite{Yessenov2021engineering,Hall2021NGVDrealizing}.
\end{enumerate}

What feature of the AD unique to ST wave packets produces these contrasts with traditional expectations? We have recently identified that the propagation angle $\varphi(\omega)$ for baseband ST wave packets is \textit{non-differentiable}; that is $\frac{d\varphi}{d\omega}$ is \textit{not} defined at $\omega\!=\!\omega_{\mathrm{o}}$ [Fig.~\ref{Fig:nondiff}(b,c)], which fuels the above-listed characteristics. It may initially appear that non-differentiable AD is an exotic or pathological condition, or perhaps even a non-physical one. This intuitive preconception is reinforced by the fact that conventional optical components such as gratings and prisms yield only differentiable AD. However, this is far from the case. Indeed, the experimental arrangement described in Section~\ref{Section:SynthesisSetup} is a universal AD synthesizer: arbitrary angular profiles $\varphi(\omega)$ can be instantiated, whether differentiable or non-differentiable \cite{Hall2021OESynthesizer}.

\begin{figure}[t!]
  \begin{center}
  \includegraphics[width=10.0cm]{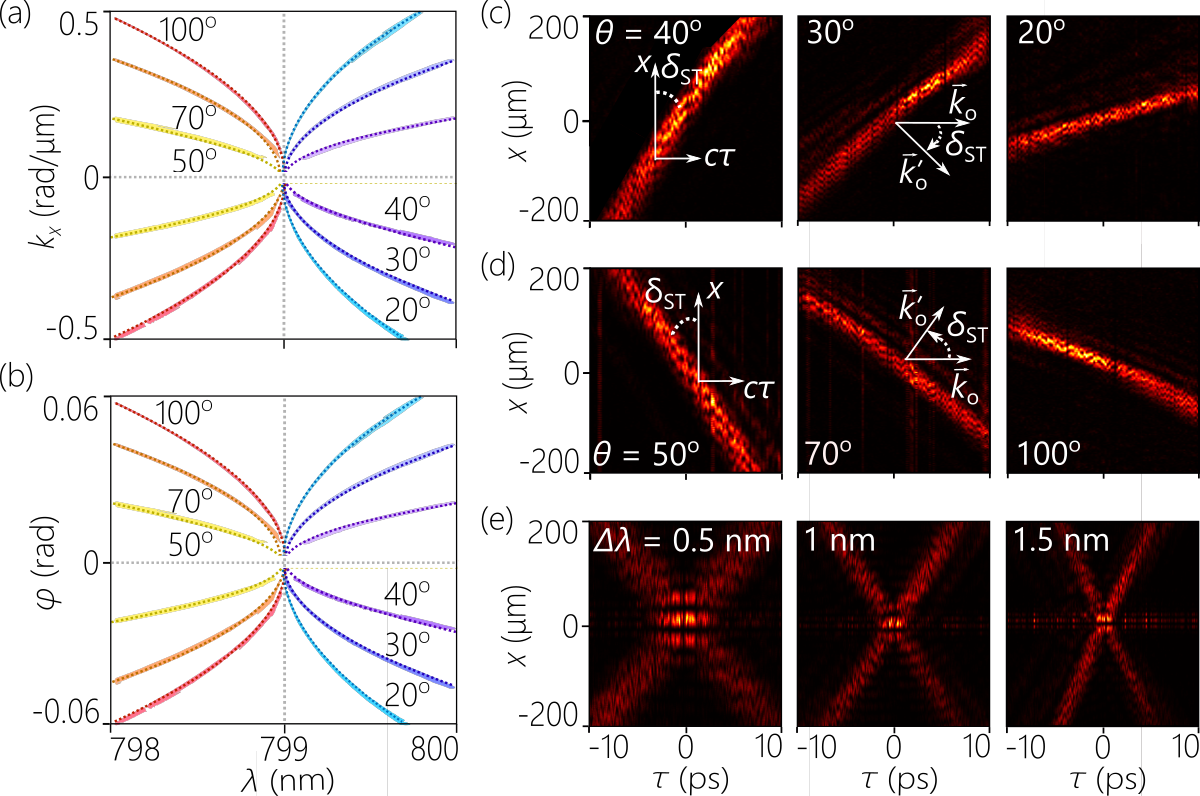}
  \end{center}
  \caption{(a) Measured spectrum $\widetilde{\psi}(k_x,\lambda)$, (b) angular dispersion $\varphi(\lambda)$, and (c-e) intensity profile $I(x;\tau)$ for baseband ST wave packets. In (c,d) we show the evolution of the pulse-front tilt angle with $\theta$ for subluminal and superluminal ST wave packets at a fixed bandwidth of $\Delta\lambda\!\approx\!1$~nm. In (e), the profiles evolve with $\Delta\lambda$ at fixed $\theta\!=\!50^{\circ}$; see \cite{Hall2021OL-AD-PFT}.}
  \label{Fig:tiltangle}
\end{figure}

The need for non-differentiability can be understood in the context of changing $\widetilde{v}$ away from $c$ on-axis (Eq.~\ref{Eq:GroupVelocityGeneral}). At $\varphi_{\mathrm{o}}\!=\!0$, $\widetilde{v}\!=\!c$ independently of $\varphi_{\mathrm{o}}^{(1)}$. However, the picture changes if $\varphi_{\mathrm{o}}^{(1)}$ is not defined. In general, for any differentiable function $f(x)$, the quantity $x\tfrac{df}{dx}\!\rightarrow\!0$ when $x\!\rightarrow\!0$, except when $f(x)$ is \textit{not} differentiable at $x\!=\!0$, whereupon $x\tfrac{df}{dx}$ can be differ from 0 even when $x\!\rightarrow\!0$; e.g., when $f(x)\!\propto\!\sqrt{x}$. From the parabolic spectral approximation for baseband ST wave packets in the vicinity of $k_{x}\!=\!0$ ($\varphi_{\mathrm{o}}\!=\!0$), $\varphi(\omega)\!\approx\!\eta\sqrt{\tfrac{\Omega}{\omega_{\mathrm{o}}}}$, which is not differentiable at $\omega\!=\!\omega_{\mathrm{o}}$, and $\widetilde{v}\!\approx\!c/(1-\omega\varphi\tfrac{d\varphi}{d\omega})\!=\!c/\widetilde{n}$, where $\eta\!=\!2(1-\widetilde{n})$ [Fig.~\ref{Fig:nondiff}(b,c)]. Thus the non-differentiability of $\varphi$ at $\omega_{\mathrm{o}}$ results in $\widetilde{v}$ deviating from $c$ on-axis. Furthermore, $\widetilde{v}$ is independent of frequency, indicating that all dispersion orders $k_{z}^{(n)}$ for $n\!\geq\!2$ are absent despite the presence of AD.

As noted above, in the presence of AD, the pulse front is tilted by an angle $\delta_{\mathrm{o}}^{(1)}$, where $\tan{\delta_{\mathrm{o}}^{(1)}}\!=\!\omega_{\mathrm{o}}\varphi_{\mathrm{o}}^{(1)}$. This is considered to be a universal relationship \cite{Hebling1996OQE}: it is independent of the origin of AD, and is also independent of the pulse shape and bandwidth. We recently demonstrated experimentally that for baseband ST wave packets in free space, the angle of the pulse-front tilt changes with the bandwidth, $\tan{\delta_{\mathrm{ST}}^{(1)}}\!\propto\!\tfrac{1}{\sqrt{\Delta\omega}}$. This can be shown to be a consequence of the non-differentiability of $\varphi(\omega)$ at $\omega\!=\!\omega_{\mathrm{o}}$ [Fig.~\ref{Fig:tiltangle}(a,b)]. Because the spectrum cannot extend beyond the non-differentiable frequency $\omega_{\mathrm{o}}$, it represents a natural terminus for the spectrum \cite{Hall2021OEConsequences}: it is the maximum possible frequency for subluminal ST wave packets and the minimum for superluminal ST wave packets. As such, changing the bandwidth involves changing the central frequency of the spectrum. The AD changes very rapidly in the vicinity of $\omega_{\mathrm{o}}$, resulting in changes in the tilt angle of the pulse front even for small bandwidths [Fig.~\ref{Fig:tiltangle}(c-e)].

\begin{figure}[t!]
  \begin{center}
  \includegraphics[width=12.6cm]{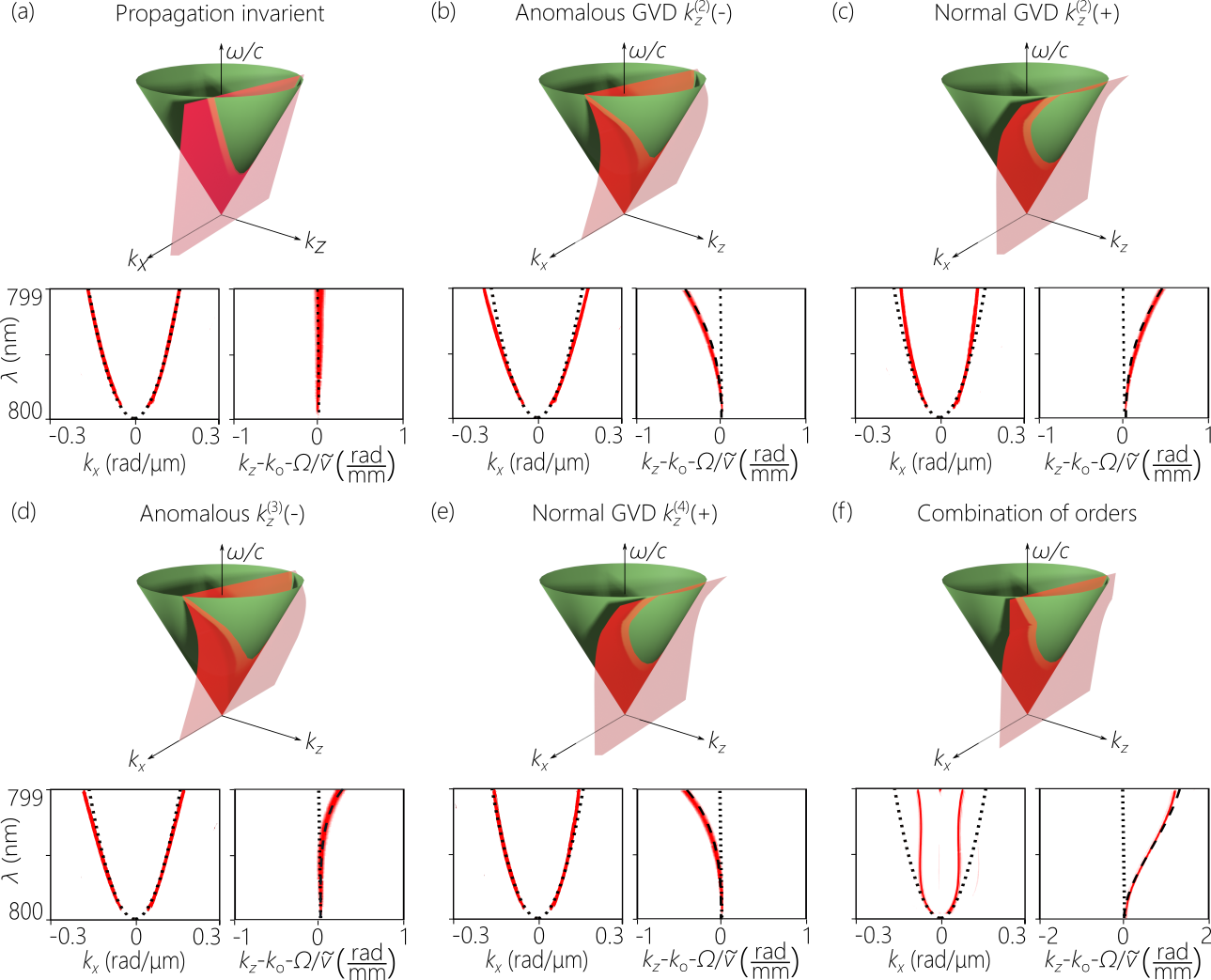}
  \end{center}
  \caption{Dispersive ST wave packets in free space. For each wave packet we plot the spectral support domain on the light-cone surface, and the spectral projections onto the $(k_{x},\lambda)$ and $(k_{z},\lambda)$ planes; the spectral tilt angle is $\theta\!=\!50^{\circ}$ throughout. (a) Non-dispersive ST wave packet $k_{z}^{(2)}\!=\!0$; (b) anomalous GVD $k_{z}^{(2)}\!=\!-100$~ps$^2$/m; (c) normal GVD $k_z^{(2)}\!=\!100$~ps$^2$/m; (d) third-order dispersion $k_z^{(3)}\!=\!-80$~ps$^3$/m; (e) fourth-order dispersion $k_z^{(4)}\!=\!40$~ps$^4$/m; and (f) a combination of multiple dispersion orders: $k_{z}^{(2)}\!=\!700$~ps$^2$/m, $k_{z}^{(3)}\!=\!-500$~ps$^3$/m, and $k_{z}^{(4)}\!=\!100$~ps$^4$/m; see \cite{Yessenov2021engineering,Hall2021NGVDrealizing}.}
  \label{Fig:measureGVD}
\end{figure}

With respect to GVD, it can be shown that the following propagation angle:
\begin{equation}
\sin{\{\varphi(\omega)\}}=\eta\sqrt{\frac{\Omega}{\omega_{\mathrm{o}}}}\;\;\frac{\omega_{\mathrm{o}}}{\omega}\;\;\sqrt{\left\{1+\frac{1+\widetilde{n}}{2}\frac{\Omega}{\omega_{\mathrm{o}}}+\frac{\sigma}{2}\left(\frac{\Omega}{\omega_{\mathrm{o}}}\right)^{2}\right\}\left\{1-\frac{\sigma}{1-\widetilde{n}}\frac{\Omega}{\omega_{\mathrm{o}}}\right\}},
\end{equation}
produces an axial wave number of the form $k_{z}(\omega)=k_{\mathrm{o}}+\frac{\Omega}{\widetilde{v}}+\frac{1}{2}k_{2}\Omega^{2}$, where $k_{2}$ is the axial GVD coefficient, and $\sigma\!=\!\tfrac{1}{2}k_{2}\omega_{\mathrm{o}}c$. This expression is agnostic with respect to the sign of $k_{2}$; i.e., both normal \textit{and} anomalous GVD are accessible. Furthermore, the expression for $k_{z}(\omega)$ terminates exactly at the second-order term in $\Omega$; all higher-order dispersion terms are eliminated. However, $\varphi(\omega)$ is once again non-differentiable because of the $\sqrt{\Omega}$ term.

It is now clear that the classical result in \cite{Martinez1984JOSAA} makes two assumptions: on-axis propagation $\varphi_{\mathrm{o}}\!=\!0$ and differentiable AD (i.e., $\varphi_{\mathrm{o}}^{(n)}$ for all $n$ are finite and well-defined). Therefore, in the presence of non-differentiable AD, normal GVD can be produced in an on-axis ST wave packet \cite{Hall2021OL-AD-PFT,Yessenov2021engineering}, or by sculpting off-axis fields \cite{Porras2003PREBessel-X}. Both classes of ST wave packets exhibiting normal GVD in free space have been recently realized \cite{Hall2021NGVDrealizing}. Examples of particular dispersive ST wave packets are plotted in Fig.~\ref{Fig:measureGVD}, including ones endowed with anomalous and normal GVD while eliminating all higher-order dispersion terms [Fig.~\ref{Fig:measureGVD}(b,c)], wave packets endowed with a third- or fourth-order dispersion term while eliminating all other terms including second order [Fig.~\ref{Fig:measureGVD}(d,e)], and a wave packet whose dispersion profile is a weighted superposition of terms of different orders [Fig.~\ref{Fig:measureGVD}(f)]. In these cases, the \textit{plane} $\mathcal{P}_{\mathrm{B}}(\theta)$ for a propagation-invariant ST wave packet [Fig.~\ref{Fig:measureGVD}(a)] is replaced by a \textit{planar curved surface} that is parallel to the $k_{x}$-axis, so that the projection onto the $(k_{z},\tfrac{\omega}{c})$-plane is a prescribed curve whose parameters can be tuned experimentally.

The earliest reported experiment in the development of localized waves was performed by the Saari group with the goal of demonstrating dispersion-free propagation in a dispersive medium \cite{Sonajalg1996OL,Sonajalg1997OL} by modifying the structure of an X-wave via a lensacon (a combination of an axicon and a converging lens). The synthesized wave packet was \textit{not} baseband, and pulses of width $\sim\!250$~fs were used, so that the wave packet was basically separable in space and time (no characterization of the wave packet 3D structure was provided); see Fig.~\ref{Fig:subluminal_xtoo}(a). The path is now paved for observing dispersion-free propagation in dispersive media of any kind. There is a significant theoretical literature on such baseband ST wave packets \cite{Porras2001OL,Porras2003OL,Porras2003PREBessel-X,Longhi2003PRE,Porras2004,Longhi2004OL,Porras2007JOSAB,Malaguti2008OL,Malaguti2009PRA}. Prior to our recent experiments, only nonlinear optical interactions produced such wave packets in dispersive media \cite{DiTrapani2003PRL,Porras2005OL,Porras2007JOSAB,Porras2007PRA}. The ST wave packets produced for propagation in anomalous \cite{Dallaire2009OE} and normal \cite{Jedrkiewicz2013OE} GVD were not eventually coupled to dispersive media and were only examined in free space. We therefore expect this area to witness significant growth in the next few years.

\subsection{Axial spectral encoding}\label{Sec:AxialSPectralEncoding}

\begin{figure}[t!]
  \begin{center}
  \includegraphics[width=8cm]{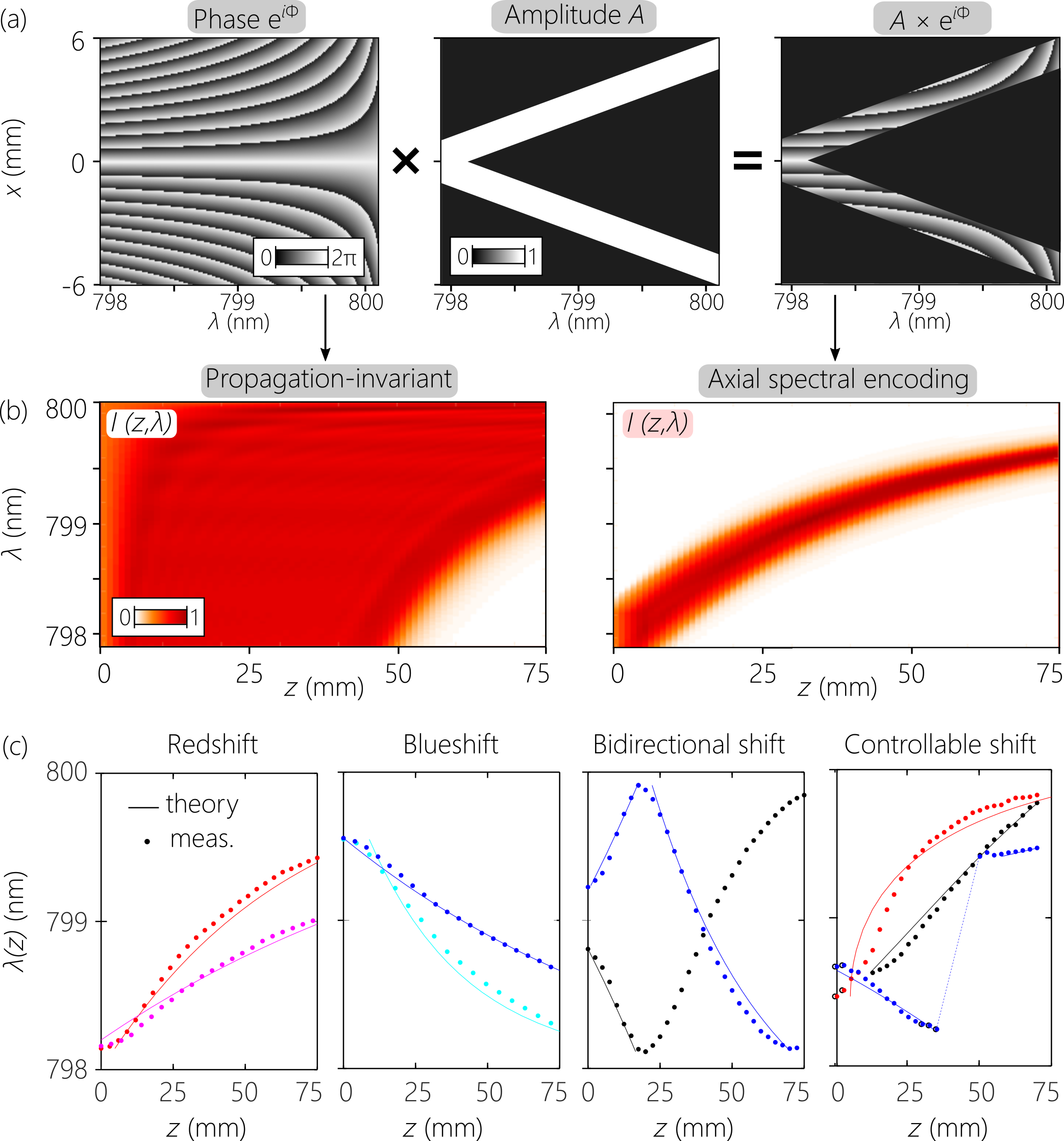}
  \end{center}
  \caption{(a) Imposing an amplitude mask on the SLM phase distribution for a propagation-invariant ST wave packet ($\theta\!=\!50^{\circ}$ and $\widetilde{v}\!\approx\!1.19c$) produces an amplitude masked phase distribution that yields a ST wave packet with tailored axial spectral encoding but the same $\widetilde{v}$. (b) Calculated spectral evolution $I(z,\lambda)$ at the wave packet center for the ST wave packets in (a) with and without amplitude masking. (c) Examples of the axial dynamics along $z$ of the center of the spectrum for axial spectrally encoded ST wave packets: red-shifting, (b) blue-shifting, (c) bidirectional spectral shifting, and (d) controllable axial spectral encodings; see \cite{AllendeMotz2020PRA}.}
  \label{Fig:AxialEncoding}
\end{figure}

The on-axis spectrum of a ST wave packet can be controlled along the propagation axis without impacting -- thereby realizing a particular \textit{axial spectral encoding} -- its other characteristics \cite{Motz2021PRAaxila-encoding}. By axial spectral encoding we refer to producing a controlled evolution along $z$ of the on-axis ($x\!=\!0$) spectrum of the ST wave packet. This is achieved by combining spectral-\textit{amplitude} modulation along with the spectral-\textit{phase} modulation described in Fig.~\ref{Fig:DesignOfThePhaseDistribution}. An example is shown in Fig.~\ref{Fig:AxialEncoding}(a) where the phase distribution needed for the synthesis of a baseband ST wave packet with a spectral tilt angle $\theta$ is then spatially filtered with an amplitude mask, resulting in a new complex amplitude spectral modulation. In the specific case shown in Fig.~\ref{Fig:AxialEncoding}(a), the complex spectral-amplitude modulation yields a wave packet whose on-axis spectrum red-shifts. In absence of the amplitude mask, the on-axis spectrum is broad and constant along $z$ [Fig.~\ref{Fig:AxialEncoding}(b)]. After adding the amplitude mask, the on-axis spectrum is narrowed, and the central wavelength of this spectrum gradually shifts to longer wavelengths with free propagation [Fig.~\ref{Fig:AxialEncoding}(c)]. 

The underlying methodology can be understood as follows. The complex amplitude imparted to the field of temporal frequency $\omega$ at a particular column on the SLM is $\approx\!e^{i\pm k_{x}(\omega)x}\mathrm{rect}(\tfrac{x'\pm x_{\mathrm{o}}(\omega)}{W_{\mathrm{o}}(\omega)})$; where $\mathrm{rect}(\tfrac{\cdot}{W})$ is a function having a constant unit amplitude over a width $W$, $x_{\mathrm{o}}(\omega)$ is the center of the amplitude modulation, which is selected to deliver this frequency $\omega$ to the axial position $z(\omega)\!\sim\!\tfrac{k}{k_{x}}x_{\mathrm{o}}(\omega)$, and the axial extent of this frequency centered at $z(\omega)$ is $W(z,\omega)\!\sim\!W_{\mathrm{o}}(\omega)\sqrt{1+(z/z_{\mathrm{o}}(\omega))^{2}}$, where $z_{\mathrm{o}}(\omega)\!=\!\pi W_{\mathrm{o}}^{2}/\lambda$ \cite{Motz2021PRAaxila-encoding}. We can thus reverse engineer the SLM amplitude mask by determining the values of $x_{\mathrm{o}}(\omega)$ and $W_{\mathrm{o}}(\omega)$ to localize the frequency $\omega$ at the target position or even multiple positions along $z$. Using this methodology, several axial spectral encodings are plotted in Fig.~\ref{Fig:AxialEncoding}(c-f), including: red-shifting [Fig.~\ref{Fig:AxialEncoding}(c)] and blue-shifting [Fig.~\ref{Fig:AxialEncoding}(d)] spectra at different rates; bi-directional spectra [Fig.~\ref{Fig:AxialEncoding}(e)] that red-shift and then blue-shift, and vice versa; and arbitrary axial spectral encodings [Fig.~\ref{Fig:AxialEncoding}(f)]. Time-resolved measurements reveal that the spatio-temporal profiles are minimally impacted \cite{Motz2021PRAaxila-encoding}.

This approach is independent, in principle, of the central wavelength or bandwidth, it minimally impacts the propagation invariance of the ST wave packet, and does not affect the group velocity of the underlying wave packet -- whether subluminal or superluminal. By controlling the encoding bandwidth and rate of axial change, ST wave packets endowed with axial spectral encoding may find applications in microscopy and remote sensing by providing a pathway to axial range finding through spectral stamping.

\subsection{Accelerating ST wave packets}\label{Sec:Acceleration}

\begin{figure}[t!]
  \begin{center}
  \includegraphics[width=9.0cm]{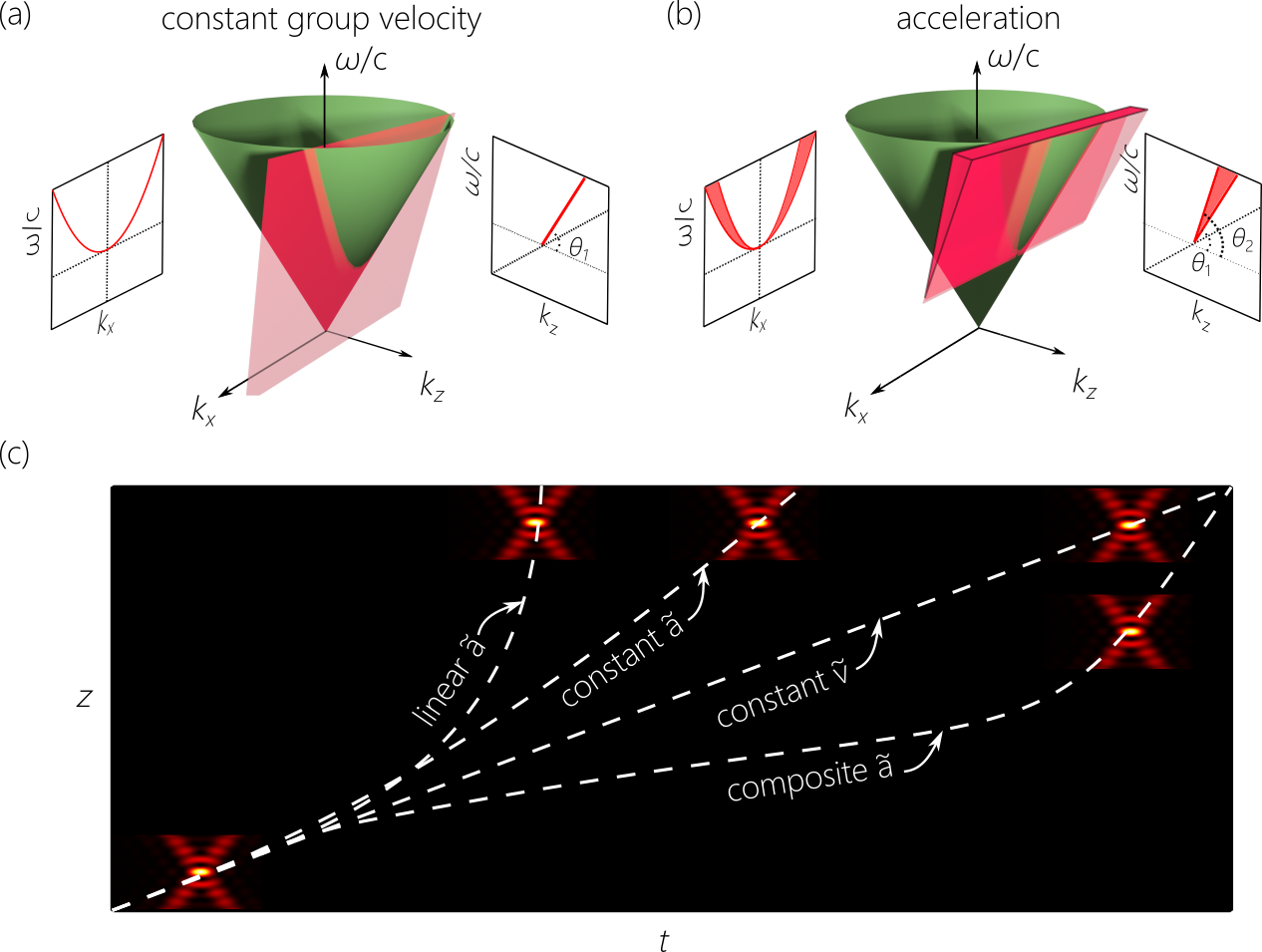}
  \end{center}
  \caption{Accelerating ST wave packets. (a) Representation of a propagation-invariant ST wave packet and (b) an accelerating ST wave packet on the light-cone surface. (c) Illustration of ST wave packets that propagate at constant-$\widetilde{v}$, constant acceleration $\widetilde{a}$, linearly varying $\widetilde{a}$, and composite $\widetilde{a}$.}
  \label{Fig:Acc_concept}
\end{figure}

\begin{figure}[t!]
  \begin{center}
  \includegraphics[width=9cm]{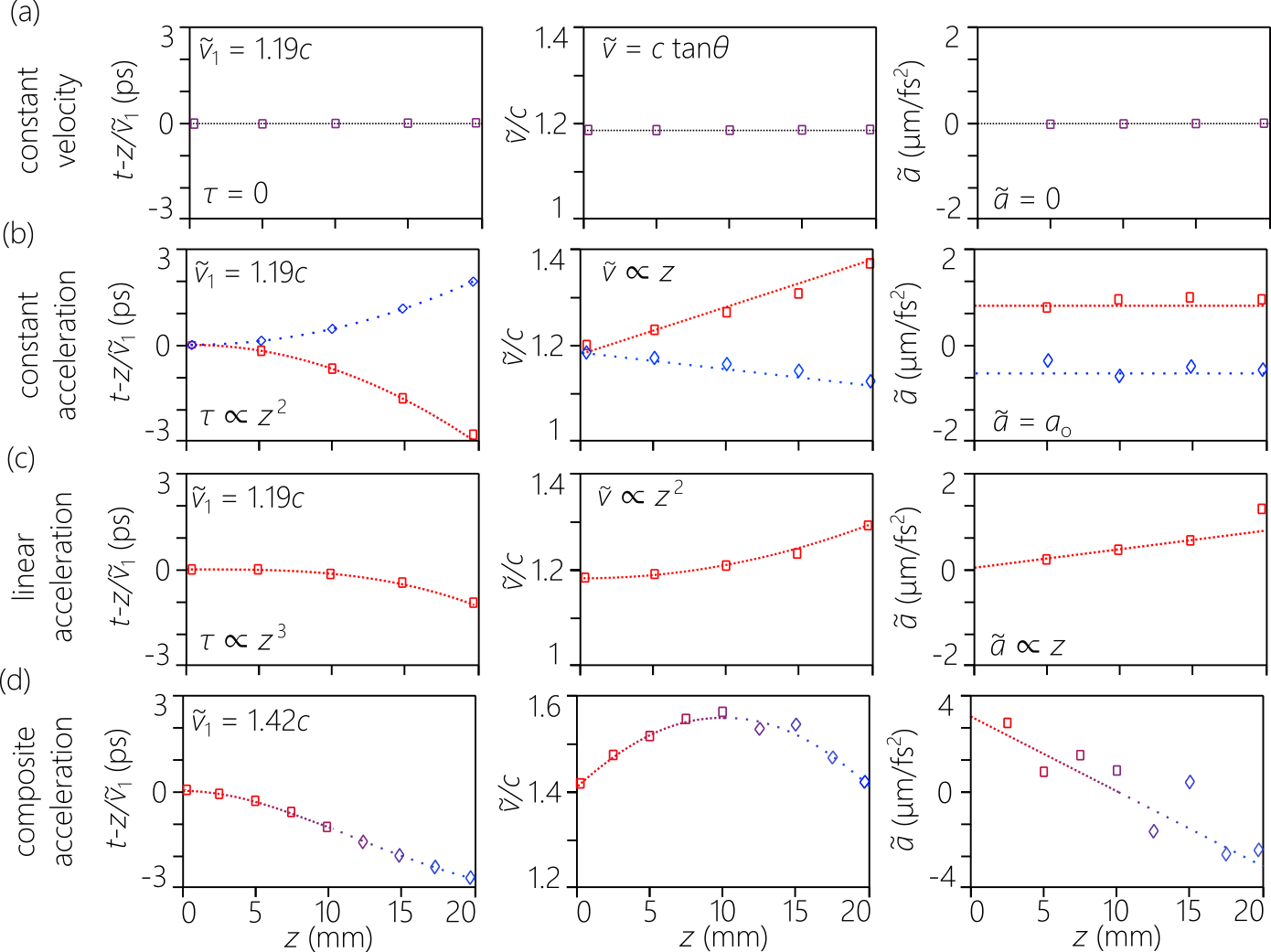}
  \end{center}
  \caption{Measured group delay $\tau\!=\!t-z/\widetilde{v}_{1}$, group velocity $\widetilde{v}$, and acceleration $\widetilde{a}$ along $z$ for accelerating ST wave packets. The data corresponds to ST wave packets having (a) constant-$\widetilde{v}$ ST wave packet with $\theta\!=\!50^{\circ}$; (b) constant-$\widetilde{a}$; (c) linear acceleration $\widetilde{a}\!\propto\!z$; and (d) composite acceleration; see \cite{Hall2021OLAccel}.}
  \label{Fig:Acc_measure}
\end{figure}

The axial distribution of $\widetilde{v}\!=\!\widetilde{v}(z)$ can also be controlled by engineering the spectral support domain on the light-cone surface to produce accelerating or decelerating wave packets. To date, two strategies have been exploited to produce optical wave packets that accelerate axially: accelerating Airy pulses in a dispersive medium \cite{Chong2010NP}, and wave-front-modulated X-waves \cite{Clerici2008OE,ValtnaLukner2009OE}. In both cases, \textit{only minute changes} in the group velocity have been reported (typically $\Delta\widetilde{v}\!\sim\!10^{-4}c$), and \textit{only constant accelerations} have been produced. Synthesizing ST wave packets that accelerate axially follows a different strategy. Rather than the precise association between $k_{x}$ and $\omega$ necessary for propagation invariance, acceleration requires that we associate with each frequency $\omega$ a \textit{finite} spatial spectrum whose bandwidth and central spatial frequency are both wavelength-dependent \cite{Yessenov2020PRLaccel}. This can be understood through the following heuristic argument. Consider two propagation-invariant ST wave packets at group velocities $\widetilde{v}_{1}\!=\!c\tan{\theta_{1}}$ and $\widetilde{v}_{2}\!=\!c\tan{\theta_{2}}$, whose spectral support domains are each a 1D trajectory [Fig.~\ref{Fig:Acc_concept}(a)]. Any intermediate group velocity corresponds to an intermediate spectral tilt angle $\theta_{1}\!<\!\theta\!<\!\theta_{2}$. It thus appears reasonable that a ST wave packet whose 2D support domain extends between the 1D trajectories associated with $\theta_{1}$ and $\theta_{2}$ will have a group velocity that varies between $\widetilde{v}_{1}$ and $\widetilde{v}_{2}$ [Fig.~\ref{Fig:Acc_concept}(b)], potentially with an arbitrary axial distribution of the group velocity [Fig.~\ref{Fig:Acc_measure}(c)].

This intuition can be placed on firmer ground by considering an axial group-velocity profile $\widetilde{v}(z)\!=\!\widetilde{v}_{1}+\tfrac{1}{m}\Delta\widetilde{v}(\tfrac{z}{L})^{m}\!=\!c/\widetilde{n}(z)$, where $L$ is the maximum axial propagation distance, $\widetilde{v}_{1}\!=\!\widetilde{v}(0)$ is the initial group velocity, $\widetilde{v}_{2}\!=\!\widetilde{v}(L)$ is the terminal group velocity, $\Delta\widetilde{v}\!=\!\widetilde{v}_{2}-\widetilde{v}_{1}$, and the exponent $m$ is \textit{not} necessarily an integer. The acceleration is $\widetilde{a}(z)\!=\!\tfrac{d\widetilde{v}}{dz}\!=\!\tfrac{\Delta\widetilde{v}}{L}(\tfrac{z}{L})^{m-1}$. We take the constant-group-velocity condition at $m\!=\!0$ to be $\widetilde{v}(z)\!=\!\widetilde{v}_{1}$; constant acceleration is associated with $m\!=\!1$, $\widetilde{v}(z)\!=\!\widetilde{v}_{1}+\Delta\widetilde{v}\tfrac{z}{L}$; linear acceleration with $m\!=\!2$, $\widetilde{v}(z)\!=\!\widetilde{v}_{1}+\tfrac{1}{2}\Delta\widetilde{v}(\tfrac{z}{L})^{2}$; and so on [Fig.~\ref{Fig:Acc_concept}(c)]. Using this model, one can establish an algorithm for constructing the SLM phase distribution, which we have recently exploited to produce ST wave packets traveling at a constant acceleration $m\!=\!1$ of record-high $\Delta\widetilde{v}\!\sim\!c$ \cite{Yessenov2020PRLaccel}, and also arbitrary acceleration rates at $m\!=\!2$, $m\!=\!3$, and even fractional exponents such as $m\!=\!0.5$ \cite{Hall2022OLArbAccel}. Additionally, we have realized a composite acceleration profile in which the $z$-axis is segmented into different acceleration domains: the wave packet accelerates from $\widetilde{v}_{1}$ to $\widetilde{v}_{2}$ then decelerates from $\widetilde{v}_{2}$ back to $\widetilde{v}_{1}$ \cite{Hall2022OLArbAccel}; see Fig.~\ref{Fig:Acc_measure}. 

Recent theoretical work by Li \textit{et al}. \cite{Li2020SR,Li2020CP,Li2021CP} has introduced new ideas exploiting carefully sculpted angular dispersion to produce wave packets that accelerate with arbitrary axial profiles of the group velocity that complement the experimental approach in \cite{Yessenov2020PRLaccel,Hall2021OLAccel}.

\subsection{The ST Talbot effect}

\begin{figure}[t!]
  \begin{center}
  \includegraphics[width=10.0cm]{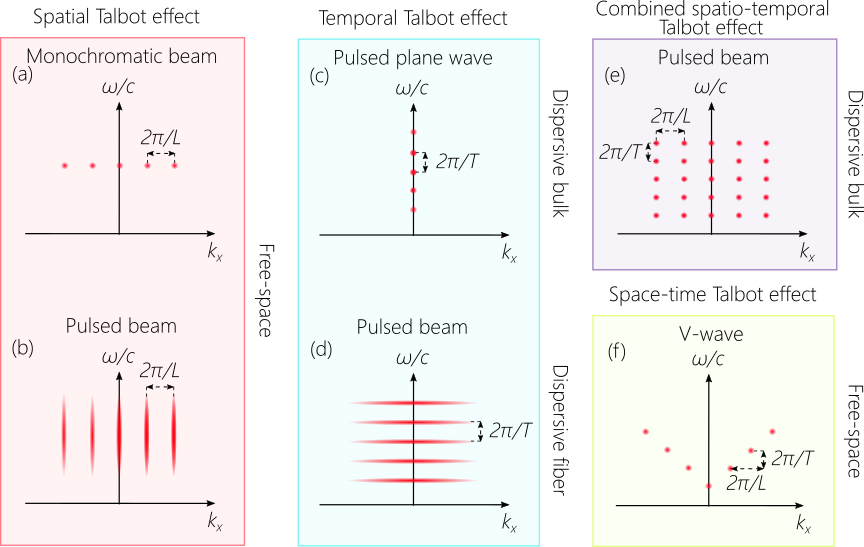}
  \end{center}
  \caption{The spectral representation in the $(k_{x},\tfrac{\omega}{c})$-plane of the Talbot effect in space, time, and space-time. (a) In the spatial Talbot effect, the spatial spectrum of a monochromatic beam is sampled periodically along $k_{x}$. (b) The spatial Talbot effect when implemented with a \textit{pulsed} beam. (c) In the temporal Talbot effect, the temporal spectrum of a plane-wave pulse is sampled periodically along $\omega$. (d) The temporal Talbot effect when implemented with a pulsed \textit{beam}. (e) Combined spatial and temporal Talbot effects, in which the spatial and temporal spectra are both sampled periodically. The spatio-temporal spectra in (a-e) are all separable with respect to $k_{x}$ and $\omega$. (f) The ST Talbot effect makes use of a V-shaped spatio-temporal spectrum, $\omega-\omega_{\mathrm{o}}\!=\!\alpha c|k_{x}|$, sampled periodically along $k_{x}$ \textit{and} $\omega$.}
  \label{Fig:Talbot_theory}
\end{figure}

The spatio-temporal spectra considered above were all continuous in terms of $k_{x}$ and $\omega$. The question arises as to the effect of discretizing the spectra of ST wave packets. Discretizing the spatial spectrum of a monochromatic field results in a periodic transverse profile, which gives rise to the Talbot effect \cite{PATORSKI1989PO,Wen2013AOP}. Reported for the first time in 1836 \cite{Talbot1836PM}, the Talbot effect is a wave interference phenomenon whereby a monochromatic field at a wavelength $\lambda_{\mathrm{o}}$ having a periodic transverse spatial profile of period $L$ (the spatial spectrum is discretized at multiples of $\tfrac{2\pi}{L}$) first diffracts before undergoing periodic axial revivals in planes separated by the spatial Talbot length $z_{\mathrm{T},x}\!=\!\tfrac{2L^{2}}{\lambda_{\mathrm{o}}}$ \cite{Rayleigh1881,Montgomery1967JOSA,Berry1996JMO}; see Fig.~\ref{Fig:Talbot_theory}(a). Such spatial reconstruction can be observed even in the presence of a finite temporal bandwidth \cite{Packross1984,Lancis1995,Guerineau2000}; see Fig.~\ref{Fig:Talbot_theory}(b).

Because the only ingredients in the Talbot effect are wave interference of a discretized spatial spectrum and diffraction, a temporal analog \cite{Kolner1994IEEEJQE,vanHowe2006} -- the temporal Talbot effect -- exists whereby a periodic pulse train of period $T$ (the temporal spectrum is discretized at multiples of $\tfrac{2\pi}{T}$) traveling in a dispersive medium with GVD coefficient $k_{2}$ exhibits axial revivals \cite{Jannson1981JOSA,Andrekson1993,Arahira1998,Shake1998,Azana1999,Atkins2003,FernandezPousa2004}; see Fig.~\ref{Fig:Talbot_theory}(c). The pulses in the periodic train first disperse before reconstituting themselves periodically along the propagation axis at planes separated by the temporal Talbot length $z_{\mathrm{T},t}\!=\!\tfrac{T^{2}}{\pi|k_{2}|}$. However, this effect has never been observed in a bulk dispersive medium. Instead, it has been only realized in single-mode optical fibers. The fiber mode has a finite spatial profile that is separable from the temporal spectrum \cite{Mitschke1998OPN,Fatome2004}, and thus does not mask the observation of this effect in time; see Fig.~\ref{Fig:Talbot_theory}(d).

\begin{figure}[t!]
  \begin{center}
  \includegraphics[width=10.0cm]{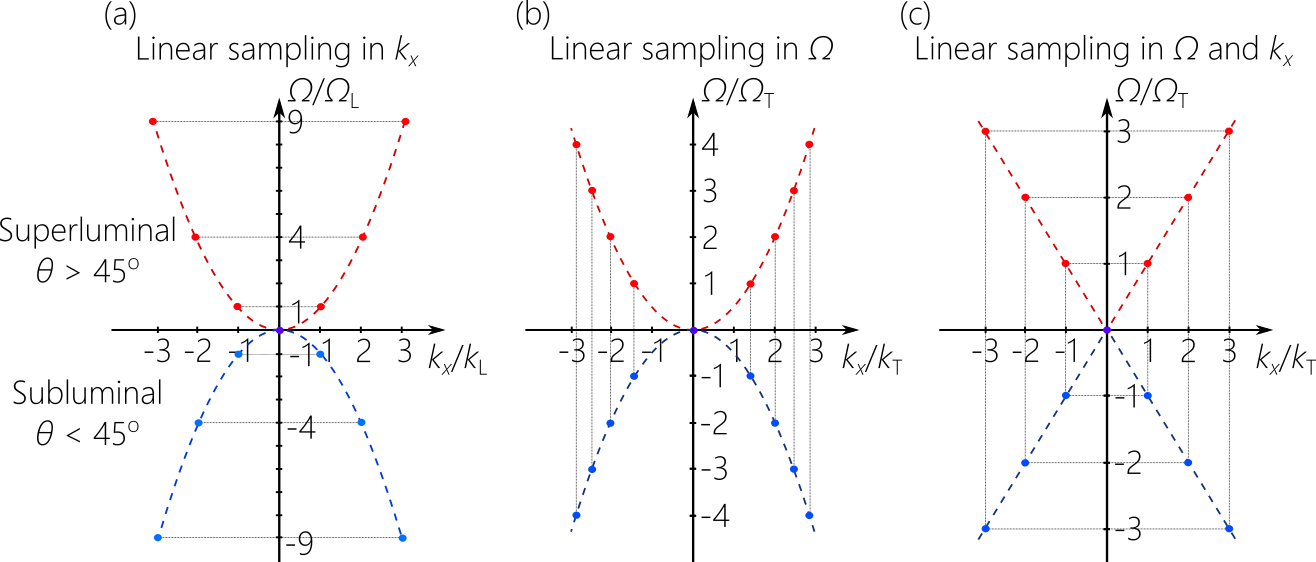}
  \end{center}
  \caption{(a) The spectra of superluminal and subluminal propagation-invariant ST wave packets discretized periodically in $k_x$ and (b) in $\Omega$. (c) A luminal, dispersive V-wave is discretized periodically in both $k_{x}$ and $\Omega$.}
  \label{Fig:discrete_STWP}
\end{figure}

Can the Talbot effect be observed in space \textit{and} time simultaneously? This requires producing a pulsed field in the form of a spatio-temporal lattice that is periodic in space and time propagating in a bulk dispersive medium. The spatial and temporal spectra need to be discretized along a 2D mesh as shown in Fig.~\ref{Fig:Talbot_theory}(e). If the parameters of the field lattice and the properties of the medium are carefully selected to equalize the spatial and temporal Talbot lengths $z_{\mathrm{T},x}\!=\!z_{\mathrm{T},t}\!=\!z_{\mathrm{T}}$, then the spatio-temporal lattice first degrades in space and time under the joint impact of diffraction and dispersion before axial revivals occur at planes separated by $z_{\mathrm{T}}$. However, the likelihood of realizing independent spatial and temporal Talbot effects simultaneously is low. Because diffraction and dispersion are independent physical phenomena that unfold on different length scales, the spatial and temporal Talbot lengths are usually incommensurate. Whereas the spatial Talbot effect has been studied in free space, the temporal Talbot effect has been observed only in single-mode optical fibers. The prospects are therefore not promising for realizing a Talbot effect in space \textit{and} time.

Consider the discretization of the spatio-temporal spectrum of a propagation-invariant ST wave packet. Because of the tight association between $k_{x}$ and $\omega$ in such a ST wave packet, discretizing $k_{x}$ implies a discretization of $\omega$, and vice versa [Fig.~\ref{Fig:discrete_STWP}]. If the spatial spectrum is discretized periodically along $k_{x}$ (thus producing a periodic transverse spatial profile), the temporal spectrum is \textit{not} discretized periodically [Fig.~\ref{Fig:discrete_STWP}(a)], and vice versa [Fig.~\ref{Fig:discrete_STWP}(b)]. Nevertheless, periodically discretizing $k_{x}$ has led to the observation of the spatial Talbot effect in the temporal domain as a manifestation of `time-diffraction' \cite{Yessenov2020PRLveiled}.

However, GVD is needed to observe the temporal Talbot effect. As described in Section~\ref{Sec:DispersiveWPs}, GVD is readily exhibited by dispersive ST wave packets in free space by modifying the angular-dispersion profile of the ST wave packet. Introducing GVD with extremely large values of $k_{2}$ of either sign produces ST wave packets whose dispersion length \textit{in free space} is extremely small. By periodically discretizing $\omega$, the on-axis pulse train displays the temporal Talbot effect in free space over short distances, leading to the first observation of the temporal Talbot effect in a freely propagating field not confined to a single-mode fiber \cite{Hall2021OLtempTalbot}.

\begin{figure}[t!]
  \begin{center}
  \includegraphics[width=10.0cm]{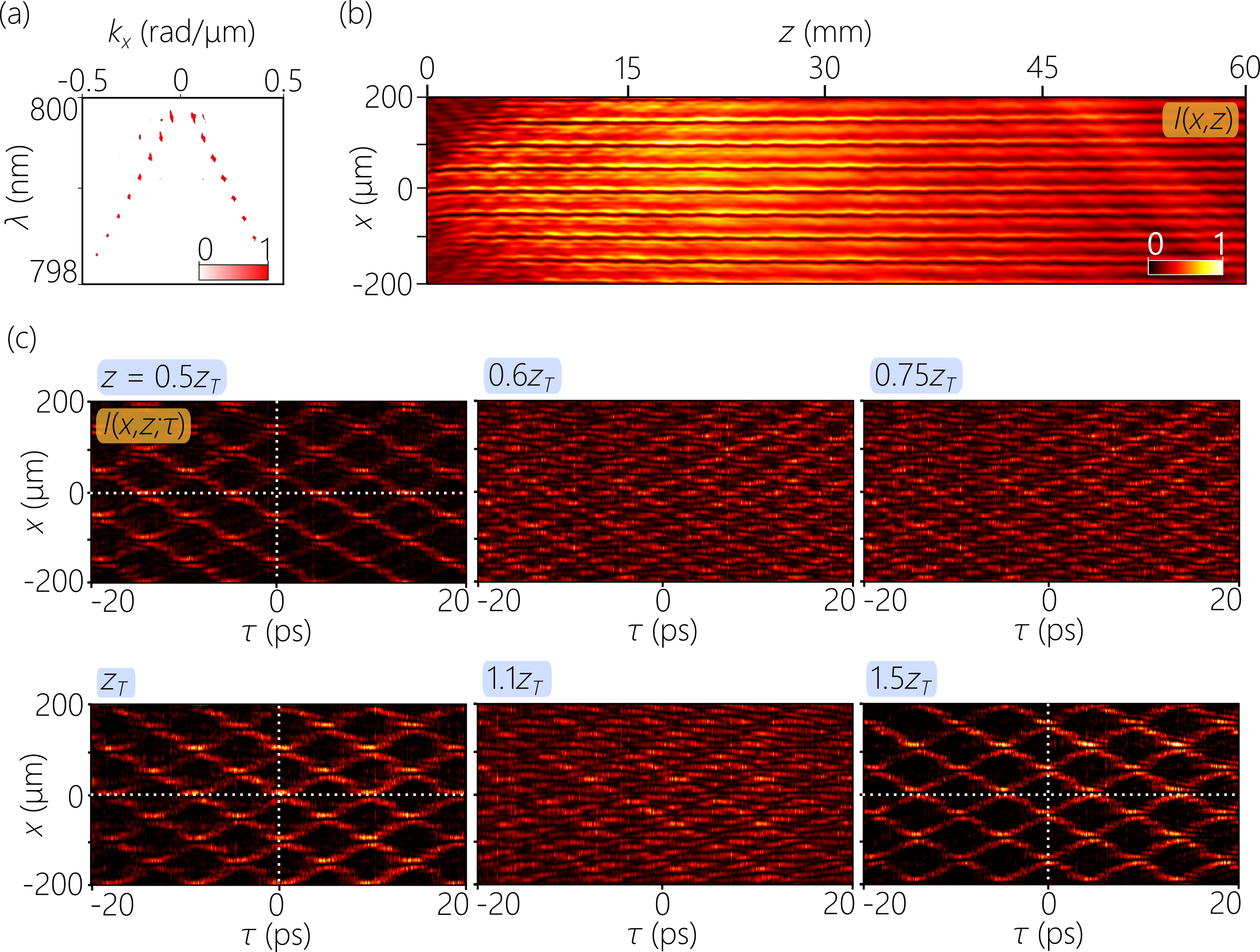}
  \end{center}
  \caption{Measurements of the ST Talbot effect with $\alpha\!=\!0.033$, $\lambda_{\mathrm{o}}\!=\!800$~nm, $\Delta\lambda\!=\!2$~nm, $L\!=\!100$~$mu$m, $T\!=\!10$~ps, and $z_{\mathrm{ST}}\!=\!25$~mm. (a) The time-averaged intensity $I(x,z)$ is diffraction-free. (b) The spatio-temporal spectrum of the ST field in $(k_{x},\lambda)$-space. (c) The spatio-temporal profile $I(x,z;\tau)$ measured at $0.5z_{\mathrm{ST}}$, $0.6z_{\mathrm{ST}}$, $0.75z_{\mathrm{ST}}$, $z_{\mathrm{ST}}$, $1.1z_{\mathrm{ST}}$, and $1.5z_{\mathrm{ST}}$; see \cite{Hall2021APLSTTalbot}.}
  \label{Fig:Talbot_measure}
\end{figure}

The need to discretize both $k_{x}$ \textit{and} $\omega$ periodically to observe the ST Talbot effect can be achieved by exploiting a particular dispersive ST wave packet where $k_{x}$ and $\omega$ are linearly related: $\Omega\!=\!\alpha c|k_{x}|$; see Fig.~\ref{Fig:discrete_STWP}(c). Because of the resulting V-shaped spatio-temporal spectrum [Fig.~\ref{Fig:Talbot_measure}(a)], this dispersive ST wave packet has been dubbed a `V-wave' \cite{Hall2021PRA-Vshape}. The underlying angular dispersion is differentiable, the resulting GVD is anomalous $c\omega_{\mathrm{o}}k_{2}\!=\!-\tfrac{1}{\alpha^{2}}$, and the group velocity is always $\widetilde{v}\!=\!c$. In addition to guaranteeing that $k_{x}$ and $\omega$ are periodically sampled simultaneously, the V-wave has a unique characteristic: the diffraction and dispersion lengths are intrinsically equal. This can be understood by observing the `Janus'-like behavior of its spectral phase $e^{ik_{z}z}$. Because $\Omega\!=\!\alpha c|k_{x}|$ in a V-wave, it can be shown that $k_{z}\!\approx\!k_{\mathrm{o}}+\tfrac{\Omega}{c}-\tfrac{k_{x}^{2}}{2k_{\mathrm{o}}}\!=\!k_{\mathrm{o}}+\tfrac{\Omega}{c}+\tfrac{1}{2}k_{2}\Omega^{2}$ for a V-wave, with $\alpha\!=\!\tfrac{1}{c\sqrt{-k_{2}k_{\mathrm{o}}}}$. In other words, when $k_{z}$ for a V-wave is expanded in terms of $\Omega$, it represents a pulse propagating in a dispersive medium undergoing GVD, and when expanded in terms of $k_{x}$, it represents a beam diffracting in free space. Consequently, the diffraction and dispersion lengths of a V-wave uniquely are intrinsically equal, not due to any careful tuning of material and field characteristics, but due to the structure of the wave packets' spatio-temporal spectrum. Discretizing $k_{x}$ and $\omega$ periodically results therefore in equal lengths for spatial and temporal Talbot revivals, thus leading to the ST Talbot effect. Whereas the time-averaged intensity has a spatially periodic profile that remains diffraction-free along $z$ and does not exhibit any axial dynamics [Fig.~\ref{Fig:Talbot_measure}(b)], veiled beneath this profile is complex spatio-temporal dynamics [Fig.~\ref{Fig:Talbot_measure}(c)] in which an initial periodic spatio-temporal lattice undergoes diffraction and dispersion, only for the initial lattice to undergo revivals at planes separated by the ST Talbot length $z_{\mathrm{T}}\!=\!\tfrac{L^{2}}{2\lambda_{\mathrm{o}}}\!=\!\tfrac{T^{2}}{\pi|k_{2}|}$; see \cite{Hall2021APLSTTalbot}. The phenomena associated with the conventional spatial and temporal Talbot effects (such as rate multiplication) can be observed in the joint spatio-temporal domain [Fig.~\ref{Fig:Talbot_measure}(c)].

Previous work on self-imaging of ST wave packets examined theoretically the effect with wave packets that lack themselves a periodic structure (and thus retain a continuous spectrum). In this case, the wave packet is constructed by superposing wave packets of different phase velocities \cite{Reivelt2002OE}. In another case, a new self-imaging effect was realized in space by superposing pulsed Bessel beams that travel obliquely with respect to the propagation axis \cite{Bock2017OL}. In both cases, the investigated self-imaging effects are -- strictly speaking -- unrelated to the conventional spatial Talbot effect.

\section{Refraction of ST wave packets}\label{Section:Refraction}

New and counter-intuitive refractive phenomena arise when ST wave packets impinge on a planar interface between two non-dispersive, isotropic, and homogeneous dielectrics, whether at normal or oblique incidence. The unique spectral structure of ST wave packets leads to fascinating consequences upon refraction, such as group-velocity invariance or group-velocity inversion across the interface, anomalous refraction (increase in the group velocity upon transition to a higher-index medium), and changes in the group velocity of the transmitted wave packet with incident angle. In general, striking refractive phenomena arise for baseband ST wave packets, whereas the refraction of X-waves and sideband ST wave packets are less interesting. We will thus focus on the former, and then explain briefly why the latter follow the same basic trend of conventional wave packets.

\begin{figure}[t!]
  \begin{center}
  \includegraphics[width=10cm]{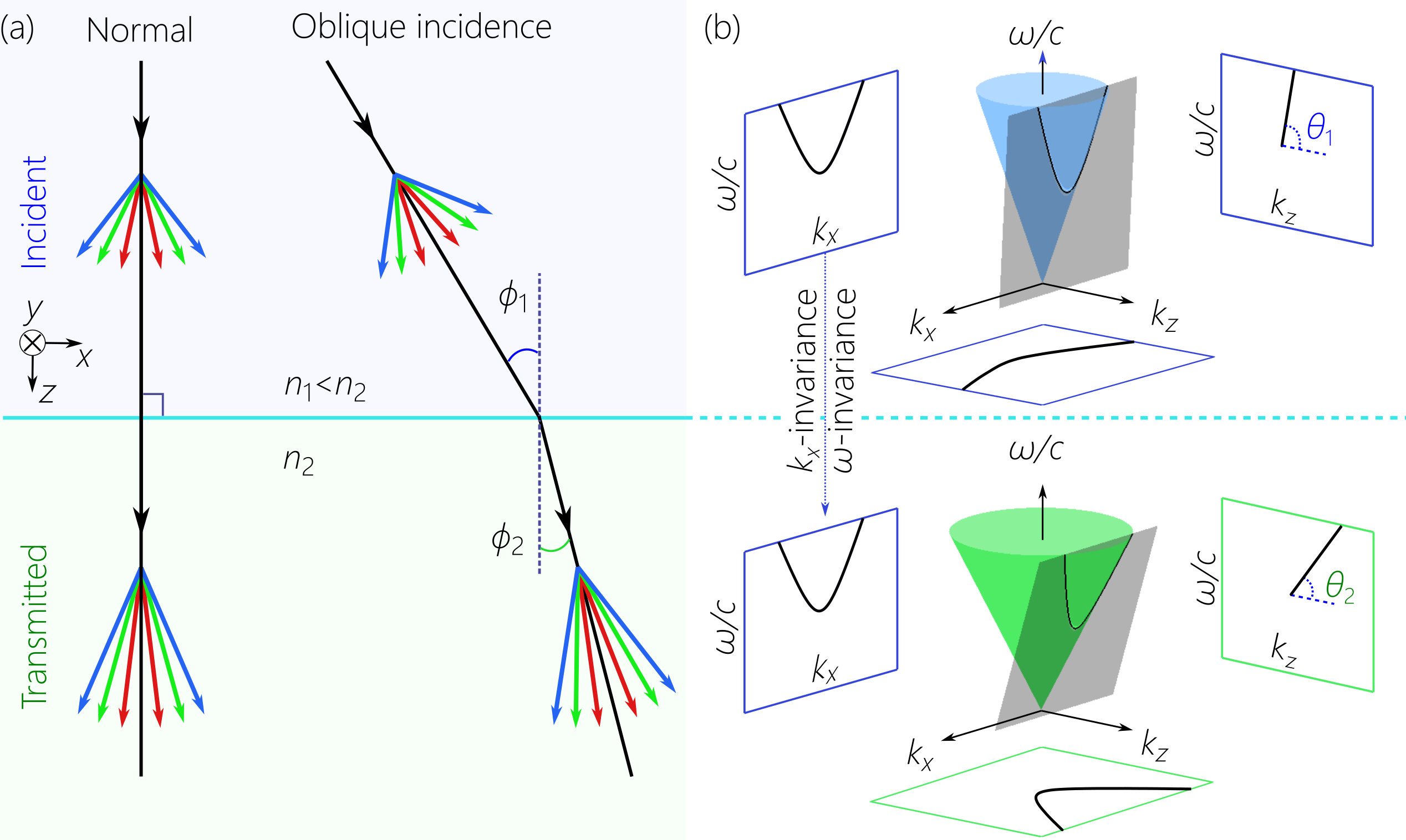}
  \end{center}
  \caption{Refraction of ST wave packets. (a) A ST wave packet is incident normally or obliquely at a interface. (b) The light-cone opening angle changes with the refractive index. At normal incidence, the projection onto the $(k_{x},\tfrac{\omega}{c})$-plane is invariant, enforcing a change in the spectral tilt angle from $\theta_{1}$ to $\theta_{2}$.}
  \label{Fig:RefractionOverview}
\end{figure}

Refraction at a planar interface between two homogeneous, isotropic, non-dispersive dielectrics (refractive indices $n_{1}$ and $n_{2}$) is governed by the conservation of transverse momentum ($k_{x}$ is invariant across the interface), and conservation of energy ($\omega$ is invariant across the interface); see Fig.~\ref{Fig:RefractionOverview}(a). These constraints for a monochromatic plane-wave give rise to Snell's law $n_{1}\sin{\phi_{1}}\!=\!n_{2}\sin{\phi_{2}}$, where $\phi_{1}$ and $\phi_{2}$ are the angles made with respect to the normal to the interface by the incident and transmitted waves, respectively. The same relationship applies to plane-wave pulses \textit{in absence of dispersion}, whereupon the group velocities in the two media of the incident and transmitted wave packets are $\widetilde{v}_{1}\!=\!c/n_{1}$ and $\widetilde{v}_{2}\!=\!c/n_{2}$, respectively, which are thus determined by the local optical environment. The general expectation is thus that the group velocity will drop when traversing the interface from a low-index medium to a high-index one, $n_{2}\!>\!n_{1}$, and that changing the incident angle does not impact $\widetilde{v}_{2}$. Whereas Snell's law governs the dynamics of an external DoF (the propagation angle), new laws of refraction for ST wave packets can be formulated in terms of the spectral tilt angle $\theta$, which is an internal DoF that determines the global structure of the ST wave packet and is in principle independent of the extrinsic DoFs (profile shape, central wavelength, beam width, temporal bandwidth, etc.). 

The origin of the remarkable refraction phenomena exhibited by ST wave packets can be understood as follows. Consider an incident ST wave packet with group index $\widetilde{n}_{1}\!=\!\cot{\theta_{1}}$, where $\theta_{1}$ is the spectral tilt angle in the first medium. Upon transmission, where the opening angle of the light-cone increases because $n_{2}\!>\!n_{1}$, the invariance of $k_{x}$ and $\omega$ results in a ST wave packet with a new spectral tilt angle $\theta_{2}$, $\widetilde{n}_{2}\!=\!\cot{\theta_{2}}$, and the group velocity may thus change in unexpected ways. By identifying a new optical refraction invariant that we call the `spectral curvature', we formulate an equation for the change in $\theta$ across the planar interface, which thus represents a new law of refraction unique to ST wave packets. 

\begin{figure}[t!]
  \begin{center}
  \includegraphics[width=9.5cm]{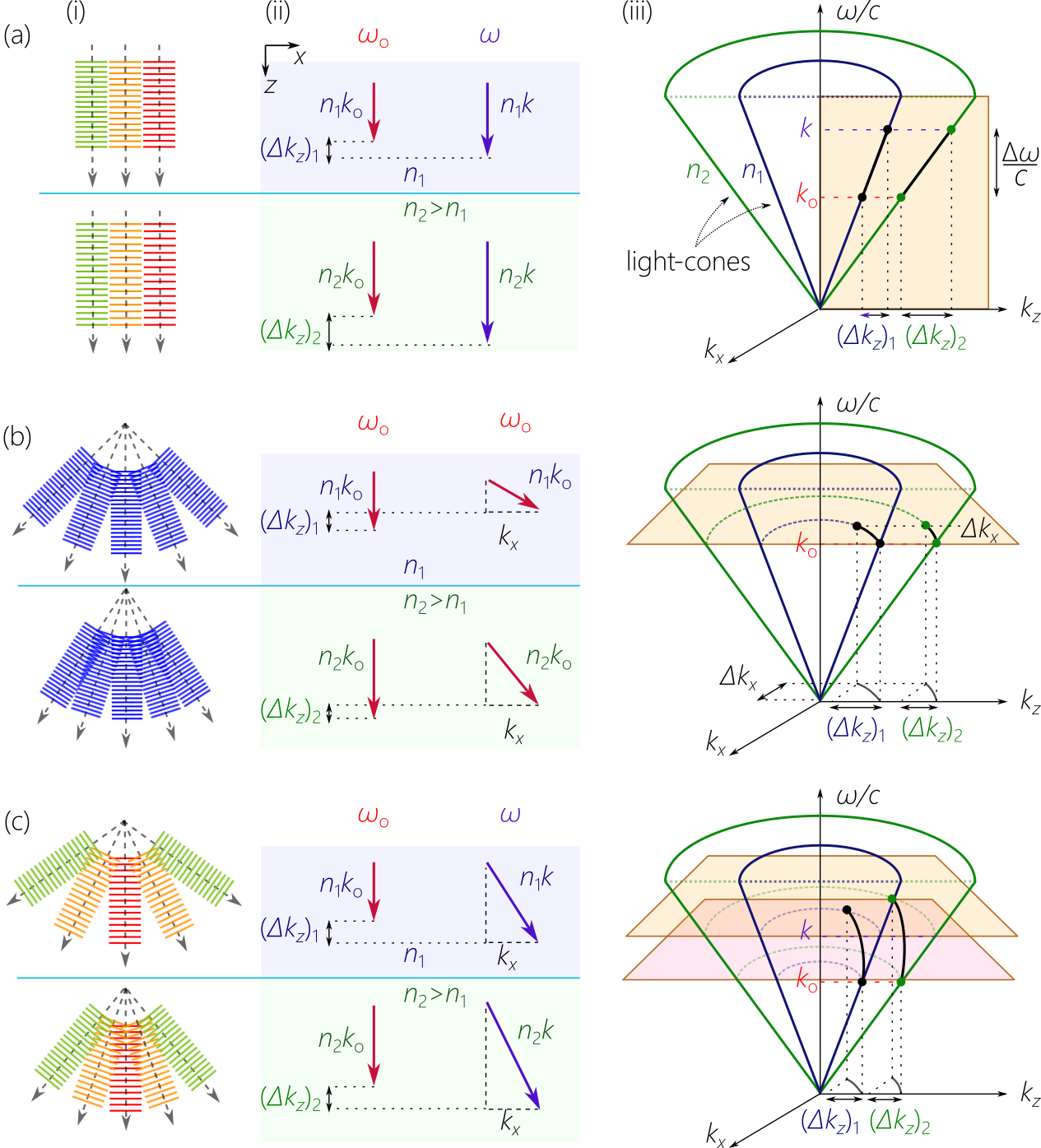}
  \end{center}
  \caption{Refraction of monochromatic plane waves from a medium of index $n_{1}$ to another of index $n_{2}\!>\!n_{1}$. The third column depicts the plane waves as points on the light-cones associated with the two media. (a) Two plane waves at frequencies $\omega_{\mathrm{o}}$ and $\omega\!>\!\omega_{\mathrm{o}}$ are normally incident on the interface, and $(\Delta k_{z})_{2}\!>\!(\Delta k_{z})_{1}$ because of light-cone inflation. (b) Two plane waves of the same frequency $\omega_{\mathrm{o}}$ are incident normally and obliquely on the interface, and $(\Delta k_{z})_{2}\!<\!(\Delta k_{z})_{1}$ despite light-cone inflation. (c) Two plane waves at frequencies $\omega_{\mathrm{o}}$ and $\omega\!>\!\omega_{\mathrm{o}}$ are incident normally and obliquely, respectively, and we may have either $(\Delta k_{z})_{2}\!<\!(\Delta k_{z})_{1}$ or $(\Delta k_{z})_{2}\!>\!(\Delta k_{z})_{1}$.}
  \label{Fig:RefractionConcept}
\end{figure}

The peculiar nature of the refraction of ST wave packets points to non-trivial refraction-driven dynamics exhibited by their spectral support domain. To understand this behavior quantitatively, we examine first the refraction of two simpler field configurations: a plane-wave pulse and a monochromatic beam. First, consider the refraction from a low-index medium $n_{1}$ to a high-index medium $n_{2}\!>\!n_{1}$ (the light-cone-angle in the second medium is larger) of a plane-wave pulse at normal incidence [Fig.~\ref{Fig:RefractionConcept}(a)], whose spectral support domain lies on the light-line ($k_{x}\!=\!0$). The surface of the light-cone `inflates' and the distance between any two points on the surface of the light-cone in the first medium will in general increase in the second. However, the temporal bandwidth $\Delta\omega$ is invariant, so light-cone inflation stretches $\Delta k_{z}$ from $(\Delta k_{z})_{1}$ to $(\Delta k_{z})_{2}\!>\!(\Delta k_{z})_{1}$, and $\widetilde{v}$ drops $\widetilde{v}_{2}\!=\!\tfrac{\Delta\omega}{(\Delta k_{z})_{2}}\!<\!\widetilde{v}_{1}\!=\!\tfrac{\Delta\omega}{(\Delta k_{z})_{1}}$ as expected. In other words, $\widetilde{n}_{1}\!=\!n_{1}$ and $\widetilde{n}_{2}\!=\!n_{2}$, and light-cone inflation reduces the group velocity of a normally incident plane-wave pulse.

\begin{figure}[t!]
  \begin{center}
  \includegraphics[width=13.2cm]{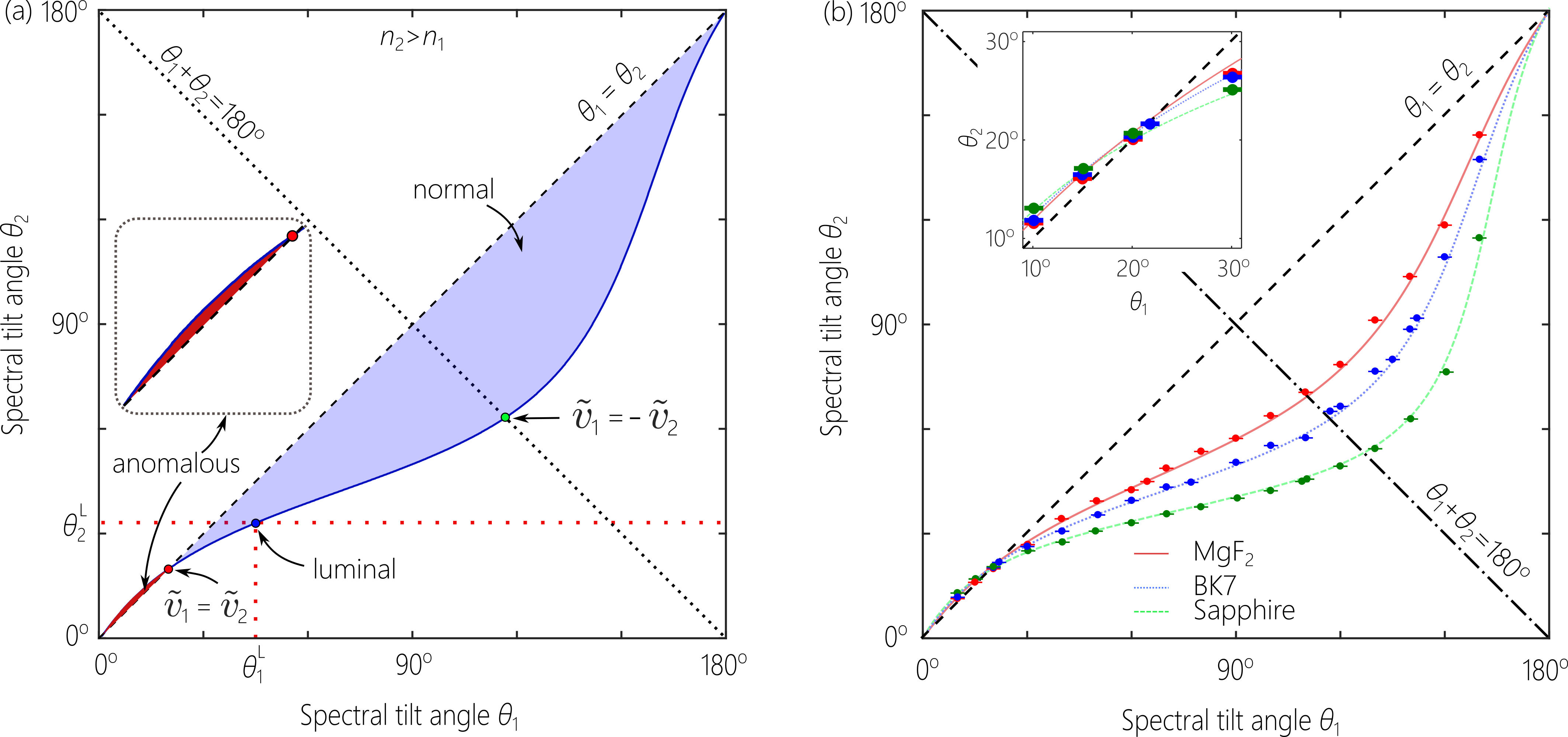}
  \end{center}
  \caption{(a) The relationship between $\theta_{1}$ and $\theta_{2}$ based on Eq.~\ref{Eq:LawOfRefractionNormal} with $n_{1}\!=\!1$ and $n_{2}\!=\!1.5$, but the overall features are generic. (b) Experimental verification of the law of refraction at normal incidence (Eq.~\ref{Eq:LawOfRefractionNormal}) from free space.}
  \label{Fig:NormalIncidenceData}
\end{figure}

Consider next a normally incident monochromatic beam whose spectral support domain is the circle at the intersection of the light-cone with a horizontal iso-frequency plane [Fig.~\ref{Fig:RefractionConcept}(b)]. Although the radius of this circle increases with light-cone inflation, the spatial bandwidth $\Delta k_{x}$ is fixed because of the conservation of transverse momentum. In the paraxial approximation, $(\Delta k_{z})_{2}\!\approx\!\tfrac{(\Delta k_{x})^{2}}{2n_{2}k_{\mathrm{o}}}\!<\!(\Delta k_{z})_{1}\!\approx\!\tfrac{(\Delta k_{x})^{2}}{2n_{1}k_{\mathrm{o}}}$, which is the opposite trend compared to that of the plane-wave pulse. Here $c\tfrac{(\Delta k_{z})_{1}}{\Delta\omega}\!=\!\tfrac{g}{n_{1}}$ and $c\tfrac{(\Delta k_{z})_{2}}{\Delta\omega}\!=\!\tfrac{g}{n_{2}}$, where $g\!=\!\tfrac{1}{2}(\tfrac{\Delta k_{x}}{k_{\mathrm{o}}})^{2}\tfrac{\omega_{\mathrm{o}}}{\Delta\omega}$ is independent of the refractive indices. In other words, the conservation of transverse momentum \textit{shrinks} the axial wave numbers rather than \textit{stretching} them as in the case of plane-wave pulses. However, because the beam is monochromatic, we cannot define a group-velocity.

Light-cone inflation thus affects these two field configurations in opposite directions: $\Delta\omega$-invariance stretches $\Delta k_{z}$ for a plane-wave pulse; whereas $\Delta k_{x}$-invariance shrinks $\Delta k_{z}$ for a monochromatic beam. In the case of a baseband ST wave packet these two effects are combined, and the two opposing trends compete [Fig.~\ref{Fig:RefractionConcept}(c)]. In each medium, the group index of the ST wave packet is $\widetilde{n}\!=\!\tfrac{c}{\widetilde{v}}\!=\!c\tfrac{\Delta k_{z}}{\Delta\omega}\!=\!n-\frac{g}{n}$, where the first term is that for a plane-wave pulse, the second is for a monochromatic beam, and the invariant quantity $g$ is defined as $g\!=\!n(n-\widetilde{n})$. Because the quantity $g$ is \textit{independent} of the refractive index, the quantity $n(n-\widetilde{n})$ is a refractive constant in any medium \cite{Ginis2020NP}. We dub $g$ the `spectral curvature' because it is related to the curvature of the parabolic spectral projection onto the $(k_{x},\tfrac{\omega}{c})$-plane. Note that $g\!>\!0$ in the superluminal regime and $g\!<\!0$ in the subluminal. For conventional wave packets $g\!\approx\!0$, so that the refractive phenomena associated with ST wave packets cannot be observed.

The spectral curvature $g$ is a new optical refraction invariant that can be used to obtain a law of refraction for ST wave packets at normal incidence:
\begin{equation}
n_{1}(n_{1}-\widetilde{n}_{1})=n_{2}(n_{2}-\widetilde{n}_{2}).
\label{Eq:LawOfRefractionNormal}
\end{equation}
We plot this relationship in Fig.~\ref{Fig:NormalIncidenceData}(a) in terms of the spectral tilt angles $\theta_{1}$ and $\theta_{2}$. This unique law of refraction for baseband ST wave packets has been verified at normal incidence \cite{Bhaduri2020NP}; see Fig.~\ref{Fig:NormalIncidenceData}(b). Intriguingly, we can prepare ST wave packets in free space ($n_{1}\!=\!1$) with group index $\widetilde{n}_{1}$ such that they subsequently travel in any medium beyond the interface ($n_{2}\!=\!n$) at the speed of light in vacuum $\widetilde{n}_{2}\!=\!1$ if $\widetilde{n}_{1}\!=\!1+n-n^{2}$ \cite{Bhaduri2019Optica}. Interestingly, the transition from positive to negative values of $\widetilde{n}_{1}$ occurs at a refractive index $n$ equal to the golden mean $\phi\!=\!\tfrac{1+\sqrt{5}}{2}\!\approx\!1.618$.

\begin{figure}[t!]
  \begin{center}
  \includegraphics[width=9cm]{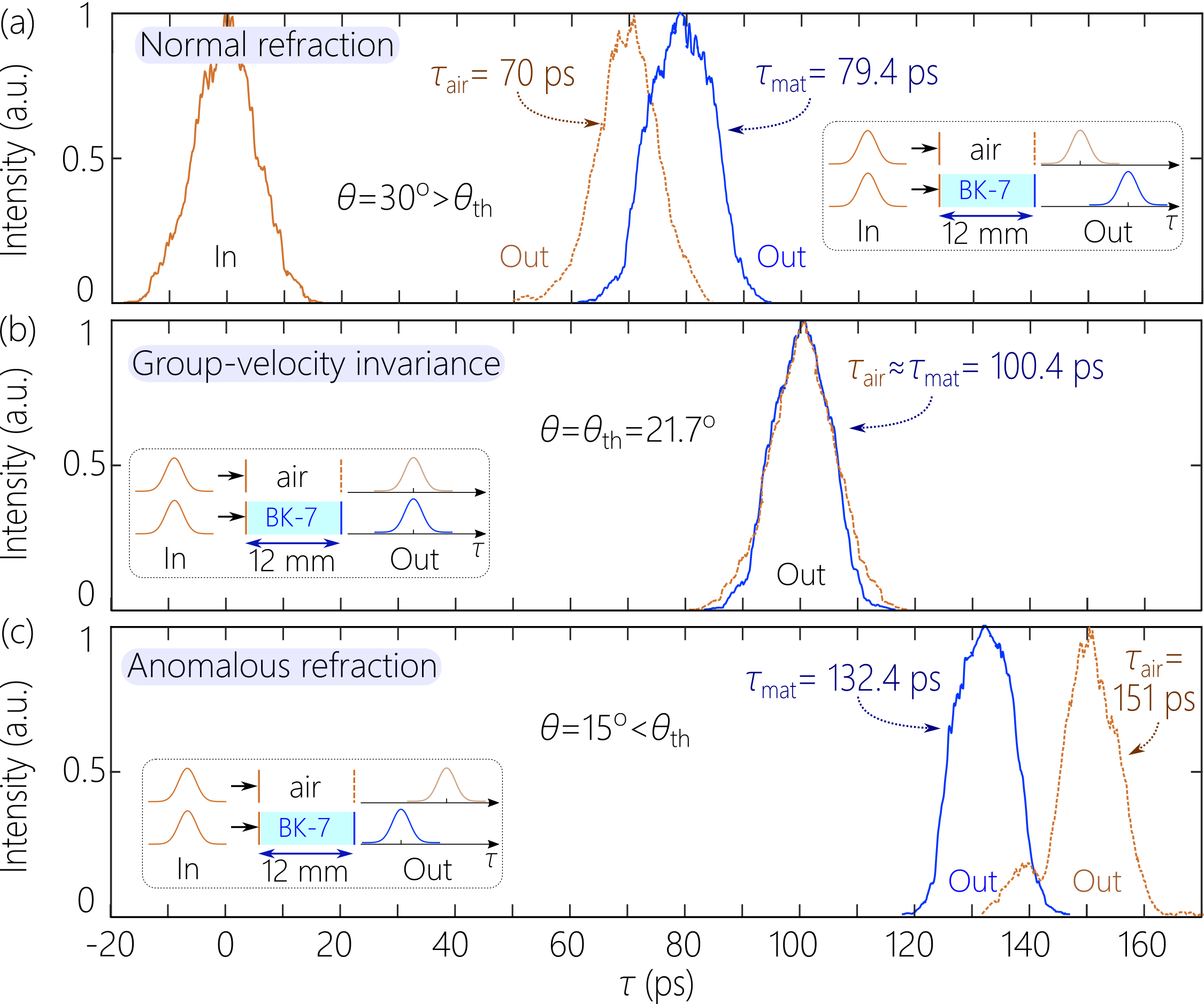}
  \end{center}
  \caption{ST wave packets after traversing 12~mm of air (group delay $\tau_{\mathrm{air}}$) and 12~mm of BK7 (group delay $\tau_{\mathrm{mat}}$) where $\theta_{\mathrm{th}}\!\approx\!21.7^{\circ}$ for normal incidence from free space. (a) Normal refraction $\tau_{\mathrm{mat}}\!>\!\tau_{\mathrm{air}}$ at $\theta_{1}\!=\!30^{\circ}\!>\!\theta_{\mathrm{th}}$; the wave packet is slower in BK7 than in free space. (b) Group-velocity invariance $\tau_{\mathrm{mat}}\!=\!\tau_{\mathrm{air}}$ at $\theta_{1}\!=\!21.7^{\circ}\!=\!\theta_{\mathrm{th}}$; the wave packet has the same speed in free space and BK7. (c) Anomalous refraction $\tau_{\mathrm{mat}}\!<\!\tau_{\mathrm{air}}$ at $\theta_{1}\!=\!15^{\circ}\!<\!\theta_{\mathrm{th}}$; the wave packet is faster in BK7 than in free space.}
  \label{Fig:NormalAnomalousRefractionData}
\end{figure}

The shape of the curve in Fig.~\ref{Fig:NormalIncidenceData}(a) is generic for any pair of refractive indices $n_{1}$ and $n_{2}$, and we can readily extract from it several surprising predictions.
\begin{enumerate}
    \item \textit{Group-velocity invariance}. Setting $\widetilde{n}_{1}\!=\!n_{1}+n_{2}$ results in $\widetilde{n}_{2}\!=\!\widetilde{n}_{1}$, which corresponds to the point at the intersection of the curve with the diagonal $\theta_{1}\!=\!\theta_{2}$. This threshold value $\widetilde{n}_{\mathrm{th}}\!=\!n_{1}+n_{2}$ occurs in the subluminal regime and represents a ST wave packet that traverses the interface with no change in group velocity -- regardless of the index contrast between the two media. We refer to this phenomenon as \textit{group-velocity invariance}.
    \item \textit{Normal and anomalous refraction}. When $\widetilde{n}_{1}\!<\!\widetilde{n}_{\mathrm{th}}$, the expected refractive behavior follows: $\widetilde{n}_{2}\!>\!\widetilde{n}_{1}$ when $n_{2}\!>\!n_{1}$, and the group velocity drops in the high-index medium, which we refer to as \textit{normal refraction}. In contrast, when $\widetilde{n}_{1}\!>\!\widetilde{n}_{\mathrm{th}}$, the opposite trend occurs and $\widetilde{n}_{2}\!<\!\widetilde{n}_{1}$ with $n_{2}\!>\!n_{1}$; i.e., the group velocity anomalously \textit{increases} in the higher-index medium, which we thus refer to as \textit{anomalous refraction}.
    \item \textit{Group-velocity inversion}. When $\widetilde{n}_{1}\!=\!n_{1}-n_{2}$, then $\widetilde{n}_{2}\!=\!n_{2}-n_{1}\!=\!-\widetilde{n}_{1}$; i.e., the group velocity of the transmitted wave packet has the same magnitude but opposite sign of that for the incident wave packet, which we refer to as \textit{group velocity inversion}. This corresponds to the intersection of the curve in Fig.~\ref{Fig:NormalIncidenceData}(b) with the anti-diagonal $\theta_{1}+\theta_{2}\!=\!180^{\circ}$. A ST wave packet traversing equal-length layers of indices $n_{1}$ and $n_{2}$ incurs zero group delay, or \textit{group-delay cancellation}.
\end{enumerate}

These phenomena of normal refraction [Fig.~\ref{Fig:NormalAnomalousRefractionData}(a)], group-velocity invariance [Fig.~\ref{Fig:NormalAnomalousRefractionData}(b)], anomalous refraction [Fig.~\ref{Fig:NormalAnomalousRefractionData}(c)], and group-velocity inversion and group-delay cancellation \cite{Bhaduri2020NP,AllendeMotz2021refraction} have all been confirmed experimentally.

\begin{figure}[t!]
  \begin{center}
  \includegraphics[width=13cm]{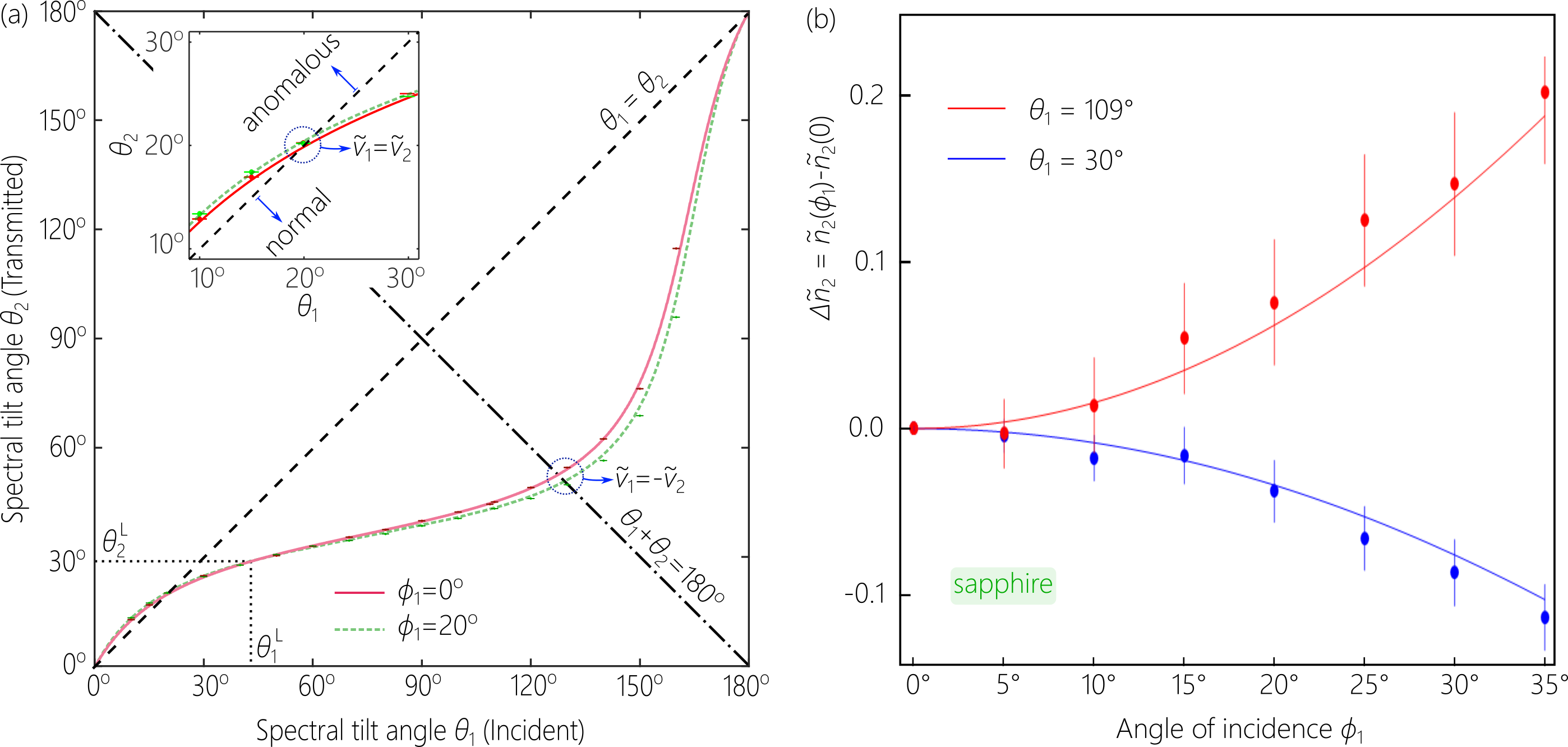}
  \end{center}
  \caption{(a) Law of refraction for ST wave packets at oblique incidence $\phi_{1}\!=\!20^{\circ}$ (Eq.~\ref{Eq:ObliqueIncidence}), compared to normal incidence (Eq.~\ref{Eq:LawOfRefractionNormal}). The inset highlights the transition from normal to anomalous refraction. (b) Measured change in $\widetilde{n}_{2}$ at oblique incidence onto sapphire ($n_{2}\!=\!1.76$) from free space with respect to normal incidence, $\Delta\widetilde{n}_{2}\!=\!\widetilde{n}_{2}(\phi_{1})-\widetilde{n}_{2}(0)$ for subluminal ($\theta_{1}\!=\!30^{\circ}$, $\widetilde{v}\!=\!0.58c$) and superluminal ($\theta_{1}\!=\!109^{\circ}$, $\widetilde{v}\!=\!-2.9c$) ST wave packets. }
  \label{Fig:ObliqueIncidenceData}
\end{figure}

Oblique incidence brings a new surprise, where the invariant spectral curvature becomes $n(n-\widetilde{n})\cos^{2}{\phi}$, and $\phi$ is the angle with respect to the normal to the interface. Therefore, the law of refraction of ST wave packets at oblique incidence becomes:
\begin{equation}\label{Eq:ObliqueIncidence}
n_{1}(n_{1}-\widetilde{n}_{1})\cos^{2}{\phi_{1}}=n_{2}(n_{2}-\widetilde{n}_{2})\cos^{2}{\phi_{2}},
\end{equation}
where $\phi_{1}$ and $\phi_{2}$ are related by Snell's law [Fig.~\ref{Fig:ObliqueIncidenceData}(a)]. An intriguing consequence of Eq.~\ref{Eq:ObliqueIncidence} is that the group index in the second medium $\widetilde{n}_{2}$ changes with $\phi_{1}$ for fixed $\widetilde{n}_{1}$. Indeed, $\widetilde{n}_{2}(\phi_{1})$ increases or the wave packet slows down with $\phi_{1}$ in the superluminal regime when $n_{2}\!>\!n_{1}$. Conversely, $\widetilde{n}_{2}(\phi_{1})$ decreases or the wave packet speeds up with $\phi_{1}$ in the subluminal regime. This unique behavior [Fig.~\ref{Fig:ObliqueIncidenceData}(b)] gives rise to new configurations for synchronization in optical communications that have been recently verified: isochronous ST wave packets \cite{AllendeMotz2021OLisochronous} and blind synchronization \cite{Yessenov2021JOSAAIII}.

\begin{figure}[t!]
  \begin{center}
  \includegraphics[width=11cm]{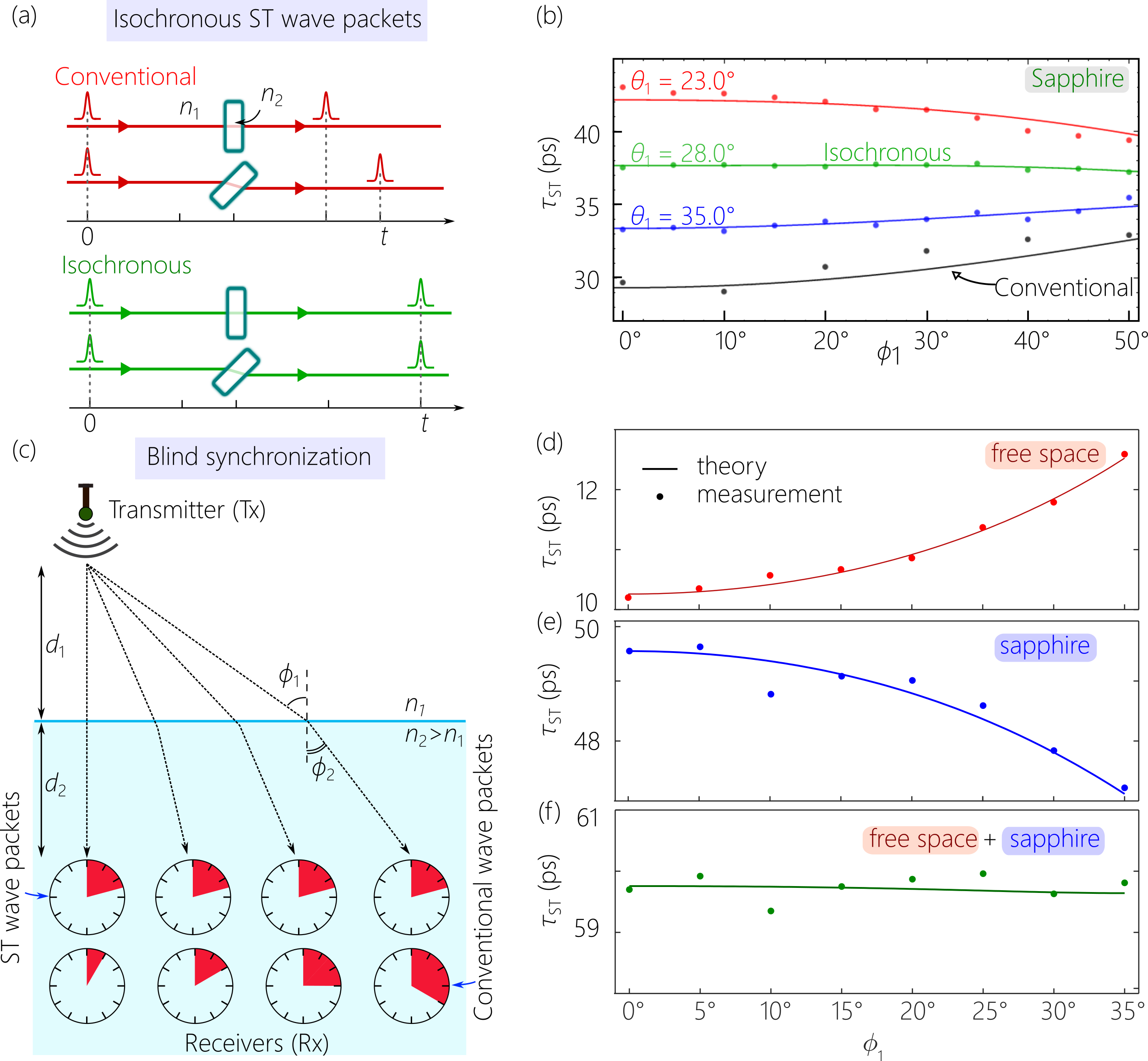}
  \end{center}
  \caption{(a) The group delay $\tau$ incurred by a conventional wave packet traversing a slab increases at oblique incidence, but is independent of the incident angle $\phi_{1}$ for an isochronous ST wave packet. (b) Measured $\tau$ with $\phi_{1}$ for different $\theta_{1}$ of incident ST wave packets from free space. (c) Multiple receivers (Rx) at \textit{a priori} unknown locations but at the same depth $d_{2}$ receive pulses from a transmitter (Tx). For conventional wave packets, $\tau$ increases with $\phi_{1}$, but can be constant for ST wave packets. (d-f) Measured $\tau$ with $\phi_{1}$ for a ST wave packet over (d) $d_{1}\!=\!1$~mm of free space; (e) $d_{2}\!=\!5$~mm of sapphire; and (f) the sum of the group delays from (d) and (e), showing an approximately constant value despite the different path lengths.}
  \label{Fig:IsochronousSynchronization}
\end{figure}

The group delay $\tau$ accrued by a conventional wave packet traversing a non-dispersive slab increases with incident angle $\phi_{1}$ because of the larger path length. For an \textit{isochronous} ST wave packet, $\tau$ is independent of $\phi_{1}$ [Fig.~\ref{Fig:IsochronousSynchronization}(a)]. This seemingly impossible task can be achieved with subluminal ST wave packets because the group velocity in the slab \textit{increases} with $\phi_{1}$ and compensates for the longer path length \cite{AllendeMotz2021OLisochronous}; see Fig.~\ref{Fig:IsochronousSynchronization}(b).

\textit{Blind synchronization} is illustrated in Fig.~\ref{Fig:IsochronousSynchronization}(c). Consider a transmitter in a medium of index $n_{1}$ broadcasting wave packets over a wide range of incident angles $\phi_{1}$ to reach receivers located at a fixed depth beyond an interface with a medium of index $n_{2}\!>\!n_{1}$, but whose positions are unknown \textit{a priori}. The group delay of a conventional pulse increases with $\phi_{1}$. However, a subluminal ST wave packet will speed up in the second medium at oblique incidence and thus has the chance to reach its receiver after accruing the same group delay as the normally incident ST wave packet. Selecting an appropriate group index $\widetilde{n}_{1}$ for the incident ST wave packet can help realize blind synchronization over a substantial angular range, as shown in Fig.~\ref{Fig:IsochronousSynchronization}(d-f). 

To the best of our knowledge, there were no previous experimental tests of the refraction of propagation-invariant optical wave packets. There have been a few theoretical studies of the refraction of X-waves \cite{Shaarawi00JASA,Attiya01PER,Shaarawi01PER,Salem12JOSAA} and FWMs \cite{Hillion93Optik,Donnelly97IEEE,Hillion98JO,Hillion99JOA}. These studies have focused on the changes in the spatio-temporal structure of the wave packets upon reflection or transmission. However, no new refractive phenomena were predicted. For X-waves, it can be shown that the quantity $n^{2}-\widetilde{n}^{2}$ is a refractive invariant at normal incidence \cite{Yessenov2021JOSAAI}. The attendant law of refraction for X-waves, $n_{1}^{2}-\widetilde{n}_{1}^{2}=n_{2}^{2}-\widetilde{n}_{2}^{2}$, implies only normal refraction. None of the phenomena we described above for baseband ST wave packets can occur in this case. A variant of this law can be derived at normal incidence for sideband ST wave packets, including FWMs, implying also that they exhibit only normal refraction \cite{Yessenov2021JOSAAI}. At oblique incidence, we can no longer define a refractive invariant. The reason is that the transverse and longitudinal wave numbers mix at oblique incidence, so that the transmitted ST wave packet changes its structure altogether and is no longer propagation-invariant. This was recognized early on in \cite{Donnelly97IEEE} for the case of FWMs.

Refraction of baseband ST wave packets in dispersive media is now being studied theoretically \cite{He2021Arxiv} and experimentally \cite{Yessenov2022OLRefDisp}. Accounting for the unique aspects of the refraction of ST wave packets will be essential for all applications where ST wave packets are synthesized in free space and then coupled into a particular medium. The results in this Section are therefore of foundational importance for future work on ST wave packets.

\section{Prospects for localizing ST wave packets in all dimensions}\label{Sec:LocalizingInAllDimensions}

Throughout this Review, we have considered ST wave packets localized along one transverse coordinate $x$, while holding the field uniform along $y$; that is, the ST wave packet takes the form of a light-sheet. This restriction can be useful in several respects: (1) it is conceptually simpler, and the reduced dimensionality enables visualization on the light-cone; (2) it is mathematically more tractable; (3) it highlights the fact that we do not need two transverse dimensions to achieve propagation invariance of a pulsed beam (Section \ref{Section:PropagationInvarianceIn1D}), whereas they are essential for diffraction-free propagation of a monochromatic beam (Section \ref{Section:DiffractionFree1D}); and (4) some applications require optical fields localized along only one transverse dimension, such as surface plasmon polaritons (Section~\ref{Sec:STSPP}) \cite{Schepler2020ACSP}, guided fields in planar waveguides (Section~\ref{Sec:Waveguides}) \cite{Shiri2020NC_Hybrid}, and light-sheet microscopy \cite{Vettenburg2014NM,Piksarv17SR}. Nevertheless, many applications require a wave packet that is localized in all dimensions, whose time-averaged intensity therefore takes the form of a `needle' rather than a `sheet'. Examples include optical communications that requires coupling to optical fibers, nonlinear optics, and laser machining. Additionally, some spatial attributes of the optical field require two dimensions for their embodiment, such as orbital angular momentum (OAM) \cite{Yao2011AOP}, and optical vortices or vector beams \cite{Bazhenov1992JMO,Basistiy1995OC,Gahagan1996OL,Zhan2009AOP,Swartzlander2014Book}. It is therefore critical to identify pathways towards synthesizing such ST fields, which we refer to as `3D ST wave packets'.

Unfortunately, the experimental strategy described in Section~\ref{Section:SynthesisSetup} cannot be directly extended to two transverse dimensions, and thus cannot produce 3D ST wave packets. One dimension of the 2D SLM at the heart of the setup in Fig.~\ref{Fig:STsetup} is dedicated to $\lambda$ and the other to $k_{x}$, so a second transverse dimension cannot be simultaneously manipulated. Indeed, this constraint applies to all spatio-temporal fields synthesized via a 2D SLM after spatially resolving the spectrum. For example, the field associated with the transverse OAM demonstrated in \cite{Jhajj2016PRX,Hancock2019Optica,Chong2020NP,Hancock2021Optica} produced by the setup in Fig.~\ref{Fig:ModulationSchemes1}(b) cannot be modified by the SLM along the transverse dimension parallel to the OAM axis (Section~\ref{Sec:OpticalVortices}). 

Moreover, cascading two such systems -- one dedicated to $k_{x}$ and the other to $k_{y}$ -- would not solve the problem of producing 3D ST wave packets because we require that each $\omega$ be related to a single \textit{radial} wave number or spatial frequency $k_{r}\!=\!\sqrt{k_{x}^{2}+k_{y}^{2}}$, which necessitates manipulating the spatial frequencies along $x$ and $y$ simultaneously rather than successively. The constraint $k_{z}\!=\!k_{\mathrm{o}}+\Omega/\widetilde{v}$ in Eq.~\ref{Eq:BasebandPlane} for baseband ST wave packets still holds for 3D ST wave packets, but we now require that $\tfrac{\Omega}{\omega_{\mathrm{o}}}\!=\!\tfrac{k_{r}^{2}}{2k_{\mathrm{o}}^{2}(1-\widetilde{n})}$. That is, we aim at producing non-differentiable \textit{radial} angular dispersion with high spectral resolution -- a task that has not been achieved in optics to date.

In this context, it is useful to remind ourselves of previously synthesized classes of 3D ST wave packets. The propagation-invariant X-waves reported in \cite{Saari1997PRL} do indeed incorporate both transverse dimensions, but X-waves are altogether free of angular dispersion (all the temporal frequencies travel at the same angle with the propagation axis). Furthermore, the results reported for the modified X-waves in \cite{Sonajalg1997OL} are inconclusive with regards to the spatio-temporal field structure achieved. On the other hand, the sole experimental report of optical FWMs made use of one transverse spatial dimension \cite{Reivelt2002PRE}, and similarly for the attempts at preparing ST wave packets for propagation invariance in dispersive media \cite{Dallaire2009OE,Jedrkiewicz2013OE}. This emphasizes that non-differentiable radial angular dispersion has not yet been realized to date. 

Recently, two experimental approaches have been devised to address this perennial challenge and yield 3D ST wave packets localized in all dimensions \cite{Yessenov2022Localized3D,Pang2021OL}. One of these strategies (developed by Alan Willner's group at USC) is suitable for implementation with a laser-comb source having discrete spectral lines \cite{Pang2021OL}. The other strategy (developed by our group at CREOL, UCF) is suitable for an optical source having a continuous temporal spectrum \cite{Yessenov2022Localized3D}. The 3D ST wave packets produced with these two techniques have similar characteristics to those of the ST wave packets described in this Review. Moreover, such 3D ST wave packets can carry OAM, and can be shaped into optical vortices and vector-beam structures, thereby opening new vistas for propagation-invariant pulsed OAM wave packets.

A recent theoretical proposal \cite{Guo2021Light} suggests an alternative nanophotonic route to the synthesis of 3D ST wave packets by making use of so-called non-local nanophotonic structures that have wave-vector-dependent -- rather than space-dependent -- transfer functions \cite{Kwon2018PRL,Guo2018Optica,Guo2018JOSAA,Wang2020ACSP,Guo2020Optica}. This approach makes use of a 2D photonic crystal whose isotropic band diagram has the required quadratic relationship $\omega\!\propto\!k_{r}^{2}$. When an ultrafast pulse is focused into this photonic crystal, the device acts as a spatio-temporal filter that extracts the target spectral structure for a subluminal baseband 3D ST wave packet from the incident generic field. This strategy therefore resembles the filtering strategy in \cite{Dallaire2009OE,Jedrkiewicz2013OE} for a reduced-dimensional ST wave packet, except that a 3D ST wave packet is produced in a single step in \cite{Guo2021Light}. It is likely, however, that this approach would suffer from low throughput, but it is nevertheless an important step towards the synthesis of ST wave packets with a single on-chip nanophotonic device. We expect that the immediate future will witness rapid developments in constructing simple, compact, yet versatile synthesis arrangements for 3D ST wave packets. 

\section{State-of-the-art of ST wave packets}\label{Sec:StateOfTheArt}

Over the past 5 years, a flurry of research has brought to the fore the unique behavior displayed by pulsed optical fields in which a precisely sculpted non-separable spatio-temporal structure has been inculcated. Within this general enterprise, ST wave packets have been recently investigated along exciting new lines of inquiry that we survey here. For example, several technical advances have been reported, including the synthesis of ST wave packets at large bandwidths \cite{Kondakci2018OE}, and at extreme wavelengths in the mid-infrared \cite{Yessenov2020OSAC}, terahertz, or X-ray regimes \cite{Tan2021AS}. On the conceptual side, several new investigations have enriched our understanding of the structure and behavior of ST wave packets. This includes examining the representation of ST wave packets on the surface of the light-cone \cite{Efremidis2017OL}, and clarifying the interpretation of the group velocity of superluminal X-waves and its relationship to the energy velocity \cite{Saari2019PRAenergyflow,Saari2020PRA}. An important advance has been the study of ST wave packets from the perspective of angular dispersion as initiated in \cite{Wong2017ACSP2,Kondakci2019ACSP}. From this perspective, ST wave packets are examples of TPFs \cite{Torres2010AOP,Fulop2010Review}, except that \textit{non-differentiable} angular dispersion undergirds them \cite{Hall2021OL-AD-PFT,Hall2021NGVDrealizing,Yessenov2021engineering,Hall2021OEConsequences,Hall2021OESynthesizer}, a result that is critical in the study of dispersion introduced into ST wave packets.

A particularly fruitful direction of study is that of implementing controllable deviation from strict propagation invariance. We have already seen in Section~\ref{Sec:ModifyingTheAxialProp} three such scenarios, including dispersive \cite{Yessenov2021engineering,Hall2021NGVDrealizing}, accelerating \cite{Yessenov2020PRLaccel,Hall2022OLArbAccel}, and axially spectrally encoded ST wave packets \cite{AllendeMotz2020PRA}, in addition to the axial dynamics associated with the ST Talbot effect \cite{Hall2021APLSTTalbot}. More recently, a new idea has been developed whereby a propagation-invariant ST wave packet is accompanied by another feature that focuses suddenly at a prescribed axial position without otherwise affecting the propagation-invariant field structure \cite{Wong2020AS}; see Fig.~\ref{fig:Other_groups1}(a). This field configuration is realized by combining two distinct spatio-temporal spectral structures: a one-to-one relationship between $k_{x}$ and $\omega$ responsible for the propagation-invariant ST wave packet structure, and a second spectral feature responsible for the focusing field. In this way, multiple types of contrasting behavior can be exhibited simultaneously along the propagation axis.

Moreover, deviation from the X-shaped profile in the paraxial regime is made possible for propagation-invariant ST wave packets through spectral-phase modulation. We have already seen (Section~\ref{Sec:PropInvarianceAiry}) that adding a cubic spectral phase produces an Airy ST wave packet whose spatio-temporal profile is no longer X-shaped but instead traces an Airy function in space-time \cite{Kondakci2018PRL}. More recently, a general approach has been systematically developed that produces a variety of spatio-temporal profiles without impacting their propagation invariance by engineering the spectral phase to introduce caustics into the field \cite{Wong21OE}. In this manner, a host of complex profiles can be exploited in lieu of the more typical X-shaped profile [Fig.~\ref{fig:Other_groups1}(b)].

\begin{figure}
    \centering
    \includegraphics[width=11cm] {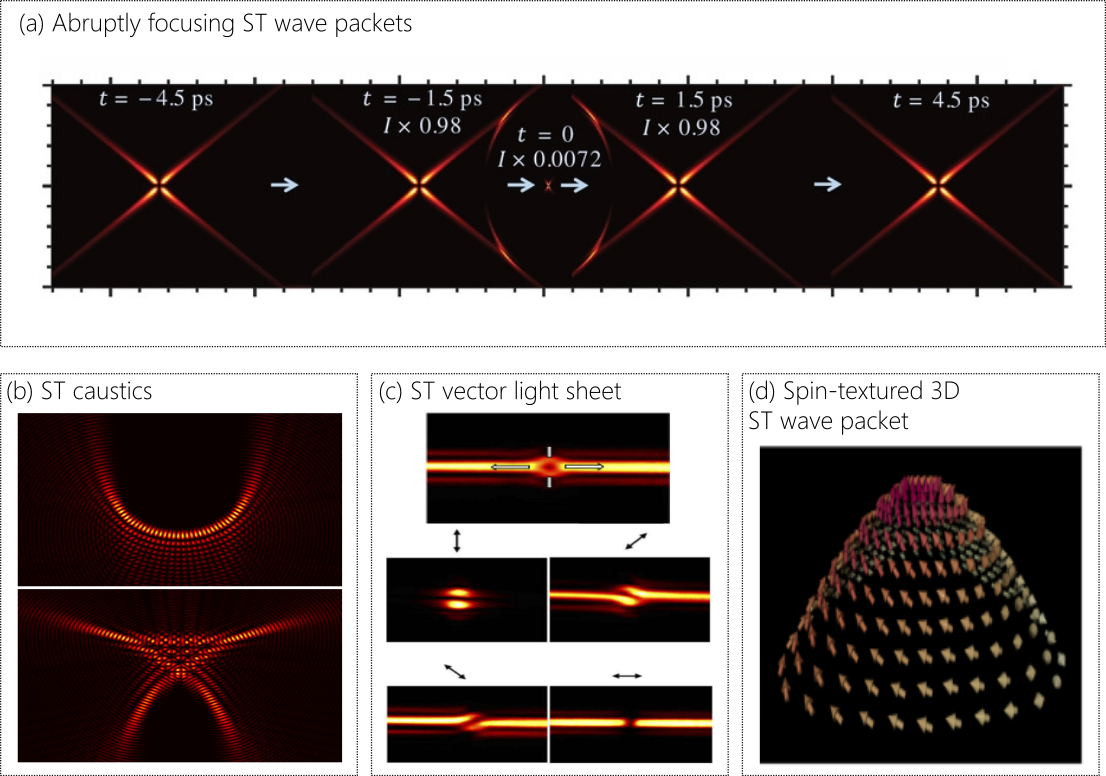} 
    \caption{(a) Abrupt focusing of a portion of the optical field accompanying a propagation-invariant ST wave packet \cite{Wong2020AS}. (b) Propagation-invariant ST light caustics \cite{Wong21OE}. (c) A vector-beam structure introduced into a ST light sheet \cite{Diouf2021OE}. (d) Spin-texture for polarization of a 3D ST wave packet in the spatio-temporal spectral space \cite{Guo2021Light}.}
    \label{fig:Other_groups1}
\end{figure}

Finally, the study of polarization in conjunction with ST wave packets has been recently initiated. Although Brittingham's FWM was a vectorial solution to Maxwell's equation \cite{Brittingham1983JAP}, work on ST wave packets since has dealt almost exclusively with scalar fields \cite{Belanger1984JOSAA,Sezginer1985JAP}. More recently, introducing polarization structures into paraxial ST wave packets has been explored \cite{Ornigotti16JO}, including vector-beam structures impressed upon ST light sheets \cite{Diouf2021OE}; Fig.~\ref{fig:Other_groups1}(c). As a further example, a recent study has demonstrated theoretically that sophisticated topological features such as Meron spin textures \cite{Guo2020PRL} can be incorporated into the polarization of 3D ST wave packets \cite{Guo2021Light}; see Fig.~\ref{fig:Other_groups1}(d). We expect the realization of 3D ST wave packets to lead to rapid developments along these lines.

\section{Novel spatio-temporal field structures}\label{Section:novelST-structures}

In addition to propagation-invariant ST wave packets, new pulsed-beam configurations have recently emerged with surprising propagation characteristics that stem from precise spatio-temporal structuring of the field. These new wave packets have a crucial characteristic in common with ST wave packets: the spatial and temporal DoFs are \textit{not} separable. Some of these spatio-temporally structured fields may share a feature with ST wave packets; e.g., they may have a controllable group velocity. In general, all these fields exhibit some form of space-time coupling \cite{Akturk2010JO}, which is usually viewed as an undesirable feature in pulsed fields, but is exploited here for useful purposes. We review here some of the prominent recent developments that have added to the vitality of this field, highlighting the commonalities and distinctions with respect to ST wave packets.

\subsection{Flying-focus}

\begin{figure}
    \centering
    \includegraphics[width=11cm] {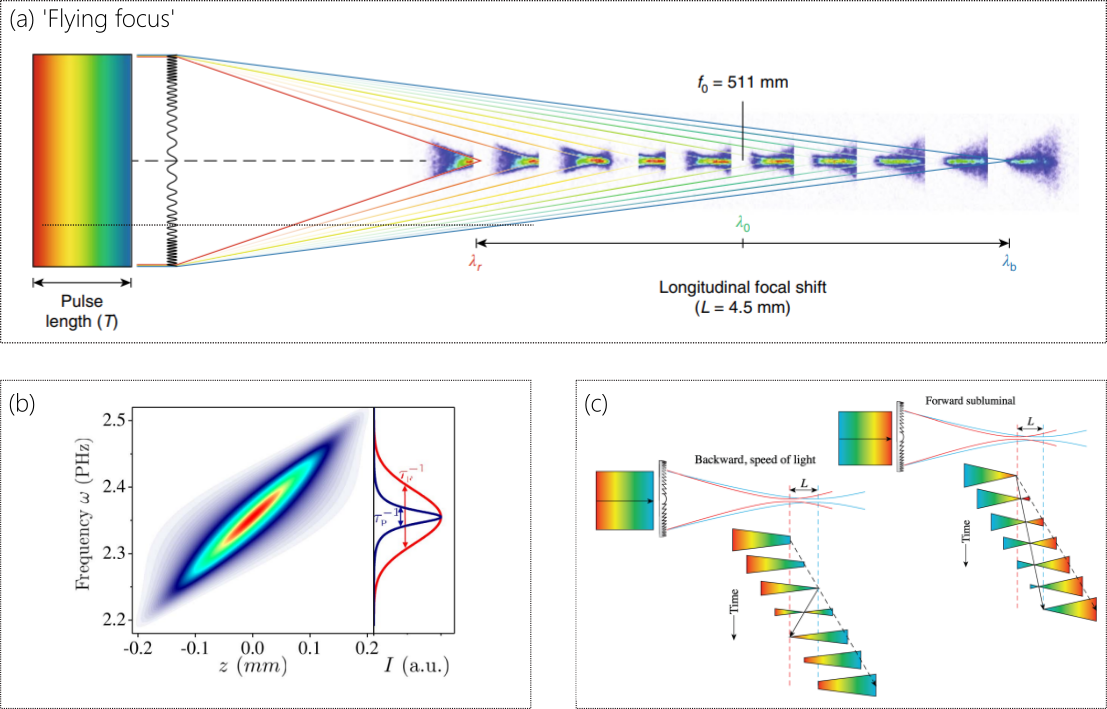} 
    \caption{A `flying-focus' as a spatio-temporally structured wave packet. (a) Controlling the group velocity of a flying focus through axial chromatism. A chirped pulse is focused with a lens endowed with chromatic aberrations \cite{Froula2018NP,Jolly2020OE}, (b) to produce a field configuration in which the on-axis spectrum is correlated with the axial position \cite{SaintMarie2017Optica}. (c) A judicious combination of chirp sign and lens focusing can create forward- or backward-propagating flying-foci \cite{Froula2019PP}.}
    \label{fig:FlyingFocus}
\end{figure}

A striking recent advance is the development of so-called `flying focus' wave packets \cite{SaintMarie2017Optica,Froula2018NP,Jolly2020OE}. These are pulsed beams whose group velocity can be tuned by introducing `axial chromatism' into the field. A plane-wave pulse is chirped to separate the wavelengths axially, and is then focused by a lens with chromatic aberrations. Consequently, different portions of the pulse spectrum are focused successively at different axial positions (determined by the dispersion introduced into the initial pulse and the particular chromatic aberrations of the lens); see Fig.~\ref{fig:FlyingFocus}(a). One can view the flying-focus as a wave packet whose on-axis wavelength $\lambda$ is tightly associated with the axial position $z$ (rather than with the axial wave number $k_{z}$ as in a propagation-invariant ST wave packet); see Fig.~\ref{fig:FlyingFocus}(b). By changing this association, forward- or backward-propagating flying-foci can be produced [Fig.~\ref{fig:FlyingFocus}(c)]. In fact, the peak of the on-axis focused pulse can be made to travel at any selected group velocity \cite{Froula2018NP}. However, despite the tunability of the group velocity, the wave packet is \textit{not} propagation-invariant. 

The range of tunability for $\widetilde{v}$ reported thus far for flying-foci \cite{Froula2018NP} is similar to that demonstrated by ST wave packets \cite{Kondakci2019NC}. Furthermore, the spectral evolution of a flying-focus resembles that of axially encoded ST wave packets (Section~\ref{Sec:AxialSPectralEncoding} and Fig.~\ref{Fig:AxialEncoding}). Despite these similarities and their lack of diffraction-free behavior,  flying-foci that are localized in all dimensions have a crucial advantage in the simplicity of their production.

The flying-focus was proposed and studied theoretically in \cite{SaintMarie2017Optica}, and subsequently demonstrated in \cite{Froula2018NP,Jolly2020OE}. The applications of flying-foci have been pursued mainly in the area of nonlinear plasma physics \cite{Froula2019PP}. For example, the ionization wave in a plasma tracks the peak of the laser-pulse intensity, so that flying-foci can produce ionization waves of arbitrary velocity \cite{Turnbull2018,Franke2019OE}, which can then help accelerate the photons in a second pulse \cite{Howard2019PRL}. Moreover, a flying-focus can result in self-acceleration and enhance the spectral broadening and intensity steepening, thereby forming an optical shock and further enhacing the spectral broadening \cite{Franke2021PRA}. Furthermore, optical nonlinearities can be exploited to tailor the arrival time of the various slices of the input pulse spectrum at different locations within the focal volume \cite{Simpson2020OE}. Indeed, cross-phase modulation in a Kerr lens can help sculpt a desired spatial profile on the flying-focus and endow it with OAM \cite{Simpson2021arxiv}.

\subsection{Toroidal pulses}

Electric and magnetic dipoles can be viewed as resulting from a pair of opposite charges and a current loop, respectively. These two elementary sources of electromagnetic radiation (and their multipolar combinations) are fundamental to our understanding of the electromagnetic properties of matter \cite{Jackson1999Book}. Less familiar are \textit{toroidal dipoles}, which are also localized electromagnetic excitations, but result from a current flowing on the surface of a torus, and represent a third independent family of elementary electromagnetic sources \cite{Papasimakis16NM}. After decades of studying toroidal dipoles across a variety of branches of physics \cite{Dubovik1990PR,Flambaum1997PRC,Ceulemans1998PRL,Afanasiev2001JPD} since their introduction by Y.~B.~Zeldovich in 1957 \cite{Zeldovich1957JETP}, they are now being carefully examined in optics -- both as an electromagnetic response of artificial photonic media endowed with particular topological structural features \cite{Papasimakis2009PRL,Kaelberer2010Science}, and as a specific toroidal field excitation propagating in free space \cite{Raybould2016OE,Raybould2017APL}. The latter, known as a toroidal pulse (and sometimes -- less descriptively -- as a flying doughnut) is particularly interesting in our context here because such wave packets are not separable with respect to their spatial and temporal DoFs \cite{Shen2021PRRes}. Indeed, the unique properties of toroidal pulses stem from the association between the spatial and temporal DoFs of the pulsed field. However, each wavelength is associated \textit{not} with a single spatial frequency, but instead with a different finite spatial bandwidth \cite{Papasimakis2018PRB}, so that toroidal pulses are \textit{not} propagation invariant.

\begin{figure}
    \centering
    \includegraphics[width=11cm] {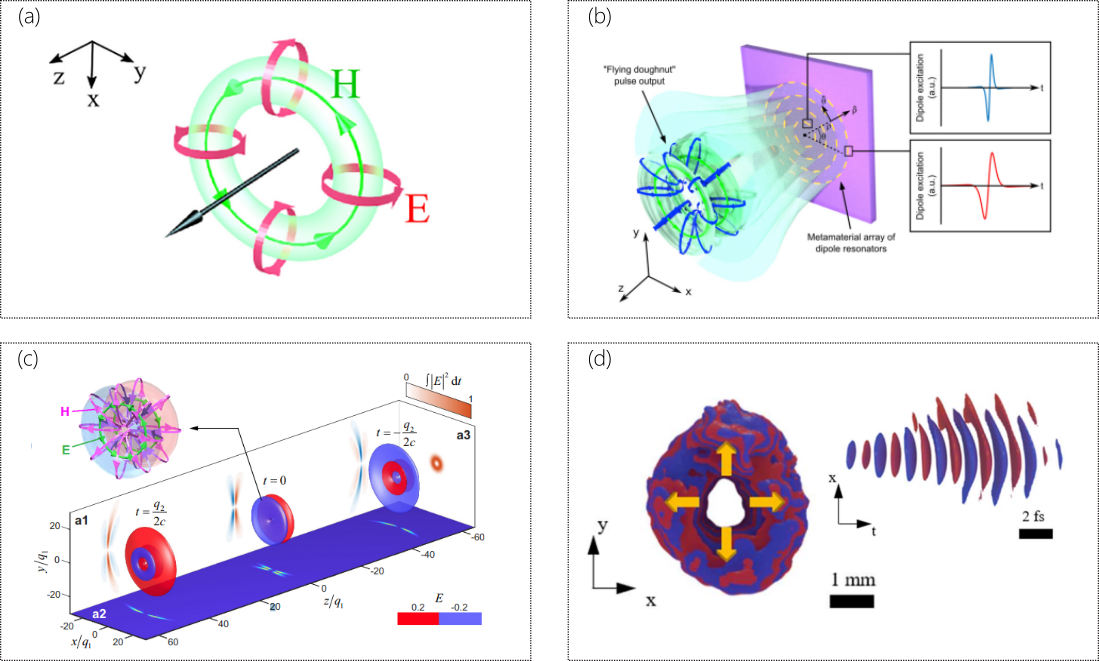} 
    \caption{Toroidal pulses as spatio-temporally structured wave packets. (a) The field structure of a toroidal pulse \cite{Raybould2017AIP}. (b) Generating a toroidal pulse from a plane-wave pulse by means of a metamaterial \cite{Papasimakis2018PRB}. (c) Axial evolution of a toroidal pulse \cite{Shen2021PRRes}. (d) Initial experimental observation of a toroidal pulse \cite{ZdagkasarXiv2021}.}
    \label{fig:Other_groups5}
\end{figure}

The first proposed toroidal pulse was a single cycle wave packet \cite{Hellwarth1996PRE}, and thus synthesizing such a field structure poses significant difficulties. A TM polarized toroidal pulse comprises an azimuthally polarized magnetic field confined in a torus-shaped region, with the electric field winding along the meridians of the torus \cite{Raybould2017APL}; see Fig.~\ref{fig:Other_groups5}(a). Viewed from a different viewpoint, each position in the toroidal-pulse wave front is associated with a prescribed temporal spectrum. Therefore, one pathway to synthesizing such a wave packet is to exploit a metasurface in which each point on the surface provides a resonance response that filters the required spectrum for the toroidal pulse at that point from a plane-wave pulse \cite{Papasimakis2018PRB,Zdagkas19Nano,Zdagkas2020PRA}; Fig.~\ref{fig:Other_groups5}(b). The spatio-temporally structured wave packet is then expected to evolve into the desired toroidal pulse at a prescribed plane [Fig.~\ref{fig:Other_groups5}(c)]. Initial results support the viability of this strategy for producing toroidal pulses \cite{Zdagkas2021arxiv}; see Fig.~\ref{fig:Other_groups5}(d). Furthermore, toroidal pulses offer a potential platform for realizing topological spin-textured polarization fields such as skyrmions \cite{Shen2021NC}. It is expected that toroidal pulses will optimize the coupling of the electromagnetic field with toroidal excitations in structured photonic environments. 

\subsection{OAM in spatio-temporally structured pulsed fields}

The propagation invariance of ST wave packets stems from the tight association between the spatial and temporal frequencies underlying the pulsed field structure. Critical to propagation invariance is that the spatial mode associated with each $\omega$ be propagation invariant itself; e.g., a plane wave or a Bessel beam. However, we may consider the association between $\omega$ and a different spatial DoF or a set of spatial modes. Although such wave packets are \textit{not} propagation invariant, they can nevertheless exhibit useful features. Indeed, one pathway to synthsizing 3D ST wave packets is by associating Bessel modes of a particular OAM order with the discrete wavelengths in a laser comb \cite{Pang2021OL} (Section~\ref{Sec:LocalizingInAllDimensions}). This suggests new ways in which OAM can be involved in the construction of spatio-temporally structured fields.

\begin{figure}
    \centering
    \includegraphics[width=11cm] {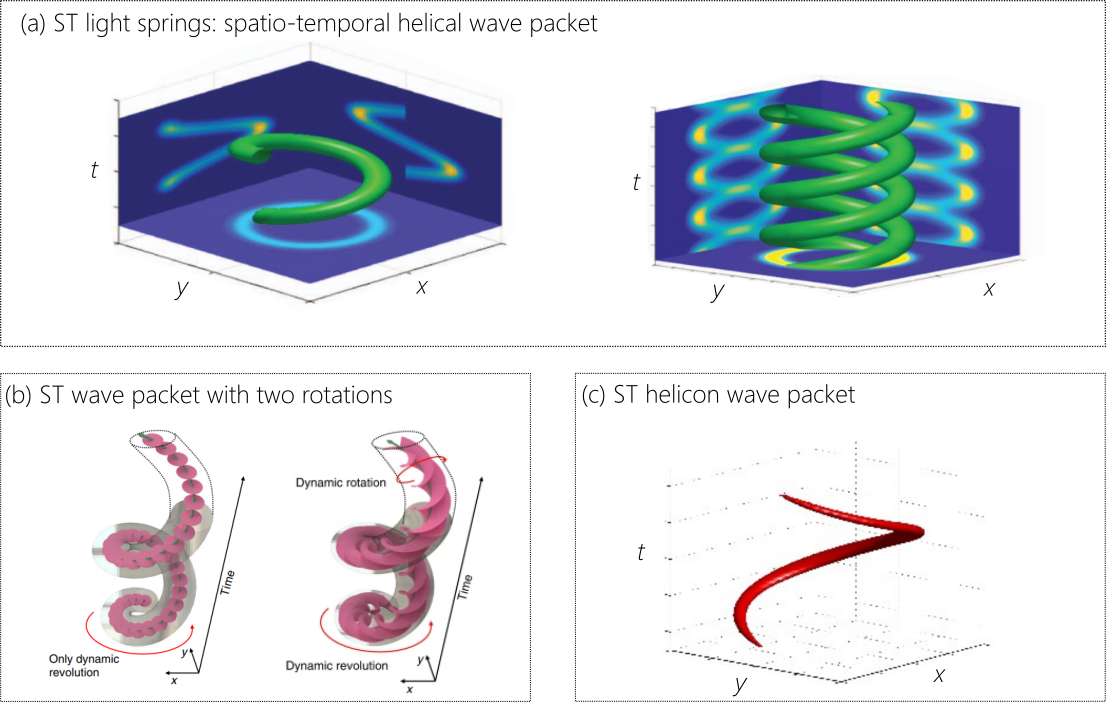} 
    \caption{Overview of some of the recent work on spatio-temporally structured wave packets. (a) Spatio-temporal encoded light springs of helical wave packets \cite{Pariente2015OL}; (b) dynamic spatio-temporal structured pulsed fields combining two orbital-angular momenta \cite{Zhao2020NC}; (c) spatio-temporal helicon in a multimode waveguide \cite{Bejot2021ACSP}.}
    \label{fig:Other_groups1}
\end{figure}

An early example of studying OAM in the context of spatio-temporally structured fields was suggested theoretically in \cite{Pariente2015OL} where a different OAM modal order was assigned to \textit{each} wavelength in a laser comb. The starting point is the recognition that combining multiple monochromatic Laguerre-Gaussian (LG) modes -- of fixed radial index but varying OAM order $\ell$ -- results in an off-axis localized spatial beam. By associating each LG mode with one temporal frequency $\ell\!=\!\ell(\omega)$, the relative phases between the LG modes become time-dependent, and the localized spatial feature rotates with time around the propagation axis in a single helix or in multiple intertwined helices; see Fig.~\ref{fig:Other_groups1}(a). Although this field structure was termed a `light spring' in \cite{Pariente2015OL}, it has since been known as helical wave packets to avoid the unintended implications of the spring `elasticity' \cite{Porras2021PRA}.

A new dimension for such helical ST structures was recently unveiled in \cite{Zhao2020NC}. Here each comb frequency line is associated \textit{not} with a single LG OAM mode of order $\ell$, but is instead associated with a prescribed superposition of LG modes having the same OAM order $\ell$ but different radial indices. The resulting wave packet combines two distinct rotation motions as manifestations of OAM. First, the spatial profile of the wave packet \textit{rotates} locally around its axis (which is absent from \cite{Pariente2015OL}), and, second, it \textit{revolves} around a central axis as in \cite{Pariente2015OL}; see Fig.~\ref{fig:Other_groups1}(b). This wave packet therefore behaves in much the same way as the Earth rotates around its axis while simultaneously revolving around the Sun.

Most recently, the association between OAM and temporal frequencies has been proposed in the context of guided ST supermodes in a multimode optical fiber as a methodology to produce `helicon' wave packets [Fig.~\ref{fig:Other_groups1}(c)]. While the structure of this wave packet resembles the helical light springs proposed in \cite{Pariente2015OL}, it nevertheless has the advantage of being propagation invariant as a result of exploiting the set of fiber modes as the spatial-mode basis. Another manifestation of OAM in a pulsed field was demonstrated in \cite{Rego2019Science}, where a pulsed beam was produced via the process of high-harmonic generation in which the OAM changes rapidly across the pulse in time, leading to a so-called `self-torque' of light. It remains to be seen whether this form of spatio-temporally structured light can be synthesized linearly without relying on a nonlinear process to produce it. Finally, a recent theoretical development indicates an intriguing relationship between pulse width (a temporal DoF) and the maximum OAM-order of the beam (a spatial DoF). It appears that the pulse width sets an upper limit on the maximum OAM-order that can be sustained by the wave packet \cite{Ornigotti2015PRL,Ornigotti2015PRA,Porras2019OL,Porras2019PRL,Agasti2021AS,Porras2021arxiv}. This limit has been examined so far in the context of X-waves and awaits to be studied for baseband ST wave packets.

\subsection{ST optical vortices}\label{Sec:OpticalVortices}

\begin{figure}
    \centering
    \includegraphics[width=10cm] {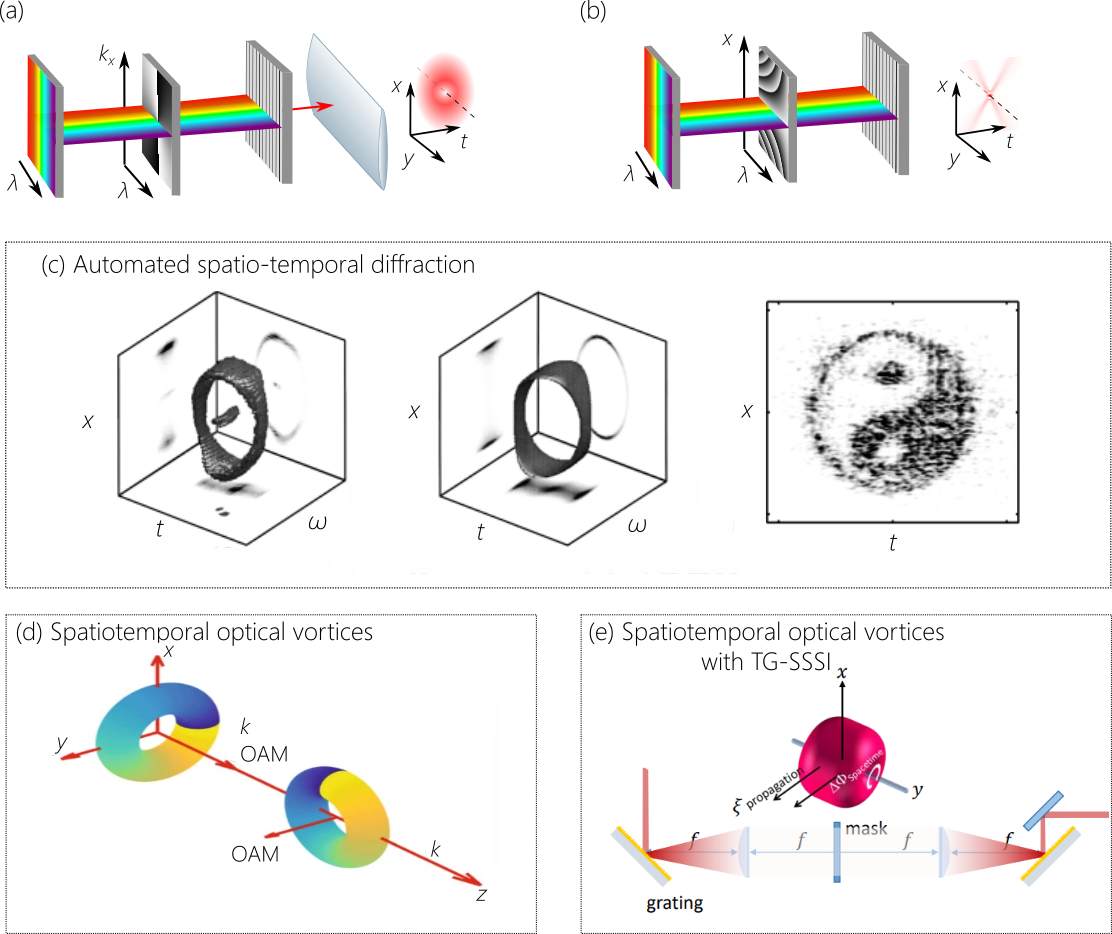} 
    \caption{(a,b) Two approaches to spatio-temporal Fourier synthesis. (c) Spatio-temporally structured fields produced using the system in (a); see \cite{Vaughan03OL}. (d) Spatio-temporal optical vortices with transverse OAM produced by the system in (a) \cite{Chong2020NP} and in (e) \cite{Hancock2019Optica}.}
    \label{fig:Other_groups3}
\end{figure}

The setup developed in \cite{Wefers1996OL,Vaughan03OL} and shown in Fig.~\ref{fig:Other_groups3}(a) was a `solution looking for a problem'. The plane of the SLM corresponds to the $(k_{x},\lambda)$ domain, whereas it corresponds to the $(x,\lambda)$ domain in the configuration shown in Fig.~\ref{fig:Other_groups3}(b) and used to produce propagation-invariant ST wave packets. In some cases, the additional lens in Fig.~\ref{fig:Other_groups3}(a) can be removed and free-space propagation is relied upon to impart the needed spatial transformation of the field. This technique was exploited in the context of quantum control of chemical interaction pathways using shaped ultrashort optical pulses after imparting spatial variation to each wavelength. Although the scheme in Fig.~\ref{fig:Other_groups3}(a) can produce extremely sophisticated pulsed beam configurations as exemplified in Fig.~\ref{fig:Other_groups3}(c), which presciently anticipated recent developments, its use was not widely adopted. Although no clear application was identified at the time, this configuration remained an intriguing possibility until it was recently exploited to produce so-called ST optical vortices (STOVs).

Optical vortices are transverse spatial field configurations (typically for a monochromatic beam) in which some optical DoF circulates around an axis, leading to a phase singularity at the vortex center \cite{Bazhenov1992JMO,Basistiy1995OC,Gahagan1996OL,Zhan2009AOP,Swartzlander2014Book}. The circulation typically occurs in the $(x,y)$-plane around the $z$-axis, and for scalar fields typically involves OAM. However, an early theoretical study proposed a spatio-temporal vortex \cite{Sukhorukov2005SPIE}, one in which the circulation occurs in the $(x,t)$ domain and spans space and time. However, this concept lay dormant until experiments by Milchberg \textit{et al}. uncovered that such STOVs occur during the nonlinear pulse collapse dynamics in self-focusing media \cite{Jhajj2016PRX}. Subsequently, it was recognized that STOVs can in fact be synthesized linearly \cite{Hancock2019Optica,Chong2020NP}, and the concept has since attracted considerable attention \cite{Wang2021Optica,Bliokh2021PRL,Chen2021arxiv,Wan2021arxiv,Cao2021PR}. It is important to note that the axis of the OAM points in the $y$ direction rather than along $z$ as in conventional OAM beams, and such a STOV is thus said to exhibit \textit{transverse} OAM. Similar to the result shown for conventional OAM \cite{Courtial1997PRA}, the second-harmonic generated from a pump with transverse OAM shows that the transverse-OAM order is doubled \cite{Gui2021NP,Hancock2021Optica}.

\section{ST photonics}\label{sec:STPhotonics}

At this early phase of development of ST wave packets and other spatio-temporally structured field structures, it is natural that interest has first focused on studying their free propagation. However, as the synthesis methodologies mature, it is expected that more efforts will be directed to investigating new opportunities for their interaction with photonic structures and devices. We survey here briefly some of the recent developments along these lines. 

\subsection{ST wave packets in waveguides}\label{Sec:Waveguides}

A much neglected area of research regarding ST wave packets is their propagation in optical fibers and waveguides. In general, because monochromatic waveguide modes by definition propagate invariantly, there has been little impact from the field of diffraction-free beams on waveguide optics. In contrast, ST wave packets offer the potential for dispersion-free pulsed modes and supermode configurations in optical waveguides. Indeed, early theoretical investigations of localized waves examined the propagation of FWMs in optical fibers \cite{Vengsarkar1992JOSAA} and X-waves in circular and coaxial waveguides \cite{ZamboniRached2001PRE,ZamboniRached2002PRE,ZamboniRached2003PRE}. Interest has been recently revived regarding the propagation of ST wave packets in waveguide structures. In general, guided ST wave packets can be classified into two broad families: hybrid ST modes in planar waveguides \cite{Shiri2020NC_Hybrid} and ST supermodes in planar waveguides \cite{Guo2021PRR,Shiri2022STsupermodes} or conventional optical fibers and waveguides \cite{Kibler2021PRL,Ruano2021JO,Bejot2021ACSP}.

\subsubsection{Hybrid ST modes}

\begin{figure}[t!]
  \begin{center}
  \includegraphics[width=8.0cm]{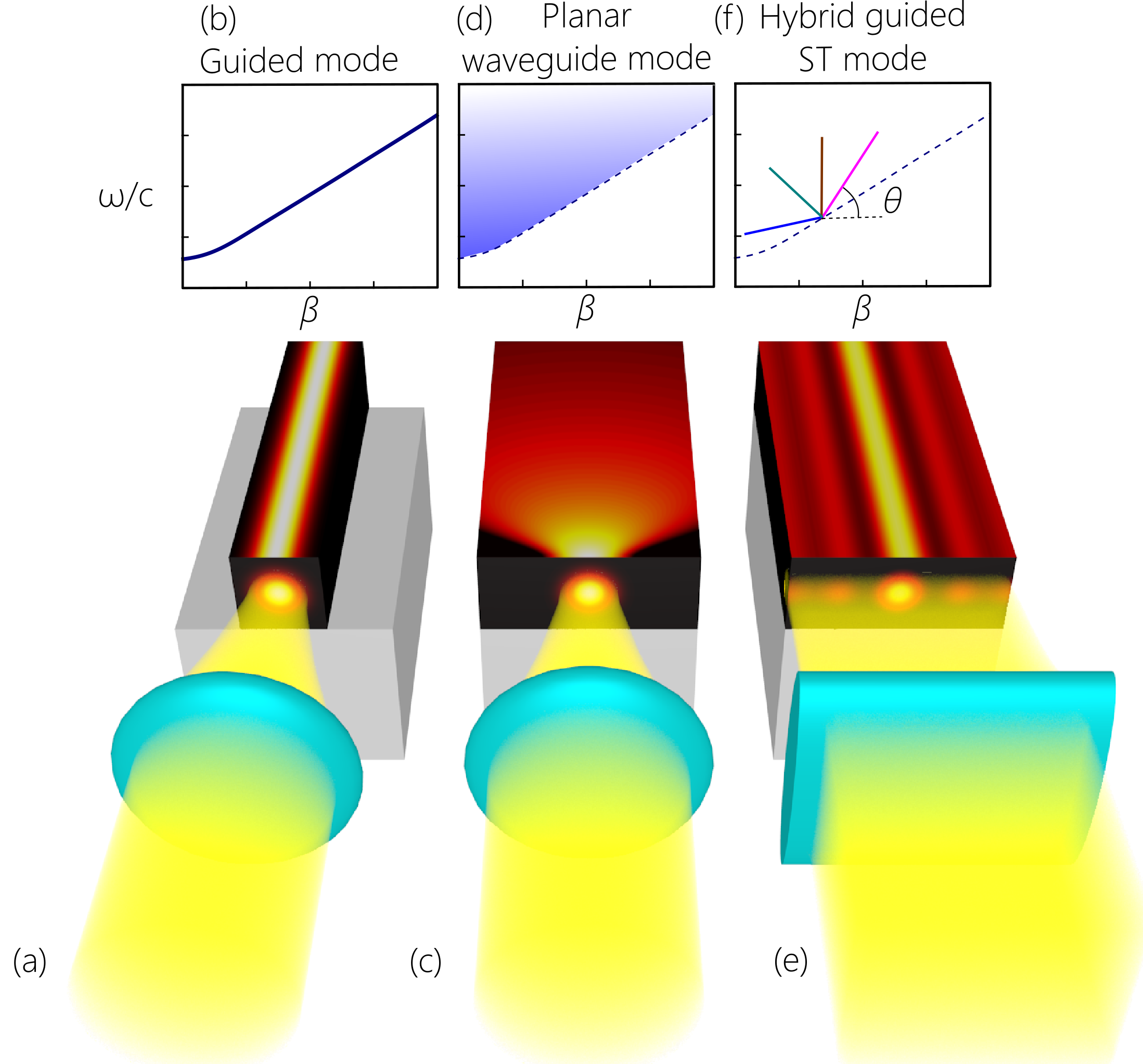}
  \end{center}
  \caption{Hybrid ST modes in a planar waveguide. (a) A conventional guided mode and (b) its dispersion relationship. (c) A planar-waveguide mode confined along $y$ but diffracts along the unbounded direction $x$, and  (d) its dispersion relationship. (e) A ST wave packet confined along $x$ and extended along $y$ is coupled via a cylindrical lens into the planar waveguide, whereupon it becomes also confined along $y$. (f) The dispersion relationship for hybrid ST modes can be tailored independently of the waveguide boundary conditions.}
  \label{Fig:HybridSTmode}
\end{figure}

In conventional waveguides, light can be confined in both transverse dimensions by a variety of optical guiding mechanisms [Fig.~\ref{Fig:HybridSTmode}(a)]; e.g., index guiding \cite{Saleh2007Book}, a photonic bandgap \cite{Russell2003Science,Joannopoulos2008Book}, gain guiding \cite{Siegman2003JOSAA,Siegman2007JOSAB}, among other options \cite{Duguay1986APL,Almeida2004OL,Wang2011OL,Pryamikov2011OE}. The dispersion relationship for a guided mode in the $(\beta,\tfrac{\omega}{c})$-plane is a 1D curve, where $\beta$ is the axial wave number [Fig.~\ref{Fig:HybridSTmode}(b)]. In a planar waveguide, light is confined along only one transverse dimension, and unrestricted along the other [Fig.~\ref{Fig:HybridSTmode}(c)]. The $(\beta,\tfrac{\omega}{c})$ dispersion relationship is no longer a 1D curve [Fig.~\ref{Fig:HybridSTmode}(d)], reflecting the freedom to propagate along the unconstrained dimension where the wave packet undergoes diffractive spreading. A \textit{hybrid ST mode} is a propagation-invariant ST wave packet whose pulsed field is confined along one transverse dimension by the guiding mechanism of the planar waveguide and is confined along the unbounded transverse dimension by virtue of the spatio-temporal structure of a baseband ST wave packet [Fig.~\ref{Fig:HybridSTmode}(e)]. Thus, two distinct guiding mechanisms combine to form the hybrid ST mode \cite{Shiri2020NC_Hybrid}. The spectral support domain for this hybrid ST mode lies at the intersection of a plane with the light-cone associated with a planar-waveguide mode [Fig.~\ref{Fig:HybridSTmode}(f)]. Uniquely, the intrinsic spatio-temporal structure of a hybrid ST mode allows overriding the constraints imposed by the planar waveguide boundary conditions, leading to a host of useful features: (1) the group index of a hybrid ST mode can be tuned continuously, independently of the waveguide structure; (2) hybrid ST modes are dispersion-free to all orders; and (3) each hybrid ST mode in a multimode planar waveguide can -- in principle -- be manipulated separately, and the group index of each tuned independently of those for the others \cite{Shiri2022ACSP}.

\subsubsection{ST supermodes}

The spatio-temporal structure introduced into a hybrid ST mode is parallel to the planar-waveguide interfaces. However, the configuration studied earlier by Zamboni-Rached \textit{et al.} \cite{ZamboniRached2001PRE,ZamboniRached2002PRE,ZamboniRached2003PRE} is one in which the ST wave packet interacts with the waveguiding structure. The ST field confined to the waveguide must be a discrete superposition of its modes, but with each mode associated with a \textit{single} wavelength. Critically, the wave number $\beta$ of each mode is associated with one wavelength according to the dispersion relationship characteristic of a freely propagating ST wave packet. This is the configuration considered in all the theoretical literature to date, which we denote \textit{ST supermodes}. In the earlier theoretical studies, the dispersion relationship for an X-wave was utilized, but the baseband ST wave-packet dispersion relationships have been exploited in recent invetsigations. This includes theoretical studies of ST supermodes produced within the waveguide via a time-modulated refractive index \cite{Guo2021PRR} or via nonlinear optical interactions mediated by a high-energy pump \cite{Kibler2021PRL}, and studies of the impact of OAM on the propagating field -- whether a FWM \cite{Ruano2021JO} or a baseband ST wave packet \cite{Bejot2021ACSP}. We have recently realized ST supermodes in a planar waveguide \cite{Shiri2022STsupermodes}. In light of the recent developments in synthesizing 3D ST wave packets \cite{Yessenov2022Localized3D}, especially using discrete laser combs \cite{Pang2021OL}, it is now conceivable that ST wave packets can be coupled to ST supermodes in multimode optical fibers.

\subsection{ST surface plasmon polaritons (ST-SPPs)}\label{Sec:STSPP}

\begin{figure}[t!]
  \begin{center}
  \includegraphics[width=13cm]{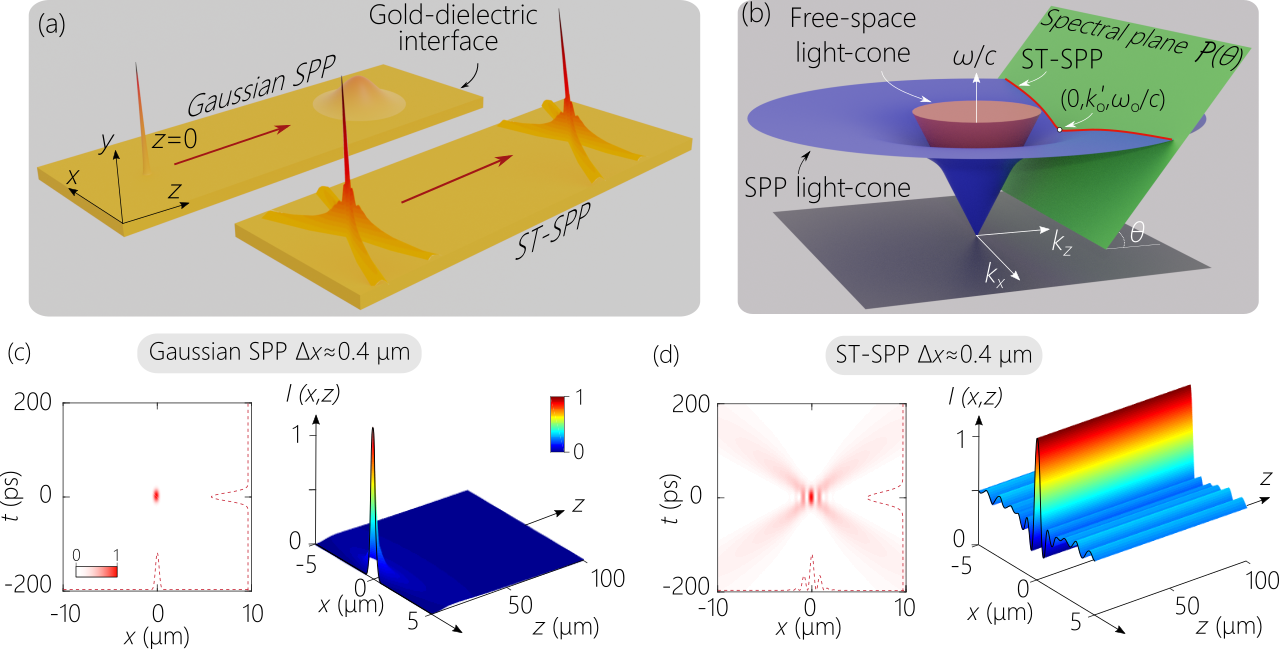}
  \end{center}
  \caption{(a) Conventional Gaussian SPP and ST-SPP wave packets are initially localized along all dimensions. Upon propagation along the surface the conventional SPP undergoes diffractive and dispersive spreading, whereas the ST-SPP is immune to both. (b) The spectral support domain for a ST-SPP at the intersection of the SPP light-cone with a plane $\mathcal{P}$ making at angle $\theta$ with the $k_{z}$-axis. (c) The initially localized SPP at $z\!=\!0$ with $\Delta x\!\approx\!400$~nm (FWHM) at $\lambda_{\mathrm{o}}\!=\!800$~nm spreads rapidly, as seen in the time-averaged intensity $I(x,z)$. The Rayleigh range is  $z_{\mathrm{R}}\!\approx\!700$~nm. (d) The X-shaped spatio-temporal intensity profile of the ST-SPP pulse with the same width as in (c) is invariant along $z$.}
  \label{Fig:ST-SPP1}
\end{figure}

Surface plasmon polaritons (SPPs) are surface waves confined to the interface between a dielectric and a metal below its plasmon resonance frequency \cite{Otto1968ZP,Economou1969PR,Kretschmann1971ZP,Raether1988Book,Maier2007Book,Zhang2012JPD}. While the SPP is confined to the metal-dielectric interface, it still undergoes diffractive spreading along the unbounded transverse dimension upon free propagation. As described in Section~\ref{Section:DiffractionFree1D}, there are no monochromatic diffraction-free beams with only one free transverse dimension, with the exception of the plane or cosine waves, and the Airy beam. Both of these beams have been realized in the context of SPPs (cosine SPPs \cite{Lin2012PRL} and Airy plasmons \cite{Salandrino2010OL,Minovich2011PRL,Li2011PRL,Zhang2011OL,Minovich2014LPR,Wang2017OE}). However, cosine waves are not confined, and Airy beams travel along a parabolic trajectory. Therefore, SPPs have yet to be demonstrated that are confined in all dimensions and travel in a straight line. Moreover, SPPs are intrinsically dispersive, such that the temporal width of a broadband SPP pulse increases with propagation. The combined effect of diffraction and dispersion on a Gaussian SPP wave packet in space and time are shown in Fig.~\ref{Fig:ST-SPP1}(a).

ST wave packets provide a platform for producing SPPs that are localized in all dimensions \cite{Schepler2020ACSP}. In contrast to the free-space dispersion relationship, the SPP dispersion relationship is:
\begin{equation}\label{SPP_dispersion}
k_{x}^{2}+k_{z}^{2}\!=\!\left(\frac{\omega}{c}\right)^{2}\frac{\epsilon_{\mathrm{m}}(\omega)\epsilon_{\mathrm{d}}}{\epsilon_{\mathrm{m}}(\omega)+\epsilon_{\mathrm{d}}}
\end{equation}
which corresponds geometrically to the SPP light-cone in Fig.~\ref{Fig:ST-SPP1}(b); here $\epsilon_{\mathrm{d}}$ and $\epsilon_{\mathrm{m}}$ are the relative permittivities of the dielectric and metal, respectively. Despite the different light-cone structure, a propagation-invariant ST-SPP is constructed in much the same way as its free-space counterpart. The spectral support domain of the ST-SPP is the intersection of the SPP light-cone with a spectral plane $\mathcal{P}(\theta)$, where the spectral tilt angle $\theta$ with respect to the $k_{z}$-axis. A ST-SPP whose spatio-temporal spectrum is confined to this intersection travels at a fixed group velocity $\widetilde{v}\!=\!c\tan{\theta}$ without dispersion at any selected wavelength, and without diffraction, almost independently of its transverse spatial width; see Fig.~\ref{Fig:ST-SPP1}(a).  

\begin{figure}
    \centering
    \includegraphics[width=8.2cm]{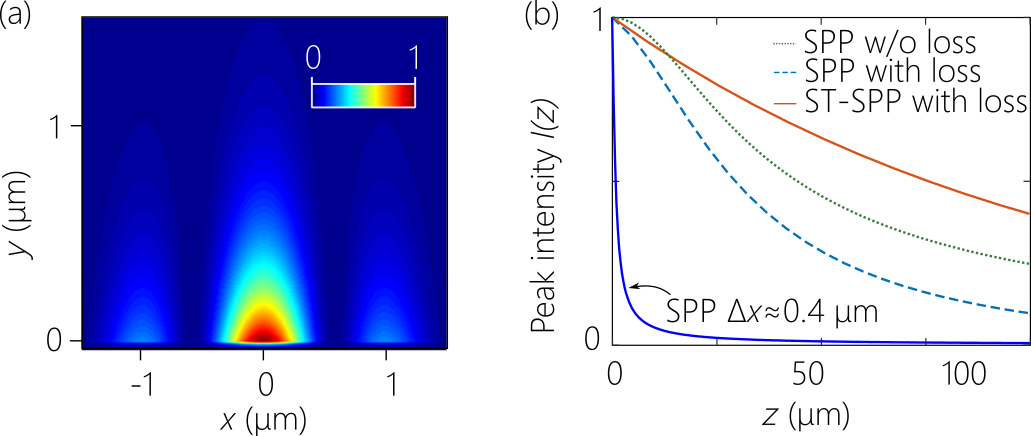}
    \caption{(a) Time-averaged intensity  of the ST-SPP in Fig.~$\ref{Fig:ST-SPP1}$ plotted in the $(x,y)$-plane. (b) Evolution of the on-axis intensity for SPPs and ST-SPPs of widths $\Delta x\!\approx\!3$~$\mu$m and $\Delta x\!\approx\!400$~nm in the presence and absence of plasmonic losses. For the SPP with $\Delta x\!\approx\!400$~nm, the plots in the presence and absence of losses coincide. For ST-SPPs, there is no observable decay in the absence of losses over the plotted length scale, and in the presence of losses the two curves for $\Delta x\!=\!3$~$\mu$m and $\Delta x\!=\!400$~nm coincide. }
    \label{fig:sub-wavelength_ST-SPP}
\end{figure}

Calculations have shown that sub-wavelength ST-SPPs can propagate invariantly for significant distances\cite{Schepler2020ACSP}. Figure~\ref{Fig:ST-SPP1}(c-d) compares a conventional Gaussian SPP and ST-SPP, both with an initial transverse width of 400~nm (FWHM) at an optical carrier wavelength of $\lambda_{\mathrm{o}}\!=\!800$~nm. The Gaussian SPP has a Rayleigh range $z_{\mathrm{R}}\!\approx\!700$~nm, whereas the ST-SPP pulse shows only a minor decrease in intensity at 2000$z_{\mathrm{R}}\!=\!1400$~$\mu$m.  The propagation distance is dictated by the spectral uncertainty $\delta\lambda$ and the tilting angle $\theta$, almost independently of $\Delta x$ \cite{Yessenov2019OE}. The transverse profile at fixed $z$ of the time-averaged intensity of a ST-SPP with $\Delta x\!\approx\!400$~nm is shown in Fig.~\ref{fig:sub-wavelength_ST-SPP}(a), showing localization in all dimensions. SPP confinement to the dielectric-metal interface with most of the intensity on the dielectric side is shown. Only a small fraction of the electric field is in the lossy metal.

However, the comparison in Fig.~\ref{Fig:ST-SPP1} does not consider the effects of dissipative electronic dynamics in the metal. By incorporating the imaginary part of the metal permittivity from the Drude model to calculate the absorption coefficient, we plot in Fig.~\ref{fig:sub-wavelength_ST-SPP}(b) the SPP and ST-SPP propagation with and without plasmonic losses for $\Delta x\!=\!400$~nm (sub-wavelength) and $\Delta x\!=\!3$~$\mu$m. For SPPs, losses compound the drop in on-axis intensity along $z$ associated with diffraction. Indeed, for $\Delta x\!=\!400$~nm, diffraction of a conventional SPP dominates over the impact of losses. For ST-SPPs, the absorption length is much smaller than the diffraction length, even for $\Delta x\!=\!400$~nm, so that diffraction and absorption decouple, and absorption alone determines the propagation distance. ST-SPPs thus have a decisive advantage over conventional SPPs for subwavelength $\Delta x$.

As of this writing, ST-SPPs have not been demonstrated experimentally. It is possible that the standard Kretschmann configuration \cite{Kretschmann1971ZP,Raether1988Book} combined with the spatio-temporal Fourier synthesis approach (Section~\ref{Section:SynthesisSetup}) can launch ST-SPPs, but new ideas are needed for realizing sub-wavelength ST-SPPs. 

\subsection{ST nanophotonics}

Nanophotonics stands to play a significant role in the area of ST wave packets and spatio-temporally structured optical fields in general. One critical task is to design nanophotonic systems that can produce a ST wave packet starting from a generic pulsed beam. It is still an open question whether a single nanostructured surface can achieve such a task. We have already mentioned the recent study showing that a nonlocal nanophotonic structure can spatio-temporally filter a subluminal ST wave packet from a focused ultrashort pulse \cite{Guo2021Light}. This approach has several interesting features; e.g., OAM of the incident pulse is maintained in the produced ST wave packet, and topological spin-texture in the field polarization can be readily introduced. More research is needed along these lines.

Along a different vein, a recent experiment made use of a metasurface to realize beam-steering at picosecond time scales \cite{Shaltout2019Science}. The metasurface in that experiment combined the functionalities of a traditional grating and lens to spatially resolve the spectrum of an incident laser frequency comb and collimate the discrete comb wavelengths at a prescribed plane. In the far field, this source produces a tilted-pulse front (TPF) because the angular dispersion inculcated into the field with this approach is differentiable. At a fixed axial plane, the passage of the TPF is recorded as a pulsed spot moving across the plane, which is interpreted as a beam being steered. 

\begin{figure}[t!]
    \centering
    \includegraphics[width=11cm] {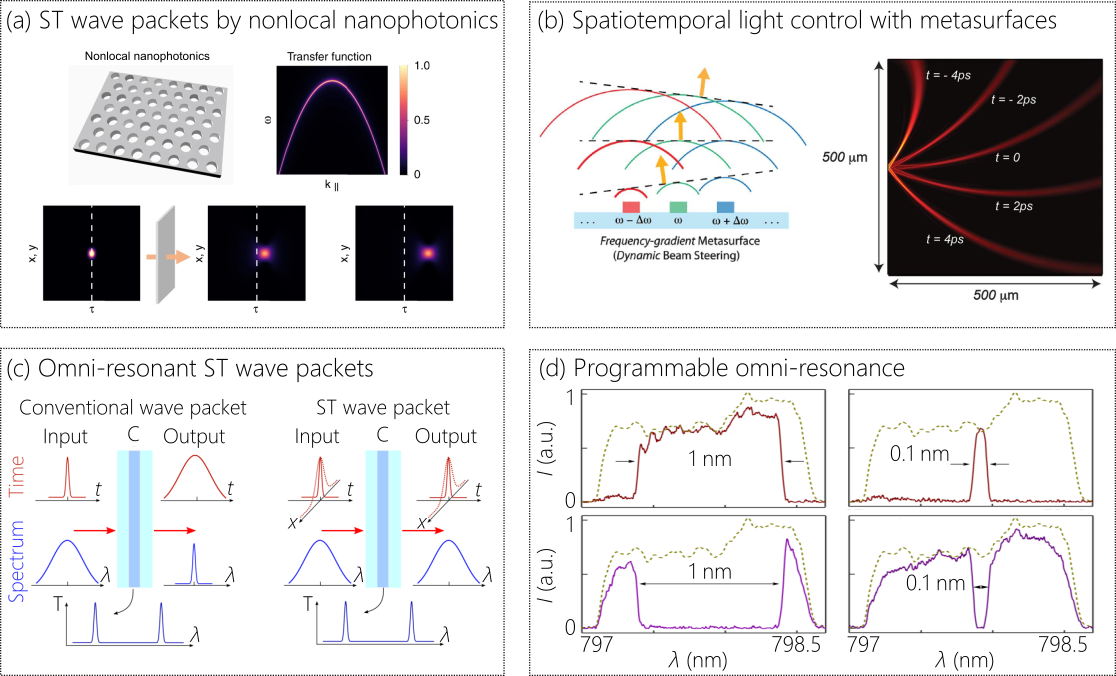}
    \caption{Overview of some recent work on spatio-temporally structured wave packets. (a) A generation of 3D space-time wave packet via nonlocal nanophotonics \cite{Guo2021Light}; (b) Spatiotemporal light control with frequency-gradient metasurfaces \cite{Shaltout2019Science}. (c) Transmission of conventional wave packets and an omni-resonant ST wave packets through a planar Fabry-P{\'e}rot cavity \cite{Shiri2020OL}. (d) Programmable omni-resonant ST wave packets with broad transmitted or rejected spectra \cite{Shiri2020APLP}.}    \label{fig:Other_groups4}
\end{figure}

\subsection{Omni-resonance and ST wave packets}

Resonant field buildup in optical cavities can enhance a variety of linear and nonlinear optical effects. One example is coherent perfect absorption (CPA) \cite{Chong2010PRL}, whereby light is guaranteed to be fully absorb in a weakly absorbing film independently of its thickness or level of intrinsic absorption by placing it in the appropriate resonant structure \cite{Wan2011Science,Zhang2012LSA,Villinger2015OL,Baranov2017NRM}. For example, CPA has been realized in a 2-$\mu$m-thick layer of polycrystalline silicon over a full octave of bandwidth (from $\sim\!800$~nm to $\sim\!1600$~nm) \cite{Pye2017OL}. However, a fundamental trade-off exists between resonant enhancement and bandwidth: increasing the cavity quality factor and thus increasing the field buildup is inevitably associated with a reduced resonant linewidth. As such, CPA cannot be harnessed for solar energy applications because most CPA realizations are based on optical resonances at discrete wavelengths and cannot provide resonantly enhanced absorption over a broad continuous spectrum.

However, this limitation can be overcome by pre-conditioning the incident field and introducing a prescribed spatio-spectral structure whereby each wavelength is assigned a particular incident angle on a Fabry-P{\'e}rot cavity \cite{Shabahang2017SR,Shabahang2019OL}. We have called this configuration `omni-resonance' because \textit{every} wavelength in a broad incident spectrum resonates with the cavity, and the resonance bandwidth is much larger than the intrinsic cavity resonant linewidth, and can even extend beyond the cavity free spectral range. This strategy allows for a useful phenomenon such as CPA to be applied to the area of solar energy by enhancing the absorption in ultrathin solar cells. We have recently exploited omni-resonant CPA for enhancing the performance of a solar cell by embedding a 200-nm-thick hydrogenated amorphous-silicon PIN diode in a planar cavity formed of dielectric Bragg mirrors designed for the NIR where silicon absorption drops \cite{Villinger2021AOM}. Once the incident solar radiation is appropriately pre-conditioned, omni-resonance over a bandwidth of $\sim\!60$~nm enhances the absorption and doubles the electric current collected compared to the bare solar cell \cite{Villinger2021AOM}. The pre-conditioning optical system is alignment-free and consists of three parallel thin planar structures: an array of polymer micro-prisms sandwiched between two gratings.

Most relevant to our perspective here, omni-resonance has been realized with ST wave packets \cite{Shiri2020OL}. The omni-resonant ST wave packet is \textit{not} spectrally filtered by the FP cavity when its bandwidth exceeds the resonant linewidth as occurs with conventional wave packets. Instead, the full bandwidth resonates with the cavity and is transmitted [Fig.~\ref{fig:Other_groups4}(c)]. Furthermore, sculpting the spatio-temporal spectrum of the ST wave packet enables programmable omni-resonant interactions with the cavity. For example, the cavity can be omni-resonant over a controllable bandwidth, can be a notch filter with tunable spectral-rejection bandwidth, or provide an arbitrary spectral response -- all while retaining the diffraction-free behavior of the transmitted ST wave packet \cite{Shiri2020APLP}; see Fig.~\ref{fig:Other_groups4}(d).

The concept of omni-resonance implemented with coherent or incoherent ST fields resolves the long-standing challenge of delivering resonant optical enhancement over continuous broad spectra, can be applied to nonlinear absorption and refraction in planar cavities, and may potentially be extended across the entire visible spectrum, thereby providing opportunities for optical signal and image processing at low light levels.

\section{What we have \textit{not} covered}
\label{Section:notcovered}

We have covered here only a portion of the rapidly developing field of ST wave packets that has emerged over the past 5 years. We have thus passed on some topics that have previously attracted interest, such as tunneling of ST wave packets \cite{Shaarawi2000PRE,Shaarawi2000JPA3,Shaarawi2002PRE}, investigating the impact of the wave packet structure on double slit interference \cite{Shaarawi94PLA,Shaarawi2011JOSAA}, and implications for plasma physics \cite{Tippett1991PRA,Ziolkowski1991PRA}. We have restricted ourselves to freely propagating or guided fields in linear, non-dispersive, isotropic, homogeneous media. We have thus \textit{not} reviewed the propagation of ST wave packets in dispersive, anisotropic, or nonlinear media, or in artificial structures such as coupled waveguide systems \cite{Christodoulides2004OL,Droulias2005OE,Lahini2007PRL,Heinrich2009PRL}. Moreover, we have focused on \textit{optical} ST wave packets. However, interest in ST wave packets in the microwave regime continues \cite{Comite2018APL,Chiotellis18PRB,Fuscaldo2019IEEE,Chiotellis2020IEEETAP}.

We describe here briefly topics of interest that we have not covered because they are still at an early stage of development.

\subsection{Relationship with special relativity}

It was recognized early on that there is an intimate connection between the unique structure of propagation-invariant ST wave packets and special relativity \cite{Belanger1986JOSAA}. Indeed, it can be shown that applying a relativistic Lorentz transformation to a monochromatic beam yields a baseband ST wave packet \cite{Saari2004PRE,Longhi2004OE,Kondakci2018PRL}! That is, if a laser emits a monochromatic beam, a relativistic observer will \textit{not} record a Doppler-shifted monochromatic beam, but rather a pulsed beam; more precisely, a baseband ST wave packet. The spectral support domain of the monochromatic beam lies along the circle at the intersection of the light-cone with a horizontal iso-frequency plane. Implementing a Lorentz transformation with Lorentz factor $\beta$ tilts this circle by an angle $\theta$ given by $\beta\!=\!\tan{\theta}$, thus producing a subluminal ST wave packet \cite{Kondakci2018PRL}. More research is needed to explore the full extent of the confluence between ST wave packets and special/general relativity, the phenomenon of time diffraction \cite{Moshinsky1952,Longhi2004OE,Saari2004PRE,Porras2017OL,Porras2018PRA}, and other aspects of special relativity impacting ST wave packets \cite{Saari2020JPC}. Furthermore, the concept of STOVs can be connected to conventional OAM of monochromatic beams via Lorentz transformations \cite{Bliokh2012PRA}. Finally, it is to be expected that ST wave packets will be relevant to recent studies on the scattering of light from moving media and structures \cite{Bahabad2014OQE,Plansinis2015PRL,Qu2016JOSAB,Leger2019AP,Leger2021arxiv}.

\subsection{Incoherent ST fields} 

The relationship between diffraction-free behavior and partial coherence (spatial and temporal) is sometimes obscure. For example, in \cite{Lumer2015Optica} a condition is derived for an Airy beam to retain the transverse acceleration of its diffraction-free profile intact in presence of spatial incoherence. However, the results in \cite{Fischer2005OE,Fischer2006JOA} throw doubt on the possibility of maintaining the diffraction-free propagation of a Bessel beam produced from sources with low degrees of spatial coherence. It is important to note that these studies have considered fields in which the the spatial and temporal DoFs are largely separable.

Although we have dealt in this Review exclusively with coherent pulsed fields, temporally incoherent fields can nevertheless be exploited to produce propagation-invariant ST fields \cite{Yessenov2019Optica}. Indeed, the first demonstrations of X-waves \cite{Saari1997PRL} and FWMs \cite{Reivelt2002PRE} made use of incoherent fields rather than coherent pulses because of the broad bandwidths needed. Previous work has examined the coherence function of ST wave packets \cite{Jedrkiewicz2006PRL,Jedrkiewicz2007PRA}, which has shown that their coherence is spatio-temporally coupled. This indicates a major distinction in propagation characteristics between conventional incoherent fields whose spatial and temporal DoFs are \textit{separable} and incoherent ST fields that are \textit{non-separable}. 

The spatio-temporal Fourier synthesis strategy employed with pulsed fields can be implemented with incoherent light from a LED \cite{Yessenov2019Optica} or SLD \cite{Yessenov2019OL}. The time-averaged intensity is indistinguishable from that resulting from a coherent pulsed laser having the same temporal spectral intensity. The spatio-temporal structure of the incoherent ST field is observed in the cross-coherence function resulting from the interference of the ST field with the initial incoherent light field. The speed of the cross-coherence function -- which has been dubbed the \textit{coherence group velocity} -- can be tuned in free space just as the group velocity of coherent ST wave packets can be \cite{Yessenov2019OL}. This is a new area for optical coherence where the conventional theory needs to be adapted to scenarios where deterministic correlations are introduced between coupled DoFs. Although previous theoretical studies have examined these questions \cite{Turunen2008OE}, it is now possible to explore them experimentally.

\subsection{ST wave packets and classical entanglement}

The unique features of ST wave packets are a consequence of the tight association between the spatial and temporal DoFs of the field. ST wave packets are therefore an embodiment of so-called `classical entanglement' \cite{Borges2010PRA,Qian2011OL,Kagalwala2013NP}, which refers to classical optical fields whose physical DoFs are non-separable and thus have the same mathematical structure of multi-partite quantum mechanical states \cite{Spreeuw1998FP,Spreeuw2001PRA}. To date, classical entanglement has been investigated solely with discretized DoFs: polarization, spatial, and time-bin modes \cite{Abouraddy2014OL,Kagalwala2015SR,Aiello2015NJP,Berg2015Optica,Qian2016PRL,Eberly2016PS}. In contrast, ST wave packets are a unique example of classical entanglement with \textit{continuous} DoFs. The relationship between the propagation distance and the Schmidt number \cite{Kondakci2019OL} (a common measure for entanglement in large-dimensional systems \cite{Ekert95AJP,Law2000PRL,Law2004PRL}) has been recently established. By adding the polarization DoF, the concepts of double-entanglement and hyperentanglement \cite{Kwiat1997JMO,Barreiro2005PRL} can be explored in a classical context \cite{Diouf2021OE}.

\subsection{ST wave packet propagation in dispersive and nonlinear media}

The study of baseband ST wave packets produced by nonlinear interaction and their propagation in dispersive media reached a high degree of maturity prior to 2010. Much of the efforts that have been directed over the past 20~years to studying the propagation of ST wave packets in dispersive and nonlinear media has been reviewed in \cite{HernandezFigueroa2008Book} (Chapters 8 and 9), in \cite{Figueroa2014Book} (Chapters 9, 12, 18 and 22), and in \cite{Faccio2007Book}. In generating ST wave packets in a nonlinear medium, a host of nonlinear processes are typically involved, including second-harmonic generation, Kerr effects, Raman effects, nonlinear absorption, among others. In general, however, some form of phase-matching of the wave vectors over a broad spectrum is visible in the measured spatio-temporal spectra, in which the different wavelengths emerging from the medium travel at different angles. Despite the many interesting results reported, there are a few drawbacks of this methodology. From a technical perspective, a high-energy pump laser is required, thus placing a heavy burden on target applications of the ST wave packets produced in this fashion. More importantly from a fundamental perspective, the spectral uncertainty $\delta\lambda$ characteristic of these wave packets is extremely large $\delta\lambda\!\sim\!1$~nm, so that the propagation-invariant distance exhibited is small (see Section~\ref{Section:PropagationDistance}). Furthermore, only minute deviations of the group velocity of these wave packets from $c$ are achievable.

Additionally, one cannot explore the full range of dynamics of ST wave packets in dispersive media. Previous theoretical work uncovered a host of fascinating phenomena in dispersive media that have gone unobserved to date, such as spherically symmetric wave packets dubbed `O-waves' that are related to Mackinnon's wave packet \cite{Mackinnon1978FP}, and which exist in \textit{both} normal and anomalous GVD regimes. Indeed, the comprehensive theoretical framework developed by Malaguti \text{et al}. \cite{Malaguti2008OL,Malaguti2009PRA} delineates a large span of striking transitions between several propagation regimes, including transitions from O-waves to X-waves to V-waves by tuning a continuous parameter such as the group velocity $\widetilde{v}$. None of these phenomena have been observed to date. The sole exception has been the observation in the context of nonlinear X-waves of a transition from an X-shaped spatio-temporal spectrum (actually a double parabola) to an O-shaped spectrum as the wavelength of the of laser pump is tuned across the zero-dispersion wavelength of water \cite{Porras2005OL}. However, the spatio-temporal profile of the wave packets associated with this transition have not been recorded. Now that dispersive ST wave packets can be readily synthesized \cite{Yessenov2021engineering,Hall2021NGVDrealizing}, we expect rapid developments to occur in this area, the predicted dynamics of ST wave packets in dispersive media can be explored, and perhaps the hypothesized O-waves will be observed. Moreover, one can potentially use ST wave packets to \textit{initiate} nonlinear interactions \cite{Porras2018OE} rather than relying on nonlinear phenomena to \textit{produce} them.

\section{Roadmap for future developments}
\label{Section:Roadmap}

The recent progress in the area of ST wave packets and related spatio-temporally structured fields (flying-foci, toroidal pulses, STOVs, etc.) paves the way to a reconceptualization of our understanding of pulsed optical fields. These efforts are currently morphing into a broader field of study that we can call \textit{space-time optics and photonics}. It is still too early to predict the applications that will emerge from this burgeoning activity. One must be cognizant of previous cycles of hype-and-bust that have occurred in the area of diffraction-free beams and propagation-invariant wave packets. The mood circa 2010 is captured in \cite{Turunen2010PO}, which details the few applications that survived for monochromatic diffraction-free beams (particularly Bessel beams) from the initial flurry of suggested possibilities. In the case of propagation-invariant wave packets, the proposed applications never materialized because of the difficulties involved in synthesizing them. In this regard, the impracticality of producing and tuning the properties of X-waves and FWMs weigh heavily on their future prospects. We therefore focus here on the emerging areas of scientific inquiry regarding ST wave packets and only make general suggestions of the potential applications whose viability will be put to the test over the next few years in this rapidly evolving field of research.

\textit{ST wave packets in dispersive media.} As noted in Section~\ref{Section:HistoricalSketch}, ST wave packets produced by linear optics have historically been of the sideband variety (e.g., FWMs and X-waves), whereas theoretical studies of propagation-invariant ST wave packets in dispersive media have focused on baseband ST wave packets. Consequently, these two lines of inquiry have not yet merged. Instead, the study of ST wave packets in dispersive media has been primarily used in exploring their nonlinear generation. We anticipate that this state of affairs will change rapidly, and that propagation invariance can now be verified in dispersive media and that the predicted phenomena of O-waves and unique spectral dynamics associated with changes in $\widetilde{v}$ can be observed.

\textit{Nonlinear and quantum optics of ST wave packets.} It has always been implicitly assumed that the angular dispersion in optical fields is differentiable. Now that it is understood that non-differentiable angular dispersion underpins the propagation invariance of ST wave packets, a potentially useful avenue of investigation is to study the consequences of introducing non-differentiable angular dispersion into pulsed fields for nonlinear and quantum optics. A host of nonlinear optical interactions are governed by phase-matching conditions, and TPFs have already been useful in various nonlinear effects (e.g., SHG \cite{Richman1998OL,Richman1999AO} and THz generation \cite{Hebling2002OE}). The versatility afforded by ST wave packets in light of the non-differentiable AD underpinning them  has yet to be exploited in nonlinear optics \cite{Yessenov2021engineering,Hall2021NGVDrealizing,Hall2021OESynthesizer}. Moreover, the recent synthesis of 3D ST wave packets is expected to lead to breakthroughs in this area in the near future. Such findings are expected to benefit our understanding of the spatio-temporal structure of entangled photon pairs produced via spontaneous parametric downconversion (SPDC). The spatio-temporal structure of such entangled photons has been examined theoretically \cite{gatti2009PRL,Caspani2010PRA,Brambilla2010PRA,Brambilla2012PRA,Horoshko2012EPJD,Gatti2012PRA,Gatti2014IJQI} and verified experimentally \cite{Jedrkiewicz2012PRL,Brambilla2014JOSAB}, and the impact of the sign of GVD on this spatio-temporal spectrum has been recently studied \cite{Spasibko2016OL,Cutipa2020OL}. Recent efforts in ST field quantization will be crucial for pushing this field forward \cite{Ornigotti2017PRA,Ornigotti2018JO}.

\textit{ST photonics.} The area of ST photonics in particular appears to be primed for rapid developments given the state-of-the-art currently achieved. This incorporates a broad range of potential opportunities, including novel synthesis methodologies using passive nanophotonic devices to produce ST wave packets from an incident pulsed field, or advanced designs for on-chip lasers or solid-state lasers that directly emit ST wave packets or other embodiments of spatio-temporally structured light. Moreover, recent work on spatio-temporal mode-locking and nonlinear effects in multimode fiber lasers \cite{Wright2015NP,Wright2015PRL,Wright2016NP,Wright2017Science} may contribute to this area. Besides continuing efforts in the area of guided ST wave packets (whether hybrid ST modes of ST supermodes), new results are anticipated in light of recent developments in the synthesis of 3D ST wave packets, especially the possibility of coupling to ST supermodes in multimode fibers and conventional waveguides. Furthermore, realizing ST-SPPs may be useful for near-field imaging, biosensing, and chemical sensing, for surface-enhanced Raman spectroscopy, or for data storage \cite{Zhang2012JPD}. Additionally, broadband resonantly enhanced interactions with Fabry-P{\'e}rot cavities mediated by ST wave packets are to be anticipated. Another avenue of interest is the potential synergy between ST wave packets and the emerging area of time-varying metasurfaces \cite{Shaltout2015OME,Shaltout2016JOSAB,Shaltout2019ScienceReview}, more recently known as space-time metasurfaces \cite{Caloz2020IEEE1,Caloz2020IEEE2,Rocca2020IEEE}.

\textit{ST wave packets and turbulence.} Finally, it will worthwhile to investigate the impact of turbulence on the propagation of ST wave packets, which is of paramount importance for optical communications over large propagation distances. To date, we have demonstrated propagation distances extending to $\sim\!100$~m outside the laboratory \cite{Bhaduri2019OL}, and further extending this distance to the kilometer-scale appears to be within reach. Along the same vein, several features of ST wave packets make them of interest for biomedical optics. Propagation invariance even with only one transverse dimension suggests them as a candidate as the excitation field in fluorescent light-sheet microscopy, and the self-healing characteristics of ST wave packets (which can be realized using incoherent light) points to their potential utility in imaging through turbid tissue.

\section{Conclusion}

We have surveyed the class of pulsed optical beams that we have denoted space-time (ST) wave packets as a centerpiece of the emerging field of ST optics and photonics. We have attempted to provide a critical evaluation of the body of work that has accumulated in this area since Brittingham discovered the FWM in 1983. In light of rapid recent developments in the synthesis of ST wave packets, it is critical to determine which portions of this literature remain a springboard for further developments and which are currently of historical value. We have argued that the better-known examples of FWMs and X-waves (in addition to sideband ST wave packets) unfortunately are not avenues for fruitful investigations in optics (but remain useful in acoustics, ultrasonics, and perhaps microwaves). Instead, the class of \textit{baseband} ST wave packets -- which has been almost neglected to date -- appears as the most promising path for further discoveries. This is attested to by the rapidly accumulating results on ST wave packets, which have verified established theoretical predictions that have gone untested for decades. Most importantly, baseband ST wave packets can be readily synthesized with narrow bandwidths in the paraxial regime, and thus may lead to new applications as the field continues to mature.

These recent breakthroughs are the culmination of a 40-year quest that began with Brittingham's theoretical discovery in 1983. After 4 decades of research, this area is now coming into its own, and many basic discoveries are anticipated over the next few years. Recent work over the past 5 years in particular constitutes a rebooting of the field of propagation-invariant wave packets by virtue of the following novel insights and breakthroughs:
\begin{enumerate}
    \item \textit{A better understanding of which wave packets are more readily synthesized in optics.} Most previous efforts were directed to X-waves and FWMs, which are unfortunately a dead end for optics because of the exorbitant requirements (in terms of bandwidth and numerical aperture) necessary for their synthesis. Baseband ST wave packets are a more versatile class of ST wave packets that can be synthesized with small numerical apertures and bandwidths. A fundamental conceptual change regarding propagation invariance has resulted from relying on geometrical representations on the light-cone surface rather than closed-form mathematical expressions, which has made this research field readily accessible, and has led to the ongoing expansion of research in this area. Crucially, the determining factors for the propagation limits of ST wave packets are now understood to be the intrinsic DoFs: the spectral uncertainty $\delta\omega$ and the spectral tilt angle $\theta$. 
    \item \textit{New experimental approaches.} A new methodology for synthesizing ST wave packets has been developed that relies on sculpting the spatio-temporal spectrum via a phase-modulating device such as a SLM or phase plate. Such an arrangement constitutes a universal angular-dispersion synthesizer that combines ultrafast-pulse shaping with the techniques of Fourier optics.
\end{enumerate}

Furthermore, a host of new spatio-temporally structured pulsed field configurations have recently emerged in which the spatial and temporal DoFs are non-separable, including flying-focus wave packets, toroidal pulses, and ST optical vortices. These novel wave packets provide a new canvas of surprising characteristics, thereby laying the foundation for new vistas of research opportunities in the area of ST optics and photonics.

To date, optics has been almost neatly divided between communities that manipulate the spatial DoFs of light (e.g., optical astronomy, microscopy, imaging) and its temporal DoFs (e.g., optical communications, ultrafast optics, laser combs). Indeed, many fundamental concepts and results in optics implicitly assume that the spatial and temporal DoFs in the optical field are separable. In contrast, ST wave packets and other spatio-temporally structured fields necessitate the joint manipulation of the spatial and temporal DoFs. In general, the recently emerging results that have been reviewed here clearly indicate the fruitful avenues of research that exploit the rich consequences of the non-separability of the spatial and temporal DoFs of a coherent pulsed beam or incoherent broadband field. The well-known quote from Hermann Minkowski regarding the space-time of special relativity bears repeating: ``Henceforth space by itself, and time by itself, are doomed to fade away into mere shadows, and only a kind of union of the two will preserve an independent reality.'' We hope to have convinced the reader that a similar process is ongoing in optical physics, whereby manipulating light jointly in space and time is opening fascinating new avenues of study, and is laying the foundations for a well-defined field of study with anticipated applications in nonlinear optics, optical signal processing, optical imaging and microscopy, and optical sensing and metrology.

\section*{Acknowledgments}
We are grateful for contributions to this research by H. E. Kondakci, B. Bhaduri, A. Shiri, A. M. Allende Motz, L. Mach, A. K. Jahromi, M. L. Villinger, C. Villinger, Y. Zhiyenbayev, and R. Aravindakshan, and by our collaborators M. A. Alonso, S. Ponomarenko, M. Meem, R. Menon, G. Atia, D. Mardani, M. B. Nasr, D. Reyes, J. Pena, S. Rostami Fairchild, M. Richardson, Q. Ru, K. L. Vodopyanov, Z. Chen, M. P. J. Lavery, J. Free, E. G. Johnson, M. Diouf, M. Harling, K. C. Toussaint, N. S. Nye, and D. N. Christodoulides. We thank A. Dogariu, M. Clerici, M. G. Raymer, F. Qu{\'e}r{\'e}, S. W. Jolly, P. Saari, M. A. Porras, and B. E. A. Saleh for valuable discussions.


\section*{Funding}
U.S. Office of Naval Research (ONR) N00014-17-1-2458, ONR MURI Award N00014-20-1-2789. 

\clearpage

\bibliography{MasterAbouraddyST,Schepler-misc.bib}

\end{document}